\documentclass[10pt]{iopart}
\usepackage{graphicx}
\RequirePackage[T1]{fontenc}
\usepackage[utf8]{inputenc}
\usepackage[english]{babel}
\usepackage{makecell}
\usepackage[dvipsnames]{xcolor}
\usepackage{float}

\usepackage[normalem]{ulem}

\usepackage[perpage]{footmisc}

\usepackage{lscape}  \usepackage{longtable}

\usepackage{hhline}
\usepackage{gensymb}
\usepackage{subcaption}

\usepackage[obeyFinal]{todonotes}
\usepackage{lineno}

\expandafter\let\csname equation*\endcsname\relax
\expandafter\let\csname endequation*\endcsname\relax

\RequirePackage{mathptmx}
\RequirePackage{mathtools}
\RequirePackage{amssymb}
\RequirePackage{multirow}
\RequirePackage{xspace}
\RequirePackage{placeins}

\usepackage{caption}
\usepackage{enumitem}

\usepackage{cite}

\usepackage{cuted}

\RequirePackage{flushend}
\RequirePackage[colorlinks,citecolor=blue,urlcolor=blue,linkcolor=blue]{hyperref}

\def\bracketbar{\smash{\hbox{\kern-7pt\raise3pt    \hbox{{\tiny(}{\lower1.5pt\hbox{\bf--}}{\tiny)}}}}}
\providecommand{\varpm}{\mathbin{\vcenter{\hbox{  \oalign{\hfil$\scriptstyle+$\hfil\cr
          \noalign{\kern-.3ex}
          $\scriptscriptstyle({-})$\cr}}}}}
\providecommand{\varmp}{\mathbin{\vcenter{\hbox{  \oalign{$\scriptstyle({+})$\cr
          \noalign{\kern-.3ex}
          \hfil$\scriptscriptstyle-$\hfil\cr}}}}}

\providecommand{\ehadrec}{\ensuremath{E_{\mathrm{had}}^{\mathrm{rec}}}\xspace}

\providecommand{\qz}{\ensuremath{q_{0}}\xspace}
\providecommand{\qq}{\ensuremath{Q^{2}}\xspace}

\providecommand{\enu}{\ensuremath{E_{\nu}}\xspace}
\providecommand{\enutrue}{\ensuremath{E_{\nu}^{\mathrm{true}}}\xspace}
\providecommand{\enureco}{\ensuremath{E_{\nu}^{\mathrm{reco}}}\xspace}
\providecommand{\enuqe}{\ensuremath{E_{\nu}^{\mathrm{QE}}}\xspace}

\providecommand{\nubar}{\ensuremath{\bar{\nu}}\xspace}
\providecommand{\nue}{\ensuremath{\nu_{e}}\xspace}
\providecommand{\nueb}{\ensuremath{\nubar_{e}}\xspace}
\providecommand{\numu}{\ensuremath{\nu_{\mu}}\xspace}
\providecommand{\numub}{\ensuremath{\nubar_{\mu}}\xspace}

\providecommand{\argon}{$^{40}$Ar\xspace}
\providecommand{\water}{H$_{2}$O\xspace}

\newcommand{\dm}[1]{\ensuremath{\Delta m^2_{#1}}\xspace}    \newcommand{\deltacp}{\ensuremath{\delta_{\rm CP}}\xspace}   

\newcommand{\maqe}{\ensuremath{M_{\mathrm{A}}^{\mathrm{QE}}}\xspace}

\newlength{\parenbarKernelHeight}

\providecommand{\nuany}{\ensuremath{\nu^{\bracketbar}}\xspace}
\providecommand{\numuany}{\ensuremath{\nuany_\mu}\xspace}
\providecommand{\nueany}{\ensuremath{\nuany_{e}}\xspace}

\usepackage{dblfloatfix}

\usepackage{threeparttable}

\begin{document}
\renewcommand{\tableautorefname}{Tab.}
\renewcommand{\figureautorefname}{Fig.}
\renewcommand{\equationautorefname}{Eq.}
\renewcommand{\sectionautorefname}{Sec.}
\renewcommand{\subsectionautorefname}{Sec.}
\renewcommand{\subsubsectionautorefname}{Sec.}

\renewcommand{\appendixautorefname}{}

\clearpage

\title[CP-violation or Nuclear Excitation?]{CP-violation or Nuclear Excitation: Reviewing the Role of Neutrino Interaction Model Uncertainties on Accelerator-Based Neutrino Oscillation Measurements}

\author{S. Dolan$^1$, L. Pickering$^2$, P. Stowell$^3$, C. Wilkinson$^4$, C. Wret$^5$}

\address{$^1$ CERN, 1211 Geneva 23, Switzerland}
\address{$^2$ STFC, Rutherford Appleton Laboratory, Harwell Oxford, United Kingdom}
\address{$^3$ University of Sheffield, School of Mathematical and Physical Sciences, Sheffield, United Kingdom}
\address{$^4$ Lawrence Berkeley National Laboratory, Berkeley, CA 94720, USA}
\address{$^5$ Imperial College London, Department of Physics, London, United Kingdom}

\ead{stephen.joseph.dolan@cern.ch, luke.pickering@stfc.ac.uk, p.stowell@sheffield.ac.uk, 
cwilkinson@lbl.gov, clarence.wret@imperial.ac.uk}

\begin{abstract}
Accelerator-based neutrino oscillation experiments have the potential to revolutionise our understanding of fundamental physics, offering an opportunity to characterise charge-parity violation in the lepton sector; to determine the neutrino mass ordering; and to explore the possibility of physics beyond three-flavour neutrino mixing. However, as more data is collected, the current and next generation of experiments will require increasingly precise control over the systematic uncertainties within their analyses. It is well known that some of the most challenging uncertainties to overcome stem from our uncertain modelling of neutrino--nucleus interactions, which also affect the event rates used to infer the oscillation probability. The sources of these uncertainties are often related to subtle details of the pertinent nuclear physics which are extremely difficult to control with sufficient precision. Confronting such uncertainties requires both state-of-the-art theoretical modelling and precise measurements of neutrino interaction event rates at experiment's near detectors,  before oscillations occur. In this work, we review the role of neutrino interaction systematic uncertainties in current and future measurements of neutrino oscillation as well as the experimental and theoretical prospects for reducing them to an acceptable level for the next generation of experiments. 

\end{abstract}
\submitto{\RPP}
\maketitle

\ioptwocol

\section{Introduction}
\label{sec:intro}
Precision measurements of neutrino flavour oscillations are a unique and powerful opportunity to study physics beyond the standard model~\cite{ParticleDataGroup:2024cfk}. Neutrino oscillations have been firmly established as a phenomenon in which neutrinos and antineutrinos produced with a defined \textit{flavour} (electron, muon or tau) have a nonzero probability of being observed as a different flavour after propagating some distance, with the oscillation probability depending on that distance $L$, and the energy of the neutrino, $E_{\nu}$~\cite{ParticleDataGroup:2024cfk}. Although neutrinos cannot be observed directly, they weakly interact, and their flavour can be identified from the outgoing charged lepton in charged-current (CC) weak interactions. Neutrino oscillation has been observed using neutrinos produced from a variety of different sources, including naturally produced solar~\cite{SNO:2002tuh,Super-Kamiokande:2001ljr} and atmospheric~\cite{Super-Kamiokande:1998kpq,IceCube:2017lak} neutrinos, as well as artificially produced neutrinos from reactors~\cite{KamLAND:2008dgz,DayaBay:2012fng,DoubleChooz:2011ymz,RENO:2012mkc,JUNO:2025gmd} and particle accelerators~\cite{Ambats:1998aa, Ahn:2006zza, Acquafredda:2009zz, NOvA:2007rmc, T2K:2011qtm}. 
Accelerator neutrino sources have the advantage of a well-defined baseline, and a controllable neutrino source whose energy, direction and initial flavour composition are relatively well understood, with the ability to produce a beam of predominantly muon neutrinos or muon antineutrinos. This review focuses on the role played by few-GeV neutrino--nucleus interaction physics in currently operating (T2K~\cite{T2K:2011qtm} and NOvA~\cite{NOvA:2007rmc}) and next-generation (Hyper-K~\cite{Hyper-Kamiokande:2018ofw} and DUNE~\cite{Abi:2020wmh}) long-baseline neutrino oscillation experiments (LBL).

Current measurements of neutrino oscillation are well described by the Pontecorvo--Maki--Nakagawa--Sakata (PMNS) three-flavour mixing paradigm~\cite{10.1143/PTP.28.870,Bilenky:1978nj,ParticleDataGroup:2024cfk}, in which neutrino mass and flavour states are related by a 3$\times$3 unitary mixing matrix. Neutrino oscillations are governed by seven parameters: three mixing angles ($\theta_{12}$, $\theta_{13}$ and $\theta_{23}$); a charge-parity (CP) violating phase, \deltacp; the squared mass differences between neutrino mass states (often referred to as \textit{mass splittings}), $\smash{\Delta m^{2}_{jk} = m^{2}_{j} - m^{2}_{k}}$; and the neutrino mass ordering (specifically, the sign of $\dm{32}$). 
The discovery of neutrino oscillations, which shows that neutrinos have mass, was awarded the 2015 Nobel prize in physics~\cite{nobel2015}. Future measurements of neutrino oscillations seek to: determine whether CP symmetry is violated in neutrino oscillations ($\sin\deltacp\neq0$); establish the ordering of neutrino masses; test whether the PMNS three-flavour paradigm is complete; and make precision measurements of all of the parameters governing oscillation. 

\begin{figure*}[htbp]
  \centering
  \includegraphics[width=0.8\textwidth]{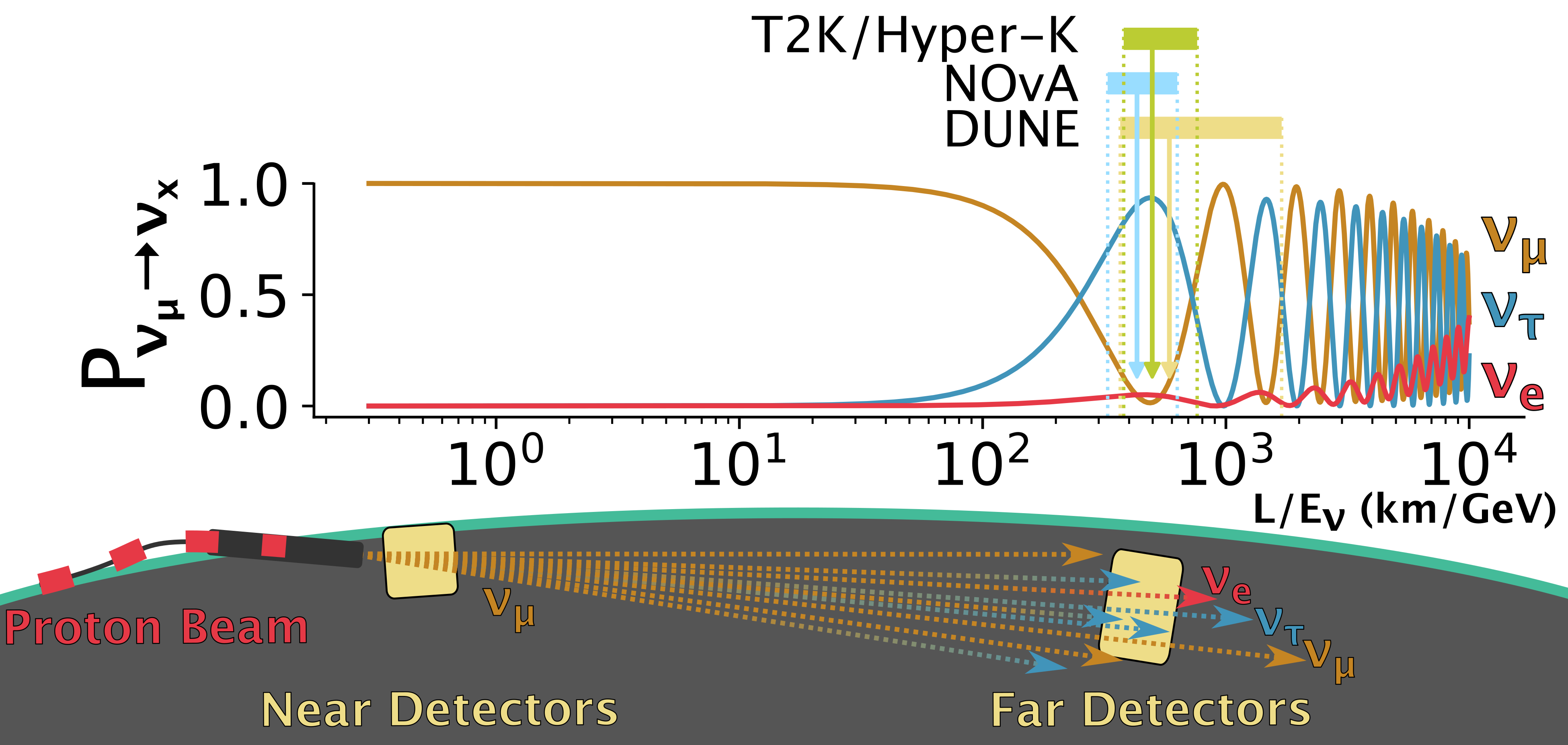}
  \caption{A schematic of an LBL neutrino oscillation experiment. The probability for a muon neutrino to oscillate to a particular flavour is shown as a function of the distance travelled divided by the neutrino energy ($L$/\enu), calculated with the NuFit 5.2 best fit parameter values~\cite{Esteban:2020cvm,nufitweb}. The span in $L$/\enu covered by current and future experiments is indicated, where $L$ is fixed at the experiments' baseline and the full width at half the maximum of \enu is used. The flux peaks are marked by arrows.}
  \label{fig:lbl}
\end{figure*}

\begin{figure}[htbp]
  \centering
  \includegraphics[width=0.98\linewidth]{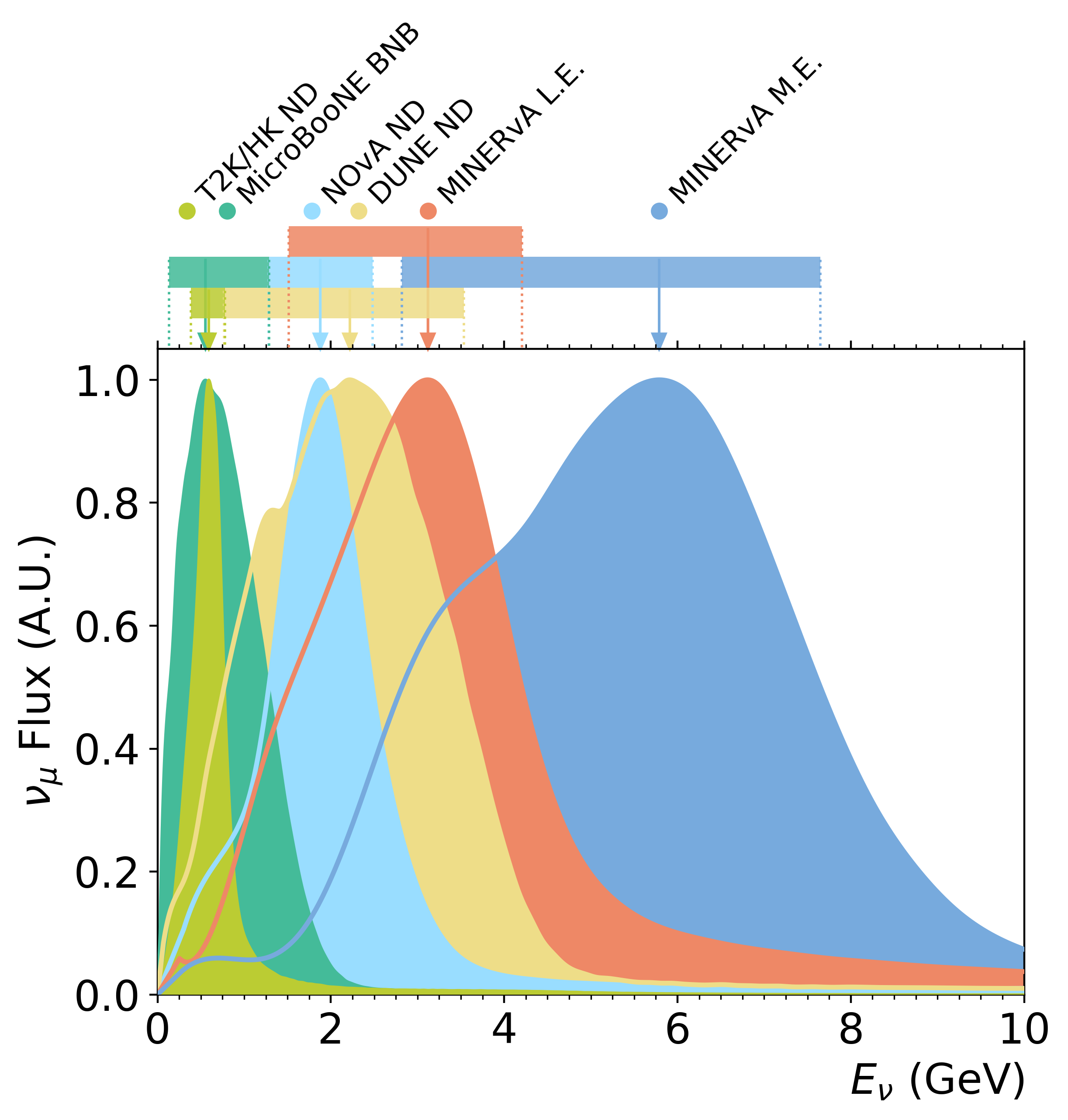}
  \caption{The shape of the predicted muon neutrino energy distributions for a range of experimental neutrino beams~\cite{MiniBooNE:2008hfu, T2K:2011qtm, MINERvA:2016iqn, MINERvA:2022vmb, Hyper-Kamiokande:2018ofw, NOvA:2007rmc, Abi:2020wmh, DUNE:2020ypp}. The coloured bands above the plot area show the full-width half-max (FWHM) ranges of the corresponding energy spectra, with the peaks marked by arrows.}
  \label{fig:exptfluxes}
\end{figure}

LBL experiments all have a similar basic design, a schematic for which is illustrated in~\autoref{fig:lbl}. A beam of $\mathcal{O}$(10--100) GeV protons is first accelerated and then collided with a fixed target to produce a secondary beam of mostly pions and kaons. Positive \textit{or} negative particles from this beam are directed into a decay volume using a series of magnetic focusing horns, which then decay to produce a beam of $\mathcal{O}$(0.1--10)~GeV predominantly muon neutrinos or antineutrinos respectively. There are percent-level contaminations from electron (anti)neutrinos and muon neutrinos of the \textit{wrong sign} (for example neutrinos contributing to the beam when negative mesons are selected). Accelerator muon neutrino flux distributions for current and planned experiments are shown in \autoref{fig:exptfluxes} (muon antineutrino spectra are typically very similar).
The neutrino beam is then sampled at least twice: once, $\mathcal{O}$(100) m from the target, before oscillation at a near detector (ND) complex; and again, $\mathcal{O}$(100--1000) km further downstream, where the flavour composition of the beam is expected to be close-to-maximally modified by neutrino oscillations at a far detector (FD). A comparison of the muon (anti)neutrino CC event rates at the ND with the  muon (anti)neutrino \textit{survival} and electron (anti)neutrino \textit{appearance} CC event rates at the FD is used to study neutrino oscillations. Whilst tau (anti)neutrino appearance would provide an interesting additional signal for neutrino oscillation studies, and is the dominant oscillation channel, the beam energies of accelerator LBL experiments usually prevent such precision measurements due to the high tau production threshold ($\enu \gtrsim 3.5$ GeV for an interaction with a free nucleon\footnote{CC tau (anti)neutrino interactions with a nucleus can have a significantly lower threshold due to nuclear effects~\cite{Graczyk:2004uy}}.). An exception was the OPERA experiment~\cite{OPERA:2018nar}, for which tau neutrino appearance was the primary channel, and which was optimised to maximise the tau neutrino CC event rate rather than the neutrino oscillation probability.

\begin{figure}[htbp]
  \centering
  \begin{subfigure}[b]{0.48\textwidth}
    \includegraphics[width=1.00\linewidth,clip,trim=0mm 0mm 0mm 17mm]{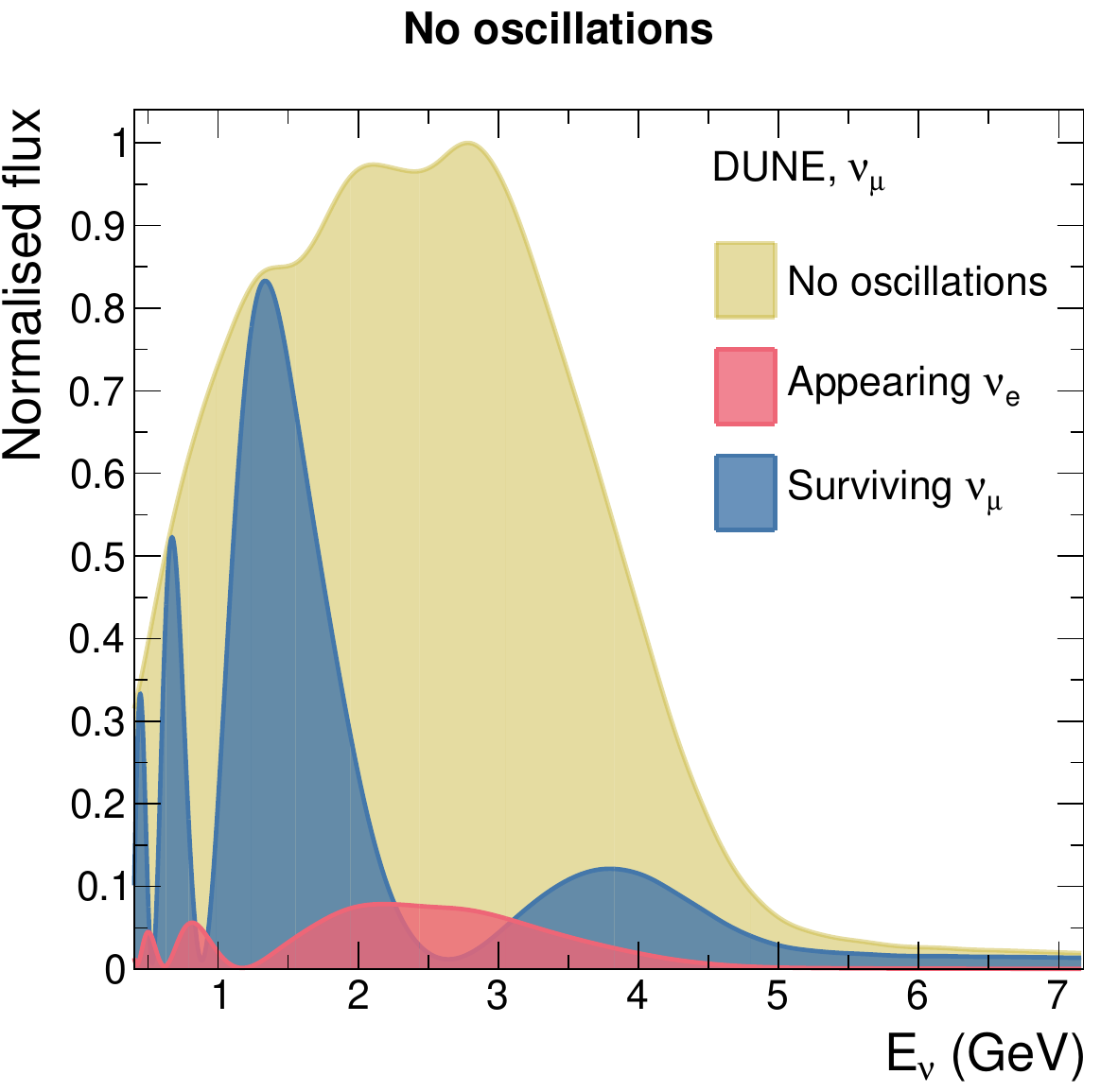}
    \caption{DUNE}
  \end{subfigure}
  \begin{subfigure}[b]{0.48\textwidth}
    \includegraphics[width=1.00\linewidth,clip,trim=0mm 0mm 0mm 17mm]{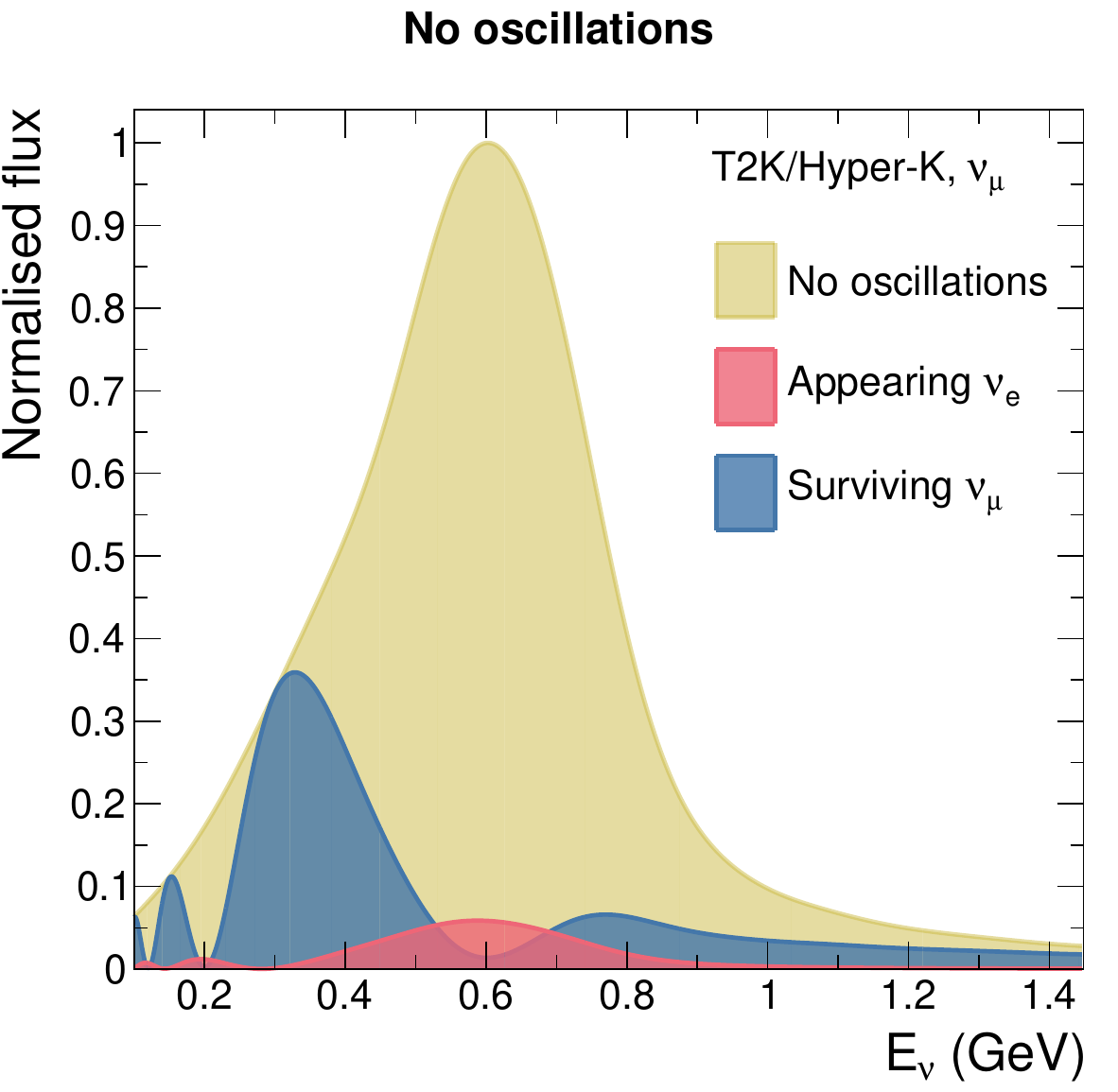}
    \caption{T2K/Hyper-K}
  \end{subfigure}
  \caption{The DUNE and T2K/Hyper-K unoscillated, $\numu \rightarrow \numu$ survival, and $\numu \rightarrow \nue$ appearance fluxes, shown as a function of neutrino energy~\cite{T2K:2011qtm, DUNE:2020ypp}. Each flux is normalised so that the maximum corresponds to 1.
  }
  \label{fig:oscfluxes}
\end{figure}

\autoref{fig:oscfluxes} shows the oscillated muon neutrino survival and electron neutrino appearance fluxes at the DUNE and Hyper-K/T2K FDs, using the best fit neutrino oscillation parameter values from the NuFit 5.2 fit to global measurements~\cite{Esteban:2020cvm,nufitweb}\footnote{Unless otherwise stated, all applications of neutrino oscillations throughout this work use the best fit neutrino oscillation parameter values from the NuFit 5.2 fit to global measurements~\cite{Esteban:2020cvm,nufitweb}, using the result assuming the normal mass ordering and not including data from Super-Kamiokande.}. By measuring the depth and position of the \textit{dip} in the muon (anti)neutrino survival probability, LBL experiments are sensitive to the mass splitting \dm{32} and the $\theta_{23}$ mixing angle. By measuring the normalisation and shape of the electron (anti)neutrino appearance probability, they are sensitive to the $\theta_{13}$ and $\theta_{23}$ mixing angles, and to \deltacp through a combination of both appearance channels. Additionally, electron (anti)neutrino appearance gives sensitivity to the \textit{octant} of $\theta_{23}$ (whether $\theta_{23}$ is larger or smaller than $\pi/4$)\footnote{The electron (anti)neutrino appearance probability is proportional at leading order to $\sin^{2}\theta_{23}$, whereas the muon (anti)neutrino survival probability is proportional at leading order to $\sin^{2}2\theta_{23}$.}.
The influence of matter effects on the electron (anti)neutrino oscillation probability~\cite{Wolfenstein:1977ue,Mikheev:1986gs} makes LBL experiments sensitive to the neutrino mass ordering, with an influence roughly proportional to the neutrino energy. Precision measurements of neutrino oscillation parameters require inference of the normalisation and shape of the oscillated flux. 

Current measurements from T2K~\cite{T2K:2023smv, T2K:2025yoy} and NOvA~\cite{NOvA:2021nfi, NOvA:2023iam, NOvA:2025tmb}, as well as a combined T2K--NOvA analysis~\cite{T2K:2025wet}, offer world-leading constraints on \dm{32} and $\sin^2\theta_{23}$, known at the 1.5--2\% and 3--6\% level respectively. Current measurements, at most, exclude CP-conserving values of \deltacp at the $\sim$2$\sigma$ level. LBL measurements, measurements of atmospheric neutrino oscillations~\cite{Super-Kamiokande:2023ahc,IceCubeCollaboration:2024ssx} and their combinations~\cite{T2K:2024wfn} provide a 1--2$\sigma$ exclusion of the inverted neutrino mass ordering. The next generation of LBL experiments, Hyper-K and DUNE, are designed to advance each of these measurements decisively, as outlined in \autoref{sec:intro}\footnote{Future measurements of the mass ordering are also possible through non-LBL experiments: JUNO, a next-generation reactor experiment~\cite{JUNO:2021vlw}; neutrino telescopes such as IceCube and KM3NeT~\cite{IceCube-Gen2:2019fet,KM3NeT:2021rkn}.}; and test whether the standard PMNS three-flavour mixing paradigm is sufficient to fully describe the neutrino sector. Future LBL measurements also complement those made by other neutrino experiments to provide a more complete picture of PMNS neutrino mixing or to look for physics beyond it by making measurements of the same parameters using different oscillation channels, baselines and energies (see, for example, Refs.~\cite{Esteban:2020cvm, Ellis:2020ehi, Capozzi:2021fjo, Denton:2023zwa, Chauhan:2023faf, Esteban:2024eli}). 

In all measurements, the oscillated neutrino flux distribution is not directly observable, instead it must be inferred from the \textit{rate} of neutrino interactions observed in the detector. \autoref{eq:rate-simple} provides a simplified general expression for the rate of neutrino interactions in neutrino oscillation experiments, neglecting backgrounds:
\begin{equation}
  \frac{dN}{d\vec{\mathbf{y}}} \propto \int{ d\enu d\vec{\mathbf{x}}\;\; \Phi(E_\nu) \; \sigma(\enu, \vec{\mathbf{x}}) \; M(\vec{\mathbf{y}} \mid \vec{\mathbf{x}}) \; P(\enu, L)},
  \label{eq:rate-simple}
\end{equation}
\noindent where the expected differential event rate as a function of observable particle kinematic variables, $\vec{\mathbf{y}}$, is proportional to the convolution of the neutrino flux $\Phi(E_\nu)$, the neutrino cross section $\sigma(\enu, \vec{\mathbf{x}})$ as a function of neutrino energy and the true outgoing particle kinematics $\vec{\mathbf{x}}$, the detector efficiency and smearing $M(\vec{\mathbf{y}}\mid\vec{\mathbf{x}})$ and the oscillation probability $P(\enu, L)$, which can be neglected at the ND. It is clear from \autoref{eq:rate-simple} that the flux, cross section and detector response must all be well understood as systematic uncertainties on each of these components propagate directly to uncertainties on inferred oscillation probabilities. A key challenge for LBL experiments is that at few-GeV energies neutrinos can interact in a variety of different ways with poorly known cross sections, producing a complex range of interaction products. Another is that due to unobserved neutral particles and detector smearing and thresholds, many different interaction products can have similar detector observables (for example, in  $M(\vec{\mathbf{y}} \mid \vec{\mathbf{x}})$, many $\vec{\mathbf{x}}$ can map to the same $\vec{\mathbf{y}}$). These two effects make it challenging to determine neutrino cross sections and to reconstruct the neutrino energy from detector observables, thereby making neutrino interaction modelling of vital importance to current and future experiments.

A major advantage for LBL experiments over those using natural sources is the ability to use an ND to sample the unoscillated beam, which allows the flux and cross-section models to be validated and constrained \textit{in situ}, benefitting from the high event rate close to the neutrino source. 
Although in principle the ND constrains the same flux and cross-section models as the FD, uncertainties do not trivially cancel between them, as by construction, and shown in \autoref{fig:oscfluxes}, there is a radical change in neutrino energy and flavour in the ND and FD spectra due to oscillations (see the integral over \enu in \autoref{eq:rate-simple}). There are also differences in the flux shape due to the spatial extent of the beam, which has some divergence at the ND due to the transverse extent of the decay pipe relative to the primary beam direction. The ND and FD also generally differ in design, often due to a need for different detector technologies to cope with different intensities, and always through their different sizes, leading to different acceptances to particles. Additionally, with very little intrinsic electron (anti)neutrino contamination in the beam, the ability to constrain the electron (anti)neutrino cross section with the ND is limited.

Although not shown in \autoref{eq:rate-simple}, there are further challenges that affect the inference of neutrino interaction probabilities in a real LBL experiment. Impurities in the beam mean that more than one neutrino source usually contributes to the event rate for any selection of neutrino interactions. For example, selections of events at the FD targeting $\nue^{\bracketbar}$ interactions are dominated by $\numu^{\bracketbar} \rightarrow \nue^{\bracketbar}$ appearance, but there are $\nue^{\bracketbar} \rightarrow \nue^{\bracketbar}$ contributions due to intrinsic electron (anti)neutrinos present in the unoscillated beam. Additionally, wrong sign contributions cannot typically be distinguished in large unmagnetised FDs. Due to imperfect event selections, wrong flavour contributions complicate matters further, and there can also be neutral current (NC) weak interactions in which no charged lepton is produced which may be mis-reconstructed as a CC interaction and included in the sample. Finally, non-beam related backgrounds may also contribute through a variety of mechanisms.

\begin{table*}[htbp]
  \centering
  {\renewcommand{\arraystretch}{1.2}
\begin{tabular}{l|l|l|l}
\hline\hline
Experiment                  & Hyper-K/T2K            & NOvA                 & DUNE             \\ \hline
Peak neutrino energy        & 0.6~GeV                & 1.9 GeV              & 2.2~GeV          \\ 
Baseline                    & 295 km                 & 810 km               & 1300 km          \\
Flux width (FWHM)           & $[0.4-0.8]$~GeV        & $[1.3-2.5]$~GeV      & $[0.8-3.5]$~GeV  \\ 
Primary ND technology       & Plastic scintillator   & Liquid scintillator  & LArTPC           \\ 
Primary ND target material(s)  & C$_{8}$H$_{8}$, H$_2$O & 67\% $^{12}$C, 16\% $^{35}$Cl, 11\% $^{1}$H, 6\% Other & \argon            \\ 
Primary FD technology       & Water Cherenkov        & Liquid scintillator  & LArTPC           \\ 
Primary FD target material(s)  & H$_2$O              & 67\% $^{12}$C, 16\% $^{35}$Cl, 11\% $^{1}$H, 6\% Other & \argon            \\ 
\hline\hline
\end{tabular}}
  \caption{\label{tab:lbldeets} An overview of the flux and detectors used for current and future accelerator LBL experiments. The flux details are provided for neutrino-mode muon-neutrino fluxes where the width is defined as the full width at half maximum (FWHM) of the flux. The full profile of the fluxes can be found in \autoref{fig:exptfluxes}. Experimental details are taken from Ref.~\cite{T2K:2011qtm, Hyper-Kamiokande:2018ofw} (Hyper-K/T2K), Ref.~\cite{NOvA:2007rmc} (NOvA) and Ref.~\cite{Abi:2020wmh, DUNE:2020ypp} (DUNE). Although the primary detector technology of the Hyper-K/T2K ND is plastic scintillator, Hyper-K also plans an intermediate water Cherenkov detector, as described in the main text.}
\end{table*}
Although the basic layout of LBL experiments is similar, they differ significantly in various design details, which are summarised for the current and future experiments in \autoref{tab:lbldeets}. While the value of $L$/\enu at which the oscillation probability is maximal is set by the value of \dm{32}, a key design choice is what the value of the baseline or neutrino beam energy should be (with the other adjusted to arrive at the desired $L$/\enu). There are two competing considerations: first, because accelerator neutrino beams cannot be focused perfectly and the flux falls as a function of $\sim$1/$L^2$, the rate reduces at longer baselines, only partially mitigated by the roughly linear increase in the neutrino--nucleus cross section as a function of \enu; second, sensitivity to the neutrino mass ordering due to the matter effect grows with \enu. These competing requirements drive the different choices in energy (see \autoref{fig:exptfluxes}) and baseline between the Hyper-K/T2K program and the NOvA/DUNE experiments. Additional strong constraints on the baseline come from the availability of accelerator complexes and locations suitable for hosting FDs.

The different energies required to be close to the oscillation maximum for different baselines drive differences in the detector technology used. With a peak neutrino energy of $\approx$0.6 GeV, neutrino interactions at Hyper-K/T2K generally have a small number of particles produced, well suited for reconstruction in water Cherenkov detectors, which scale to very large volumes in a cost effective way. However, water Cherenkov detectors are not well suited to high rate environments, leading to a different primary ND design for Hyper-K/T2K, although Hyper-K envisages the construction of an intermediate water Cherenkov detector (IWCD), placed $\sim$1 km from the neutrino beam production point (where oscillations are not expected to be significant)~\cite{Zhu:2023nsv,Hyper-Kamiokande:2025fci}. 
NOvA and DUNE have higher energies and access neutrino interactions with more complex interaction signatures, requiring high resolution detectors with low particle tracking thresholds. 
Another key difference between the experiments is the use of off-axis neutrino beams---exploiting the two-body pion decay kinematics to produce a narrower neutrino energy profile~\cite{offaxis}---at Hyper-K/T2K and NOvA and a broad on-axis flux distribution at DUNE  (\autoref{fig:oscfluxes}). DUNE thereby gains sensitivity by observing the oscillation shape over a broader region of energy (and therefore the oscillation pattern as a function of $L$/\enu)~\cite{DUNE:2020jqi, DUNE:2021mtg}, but requires precise energy reconstruction capabilities over the entire range, motivating the use of liquid argon Time Projection Chambers (LArTPCs).  Differences in the experimental approaches, target materials, reconstruction methods and neutrino flux distributions for current- and next-generation experiments render them sensitive to overlapping but not identical aspects of neutrino interaction modelling. 
This significantly increases the scope of the required program to reduce systematic uncertainties in general, with different challenges for different experiments, but also allows measurements or models of neutrino interactions focussed on one experiment to provide useful input to others.

\begin{table}[htbp]
  {\renewcommand{\arraystretch}{1.2}
  \begin{tabular}{l|cc|cc}
    \hline\hline
Experiment    & T2K & NOvA & Hyper-K     & DUNE       \\ \hline
$N_{\nu_\mu}^{\mathrm{FD}}$ ($\nu$-mode)             & 318 & 384  & $\approx$8800  & $\approx$15000 \\
$N_{\bar{\nu}_\mu}^{\mathrm{FD}}$ ($\bar{\nu}$-mode) & 137 & 106  & $\approx$12000 & $\approx$8100 \\
$N_{\nu_e}^{\mathrm{FD}}$ ($\nu$-mode)               & 108 & 181   & $\approx$2800  & $\approx$3300 \\
$N_{\bar{\nu}_e}^{\mathrm{FD}}$ ($\bar{\nu}$-mode)   & 16  & 32   & $\approx$1500  & $\approx$900 \\
\hline\hline
\end{tabular}}
  \caption{Integrated event rates in neutrino ($\nu$) and antineutrino-enhanced ($\bar{\nu}$) beam modes for the FD samples from T2K~\cite{T2K:2023smv} and NOvA~\cite{NOvA:2025tmb} oscillation analyses, as well as projected event rates for $\sim$10 years of Hyper-K~\cite{Hyper-Kamiokande:2025fci} and DUNE~\cite{DUNE:2020jqi} data taking (rounded to the nearest 100). Event rates for Hyper-K and DUNE assume $\deltacp =-\pi/2$ and normal neutrino mass ordering, except for the DUNE \numu and \numub samples which use $\deltacp=0$ (although these samples have little sensitivity to variations of $\deltacp$). Hyper-K assume a 1:3 neutrino:antineutrino beam run plan, whereas DUNE assume 1:1. DUNE only includes interactions between 0.5 and 10 GeV neutrino energy.}
\label{tab:evtrates}
\end{table}

To give a sense of the scale of systematic uncertainties that impact LBL measurements, \autoref{tab:evtrates} shows the projected FD event rates for current and next-generation LBL experiments, highlighting the order of magnitude increase which the latter offer compared to the former. 
Current systematic uncertainties on the FD muon neutrino ($N_{\nu_\mu}^{\mathrm{FD}}$) and muon antineutrino ($N_{\bar{\nu}_\mu}^{\mathrm{FD}}$) event rates are at the level of 4\% for T2K~\cite{T2K:2023smv, T2K:2025yoy}, whilst uncertainties on the FD electron neutrino ($N_{\nu_e}^{\mathrm{FD}}$) and electron antineutrino ($N_{\bar{\nu}_e}^{\mathrm{FD}}$) event rates are 5\% for T2K and 9\% for NOvA~\cite{NOvA:2021nfi}. 
In all cases, the dominant contributors to the systematic uncertainties are related to the modelling of neutrino interaction cross sections. Inferring statistical uncertainties from \autoref{tab:evtrates} shows that, although T2K and NOvA are not yet limited by systematic uncertainties, a dramatic reduction is required to avoid DUNE and Hyper-K being prematurely limited by them.

In this work, we review the challenges in neutrino interaction modelling that lead to dominant systematic uncertainties on neutrino oscillation measurements and the methods for reducing those uncertainties. We begin, in \autoref{sec:models}, with a brief review of neutrino interaction modelling and of the simulations used to describe neutrino interactions for neutrino oscillation analyses. An overview of how modern LBL oscillation analyses work and how their measurements are dependent on different aspects of neutrino interaction modelling is presented in \autoref{sec:howtooa}. Current constraints on cross-section models from global cross-section measurements are described in \autoref{sec:constraints}. A review of how different models describe the aspects of neutrino interactions most important for oscillation analyses is in \autoref{sec:issuesforoa}, before the impact of modelling uncertainties on future oscillation experiment sensitivities are discussed in \autoref{sec:impactonoa}. Novel methods, both demonstrated and proposed, to reduce systematic uncertainties are considered in \autoref{sec:novelapproach}. Our review finishes with a qualitative discussion of the challenges faced by next-generation experiments and prospects to face them in \autoref{sec:prospects}, with our conclusions in \autoref{sec:concl}.

\section{Modelling few-GeV neutrino interactions}
\label{sec:models}

This section provides a brief overview of neutrino interaction models at few-GeV energies in order to give essential context for the subsequent sections which demonstrate how mismodelling can impact neutrino oscillation analyses. More comprehensive reviews of neutrino interaction modelling can be found in Refs.~\cite{Formaggio:2012cpf,Alvarez-Ruso:2014bla,Mosel:2016cwa,Katori:2016yel,NuSTEC:2017hzk,Mahn:2018mai,Jachowicz:2021ieb,SajjadAthar:2022pjt, Ruso:2022qes, Pandey:2025vpa}.  

\subsection{Neutrino--nucleon interactions}
\label{sec:nucleon}
Neutrinos interact with nucleons in different ways, depending on the neutrino energy, whether the probe is a neutrino or antineutrino, the energy and momentum transfer to the nucleus (\qz, $\vec{q}_{3}$), the target nucleon's isospin, and whether the interaction is charged-current (CC) or neutral-current (NC). At the neutrino energy scales of LBL experiments, CC interactions at low energy transfers ($\qz \lesssim 0.3$~GeV) are dominated by quasi-elastic interactions, where a target neutron is transformed into a proton. At higher energy transfers ($\qz \gtrsim 0.3$ GeV), in the \textit{resonance region}, neutrino interactions excite nucleons into resonant states which then decay to produce pions and other mesons. At even higher energy transfers ($\qz \gtrsim 1.5$ GeV) neutrino interactions begin to resolve the quark sub-structure of nucleons. Since modern experiments which measure neutrino cross sections use relatively heavy nuclear targets, neutrino--nucleon interaction cross sections are primarily constrained using historical bubble chamber neutrino scattering data on hydrogen, deuterium, and other light targets. However, this data is often not accompanied by a complete treatment of systematic uncertainties and the treatment of nuclear effects remain a challenge even for these light targets~\cite{Meyer:2016oeg, Kabirnezhad:2017jmf, Nikolakopoulos:2022tut}.

\subsubsection{Quasi-elastic and elastic scattering}
Charged-current quasi-elastic (CCQE) and neutral-current elastic (NCE) interactions are those in which the struck nucleon remains a nucleon and no other hadrons are produced. 
In CCQE interactions, the (anti)neutrino interacts with an initial-state neutron (proton) via the exchange of a $W$ boson, producing a charged muon and proton (neutron) final state. In NCE interactions, no charge is exchanged.
In comparison with other channels, CCQE and NCE interactions with free-nucleon targets are relatively well understood. 
The cross section depends on the neutrino energy and a single kinematic variable, usually chosen to be the squared four-momentum transferred to the target, \qq. 
The nucleon response can be written in terms of form-factors which describe the nucleon as a target with finite extent~\cite{LLEWELLYNSMITH1972261}.

CCQE interactions can be broken down into vector, axial-vector, and pseudoscalar contributions. 
The vector part can be constrained by electron-scattering data, and the pseudoscalar component can be related to the axial-vector contribution through the partially conserved axial current (PCAC) hypothesis~\cite{LLEWELLYNSMITH1972261}.
In the case of CCQE scattering, the axial-vector form factor is often parameterised as a dipole:
\begin{equation}
 F_{\mathrm{A}}(\qq) = \frac{F_{\mathrm{A}}(\qq=0)}{\left( 1+\left( Q/\maqe \right)^2 \right)^2},
 \label{eq:axialff}
\end{equation}
which contains two parameters\footnote{The Fourier transform of a form factor corresponds to the distribution it maps out in physical space, so a dipole form factor corresponds to an exponential distribution.}. 
The numerator can be determined from $\beta$-decay, $F_{\mathrm{A}}(\qq=0)\approx-1.27$, whilst \maqe, the nucleon axial mass, was determined to be $\approx$1~GeV/$c^2$ from bubble chamber data~\cite{Barish:1977qk,Barish:1979ny,Miller:1982qi,Baker:1981su,Kitagaki:1983px,Kitagaki:1990vs,Bernard:2001rs}, although is often treated as a free parameter due to uncertainty over the quality of the bubble chamber data. However, the use of a dipole form factor is not well motivated, and alternatives have been introduced which go beyond the simple dipole form of \autoref{eq:axialff}\footnote{Early bubble chamber experiments experimented with different forms for the axial form factor, but settled on the dipole form, largely by analogy to the electromagnetic form factors~\cite{Baker:1981su, Radecky:1981fn}.}. These include relatively simple empirical corrections to the dipole~\cite{Bodek:2007ym}, as well as alternative parameterisations which include more freedom in the \qq-dependence of the axial-vector form factor using the \textit{z-expansion} formalism~\cite{Meyer:2016oeg}.

Recent theory developments have enabled the use of lattice quantum chromodynamics (LQCD) to directly calculate the axial form factor for CCQE interactions~\cite{Meyer:2022mix, Meyer:2026kdl}.
These calculations find a \qq-dependence which disagrees with the bubble chamber data and the dipole assumption.
Additionally, the MINERvA experiment has recently extracted the CCQE $\bar{\nu}_{\mu}$--proton cross section from their C$_{8}$H$_{8}$ target using kinematic separation techniques~\cite{MINERvA:2023avz}. This represents the only modern measurement of neutrino interactions on a free-nucleon target but, with the current uncertainties, the result is compatible with many available models.

For NCE scattering, there is an additional strange axial-vector form factor, which can be written similarly to \autoref{eq:axialff} but replacing $F_{\mathrm{A}}(\qq=0)$ with $\Delta s$~\cite{Garvey:1992cg, Formaggio:2012cpf, Giusti:2019cup}, where \maqe is the same as the CCQE form factor, and $\Delta s$ is the strange quark contribution to the spin of the nucleon. 
Attempts to measure $\Delta s$ experimentally~\cite{Ahrens:1986xe, MiniBooNE:2010xqw, MiniBooNE:2013dds, Giusti:2019cup, Pate:2024acz} or to calculate it with LQCD~\cite{QCDSF-UKQCD:2018afx} have been conducted, but there remains a large uncertainty on its value.

\subsubsection{The resonance region}
\label{sec:spp}
At higher energy transfers ($\qz\gtrsim300~\text{MeV}$), there is sufficient energy to excite the nucleon to a resonant state. 
Generally speaking, most models consider discrete resonances with masses below $2~\text{GeV}/c^2$, hereafter defined as \textit{hadronic mass} or $W$. Formally, 
\begin{equation}
  W^2=-\qq+m_{n}^2+2 q \cdot P_{n},
\end{equation}
\noindent where $\qq=-q^2$ is the squared four-momentum transferred to the nucleon ($q$) and $P_{n}$ ($m_{n}$) is the struck nucleon's four-vector (mass). 
For a stationary free nucleon target, this reduces to
\begin{equation}
  W_{\mathrm{free}}^2=-\qq+m_{n}^2+2\qz m_{n},
\end{equation}
sometimes referred to as the \textit{experimental $W$}, or $W_{\mathrm{exp}}$.
The resonant states typically decay to produce one pion and one nucleon, in what is known as resonant pion production (RPP). There is also a subpercent-level branching fraction to a single photon and nucleon~\cite{ParticleDataGroup:2024cfk}. Higher mass resonances may decay to produce other mesons, or more than one meson, in addition to a nucleon~\cite{PhysRevD.82.033001, Alvarez-Ruso:2014bla,Nakamura:2015rta}. Often the dominant resonance state is the Delta resonance at 1.232 GeV$/c^2$, the $\Delta(1232)$, which is involved in around 64\% of RPP muon neutrino interactions at 2.5~GeV neutrino energy, and 88\% at 0.6~GeV, as calculated with the NEUT interaction simulation package~\cite{Hayato:2002sd,Hayato:2009,Hayato:2021heg}.

For CC neutrino interactions, a double or single positively charged resonance such as a $\Delta^{++}$ and $\Delta^+$ can be excited, which subsequently decay as $\Delta^{++} \rightarrow p \pi^+$, and $\Delta^{+} \rightarrow n \pi^+$ or $\Delta^+ \rightarrow p \pi^0$, and leads to a charged lepton, charged or neutral pion, and a single nucleon after the interaction.
In CC antineutrino interactions, a neutral or negative resonance excitation such as $\Delta^0$ or $\Delta^-$ is possible, with correspondingly altered outgoing hadrons.
This picture readily extends to NC interactions, provided charge, isospin, and other quantum numbers are conserved in the interaction. 
However, for RPP and other resonant production modes (for example, kaon or photon production), the contributions are generally not from a single resonance. Instead, multiple resonances overlap, causing important interference effects~\cite{Rein:1980wg, Kabirnezhad:2017jmf,Nakamura:2015rta,Nikolakopoulos:2018gtf,Sobczyk:2018ghy,Kabirnezhad:2024cor}. Additionally, each resonance may have a different branching fraction to the $N \pi$ decay channel. There are also less understood but important \textit{non-resonant} contributions to the single-pion-production cross section, which may also interfere with the resonant excitations (see, for example, Ref.~\cite{Sobczyk:2018ghy}).

Resonance interactions are commonly broken down into the pure $I_{3/2}$ and mixed ($I_{3/2}$ and $I_{1/2}$) isospin channels. 
Isospin is a \textit{spin-like} quantity, in the sense that it sums like spin, reflecting the up and down quark content of hadrons, where the up quark has $I_3=1/2$ and the down quark has $I_3=-1/2$. 
For instance, the $\Delta^{++}$ consists of $uuu$ so is $I_3=3/2$, and the $\Delta^+$ is $uud$ so is $I_3=1/2$. 
The dominant RPP interaction for neutrinos is the CC$1\pi^+1p$ interaction, which is pure $I_{3/2}$ and only has contributions from three well-separated, mostly non-interfering, $\Delta$ resonances: $\Delta^{++}(1232)$, $\Delta^{++}(1600)$ and $\Delta^{++}(1920)$~\cite{ParticleDataGroup:2024cfk}.
The CC$1\pi^01p$ and CC$1\pi^+1n$ channels are both mixed isospin, and have complex contributions from 15--20 interfering resonances below a hadronic mass of $2~\text{GeV}/c^2$~\cite{ParticleDataGroup:2024cfk}.

For interactions on a nucleon, the RPP cross section can be formulated similarly to the CCQE case, but with significantly more complexity due to the multiple contributing resonances to each outgoing hadronic state, each with different form factors, and the relative interferences and phases between them. 
A dipole approximation of the form factor for each resonance contains two free parameters. Unlike the CCQE case, the values of these form factors at $\qq = 0$ are not constrained by $\beta$-decay measurements. 
Moreover, the presence of a two-particle hadronic final-state means that the cross section depends on four independent variables\footnote{A common choice is \qq, $W$, and the two so-called \textit{Adler angles}~\cite{ADLER1968189}, the polar and azimuthal angle of the direction of the outgoing meson in the hadronic rest frame, with respect to the direction of the energy transfer in the interaction.}.
Given the different properties of each contributing resonance, the kinematics of the outgoing nucleon and pion for RPP interactions in general are also not trivial to calculate,
and there is limited available data to constrain RPP or other resonant interactions.
Additionally, whilst there are relationships between CC and NC RPP interactions, and between neutrino and antineutrino interactions, these are complex. Each channel has different contributions from multiple interfering resonances, each with different form factors, and a poorly understood non-resonant background contribution. Applying constraints across different RPP channels is therefore challenging due to the resonances being difficult to separate, and their relative contributions depending on the incoming neutrino energy.
Recent and historical examples of RPP cross-section calculations can be found in Refs.~\cite{ADLER1968189,Rein:1980wg,Graczyk:2014dpa,Gonzalez-Jimenez:2016qqq,Gonzalez-Jimenez:2017fea,Kabirnezhad:2017jmf,Niewczas:2020fev,Sato:2021pco,Nikolakopoulos:2018gtf,Nikolakopoulos:2022tut,Sobczyk:2018ghy,Alvarez-Ruso:2015eva,Nakamura:2015rta,Hernandez:2007qq,Hernandez:2010bx,Martini:2009uj,Martini:2014dqa,Hernandez:2013jka,Lalakulich:2006sw,Lalakulich:2005cs, Kabirnezhad:2024cor}.

As for the CCQE case, the axial-vector contributions to the RPP cross section can be constrained by data from neutrino--deuterium bubble chamber experiments, primarily ANL~\cite{Derrick:1980xw} (peak $\enu \approx 0.8$ GeV) and BNL~\cite{Kitagaki:1986ct} (peak $\enu \approx 1$ GeV). However, tensions in the absolute normalisation of the cross sections measured by the two experiments led to re-evaluation of the data under the assumption of a mismodelled flux~\cite{Graczyk:2009qm, Hernandez:2007qq, Wilkinson:2014yfa}. Whilst the reanalysed data has been used for tuning RPP model parameters~\cite{Rodrigues:2016xjj,T2K:2023qjb,MINERvA:2019kfr,GENIE:2021zuu}, it brings into question the reliability of these historic datasets, and the techniques used to correct some of the data in the reanalyses cannot be applied to all of the variables of interest, leading to a confused overall picture. Higher energy (peak $\enu \approx 20$ GeV) neutrino--deuterium scattering data from BEBC~\cite{Allasia:1990uy} and neutrino--hydrogen from the Fermilab 15 ft bubble chamber~\cite{Bell:1978qu} have also been used to study resonances above the $\Delta(1232)$, and the behaviour of the $\Delta(1232)$ at higher neutrino energies~\cite{Kabirnezhad:2017jmf,Gonzalez-Jimenez:2016qqq,Kakorin:2021mqo,Lalakulich:2006sw,Kabirnezhad:2024cor}.  The bubble chamber data is often combined for hydrogen and deuterium targets, and the amount of each is seldom reported. This combined target is commonly treated as a free proton target, even though nuclear effects in deuterium have been shown to modify the cross section~\cite{Mann:1973pr,Singh:1971md,Alvarez-Ruso:1998ais}.
Other bubble chambers with non-hydrogen target materials have also been used to try and inform the neutrino--nucleon cross section in older studies~\cite{GargamelleNeutrinoPropane:1977hya, skat_1989}. Antineutrino measurements are sparse, especially in the few-GeV range.
LQCD calculations of the RPP form factors are possible but computationally challenging, and results using physical pion masses are not yet available (as discussed in, for example, Ref.~\cite{Barca:2022uhi}).
Whilst existing bubble chamber data has proven very useful, the RPP single-nucleon cross section is poorly understood, especially beyond the $\Delta(1232)$ region, which becomes increasingly important at higher neutrino energy ($\enu \gtrsim 1$ GeV).

\subsubsection{Shallow and deep inelastic scattering (SIS/DIS)}
At sufficiently large energy transfers, neutrinos undergo deep inelastic scattering (DIS) with nucleon targets in which the interaction resolves the internal quark structure of the nucleon, resulting in hadronic showers containing multiple mesons and hadrons. 
CC and NC DIS interaction cross sections differential in Bjorken $x$ and $y$~\cite{Bjorken:1968dy},
\begin{equation}
    \begin{aligned}
    x=\frac{\qq}{2m_{n}\qz};
    \end{aligned}
    \quad
    \begin{aligned}
    y = 1 - \frac{E_l}{\enu},
    \end{aligned}
\end{equation}
\noindent can be calculated using QCD and expressed in terms of parton distribution functions (PDFs)~\cite{LLEWELLYNSMITH1972261} that are validated against a wealth of high energy (>100 GeV) lepton nucleon scattering data~\cite{SajjadAthar:2020nvy}. The leading order GRV98 PDFs~\cite{Gluck:1998xa} are typically used for modelling neutrino--nucleon DIS cross sections.
This approach requires interactions to be in a regime where perturbative QCD applies. This is often taken to mean $W \geq 2$ GeV$/c^2$ and $\qq \geq 1$ GeV$^2/c^4$, but the exact limits are not precisely defined~\cite{Bodek:2010km}. 

\textit{Shallow} or \textit{soft} inelastic scattering (SIS) covers non-resonant meson-producing interactions with $W$ and \qq below the DIS regime, which significantly overlaps with RPP. The models used in simulation packages typically blend RPP and SIS predictions in an \textit{ad hoc} manner to transition from discrete interfering resonances to being continuous in hadronic mass in the DIS regime. In some models, SIS interactions are used to describe the non-resonant RPP contribution~\cite{GENIE:2021wox}. In this region, QCD becomes non-perturbative and the resolution of the interaction is not unambiguously at the quark level. 
The challenges associated with first-principle SIS calculations~\cite{SajjadAthar:2020nvy} mean current models heavily rely on empirical data-driven corrections such as the Bodek--Yang formalism~\cite{Bodek:2003wc,Bodek:2004pc,Bodek:2010km,Bodek:2021bde,Bodek:2022lqu}, which is an empirical scaling prescription tuned to electron-scattering data.
Precision few-GeV neutrino interaction data focused on the SIS/DIS region are challenging to obtain. 
Multi-pion measurements from bubble chamber experiments on light targets suffer from low statistics~\cite{Kitagaki:1986ct,Day:1983itd}, and recent high-statistics measurements are often on nuclear targets~\cite{MINERvA:2021owq, MINERvA:2025hzq}, making the separation of nuclear effects from neutrino--nucleon cross-section modelling challenging. 

SIS and DIS interactions lead to the production of quarks and gluons from the struck nucleon. \textit{Hadronisation} is the process by which observable hadrons are formed from these fragments. Hadronisation models predict the outgoing multiplicity or kinematic distributions of hadrons.
Dedicated simulation packages, often developed for collider physics, such as PYTHIA~\cite{Bierlich:2022pfr}, HERWIG~\cite{Bahr:2008pv} and SHERPA~\cite{Chahal:2022rid}, describe hadronisation, but are generally focussed on high $W$ and \qq. For example, PYTHIA is designed to model the production of complex high-multiplicity final states from high-energy collisions above a centre-of-mass energy of 10~GeV~\cite{Bierlich:2022pfr}. 
Simulations of few-GeV neutrino interactions typically use the relatively old PYTHIA 5.7 or 6~\cite{Sjostrand:2006za, Yang:2009zx, Bronner:2016huz} to describe hadronisation for $W\gtrsim2$ GeV and custom data-driven \textit{ad hoc} hadronisation models at lower $W$~\cite{Yang:2009zx,Bronner:2016gmz,Bronner:2016huz}. Using PYTHIA at such low $W$ or when predicting low-multiplicity final states (1--3 pions) is not well motivated, but better options remain elusive.
Hadronisation models are often tuned to the limited available data in an attempt to mitigate these issues~\cite{Bronner:2016gmz,PhysRevD.87.074029, Katori:2014fxa, GENIE:2021wox}, but assigning reasonable uncertainties is challenging.
Recently, PYTHIA collaborators have explored lower energy hadronisation modelling for cosmic-ray cascades in the atmosphere~\cite{Sjostrand:2020gyg,Sjostrand:2021dal}, but its impact on few-GeV neutrino interaction modelling has not yet been evaluated.

Ref.~\cite{SajjadAthar:2020nvy} reviews the challenges in describing SIS/DIS interactions in more detail. Overall, whilst the high $W$ and high \qq DIS total cross sections in the region of perturbative QCD are well understood, the SIS region and the transition region between resonant, SIS and DIS interactions is not. 

\subsubsection{Diffractive pion production}
\label{sec:diffractive_pi}
Non-resonant single pion production neutrino interactions with free nucleons at high-$W$ \smash{($W\gtrsim3$ GeV$/c^{2}$)} and low-\qq \smash{($\qq\lesssim 2$ GeV$^{2}/c^4$)} are known as \textit{diffractive pion production}.
This process is theorised to originate predominantly from only the axial-vector part of the weak interaction~\cite{Bartl:1977uj,Rein:1986cd}, leading to similar cross sections for neutrinos and antineutrinos. 
Its cross section is small at few-GeV neutrino energies, contributing at the percent-level to interactions which produce a single pion at $\enu \approx 1$ GeV. 
Diffractive pion production was hypothesised to explain the large CC$1\pi^+ 1p$ cross section in the high $W$ region, where resonances are expected to not significantly contribute, present in BEBC \numu--hydrogen data~\cite{Aachen-Birmingham-Bonn-CERN-London-Munich-Oxford:1985yec,Rein:1986cd}. 
Evidence for NC diffractive $\pi^{0}$ production has been shown by the MINERvA experiment using a plastic scintillator (C$_{8}$H$_{8}$) target~\cite{MINERvA:2016uck}, but a definitive measurement of diffractive pion production cross section has not yet been made.

\subsubsection{Radiative corrections}
Modifications to the interaction rate or kinematics of interactions can occur through higher order quantum electrodynamics (QED) effects such as virtual loop contributions beyond the tree-level Feynman diagram for a given process. These \textit{radiative corrections} have a percent-level effect on CC (anti)neutrino interaction cross sections. Their inclusion in cross-section calculations is important for precision studies of neutrino oscillations since their impact depends on the neutrino flavour. 
Recent work has allowed them to be precisely calculated, and their effect on the total and some differential cross sections are well characterised~\cite{Day:2012gb,Tomalak:2021hec,Tomalak:2022xup}.
Radiative corrections can lead to additional real photon emission which has the potential to complicate flavour identification in experiments, especially for electron neutrino interactions~\cite{Tomalak:2021hec}.

\subsubsection{Summary of neutrino--nucleon cross sections}

\begin{figure}[htbp]
  \centering
  \begin{subfigure}[b]{\linewidth}
    \centering
    \includegraphics[width=0.75\linewidth]{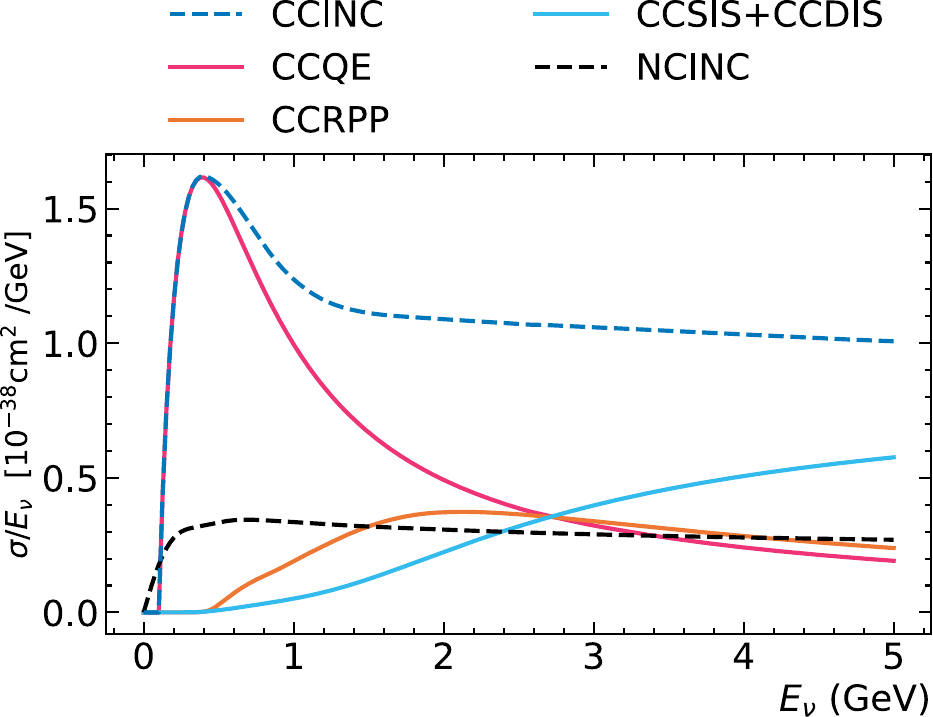}\\
    \caption{\numu--neutron}
  \end{subfigure}
  \begin{subfigure}[b]{\linewidth}
    \centering
    \includegraphics[width=0.8\linewidth]{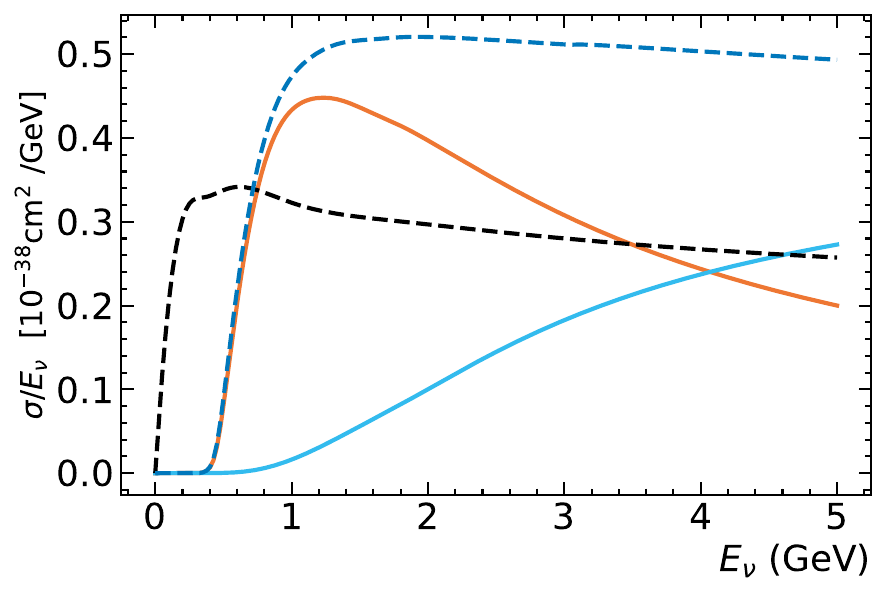}\\
        \caption{\numu--proton}
  \end{subfigure}
  \begin{subfigure}[b]{\linewidth}
    \centering
    \includegraphics[width=0.8\linewidth]{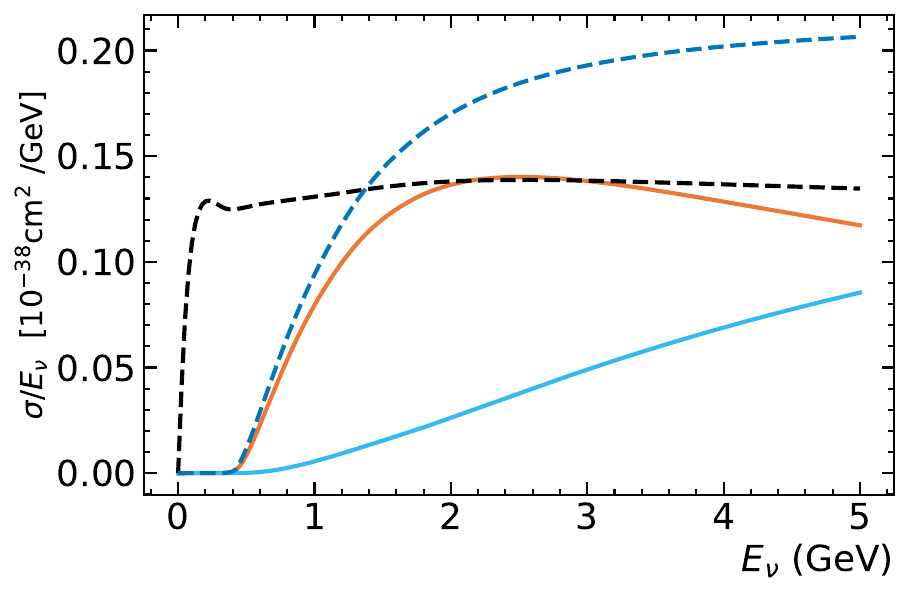}\\
    \caption{\numub--neutron}
  \end{subfigure}
  \begin{subfigure}[b]{\linewidth}
    \centering
    \includegraphics[width=0.8\linewidth]{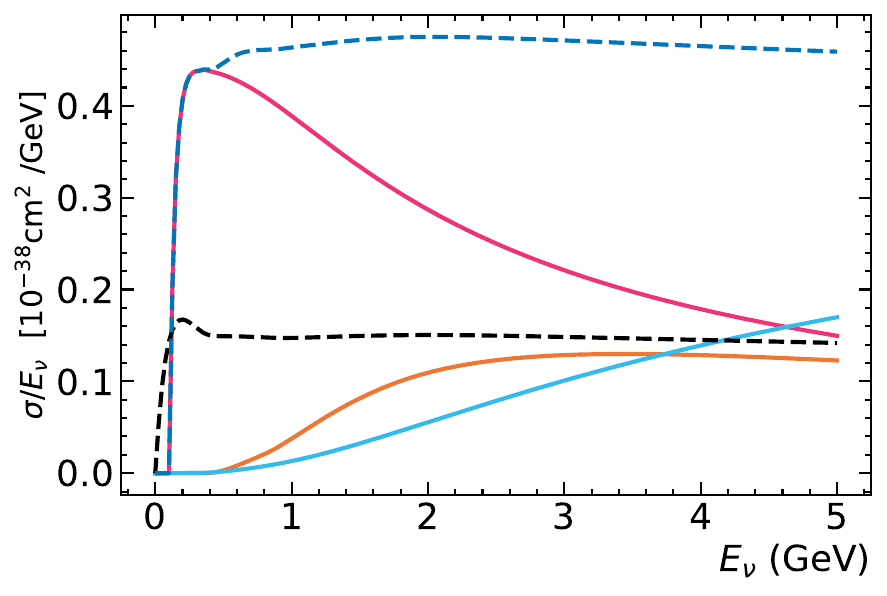}\\
        \caption{\numub--proton}
  \end{subfigure}
  \caption{The evolution of the muon neutrino and antineutrino interaction cross sections on free protons and neutrons for the different interaction channels as a function of neutrino energy as calculated with GENIE~\cite{Andreopoulos:2009rq, Andreopoulos:2015wxa}. The y-axis is presented as the cross section divided by the neutrino energy to make the cross-section evolution legible across the entire energy range shown.}
  \label{fig:sigenunucleon}
\end{figure}

The probability for an (anti)neutrino to interact with a nucleon via a particular interaction channel depends on whether the nucleon is a neutron or proton and the neutrino's energy. This is illustrated in \autoref{fig:sigenunucleon}, which shows the neutrino interaction cross section as a function of neutrino energy broken down by interaction channel. CCINC and NCINC stand for the \textit{inclusive} sum of all CC and NC interaction channels respectively. CCQE and specific RPP processes are only permitted for particular target nucleons due to charge conservation---for example, an (anti)neutrino CCQE interaction on a (neutron) proton target is not possible. Predominantly due to helicity factors, the total antineutrino cross section (considering interactions on both protons and neutrons) is lower than the neutrino cross section and evolves differently with energy. At low (anti)neutrino energies, for a target (proton) neutron, CCQE interactions dominate as only low energy transfer interactions are kinematically allowed. Once the neutrino energy is larger than the range of energy transfers for which CCQE interactions are possible ($q_0\sim1$~GeV for neutrinos), the CCQE cross section \textit{saturates}, becoming constant with neutrino energy. The same concept broadly applies to RPP interactions, once the neutrino energy is large enough to permit energy transfers (and consequently $W$) which span the full range of resonances, the RPP cross section begins to saturate. The SIS+DIS cross section, once kinematically allowed, does not saturate, as larger energy transfers permit interactions with more outgoing hadrons. Once in the DIS-dominated energy regime, the cross section rises approximately linearly with (anti)neutrino energy over the energy range of interest for this work. Due to the lower mass outgoing lepton, $\nu_e^{\bracketbar}$ interactions have a lower energy threshold, leading to a slightly larger cross section with respect to $\nu_\mu^{\bracketbar}$ interactions. Similarly, NCE interactions have a massless outgoing neutrino and therefore no energy threshold, leading to a larger cross section with respect to CC interactions at low neutrino energy. However, NC interactions in general have a lower cross section due to their mediation by a Z boson rather than a W boson.

\subsection{Nuclear effects}
\label{sec:nuclear}

When a neutrino interacts with a nuclear target, rather than with a free nucleon, there are significant alterations to its interaction cross section. 
Nuclear effects can alter the dependence of cross sections on neutrino energy, flavour and nuclear target. They also alter final-state particle kinematics and multiplicities with respect to free nucleon interactions.
In this section, we give a non-exhaustive overview of nuclear effects, with a focus on those which are important to LBL neutrino oscillation experiments. 

\subsubsection{Plane wave impulse approximation (PWIA)}

When modelling the interactions described in \autoref{sec:nucleon} within a nucleus, a common simplification is the \textit{plane wave impulse approximation} (PWIA), which has two components. The \textit{impulse approximation} treats neutrino--nucleus scattering as interactions with a single nucleon within a target nucleus. Nuclear effects can cause the properties of the target nucleon to be modified with respect to those of a free nucleon, but correlations between nucleons are not treated explicitly. The \textit{plane wave} approximation neglects interactions between an outgoing nucleon and the residual nucleus\footnote{The nucleon's wavefunction is treated as a plane wave that is not distorted by the nuclear medium through which it passes.}. 

Each of these approximations is only reasonable for a limited region of kinematic phase space. For example, collective effects from the presence of many interacting nucleons in the nucleus render the impulse approximation inaccurate at low energy transfers ($\qz \lesssim100~\text{MeV}$) and the plane wave approximation is unsuitable once an experiment is sensitive to the details of outgoing hadronic particles. Fully describing neutrino interactions without PWIA is tantamount to solving an extremely complicated quantum many-body problem. However, progress has been made in applying extensions to PWIA models to consider some effective deviations from it, which are described below.

\subsubsection{Ground state modelling}
Nucleons in a nucleus move with some \textit{Fermi motion} and are confined by the nuclear potential. Some \textit{removal energy} is required to liberate nucleons from the nucleus. 
When neutrinos interact with nucleons bound inside the nucleus, this \textit{nuclear ground state} modifies the free-nucleon cross section.

\begin{figure}[hptb]
  \centering
  \includegraphics[width=0.98\linewidth]{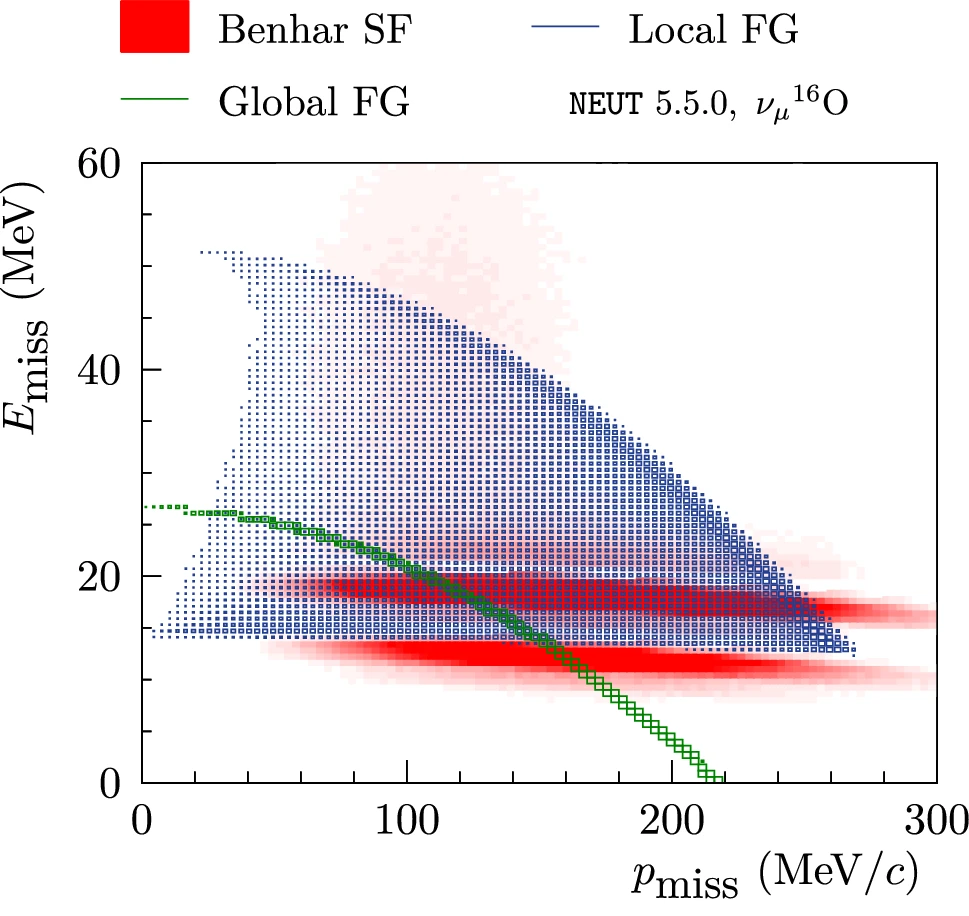}
  \caption{Nuclear spectral functions predicted by three different models in the NEUT neutrino interaction event generator, showing the removal energy ($E_{\mathrm{miss}}$) and initial state nucleon momentum ($p_{\mathrm{miss}}$) of nucleons inside an oxygen nucleus. The figure is adapted from Ref.~\cite{Hayato:2021heg}.}
  \label{fig:sfcomp}
\end{figure}
The description of the nuclear ground state differs substantially between models. \autoref{fig:sfcomp} shows the probability distribution for a nucleon to have some removal energy and initial state momentum (often called \textit{spectral functions}) for three commonly used models. Fermi Gas models describe nucleons as non-interacting fermions filling up energy states independent of position inside a global Fermi Gas (RFG)~\cite{smith1972neutrino, Bodek:1980ar} or in a radially-dependent local Fermi Gas (LFG) potential~\cite{Alvarez-Ruso:2014bla}. The latter is more realistic, accounting for the fall-off of nuclear density with radial position. The Benhar spectral function model (SF)~\cite{Benhar:1994hw, Ankowski:2014yfa} provides a significantly more sophisticated description of the nuclear ground state. It is a nuclear shell model, where the different shells can be seen as discrete peaks along the removal energy axis of \autoref{fig:sfcomp}. The two sharp peaks arise from oxygen's two p-shells and the diffuse, higher removal energy, distribution comes from oxygen's s-shell. Alternative SF models have also been produced by other authors~\cite{Ankowski:2005wi, Butkevich:2009cp}. Ground state models also often include some effective treatment for \textit{short-range correlations} (SRCs), accounting for correlated pairs of nucleons within the nucleus, which gives the model a high initial state nucleon momentum tail (evident in the Benhar SF model with respect to the Fermi Gas models in \autoref{fig:sfcomp})\cite{Bodek:1980ar,Benhar:1994hw}. Within the context of PWIA, the Fermi motion and removal energy described by a spectral function encompasses all nuclear effects. 

Nuclear ground state models have been tuned to electron--nucleus scattering measurements as a function of outgoing electron energy at fixed incoming electron energy and scattering angle~\cite{smith1972neutrino, Bodek:1980ar, JLabE91013:2003gdp}. Very generally, the SF and LFG models are able to describe electron--nucleus scattering well at energy transfers $\qz \gtrsim100~\text{MeV}$. At low energy transfers, where PWIA breaks down, these models do not describe the measurements without consideration of additional nuclear effects.

\subsubsection{Pauli blocking}
Named after the Pauli exclusion principle, Pauli blocking prevents interactions in which the outgoing nucleon would be in a quantum state that is already occupied by a nucleon within the struck nucleus. 
It thereby represents a deviation from the plane wave approximation and is accounted for with varying degrees of sophistication in models of neutrino--nucleus scattering (see the discussion in, for example, Ref.~\cite{Ankowski:2014yfa}). 
Pauli blocking is most significant when the outgoing nucleon momentum is comparable to the momentum scale of the Fermi motion of nucleons within the nucleus ($p_{n}\sim200~\text{MeV}/c$), and is most relevant for interactions with lower momentum transfers. 
Altering the treatment of Pauli blocking can change the CCQE cross section by tens of percent in some regions of kinematic phase space relevant to accelerator-neutrino experiments~\cite{Ankowski:2014yfa, Chakrani:2023htw}.
Its impact is also not always negligible at higher energies. SIS and DIS interactions generally create numerous outgoing low-momentum nucleons which may each be \textit{Pauli blocked}. Additionally, other nuclear effects may cause low momentum nucleons to be produced, even in high-energy transfer interactions, which can be Pauli blocked.

\subsubsection{Long-range correlations}
Consideration of long-range correlations between nucleons within a target nucleus, representing a deviation from the impulse approximation, can significantly reduce and reshape the neutrino interaction cross section at low energy transfers ($\qz\lesssim 100~\text{MeV}$). 
\textit{Ab initio} methods~\cite{Lovato:2023raf,Lovato:2020kba} can be employed to calculate the impact of many-body nuclear correlations on cross sections as a function of lepton kinematics. Alternatively, the random phase approximation (RPA) is often used to calculate the impact on both lepton and hadron kinematics. RPA modifies the bare weak propagator to include interactions with particle-hole pairs within the nucleus~\cite{Martini:2009uj,Nieves:2011pp,Jachowicz:2002rr,Pandey:2014tza}. 
Widely used implementations of RPA include the Valencia models~\cite{Nieves:2011pp}, the \textit{continuum random phase approximation} model (CRPA) from the Ghent group~\cite{Jachowicz:2002rr,Pandey:2014tza} and the model from Martini and collaborators~\cite{Martini:2009uj}. 
The details of what is included differ substantially between models (for example, whether $\Delta$ baryon excitations are considered) and some comparisons between them can be found in Refs.~\cite{Martini:2016eec, Dolan:2021rdd}.
The impact of RPA corrections is generally only considered relevant for CCQE interactions with low energy transfers $\qz\lesssim100~\text{MeV}$, but the effect within this region can be so large that change in the impact on the total CCQE cross section can be substantial, $\sim$10\% for a neutrino energy of 1~GeV.

At energy transfers below the threshold for single nucleon knock-out ($\qz\lesssim50~\text{MeV}$), PWIA completely breaks down. Collective excitations of large parts of the nucleus, sometimes referred to as \textit{giant resonances}, cause discrete spikes in the cross section as a function of energy transfer, as is clearly observed in electron--nucleus scattering measurements~\cite{Pandey:2014tza,Jachowicz:2021ieb}. 
Their exact strength and shape depends on both ground state and RPA treatments. Whilst broad-band neutrino beams smear out the impact of these low-energy excitations, their collective impact on the cross section could be relevant for interactions at low energy transfer. 

\subsubsection{Final-state interactions (FSI)}
\label{subsec:FSI}

Outgoing hadrons produced in neutrino--nucleus interactions need to propagate out of the dense nucleus environment before they are observable (and are \textit{final state} particles). As they propagate, they experience the nuclear potential and may interact in a variety of ways, breaking the assumptions of the plane wave approximation. These \textit{final-state interactions (FSI)} can modify the number and type of hadrons which escape the nucleus, as well as change their four-momenta with respect to the plane wave case, obscuring the initial interaction process.
A complete description of FSI requires detailed consideration of a quantum many-body problem which is currently computationally intractable, so data-driven methods are usually employed.

A common approach to simulating FSI is to use semi-classical intranuclear cascade models, some of which are reviewed in Ref.~\cite{Dytman:2021ohr}. In these models, the outgoing hadrons are individually stepped through the nucleus until they reach some radius where they are considered to have escaped the nuclear potential. 
At each step, the probability of different types of interaction are evaluated as a function of the outgoing hadron's energy. Numerical Monte Carlo methods are used to determine whether or not an interaction occurs and if so which interaction happens. These interactions include elastic and inelastic scattering, charge-exchange, pion production and pion absorption. Such intra-nuclear scattering of hadrons is related to hadron--nucleus scattering, which can be measured experimentally and used to tune or validate FSI models~\cite{PinzonGuerra:2018rju}. However, the relationship between the two is complex~\cite{Salcedo:1987md, Oset:1987re, Golan:2012wx}. This is partly because hadrons, unlike neutrinos, interact via the strong nuclear force, so they preferentially scatter from nucleons close to the surface of the nucleus. Electron--nucleus scattering data can be used to tune integrated properties of FSI, such as the probability for a hadron to escape without re-interacting (nuclear transparency)~\cite{Dytman:2021ohr,Niewczas:2019fro}.
Photon--nucleus scattering data can also be used to probe FSI in a similar way to electron-scattering data, with the advantage that the photon deposits all of its energy in the nucleus~\cite{Krusche:2004uw, Krusche:2004zc, Krusche:2011we, Buss:2011mx, Mosel:2016cwa}.

More sophisticated models such as INCL~\cite{boudard2013new,mancusi2014extension,rodriguez2017improvement,hirtz2018parametrization} and GiBUU~\cite{Buss:2011mx} propagate an ensemble of hadrons within a nucleus in steps of time, rather than single hadrons in steps of space as considered in the cascade models discussed thus far, stopping after a preset number of steps.
INCL includes additional scattering channels such as nuclear cluster production (for example, simulating the emission of alpha particles from FSI). 
GiBUU~\cite{Buss:2011mx} utilises quantum-kinetic transport theory to simulate the propagation of outgoing hadrons through the nucleus. The FSI model in GiBUU follows the evolution of each hadron’s phase space density in time while applying the same mean-field potential used for the nuclear ground state in cross-section calculations. 

Neutrino interactions almost always leave the residual nucleus in an excited state, and its subsequent de-excitation can carry away a non-negligible fraction of an interaction's energy transfer. Until recently, nuclear de-excitation has been neglected in most interaction simulation packages used by oscillation experiments, largely because the resulting low-energy products are difficult to observe in detectors. However, this process is considered within INCL, which passes the final nuclear state to the ABLA nuclear de-excitation routine~\cite{Kelic:2009yg,Ershova:2023dbv}. A detailed evaluation of INCL+ABLA compared to a semiclassical cascade FSI model (from the NuWro package~\cite{Golan2012nuwro}) is provided in Refs.~\cite{ershova2022study,Ershova:2023dbv}. 

All of the approaches to FSI modelling described so far factorise the treatment of the outgoing hadronic final state from the total cross section, and therefore leave the cross section as a function of the primary outgoing lepton kinematics untouched.
Some neutrino interaction theory calculations take an alternative approach, and do not factorise FSI from the initial interaction. These treat FSI as a distortion of the outgoing nucleon wave function by the nuclear potential, in a consistent quantum mechanical treatment (see, for example, Refs.~\cite{gonzalez2019nuclear,Nikolakopoulos:2022qkq,Ankowski:2014yfa}). In these models, the role of FSI, with respect to a plane wave approach, is to add a potential term when solving the wave equation used to determine the final-state nucleon wave function.
Such approaches change the cross section as a function of both lepton and hadron kinematics, typically substantially reducing the cross section at low energy transfers with respect to the plane wave case~\cite{Franco-Patino:2022tvv,Franco-Patino:2023msk,Nikolakopoulos:2024mjj,Dolan:2021rdd}.

The distorted wave function approach offers a more consistent calculation, but can only describe the impact of FSI on a particular final state (for example, a single proton final state). It is able to account for migrations to other final states, but does not provide explicit details about them.
Comparisons have been made between calculations using a distorted wave function and those using cascade models for interactions which leave a single nucleon and no other hadrons in the final state~\cite{Nikolakopoulos:2022qkq,Nikolakopoulos:2023pdw,Franco-Patino:2022tvv,Franco-Patino:2023msk,Dolan:2019bxf,McKean:2025khb}. These show that the two approaches are only consistent when the outgoing nucleon's kinetic energies are above 100 MeV.
As a consequence, the authors of Ref.~\cite{Nikolakopoulos:2022qkq} emphasise the need for caution in applying cascade models to interactions in which the outgoing nucleon energy is small.

\subsubsection{Meson-exchange currents and two-particle two-hole (2p2h) interactions}
Neutrino--\textit{nucleus} scattering introduces interaction channels with no free-nucleon analogue where, instead of interacting with a single nucleon, neutrinos interact directly with bound states of two or more nucleons. Such interactions are a direct departure from the impulse approximation. The dominant contributions are from \textit{2p2h} (two-particle two-hole) interactions, in which two outgoing nucleons can be produced from a variety of Feynman diagrams where the weak-propagator interacts at different points with two nucleons bound via a \textit{meson-exchange current}. The interactions between nucleons are mostly mediated by pions, with smaller contributions from $\rho$ and other heavier mesons. The Feynman diagrams include important contributions from interactions in which nucleons are promoted to virtual $\Delta$-resonances.

The most common 2p2h models used when simulating few-GeV neutrino interactions are the Valencia~\cite{Nieves:2011pp,Sobczyk:2020dkn,Sobczyk:2024ecl} and SuSA-MEC~\cite{Gonzalez-Jimenez:2014eqa,Megias:2016fjk,RuizSimo:2016rtu,RuizSimo:2016ikw} models\footnote{Implementations of the Valencia model in simulation packages have not, at the time of writing, incorporated the latest model updates\cite{Sobczyk:2024ecl}.}. Additionally, the model from Martini et al.~\cite{Marteau:1999kt, Martini:2009uj} has previously been used to gauge plausible variation in 2p2h modelling~\cite{T2K:2023smv}. Although based on similar underlying calculations, they differ in their treatment of relativistic approximations, the exact Feynman diagrams considered (including whether heavier meson propagators beyond the pion are included) and the treatment of the virtual $\Delta$ baryon (see Ref.~\cite{Megias:2018ujz} for some discussion of this issue). 
Some models, despite being referred to as 2p2h, also include a subset of possible 3p3h diagrams.
These details lead to significant changes to the cross sections, impacting the momentum transfer distributions, normalisations (which differ by more than 50\%), and the possible nucleon--nucleon final states~\cite{Russo:2025oph, Dolan:2018sbb, Dolan:2019bxf,Martini:2009uj}. These variations are larger for neutrinos than antineutrinos.

Most 2p2h calculations only provide cross section calculations for the outgoing lepton kinematics and the relative fraction of different nucleon--nucleon final states\footnote{Proton--neutron or neutron--neutron (proton--proton) for (anti)neutrino interactions, with the latter being substantially larger than the former.}. The calculations tend not to be able to derive cross sections as a function of the outgoing nucleon kinematics, leading to effective approaches being used to produce full event predictions in neutrino event generators~\cite{Russo:2025oph,Dolan:2019bxf, Schwehr:2016pvn}. However, state-of-the-art theory has recently provided the first 2p2h cross section calculations as a function of the outgoing nucleon kinematics~\cite{Martinez-Consentino:2023hcx,Sobczyk:2020dkn,Sobczyk:2024ecl}.

Compelling evidence for 2p2h interactions has been provided by electron--nucleus scattering measurements~\cite{Megias:2016lke}, and they are recognised as central for neutrino--nucleus interaction modelling after providing a convincing resolution of the MiniBooNE \textit{axial-mass puzzle}~\cite{Nieves:2011yp, Martini:2009uj, Martini:2010ex}. Additional 2p2h channels that also lead to real pion production may also be relevant, but are generally neglected. Similarly, higher-order contributions (starting with 3p3h) have been calculated to reveal a small but non-negligible contribution to the cross section~\cite{Martini:2009uj,Sobczyk:2020dkn,Sobczyk:2024ecl}.

\subsubsection{Coherent pion production}
Coherent pion production (CPP) describes the production of a pion through a coherent interaction with the entire nucleus in an analogous way to diffractive neutrino--nucleon pion production, but with a significantly larger cross section at few-GeV energies.
CPP interactions produce a lepton and pion final state without any nucleons, alongside a recoiling nucleus; a $l^-\pi^+$ for $\nu_l$ CCCPP; and a $\nu_l \pi^0$ for $\nu_l$ NCCPP. Since the interaction is with the nucleus as a whole, CPP interactions are not modified by Pauli blocking or pion FSI.
The lepton and pion are likely to be nearly collinear and forward-going, and the energy transfer to the recoiling nucleus is very small.
The CPP cross section is small compared to other interactions creating a $1l 1\pi$ final state, but is a significant contribution in the low-\qq region.
At few-GeV energies, CCCPP contributes $\sim$10--15\% of the  $\qq < 0.1~\textrm{GeV}^2$ CC single pion production cross section.
Theoretical models for CPP~\cite{Rein:1982pf, Kelkar:1996iv, Paschos:2005km, Alvarez-Ruso:2007rcg, Berger:2008xs, Amaro:2008hd, Paschos:2009ag, Sogarwal:2022xle} either use Adler's Partially Conserved Axial Current (PCAC) theorem~\cite{Piketty:1970sq,Adler:1964yx} and pion--nucleus scattering data, or the microscopic sum of resonances over all nucleons in the nucleus.

Other coherent neutrino--nucleus interactions are possible. Elastic coherent interactions are an important field of study~\cite{Abdullah:2022zue}, but can be neglected at few-GeV energies. 
Additionally, models for coherent interactions that produce a kaon instead of a pion have been developed~\cite{Alvarez-Ruso:2012kmi}.

\subsubsection{Additional nuclear effects at high energy transfer}
Although nuclear effects have the largest impact at interactions when the energy transfer is comparable to nuclear energy scales ($\mathcal{O}$(100) MeV), there are other substantial nucleus-dependent effects to consider for neutrino scattering at higher energies. 
Notably, the DIS interaction is modified by processes such as the \textit{EMC} effect~\cite{Malace:2014uea}, \textit{(anti-)shadowing}~\cite{Piller:1995kh,Kopeliovich:2012kw,Armesto:2006ph}, and Fermi motion~\cite{Bodek:1980ar,Arneodo:1992wf}. 
Analyses using both neutrino-- and lepton--nucleus DIS scattering to extract parton distribution functions (PDFs) show tension between the two~\cite{Kovarik:2010uv,Muzakka:2022wey}, especially at $x<0.1$, where nuclear shadowing effects may occur. However, it is not clear if this is an effect from physics or erroneous data~\cite{Paukkunen:2013grz,Eskola:2016oht}.
To resolve this issue, model-independent explorations of DIS using neutrino beams have been suggested, and are explored in detail in Refs.~\cite{Morfin:2021ujm,SajjadAthar:2020nvy}.
For SIS and DIS interactions, a phenomenological approach is sometimes used to parametrise modifications to PDFs in which global QCD data are fitted to determine either nuclear modification factors to free proton PDFs or effective nuclear PDFs~\cite{Kovarik:2015cma, NuSTEC:2017hzk}. 
Recent studies of incorporating a \textit{common} nuclear effect, short-range correlated pairs (SRCs), into nuclear parton distribution functions show promising results~\cite{nCTEQ:2023cpo}.
Modifications with respect to the free nucleon case are kinematic, energy, and nucleus dependent, and can exceed 10\% in regions of phase space relevant for neutrino oscillation experiments~\cite{Malace:2014uea,Peroni:1992hc}.

\subsection{Simulating neutrino--nucleus interactions with event generators and associated challenges}
\label{sec:evgen}
Neutrino \textit{event generators} are essential simulation tools for few-GeV neutrino experiments. In the context of neutrino oscillation measurements, generators are required to simulate all aspects of neutrino interactions that experiments are sensitive to. Crucially, they are used to predict the outcome of experimental event selections and the mapping of observable detector quantities to the high-level physics of interest. This requires simulating the aforementioned neutrino interaction channels over a wide range of neutrino energies (see \autoref{fig:exptfluxes}) and nuclear targets, including relevant nuclear effects and providing the full set of outgoing particle kinematics for each interaction, or \textit{event}. 

\begin{figure}[tbp]
  \centering
  \includegraphics[width=0.98\linewidth]{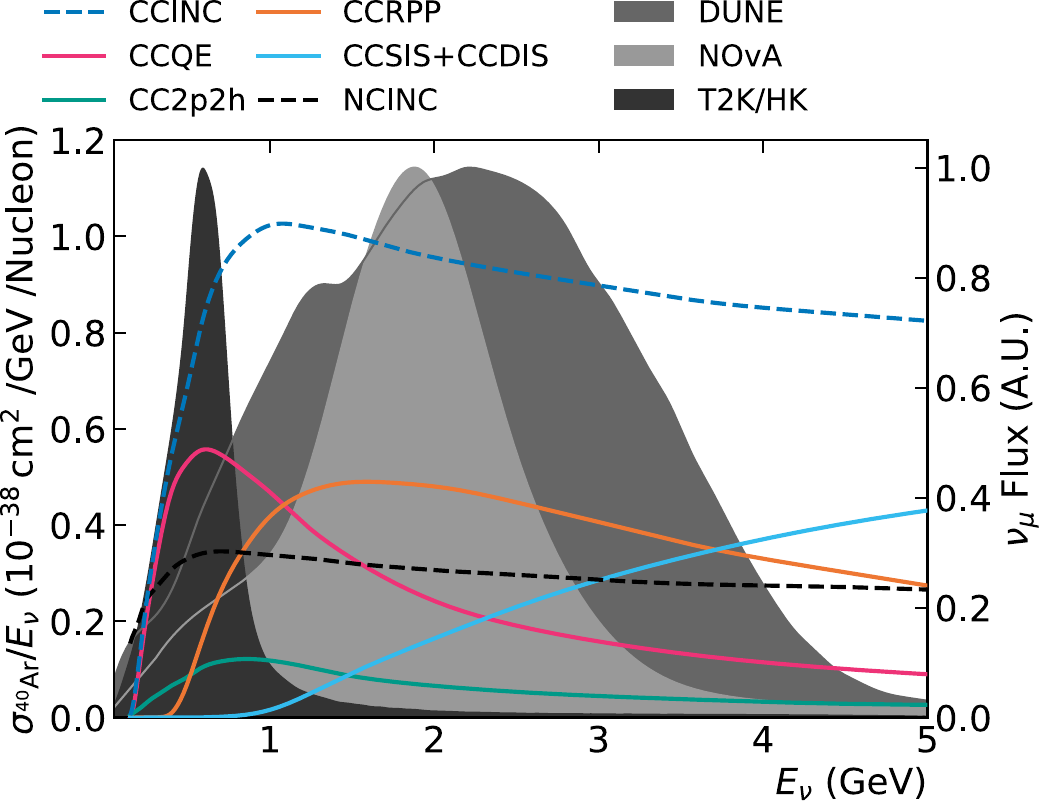}
  \caption{The \numu--\argon interaction cross section as a function of neutrino energy, divided by the neutrino energy, for various channels as predicted by the GENIE event generator in the 10a configuration. The shape of the neutrino fluxes for currently running and planned LBL neutrino oscillation and neutrino interaction experiments are overlaid.}
\label{fig:sigenu}
\end{figure}

A prediction of the total cross section of muon-neutrino interactions on an argon (\argon) nucleus target simulated using one configuration of the GENIE event generator~\cite{Andreopoulos:2009rq, Andreopoulos:2015wxa} is shown in \autoref{fig:sigenu}. To produce this, GENIE uses a subset of the models described in \autoref{sec:nucleon} and \autoref{sec:nuclear}, but requires some approximations to connect them (forming GENIE's \textit{10a} configuration~\cite{GENIE:2021zuu}). \autoref{fig:sigenu} also shows the different neutrino fluxes used by neutrino oscillation experiments, illustrating how the broad-band neutrino beams they use span regions dominated by a variety of interaction channels. This is shown quantitatively in \autoref{tab:channels} and demonstrates the requirement on generators to model neutrino interactions from a range of channels.

\begin{table}[htbp]
  \centering
  {\renewcommand{\arraystretch}{1.2}
  \begin{tabular}{c|c|c|c|c}
\hline\hline
\multicolumn{2}{c|}{Channel} & Hyper-K/T2K & NOvA & DUNE \\ \hline
\multirow{4}{*}{$\nu_\mu$} & CCQE            & 42.0\%        & 17.2\%       & 19.6\%  \\ 
& CC2p2h          & 8.3\%        & 4.6\%       & 5.3\%  \\ 
& CCRPP           & 35.1\%        & 41.6\%       & 40.1\%  \\
& CCSIS/DIS       & 13.5\%        & 34.7\%      & 33.3\%  \\ \hline
\multirow{4}{*}{$\bar{\nu}_\mu$} & CCQE            & 54.6\%        & 31.3\%       & 23.9\%  \\ 
& CC2p2h          & 11.2\%        & 7.1\%       & 8.6\%  \\ 
& CCRPP           & 25.9\%        & 38.5\%       & 44.5\%  \\
& CCSIS/DIS       & 6.2\%        & 20.0\%      & 21.1\%  \\ \hline\hline
\end{tabular}
}
\caption{The proportion of muon-(anti)neutrino CC interactions broken down by interaction channel, as predicted by the GENIE event generator in the 10a configuration, for simulations using the neutrino flux at the Hyper-K/T2K, NOvA and DUNE NDs. The simulated target materials are C$_8$H$_8$ for Hyper-K/T2K, CH$_2$ for NOvA and \argon for DUNE. The columns do not sum to 100\% since small contributions, such as CPP or heavy meson production, are not included.}
\label{tab:channels}
\end{table}

\begin{figure}[htbp]
  \centering
  \begin{subfigure}[b]{\linewidth}
    \centering
    \includegraphics[width=0.8\linewidth]{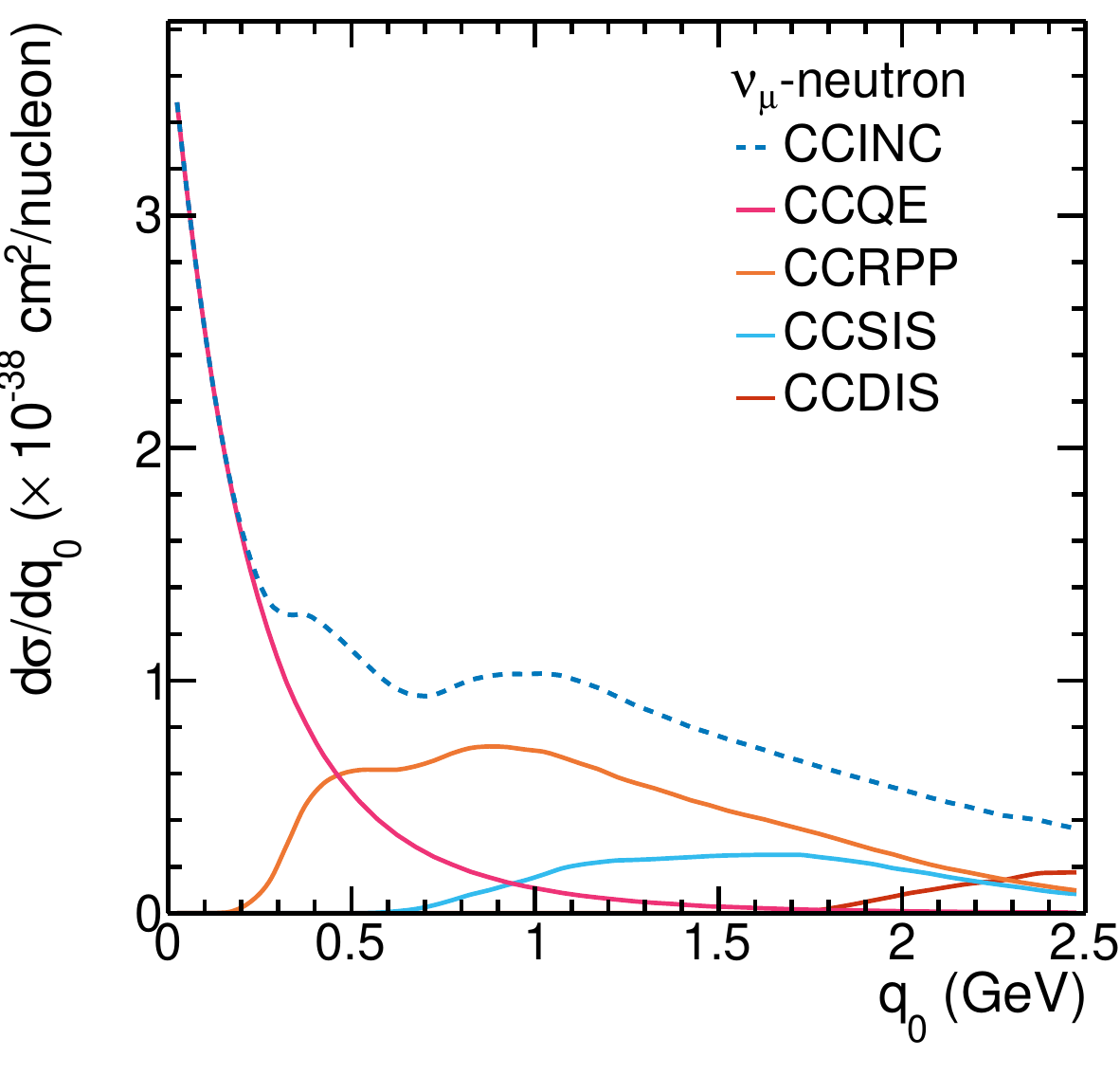}\\
    \caption{\numu--neutron}
  \end{subfigure}
  \begin{subfigure}[b]{\linewidth}
    \centering
    \includegraphics[width=0.8\linewidth]{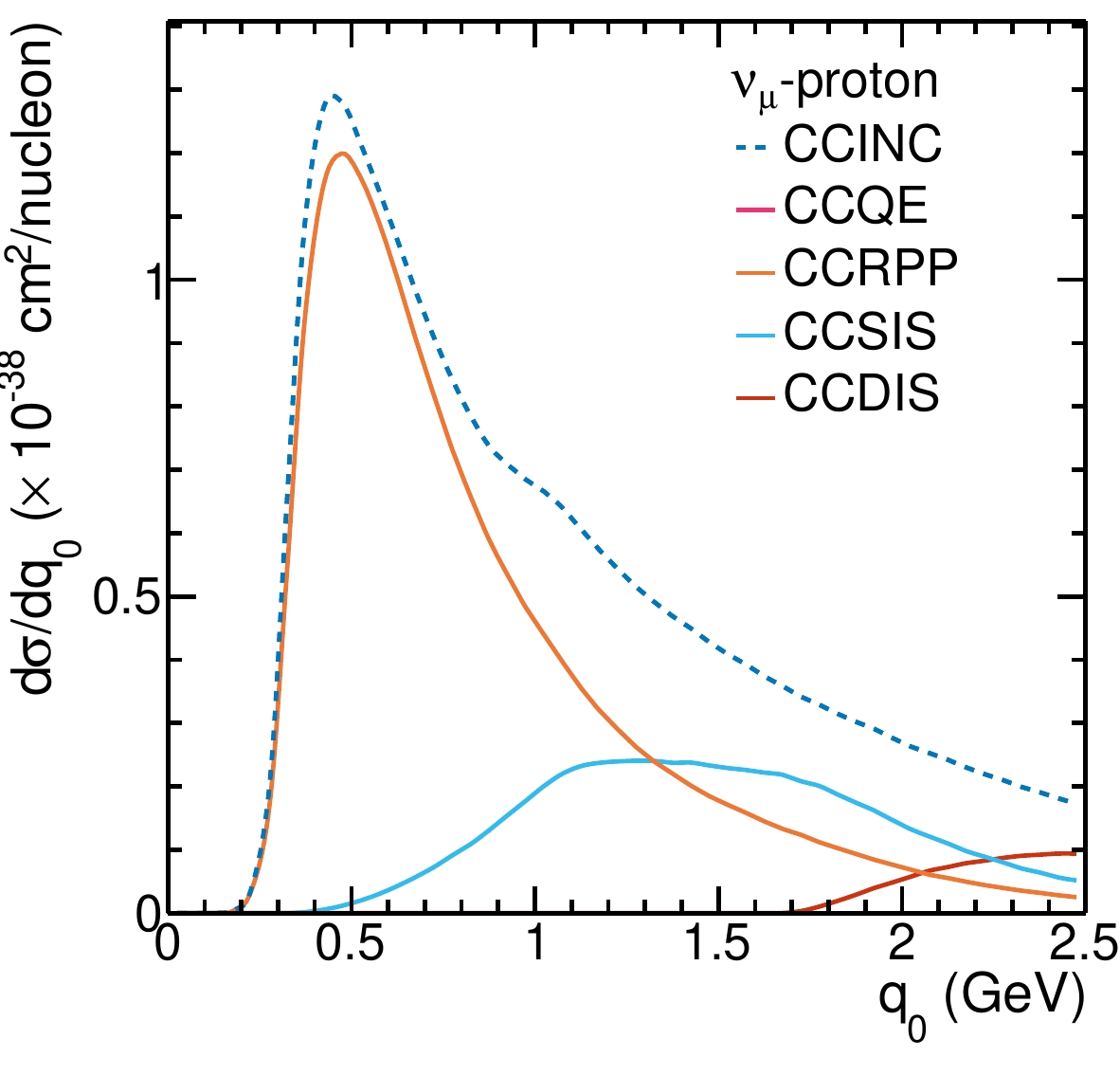}\\
        \caption{\numu--proton}
  \end{subfigure}
  \begin{subfigure}[b]{\linewidth}
    \centering
    \includegraphics[width=0.8\linewidth]{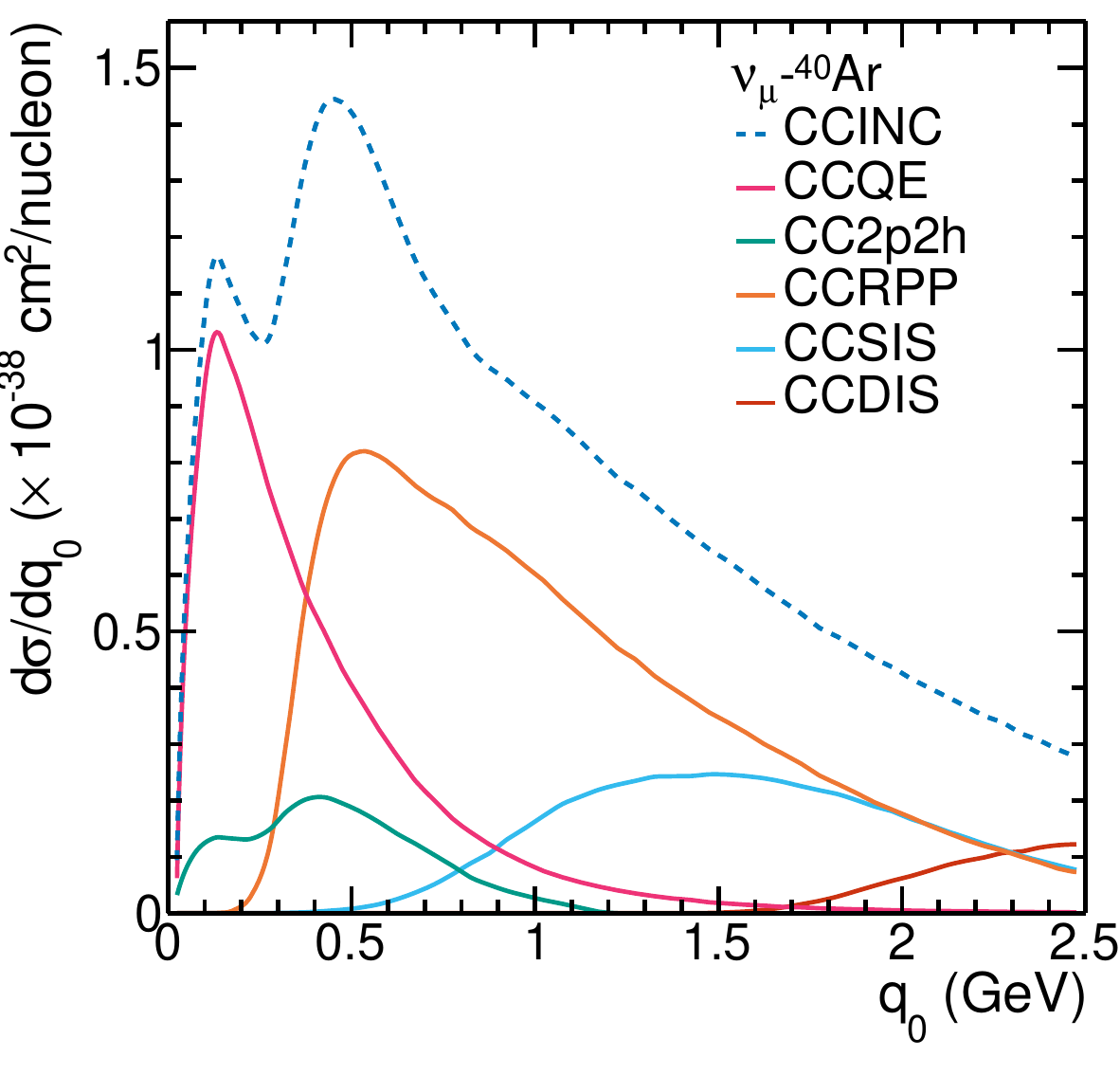}
    \caption{\numu--\argon}
    \end{subfigure}
  \caption{The evolution of the cross section for various CC channels as a function of energy transfer, \qz, shown for muon neutrinos interacting on various targets, all using the nominal DUNE flux~\cite{DUNE:2020jqi,DUNE:2021mtg} and the GENIE 10a generator. Small contributions from diffractive and coherent scattering are included in the CCINC distribution, but are not shown separately.}
  \label{fig:q0_breakdown}
\end{figure}

Beyond the total cross section, generators need to predict other interaction-level and outgoing particle kinematic distributions. An example is shown in \autoref{fig:q0_breakdown}, which compares GENIE predictions for the differential cross section as a function of energy transfer for muon neutrino interactions on free protons or neutrons to interactions on argon nuclei for a selection of CC interaction channels, averaged over the incoming neutrino energy distribution from the DUNE experiment~\cite{DUNE:2020jqi,DUNE:2021mtg} (the DUNE \textit{flux-averaged cross section}). 

\autoref{fig:q0_breakdown} serves the dual purpose of illustrating a complete generator prediction and, in the context of one particular collection of models, summarising the content of \autoref{sec:nucleon} and \autoref{sec:nuclear}. For neutrino interactions with neutrons or argon nuclei the dominance and drop off of CCQE for $\qq \leq 400$ MeV$^2$ is clear, followed by a broad range of energy transfer dominated by CCRPP interactions, until $\qq\gtrsim$2~GeV$^2$, leading to a CCSIS and CCDIS dominated region. The wide range and multi-peaked structure of RPP interactions seen for the neutrino--neutron case follows from the contributions of multiple resonances which exist in the \textit{mixed-isospin} channel. For \numu--proton interactions, CCQE scattering is not possible, and the CCRPP cross section is strongly dominated by $\Delta(1232)$ production, as this is a pure 3/2 isospin channel. 
The \numu--\argon cross section has contributions from both neutron and proton channels and demonstrates how nuclear effects change the interaction cross section considerably from a na\"ive sum of the two. 
Most obviously, there are contributions from CC2p2h, which has no equivalent in the free-nucleon case\footnote{CCCPP is not shown explicitly, but is an additional small contribution to the CCINC cross section as the energy transfer approaches zero.}.
The CCQE distribution has also changed considerably, largely due to effects from removal energy and Pauli blocking, removing much of the very low-energy transfer cross section, whilst RPA provides an additional suppression. The distributions of energy transfer for CCQE and CCRPP interactions are also broadened with respect to the free nucleon case---which would be more apparent were the figures restricted to a fixed outgoing lepton angle.

Other than GENIE, commonly used neutrino event generators include NEUT~\cite{Hayato:2002sd,Hayato:2009,Hayato:2021heg}, NuWro~\cite{Golan:2012wx,Golan2012nuwro} and GiBUU~\cite{Buss:2011mx}, which are described in \autoref{appen:gen}\footnote{Beyond those described in \autoref{appen:gen}, other neutrino event generators include ACHILLES~\cite{Isaacson:2022cwh}, FLUKA~\cite{Ferrari:2005zk,Bohlen:2014buj} and NUANCE~\cite{Casper:2002sd}.}. Each generator faces the formidable challenge of assembling disparate theory calculations into a complete simulation, requiring choices about which models to use, how to connect them and how to deal with cases when the models are not able to predict the full set of outgoing particle kinematics. Since there are a wide variety of models and numerous ways of connecting them, different plausible choices would lead to significantly different versions of \autoref{fig:sigenu} and \autoref{fig:q0_breakdown}. A consequence of this challenge is that, within each of these generators, different configuration options exist which use different sets of models for cross-section calculations and nuclear effects.

An example challenge in connecting models can be found in the SIS transition region between nucleon-level RPP interactions and quark-level DIS interactions (the region $1\lesssim \qz \lesssim 2$ GeV in \autoref{fig:q0_breakdown}). Since models usually lack a reliable description of this region, generators typically implement some \textit{ad hoc} blending or discrete cut off between their RPP, SIS and DIS models as a function of the hadronic invariant mass of the interaction. 

Generators face further difficulties of consistency when combining models of some interaction channels with models of other nuclear effects with models for the neutrino--nucleon interaction channel. For example, models of SRCs (usually applied as a nuclear modification to CCQE interactions) and 2p2h, both produce multi-nucleon final states and overlap in the physics they cover. This results in potential double counting if a generator uses calculations from different underlying theories. Furthermore, the treatment of nuclear effects is not trivial to keep consistent between interaction channels due to differences in available theory inputs. For example, generators may use different nuclear ground states or Pauli blocking treatments for different channels.

Additional approximations arise since, due to both theoretical and computational limitations, the input cross-section calculations to event generators are often \textit{inclusive} (they can only predict the outgoing lepton's kinematics) or \textit{semi-inclusive} (they can also predict \emph{some} of the outgoing hadrons' kinematics), but almost never \textit{fully exclusive} (they can predict \emph{all} of the outgoing particles' kinematics). 
This applies, for example, to many 2p2h and RPP models used in generators, which are often only capable of inclusive calculations. For these interactions, most generators introduce several approximations, that are not directly predicted by the theoretical framework of the primary interaction, to simulate the outgoing hadron kinematics. For example, in the 2p2h case, nucleon kinematics are generated from the inclusive input model with factorised models~\cite{Sobczyk:2012ms, Prasad:2024gnv}. A model which is used by many generators assumes that~\cite{Sobczyk:2012ms}: the nucleons share the four-momentum transfer evenly; their initial-state kinematics are not correlated with each other; and their initial-state kinematics are independent of the four-momentum transfer of an interaction. Some details of the approximations implicit in the approach of building exclusive predictions from semi-inclusive or inclusive input models are discussed in Refs.~\cite{Nikolakopoulos:2023pdw, Dolan:2019bxf, Dolan:2021rdd,Schwehr:2016pvn,Russo:2025oph}. 

Another important approximation inside generators is the factorisation of the primary neutrino interaction and FSI. Most cross-section calculations in event generators do not include FSI effects and so the generators pass the outgoing hadrons from the initial neutrino interaction to a separate routine to simulate their propagation through the nucleus, usually a cascade as introduced in \autoref{sec:nuclear}. The treatment of FSI therefore determines the relationship between neutrino interaction channels and observable \textit{topologies} (the particles exiting the nucleus). For example, pion absorption FSI can leave CCRPP interactions with no mesons in the final state (a \textit{CC0$\pi$} topology) and nucleon FSI can lead to CCQE interactions with multiple nucleons in the final state. Whilst cascade models of FSI alter the kinematics and particle content of the \emph{hadronic} system, they typically leave the total cross section and the lepton kinematics unchanged. As different generators use different FSI models, the mapping between the outgoing lepton kinematics and the outgoing hadronic system (kinematics and topology) within them, even when the baseline cross-section model is the same, can be very different.

This factorisation is an artificial separation of a coupled quantum mechanical process, which implies that FSIs do not impact the cross section as a function of neutrino energy or lepton kinematics. This implication is known not to hold at low energy transfers (see \autoref{sec:nuclear}). Although some models in event generators break elements of this approximation~\cite{Ankowski:2014yfa,Dolan:2019bxf, Dolan:2021rdd}, these are usually only implemented as inclusive calculations. A single model able to calculate the full exclusive one-nucleon knockout cross section, consistently accounting for the impact of FSI on the inclusive and exclusive cross sections, was recently implemented in an event generator~\cite{McKean:2025khb}. However, there are significant challenges to extending this approach to more calculations and to more interaction channels.

\subsubsection{Model spread as an uncertainty metric}

Due to all of the complications in cross-section modelling discussed, there is no comprehensive set of systematic uncertainties that cover few-GeV neutrino--nucleus cross sections. In the absence of such a comprehensive uncertainty evaluation, a useful first step can be investigating the \emph{spread} of plausible cross-section model predictions. These make different choices and impose different approximations, and so the ensemble can give a useful lower bound on the uncertainty associated with neutrino interaction modelling. Throughout this work, a variety of neutrino interaction event generator configurations are used to this end. The generators and models are selected to cover a representative range of neutrino interaction physics and/or configurations widely used by ongoing or upcoming experiments. A summary of the configurations used is presented in \autoref{tab:models}, with further details in \autoref{appen:gen}. Even when generators use similar underlying theory calculations as an input, the implementation choices can still cause them to produce different results (see Refs.~\cite{Avanzini:2021qlx,Betancourt:2018bpu}). 

An important deficiency when looking at the \textit{model spread} from an ensemble of generator configurations is that regions of agreement do not necessarily show that the physics is well understood, it could instead be that the generators make similar choices due to a lack of available options, or that generator developers tune models to a very limited body of available data. Conversely, a region of disagreement may highlight the region in which the most work has been done, and the different choices may truly be indicative of a plausible range of possible predictions. Nevertheless, the model spread approach is useful as a tool for experimentalists to explore how different a model prediction could be if a different simulation package were used in an analysis.

\begin{table*}[htbp]
\centering
\footnotesize
{\renewcommand{\arraystretch}{1.2}
\begin{tabular}{c|c|c|c|c|c|c|c}
\hline\hline
Generator & Ground                                 & QE                                        & 2p2h                                                          & RPP                                                         & SIS and                                                     & DIS and~\cite{Gluck:1998xa,Sjostrand:2006za}                                                                           & FSI~\cite{Dytman:2021ohr} \\ 
model     & state                                  &                                           &                                                               &                                                             & hadronisation                                                                     & hadronisation          &  \\ \hline
NEUT        & LFG/RFG                              & \makecell{Valencia\\\cite{Nieves:2011pp}} & \makecell{Valencia   \\\cite{Nieves:2011pp}}                  & \makecell{Berger--Sehgal\\\cite{Berger:2007rq}}              & \makecell{BY-Corr.~\cite{Bodek:2003wc} \\ $W>1.3$ GeV: Ref~\cite{NuSTEC:2020nsl}} & \makecell{GRV98 \\ $W>2.0$ GeV: Pyth.5}                    & NEUT Casc.      \\ \hline
NEUT DCC    & LFG/RFG                              & \makecell{Valencia\\\cite{Nieves:2011pp}} & \makecell{Valencia\\\cite{Nieves:2011pp}}                     & \makecell{DCC\\\cite{Nakamura:2015rta}}                     & \makecell{BY-Corr.~\cite{Bodek:2003wc} \\ $W>1.3$ GeV: Ref~\cite{NuSTEC:2020nsl}} & \makecell{GRV98 \\ $W>2.0$ GeV: Pyth.5}                    & NEUT Casc.      \\ \hline
NuWro 19    & LFG                                  & Ref~\cite{Graczyk:2003ru}                 & \makecell{Valencia\\\cite{Nieves:2011pp}}                     & \makecell{Adler\\\cite{Adler:1975mt}}                       & \makecell{Use Adler\cite{Adler:1975mt} for \\ $W<1.6$ GeV}                                            & \makecell{GRV98 \\ $W>1.4$ GeV: Pyth.6}                    & NuWro Casc.      \\ \hline
NuWro 25    & LFG                                  & Ref~\cite{Graczyk:2003ru}                 & \makecell{Valencia 20\\\cite{Sobczyk:2020dkn,Prasad:2024gnv}} & \makecell{Hybrid\\\cite{Yan:2024kkg}}                       & \makecell{Use Hybrid\cite{Yan:2024kkg} for \\ $W<1.6$ GeV}                                            & \makecell{GRV98 \\ $W>1.6$ GeV: Pyth.6}                    & NuWro Casc.      \\ \hline
GiBUU       & LFG~\cite{Leitner:2008ue}            & Ref~\cite{Leitner:2008ue}                 & Ref~\cite{Gallmeister:2016dnq}                                & \makecell{Refs.\\\cite{Lalakulich:2012cj,Lalakulich:2010ss}}& \makecell{Ref.~\cite{Mosel:2023zek}    \\ $W>2.0$ GeV: Pyth.6}                  & \makecell{GRV98 \\ $W>3.0$ GeV: Pyth.6}                    & Ref.~\cite{Buss:2011mx} \\ \hline
GENIE 10a   & \makecell{LFG\\fixed $E_{\mathrm{rmv}}$} & \makecell{Valencia\\\cite{Nieves:2011pp}} & \makecell{Valencia\\\cite{Nieves:2011pp}}                 & \makecell{Berger--Sehgal\\\cite{Berger:2007rq}}              & \makecell{BY-Corr.~\cite{Bodek:2003wc} \\ $W>2.3$ GeV: Ref~\cite{Yang:2009zx}}   & \makecell{GRV98 \\ $W>3.0$ GeV: Pyth.6}                    & hA \\ \hline
GENIE 10b   & \makecell{LFG\\fixed $E_{\mathrm{rmv}}$} & \makecell{Valencia\\\cite{Nieves:2011pp}} & \makecell{Valencia\\\cite{Nieves:2011pp}}                 & \makecell{Berger--Sehgal\\\cite{Berger:2007rq}}              & \makecell{BY-Corr.~\cite{Bodek:2003wc} \\ $W>2.3$ GeV: Ref~\cite{Yang:2009zx}}   & \makecell{GRV98 \\ $W>3.0$ GeV: Pyth.6}                    & hN Casc.   \\ \hline
GENIE 10c   & \makecell{LFG\\fixed $E_{\mathrm{rmv}}$} & \makecell{Valencia\\\cite{Nieves:2011pp}} & \makecell{Valencia\\\cite{Nieves:2011pp}}                 & \makecell{Berger--Sehgal\\\cite{Berger:2007rq}}              & \makecell{BY-Corr.~\cite{Bodek:2003wc} \\ $W>2.3$ GeV: Ref~\cite{Yang:2009zx}}   & \makecell{GRV98 \\ $W>3.0$ GeV: Pyth.6}                    & INCL Casc. \\ \hline
GENIE CRPA  & \makecell{LFG\\$q_3$ dep. $E_{\mathrm{rmv}}$\\~\cite{Dolan:2019bxf}}  & \makecell{CRPA\\\cite{Jachowicz:2002rr}} & \makecell{SuSAv2\\\cite{RuizSimo:2016rtu}}                   & \makecell{Berger--Sehgal\\\cite{Berger:2007rq}}              & \makecell{BY-Corr.~\cite{Bodek:2003wc} \\ $W>2.3$ GeV: Ref~\cite{Yang:2009zx}}    & \makecell{GRV98 \\ $W>3.0$ GeV: Pyth.6}                    & hN Casc.   \\ \hline\hline
\end{tabular}}
\normalsize
\caption{An overview of the event generator models used within this work. \textit{Casc.} stands for intranuclear cascade, $W$ is the invariant mass of the hadronic system, \textit{BY-Corr.} stands for Bodek--Yang corrections and \textit{Pyth.5/6} stands for PYTHIA 5 or 6 respectively. The NEUT, NuWro and GENIE hN cascades are all similar, based on the Salcedo--Oset model~\cite{Salcedo:1987md,Oset:1987re} for pion FSI, with differences in implementation details and tuning procedures. The SIS and DIS columns show the cross-section model on top with the hadronisation model below, alongside an indicative range of $W$ for which the hadronisation model is employed (some generators employ interpolation between models). The kinematic limits to determine when the cross section and hadronisation models transition from SIS to DIS vary. More details can be found in \ref{appen:gen}.}
\label{tab:models}
\end{table*}

\begin{figure*}[htbp]
  \centering
  \begin{subfigure}[b]{0.48\linewidth}
    \centering
    \includegraphics[width=\linewidth]{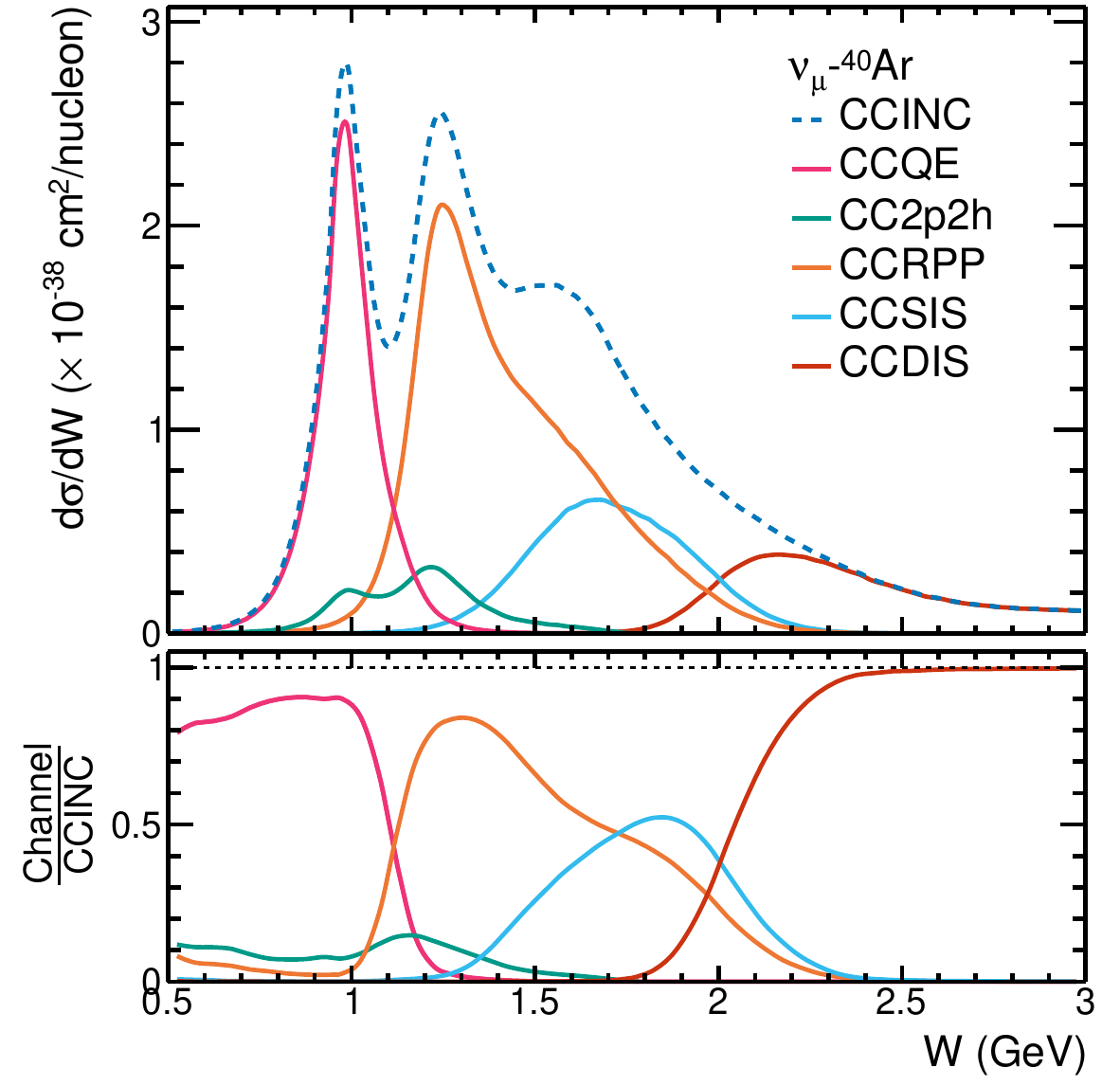}
    \caption{GENIE 10a}
  \end{subfigure}
  \begin{subfigure}[b]{0.48\linewidth}
    \centering
    \includegraphics[width=\linewidth]{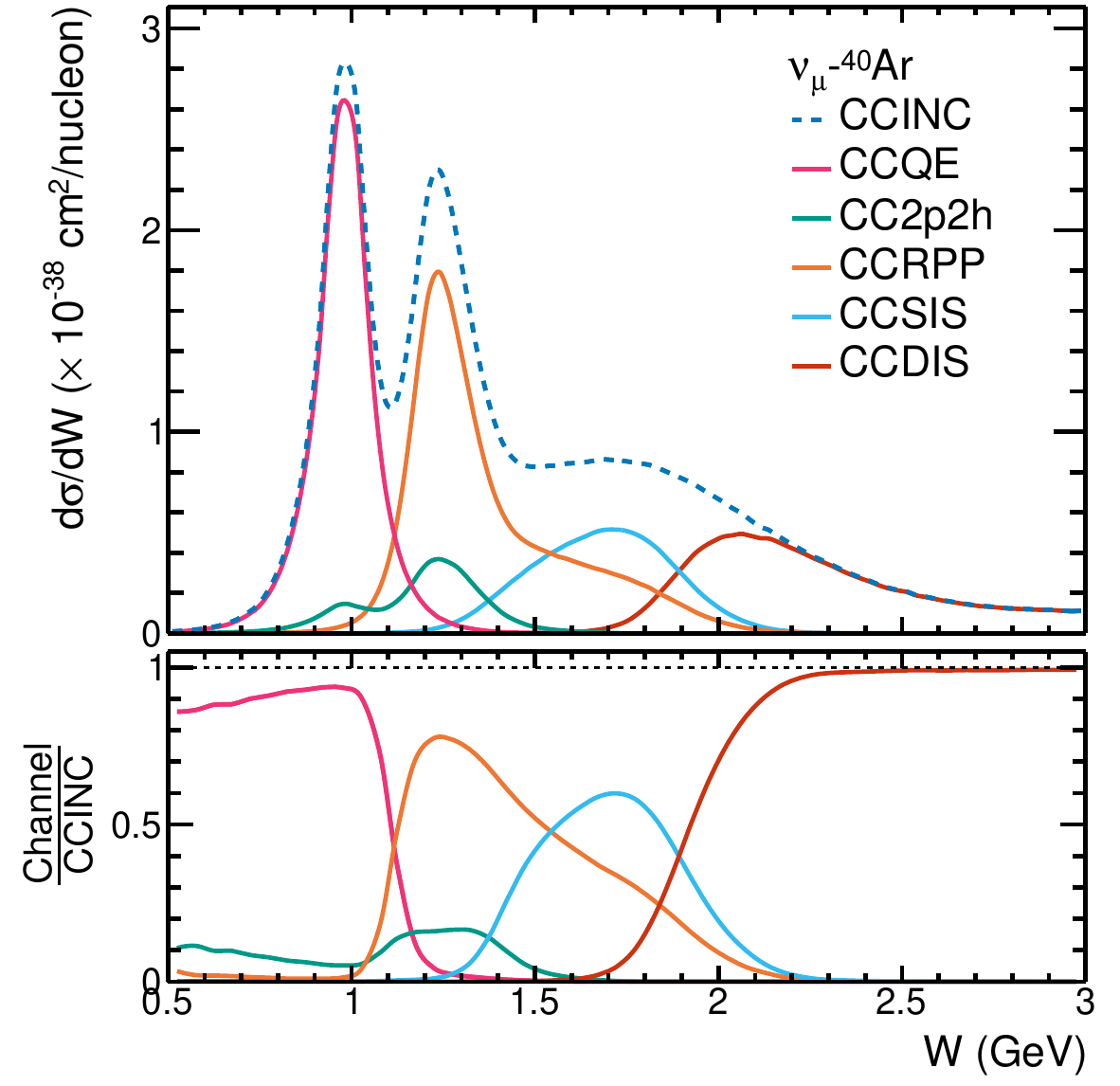}
    \caption{NuWro 25}
  \end{subfigure} 
  \caption{The differential \numu--\argon CCINC cross section as a function of the hadronic invariant mass integrated over the DUNE neutrino-enhanced ND flux, broken down by interaction channel for the GENIE 10a and NuWro 25 event generator configurations.}
  \label{fig:WCompByMode}
\end{figure*}

Model spread can be used to illustrate some of the aforementioned challenges generators face. \autoref{fig:WCompByMode} shows the differential CCINC \numu--\argon cross section averaged over DUNE's ND flux as a function of the hadronic invariant mass for two different generators (GENIE 10a and NuWro 25) broken down by contributing interaction channel\footnote{The assignment of RPP and SIS labels to generated events at intermediate $W$ ($\sim$1.3--2.0~GeV) is not consistently defined in every generator. NuWro 25 does not include an SIS interaction label. In this work, NuWro 25 interactions flagged as coming from the Hybrid~\cite{Yan:2024kkg} RPP/SIS model are assigned as RPP if they have one pion and no other mesons before FSI and SIS otherwise.}. The CCINC distributions from both models show a peak at the nucleon mass for CCQE interactions, and another peak at the $\Delta$(1232) mass for CCRPP. However, they differ in their treatment of other resonances and the CCSIS transition region between CCRPP and CCDIS interactions. As detailed in \ref{appen:gen}, NuWro 25 considers resonances only up to $W \sim$1.5 GeV and employs a DIS cross section calculation down to $W=1.6$~GeV, whilst GENIE employs a wider range of resonances up to higher $W$ in addition to low $Q^2$ Bodek--Yang corrections to the DIS cross section. These different \textit{ad hoc} approaches result in both a different CCINC cross section and a different distribution of hadronic final states. 

\begin{figure*}[htbp]
  \centering
  \begin{subfigure}[b]{0.48\linewidth}
    \centering
    \includegraphics[width=\linewidth]{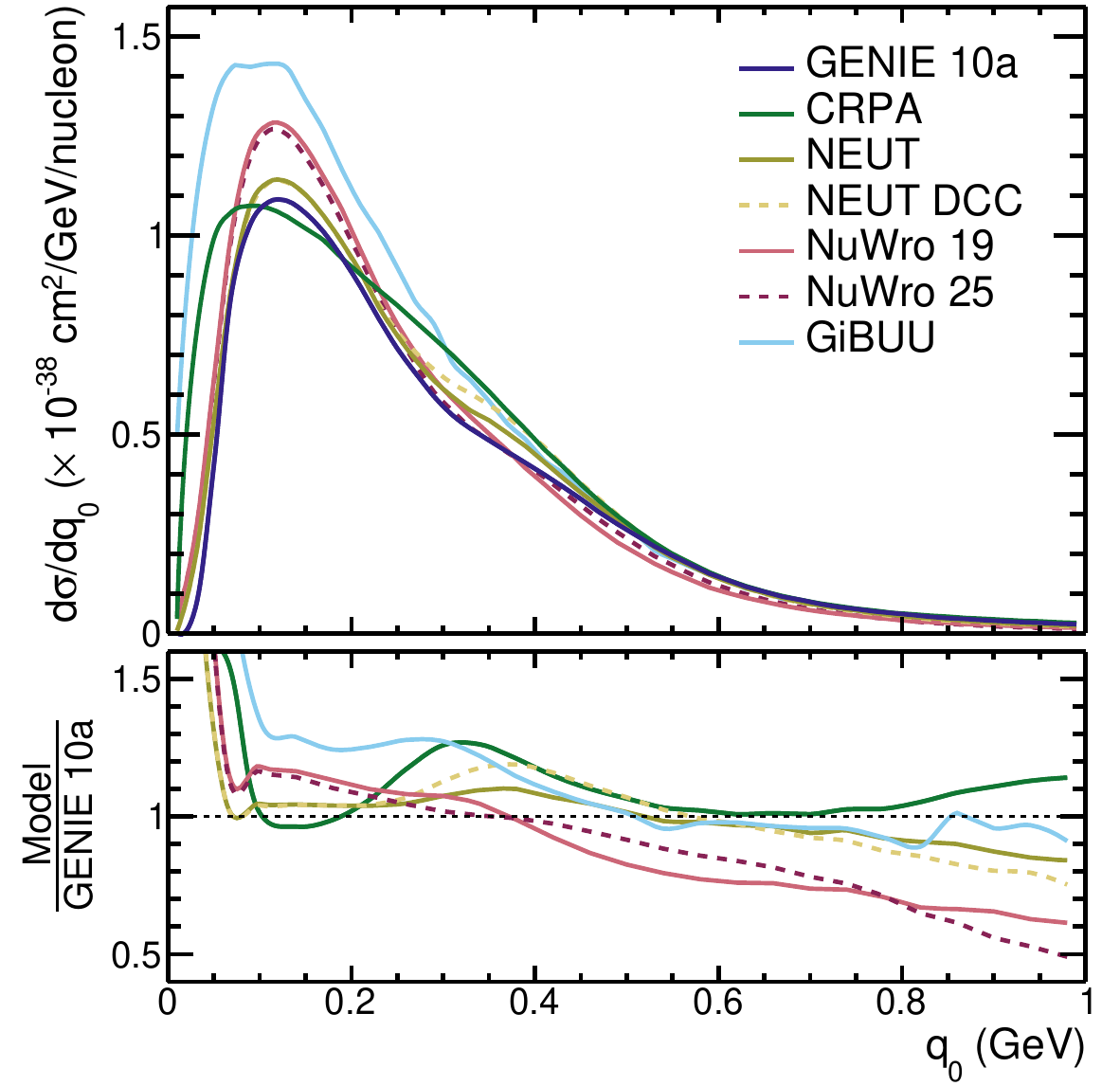}
    \caption{Hyper-K/T2K ND \numu--\water CC0$\pi$}
  \end{subfigure}
  \begin{subfigure}[b]{0.48\linewidth}
    \centering
    \includegraphics[width=\linewidth]{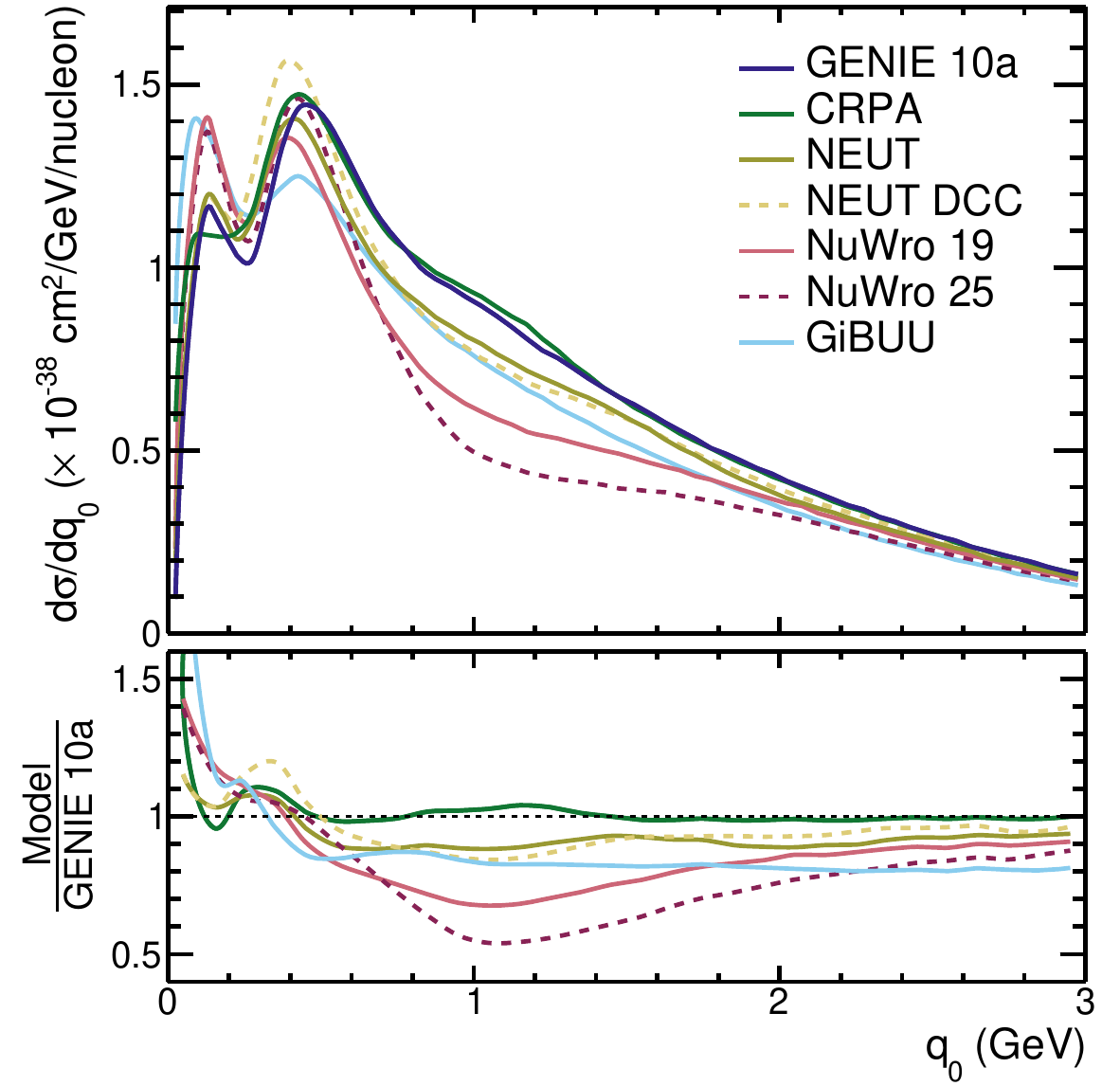}
    \caption{DUNE ND \numu--\argon CCINC}
  \end{subfigure}
  \caption{The differential cross section as a function of the energy transfer, \qz, shown for \numu--\water CC0$\pi$ interactions averaged over the Hyper-K/T2K neutrino-enhanced ND flux, and for \numu--\argon CCINC interactions averaged over the DUNE neutrino-enhanced ND flux, for a variety of generator configurations.}
  \label{fig:q0GenComp}
\end{figure*}

Generators use the widest variety of theory models in the QE-dominated low-energy transfer region, where the impulse approximation breaks down (see~\autoref{tab:models}). In this region, generators employ a wider range of plausible models using a variety of approaches. \autoref{fig:q0GenComp} shows the evolution of the cross sections as a function of energy transfer for the DUNE and Hyper-K/T2K muon neutrino fluxes on argon and water targets respectively. For DUNE, CCINC events are included, whilst for Hyper-K/T2K only CC0$\pi$ events are shown, corresponding to the primary signal channels each experiment uses to study neutrino oscillations. Since the cross section as a function of energy transfer is not sensitive to FSI variations, generator predictions where only FSI is changed are not shown. For Hyper-K, models are fairly similar for $\qz \gtrsim 100$~MeV but differ dramatically in the low-energy transfer region below this, which is particularly sensitive to the details of nuclear effect modelling (see \autoref{sec:nuclear}). In the DUNE case, large differences are also seen at low-energy transfer, in addition to other regions. Generators differ substantially in their simulation of the relative size of the two peaks (due largely to CCQE and CCRPP interactions) as well as in the CCSIS region ($1\lesssim \qz \lesssim 2$~GeV, see \autoref{fig:q0_breakdown}).

\begin{figure*}[htbp]
  \centering
  \begin{subfigure}[b]{0.48\linewidth}
    \centering
    \includegraphics[width=\linewidth]{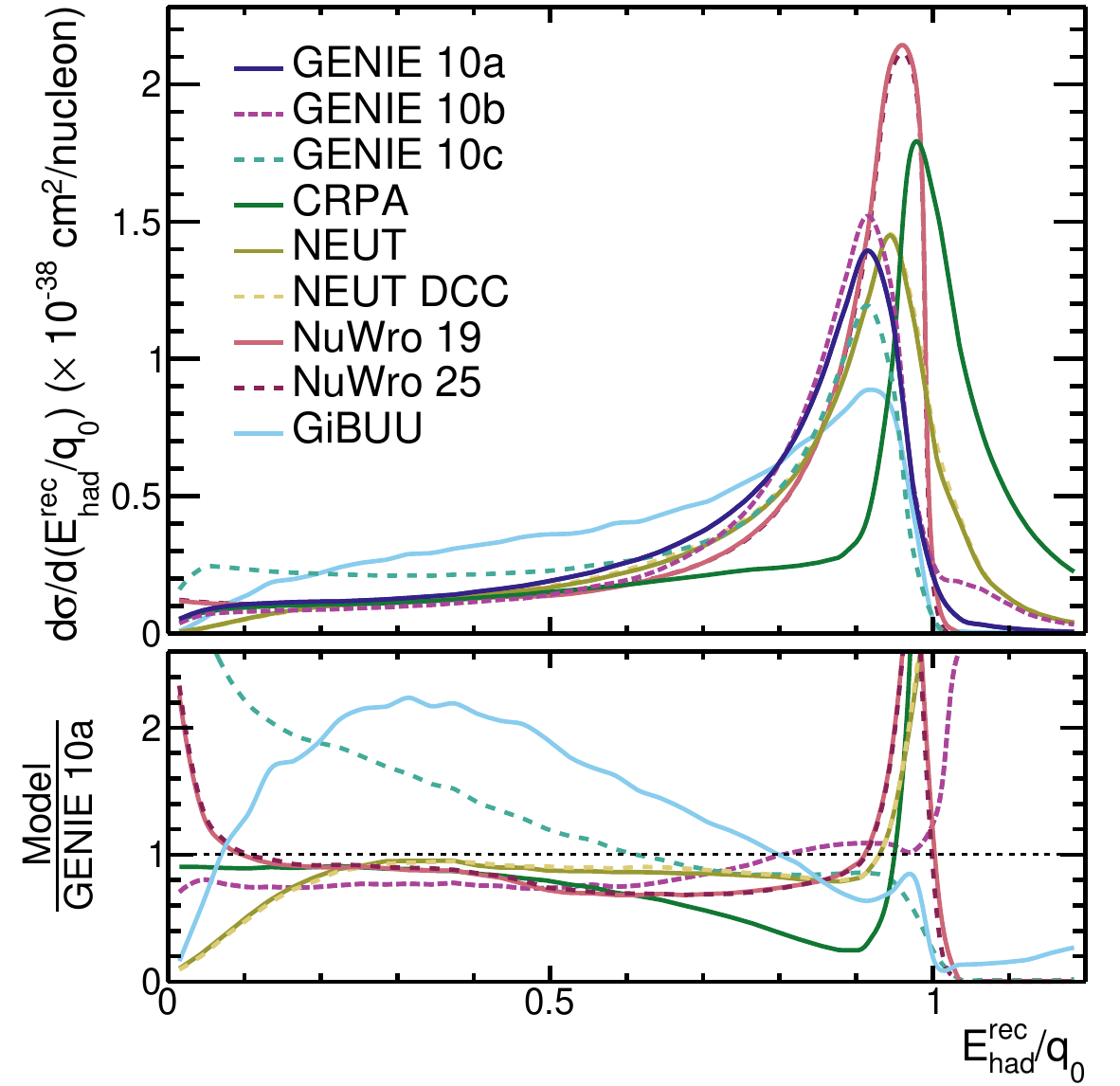}
    \caption{Hyper-K/T2K ND \numu--\water CC0$\pi$}
  \end{subfigure}
  \begin{subfigure}[b]{0.48\linewidth}
    \centering
    \includegraphics[width=\linewidth]{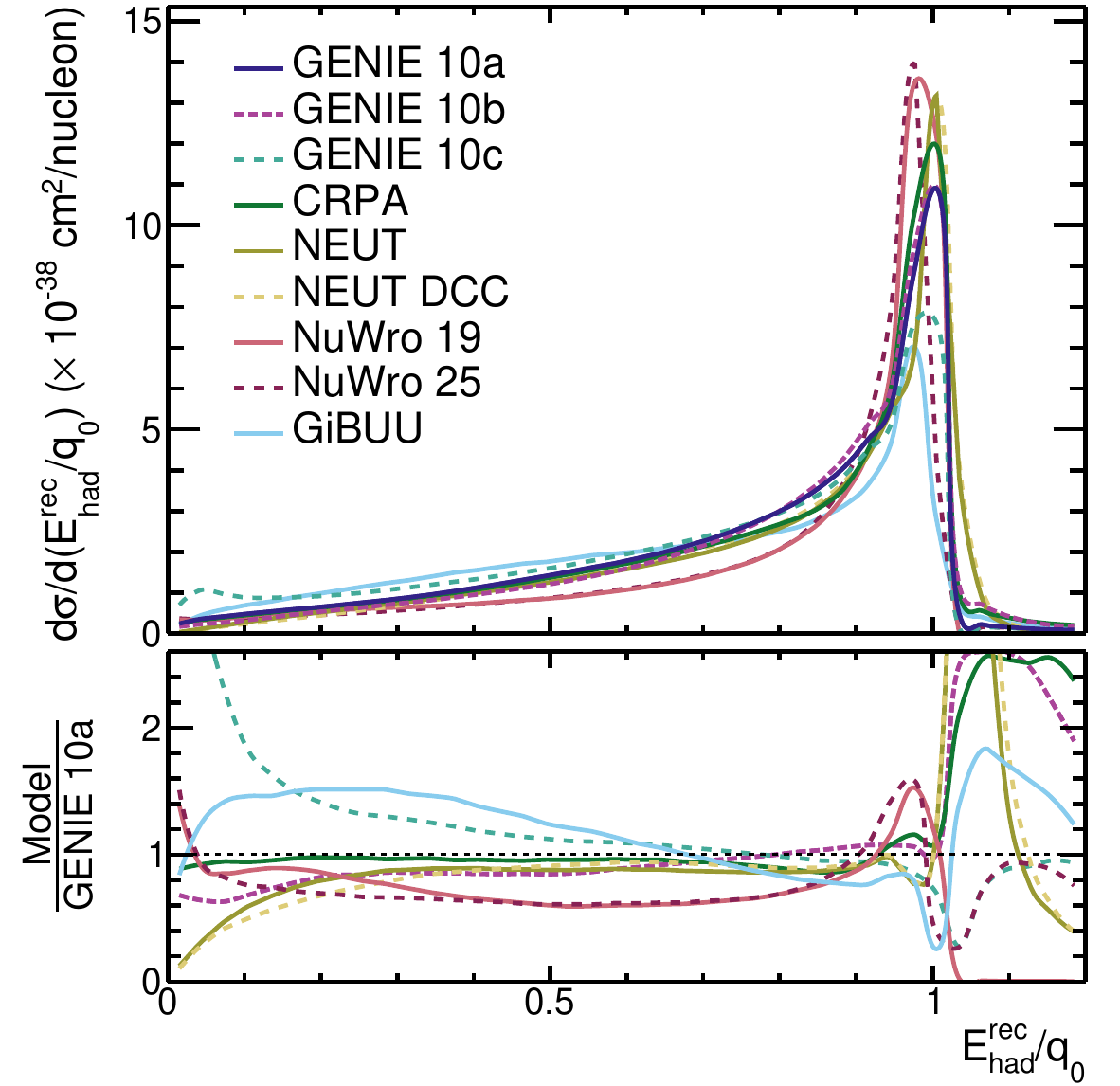}
    \caption{DUNE ND \numu--\argon CCINC}
  \end{subfigure}
  \caption{The differential muon-neutrino cross section as a function of the ratio between the visible hadronic energy (see~\autoref{eq:enuhad}) and the true energy transfer, shown for \numu--\water CC0$\pi$ interactions averaged over the Hyper-K/T2K neutrino-enhanced ND flux, and for \numu--\argon CCINC interactions averaged over the DUNE neutrino-enhanced ND flux, for a variety of generator configurations.}
  \label{fig:q0EhadMap}
\end{figure*}

Generators further differ in how energy transferred to the hadronic system maps to energy that is visible within detectors, which is particularly sensitive to the treatment of FSI modelling, as well as to approximations made to simulate outgoing hadron kinematics using inclusive input calculations. As will be discussed in \autoref{sec:howtooa}, this is of critical importance for experiments such as DUNE, which use the total visible hadronic energy in their estimate of the neutrino energy. The visible hadronic energy for an idealised detector, \ehadrec, is derived from the sum of the energies of all final-state hadrons other than neutrons, which are rarely visible at the energies of interest for few-GeV neutrino experiments. The ratio \ehadrec over \qz is shown in \autoref{fig:q0EhadMap}, for the same Hyper-K/T2K and DUNE cases considered in \autoref{fig:q0GenComp}. If there are no neutrons in the final state then $E_{\mathrm{rec}}^{\mathrm{had}}$ differs from \qz primarily due to the nuclear removal energy spent to liberate hadrons from the nuclear potential, resulting in a narrow distribution between $\sim$0.9 and 1.0. Conversely, the emission of neutrons can cause ratios much less than 0.9. \autoref{fig:q0EhadMap} shows that different treatments of nuclear removal energies, as well as neutron multiplicities and kinematics (largely driven by FSI), cause large differences between event generator predictions\footnote{Values above one are not physical (energy is not conserved), yet the approximations introduced in some event generator implementations cause non-negligible populations here. For example: the GENIE CRPA implementation uses a momentum-transfer dependent sampling of the initial state nucleon which sometimes is less than the kinetic energy nucleons have from Fermi motion; NEUT does not implement a removal energy for RPP, SIS or DIS interactions; and some GENIE FSI models allow outgoing nucleons to gain energy.}. 

\section{Neutrino interaction modelling and neutrino oscillation measurements}
\label{sec:howtooa}

As introduced in \autoref{sec:intro}, to study neutrino oscillations, LBL experiments create a predominantly muon (anti)neutrino beam and measure it twice. First at the ND before oscillation, where events constrain models of the neutrino flux, interaction cross section and detector effects, and again at the FD, where the shape and normalisation of muon and electron (anti)neutrino events are used to extract the neutrino oscillation probabilities. Although the general rate equation was given in \autoref{eq:rate-simple}, it is worth explicitly writing expressions for the number of events observed at the ND ($N^{\mathrm{ND}}$) as well as in the electron and muon neutrino channels at the FD ($N_{\nu_\mu}^{\mathrm{FD}}$ and $N_{\nu_e}^{\mathrm{FD}}$ respectively):
\begin{equation}
\begin{split}
    \frac{dN^{\mathrm{ND}}}{d\vec{\mathbf{y}}} = N_{\mathrm{tgts}}^{\mathrm{ND}} \int & d\enu d\vec{\mathbf{x}}\;\; \Phi_{\nu_\mu}^{\mathrm{ND}} (\enu) \\
    & \sigma_{\nu_\mu}(\enu, \vec{\mathbf{x}}) \; M^{\mathrm{ND}}(\vec{\mathbf{y}} \mid \vec{\mathbf{x}}) + B^{\mathrm{ND}},
    \label{eq:evtratend}
\end{split}
\end{equation}
\begin{equation}
\begin{split}
    \frac{dN_{\nu_\mu}^{\mathrm{FD}}}{d\vec{\mathbf{y}}} = N_{\mathrm{tgts}}^{\mathrm{FD}} \int & d\enu d\vec{\mathbf{x}}\;\; P_{\nu_\mu \rightarrow \nu_\mu}(E_{\nu}) \; \Phi_{\nu_\mu}^{\mathrm{FD}} (\enu) \\
    & \sigma_{\nu_\mu}(\enu, \vec{\mathbf{x}}) \; M^{\mathrm{FD}}(\vec{\mathbf{y}} \mid \vec{\mathbf{x}}) + B^{\mathrm{FD}}_{\numu},
    \label{eq:evtratenumu}
\end{split}
\end{equation}
\begin{equation}
\begin{split}
    \frac{dN_{\nu_e}^{\mathrm{FD}}}{d\vec{\mathbf{y}}} = N_{\mathrm{tgts}}^{\mathrm{FD}} \int & d\enu d\vec{\mathbf{x}}\;\; P_{\nu_\mu \rightarrow \nu_e}(E_{\nu}) \; \Phi_{\nu_\mu}^{\mathrm{FD}} (\enu) \\
    & \sigma_{\nu_e}(\enu, \vec{\mathbf{x}}) \; M^{\mathrm{FD}}(\vec{\mathbf{y}} \mid \vec{\mathbf{x}}) + B^{\mathrm{FD}}_{\nue},
    \label{eq:evtratenue}
\end{split}
\end{equation}
\noindent in each case, the expected differential event rate as a function of observable particle kinematic variables, $\vec{\mathbf{y}}$, is shown as a function of the number of targets $N_{\mathrm{tgts}}$, the neutrino flux $\Phi(E_\nu)$, the neutrino cross section $\sigma(\enu, \vec{\mathbf{x}})$ as a function of neutrino energy and the true outgoing particle kinematics $\vec{\mathbf{x}}$, the detector efficiency and smearing $M(\vec{\mathbf{y}}\mid\vec{\mathbf{x}})$ and the oscillation probability $P(\enu)$, which can be neglected at the ND. Additionally, there is a background term, $B$, which represents contributions from other flavours, NC events which leak into the relevant CC sample, and non-beam backgrounds. 

Comparing \autoref{eq:evtratend} to \autoref{eq:evtratenumu} and \autoref{eq:evtratenue} demonstrates how measurements at the ND are able to constrain much of the physics that must be understood in order to infer neutrino oscillation probabilities from FD event rates. However, there are subtle but important differences which limit how well these constraints can be \textit{extrapolated} to the FD: the relationship between detector observables $\vec{\mathbf{y}}$ and the true particles produced at the vertex $\vec{\mathbf{x}}$ are different, and by construction the flux over which the integral runs is radically different due to oscillations. The backgrounds also differ significantly between the ND and FD as well as between FD flavours. These differences mean that an \textit{a priori} understanding of the flux, detector effects and cross-section models are essential in addition to ND constraints.

The neutrino flux can be constrained using dedicated measurements from well-controlled hadron production experiments such as NA61/SHINE~\cite{NA61:2014lfx} or EMPHATIC~\cite{EMPHATIC:2019xmc}, as well as using \textit{in situ} standard candles, both of which are described in \autoref{sec:novelapproach}. 
Additionally, proton, muon, and neutrino beam monitors are commonly used to understand different aspects of the beam profile.
Similarly, detector responses and their uncertainties can be constrained through test-beam measurements, for example using ProtoDUNE~\cite{DUNE:2020ypp,DUNE:2017pqt,DUNE:2020cqd} and the Water Cherenkov Test Experiment~\cite{WCTE:2857041}, as well as \textit{in situ} detector calibrations. 

Conversely, as described in \autoref{sec:models}, the simulation of few-GeV neutrino interactions requires the modelling of a variety of nuclear physics processes which can so far only be confronted using approximate methods. Whilst these models can be benchmarked against neutrino cross-section measurements (the topic of \autoref{sec:constraints}), these are averaged over broad fluxes and cannot isolate individual processes. Out of all sources of systematic uncertainty affecting LBL experiments, the path to controlling neutrino interaction uncertainties is the least clear.

\subsection{Extrapolating constraints between the near and far detectors}
\label{subsec:NDtoFD}

Na\"{\i}vely, one might expect that by measuring the neutrino event rates at both the ND (\autoref{eq:evtratend}) and an FD (\autoref{eq:evtratenumu} or \autoref{eq:evtratenue}) of an accelerator neutrino experiment, systematic uncertainties would be trivially cancelled in a ratio of the two. This is not the case due to four main reasons:

\begin{enumerate}
\item {\bf Different ND and FD target materials.} As the ND and FD operate in very different event rate environments, it is often not practical to construct detectors with exactly the same technology. This is the case for Hyper-K/T2K, where the FD is a water Cherenkov detector whilst the NDs use a mix of target materials (see \autoref{tab:lbldeets}).

\item {\bf Different ND and FD detector responses.} Fundamentally different ND and FD detector technologies and geometries are often used to cope with their different event rate environments, leading to little correspondence in the detector response. This is the case for Hyper-K/T2K with a water Cherenkov FD and a mix of technologies at the ND, as well as for DUNE which uses time projection chambers with long drift lengths and times at the FD, but require much faster segmented detectors at the ND. 
In general, it is also the case that detector responses differ considerably for muons and electrons, which adds further complications when extrapolating from the ND to FD electron (anti)neutrino event samples. 

\item {\bf Integration over different neutrino fluxes.} Because of neutrino oscillations, the neutrino flavour content of the beam and the neutrino energy distribution for each flavour is radically different at the ND and FD (\autoref{fig:oscfluxes}). As neutrino--nucleus cross sections vary with neutrino energy, the total averaged cross section and the relative contributions from different interaction channels differ between ND and FD, as shown for Hyper-K/T2K in \autoref{fig:oscxsecfluxcomp} and quantified for both Hyper-K/T2K and DUNE in \autoref{tab:channelsoscnoosc}. In general, the disappearance of muon neutrino interactions around the unoscillated flux peak accentuates the importance of high-energy flux tails to give a larger fraction of CCRPP events for Hyper-K/T2K and of CCSIS or CCDIS events for DUNE. Additionally, the range of observable interactions depends on detector response, leading to further ND--FD differences. 

There are also differences in the unoscillated neutrino fluxes at the ND and FD due to the production of neutrinos from meson decay in a large decay volume. The ND integrates the flux of neutrinos produced with a range of outgoing angles from the decay (a line source), whereas the neutrino beam can be treated as a point source at the FD.

\begin{figure}[tb]
  \centering
  \includegraphics[width=0.98\linewidth]{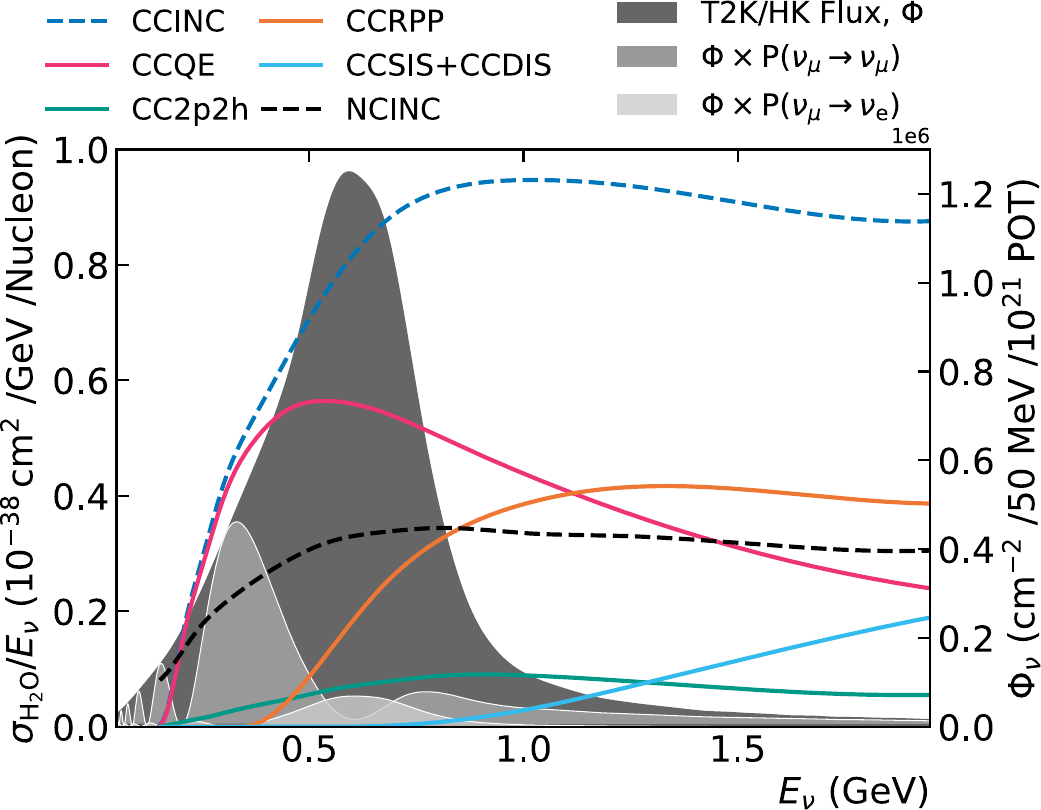}
  \caption{The cross section for neutrino--water interactions for a variety of interaction channels as simulated by the NEUT event generator. Also shown is the unoscillated neutrino-mode Hyper-K/T2K FD \numu flux, as well as the $\numu\rightarrow\numu$ and $\numu\rightarrow\nue$ oscillated fluxes.
  }
  \label{fig:oscxsecfluxcomp}
\end{figure}

\begin{table}[htbp]
  \small
  \centering
  {\renewcommand{\arraystretch}{1.2}
  \begin{tabular}{c|c|c|c|c|c}
\hline\hline
\multicolumn{2}{c|}{}        & \multicolumn{2}{c|}{Hyper-K/T2K} & \multicolumn{2}{c}{DUNE} \\ 
\multicolumn{2}{c|}{Channel} & \multicolumn{2}{c|}{(CC0$\pi$)}    & \multicolumn{2}{c}{(CCINC)} \\ \cline{3-6}
\multicolumn{2}{c|}{}        &  w/o osc.     & w/ osc.          & w/o osc.    & w/ osc. \\ \hline
\multirow{4}{*}{$\nu_\mu$}       & CCQE       & 74.8\%        & 68.1\%           & 18.4\%      & 16.1\% \\ 
                                 & CC2p2h     & 14.4\%        & 14.8\%           & 5.0\%       & 4.3\% \\ 
                                 & CCRPP      & 10.1\%        & 14.9\%           & 38.6\%      & 29.9\% \\
                                 & CCSIS/DIS  & 0.3\%         & 0.7\%            & 36.3\%      & 48.5\% \\ \hline
\multirow{4}{*}{$\bar{\nu}_\mu$} & CCQE       & 79.0\%        & 73.6\%           & 22.7\%      & 20.4\% \\ 
                                 & CC2p2h     & 14.9\%        & 15.5\%           & 8.2\%       & 7.4\% \\ 
                                 & CCRPP      & 5.7\%         & 9.5\%            & 44.2\%      & 38.0\% \\
                                 & CCSIS/DIS  & 0.1\%         & 0.3\%            & 23.0\%      & 32.6\% \\ \hline\hline
\end{tabular}
}
\caption{The proportion of each interaction channel contributing to $\numu^{\bracketbar}$--\water CC0$\pi$ interactions using the (anti)neutrino-enhanced flux at the Hyper-K/T2K FD, and $\numu^{\bracketbar}$--\argon CCINC and the (anti)neutrino-enhanced flux at the DUNE FDs, as simulated by GENIE 10a. This is shown before and after neutrino oscillations are applied. The columns do not sum to 100\% as small contributions, such as CPP and heavy meson production, have been omitted.}
\label{tab:channelsoscnoosc}
\end{table}

\item {\bf $\nue/\numu$ and $\nueb/\numub$ cross section differences.} The ND measures predominantly muon (anti)neutrino interactions whilst the FD measures oscillations through muon and electron (anti)neutrino channels. Assuming lepton universality, the only differences in CC electron and muon neutrino interaction cross sections are due to the outgoing lepton mass. However, the way in which this couples to nuclear effects~\cite{Dieminger:2023oin,Nikolakopoulos:2019qcr,Ankowski:2017yvm,Martini:2023kem}, as well as to radiative corrections to neutrino cross sections~\cite{Tomalak:2021hec,Tomalak:2022xup,Day:2012gb}, introduces differences which are not trivial to calculate. For example, the inclusion of FSI in microscopic models can cause the muon (anti)neutrino cross section to be larger than the electron (anti)neutrino cross section in some regions of kinematic phase space~\cite{Nikolakopoulos:2019qcr}. 
For CCQE interactions, muon and electron (anti)neutrino cross sections differ most when the energy transfer is comparable to the lepton mass difference. 
Lepton mass differences can also alter cross sections close to kinematic thresholds. For RPP interactions this leads to large differences in the ratio of electron and muon (anti)neutrino cross sections near the threshold energy of prominent resonances (for example the $\Delta(1232)$). This is discussed in detail in~\ref{appen:rppnuenumu}.
\end{enumerate}

\noindent Overall, neutrino interaction models are relied on to correct for differences in the interactions observed at the ND and FD resulting from each of these four effects, inducing a source of systematic uncertainty for LBL experiments.

\subsection{Neutrino energy reconstruction}
\label{subsec:enurec}

As a neutrino energy-dependent effect, measurements of neutrino oscillation rely on the ability of experiments to infer neutrino energy. As described by~\autoref{eq:evtratend}--\autoref{eq:evtratenue}, experiments perform neutrino oscillation analyses by measuring event rates as a function of some detector observable variable(s) ($\vec{\mathbf{y}}$), which provide an estimator for neutrino energy.
Its form depends on the detector technology and the dominant interaction channels given the neutrino energy spectrum in the experiment.

As the Hyper-K and T2K neutrino fluxes are mostly below 1 GeV, they are dominated by interactions with CC0$\pi$ final-state topologies (see \autoref{fig:oscxsecfluxcomp}). Furthermore, their water-Cherenkov FDs have high detection thresholds for protons and so a neutrino energy proxy ($E^{\mathrm{QE},\;\mathrm{rec}}_{\nu}$) is built using the outgoing lepton kinematics alone:
\begin{equation}
    E^{\mathrm{QE},\;\mathrm{rec}}_{\nu} = \dfrac{m_p^2-m_\ell^2-(m_n-E_{\mathrm{b}})^2+2E_\ell(m_n-E_{\mathrm{b}})}{2(m_n-E_{\mathrm{b}}-E_\ell+p_\ell^z)},
    \label{eq:enuqe}
\end{equation}
\noindent where $m_{p/\ell/n}$ is the mass of a proton, outgoing lepton or neutron (the proton and neutron masses should be swapped for antineutrino interactions); $E_l$ and $p_l^z$ are the outgoing lepton energy and momentum projected along the direction of the incoming neutrino, respectively; and $E_{\mathrm{b}}$ accounts for the nuclear binding energy of the struck nucleon, assuming a constant value (27 MeV for oxygen in this work). 
This formula is derived assuming CCQE scattering off a stationary nucleon target. It is the most common example of \textit{kinematic} neutrino energy reconstruction, where the neutrino energy is estimated from the kinematics of a subset of particles in the final state\footnote{Other examples of kinematic energy reconstruction methods can be found in Refs.~\cite{Furmanski:2016wqo, Baudis:2023tma}.}.
\autoref{eq:enuqe} is usually only applied to event samples reconstructed to have CC0$\pi$ final states, as the quantity does not map well to neutrino energy when applied to non-CCQE interactions\footnote{Recent T2K analyses use a modified formula for the subdominant samples that have a single observable pion in the final state, built assuming an RPP interaction via the excitation of an on-shell $\Delta(1232)$ resonance~\cite{T2K:2021xwb}, which is the dominant resonance at T2K energies.
In this case, the proton mass in \autoref{eq:enuqe} is replaced by the $\Delta(1232)$ mass. 
The kinematics of the outgoing pion can also be included in the energy estimator, which removes the requirement of assuming an on-shell $\Delta(1232)$ resonance, but adds the assumption of a single outgoing nucleon.}.

Experiments with detectors which are more capable of calorimetry, such as DUNE and NOvA, commonly use the entire observed hadronic final state to infer the neutrino energy. 
This is referred to as a \textit{calorimetric} neutrino energy reconstruction and, conversely to the kinematic method, can be applied to CCINC interactions.
Although it is difficult to reliably map the \textit{visible} energy observed in the detector to a variable that can be unambiguously defined from the kinematics of the final-state particles, one frequently used proxy variable is:

\begin{equation}
  E^{\mathrm{rec},\;\mathrm{avail}}_{\nu} = E_{\ell} + E_{\mathrm{had}}^{\mathrm{avail}} = E_{\ell} + \sum_{i=p,\pi^{\pm}}{T_i} + \sum_{i=\pi^{0}, \gamma, h}{E_i},
  \label{eq:enuavail}
\end{equation}
\noindent in which the reconstruction neutrino energy is the sum of: the outgoing lepton energy, $E_\ell$; the kinetic energies, $T$, of all final-state protons and charged pions; and the total energy, $E$, of all neutral pions, photons, and other hadrons (typically a small contribution at few-GeV energies). 
The latter two terms are the reconstructed \textit{available} hadronic energy, \ehadrec, as introduced in Ref.~\cite{MINERvA:2015ydy}.  Only the kinetic energy of protons is used as their rest-mass energy is also present in the initial state of few-GeV neutrino interactions. Neutrons are assumed to not leave any visible energy deposits within the detector. The kinetic energy of charged pions is used because, although they are created within an interaction, their rest-mass energy is largely lost to neutrinos via the pion decay. Conversely, the total energy of neutral pions is included as the two photons produced in the dominant decay mode more readily leave their energy within a detector.

More capable detectors may be able to identify charged pions, and so a more optimistic proxy can include a correction for the missing pion masses:
\begin{equation}
  E^{\mathrm{rec},\;\mathrm{had}}_{\nu} = E_{\ell} + \ehadrec = E_{\ell} + \sum_{i=p}{T_i} + \sum_{i=\pi^{\pm}, \pi^{0}, \gamma, h}{E_i},
  \label{eq:enuhad}
\end{equation}

These two estimators are further discussed and compared in Refs.~\cite{Wilkinson:2022dyx, Dolan:2026dfu}. Both neglect the impact of quenching effects, such as Birks' quenching in scintillator detectors or recombination effects in LArTPCs~\cite{Birks:1951boa,GERRITSEN1948407,ICARUS:2004koz,Friedland:2018vry,recomChem}. The size of such quenching effects depends on the distribution of kinetic energy among final-state hadrons which couple to misreconstruction and detector threshold effects in non-trivial ways, and requires a correction which depends on the neutrino interaction model used~\cite{Lachner:2026ufu}. They also do not account for any possibility of correcting for portions of the missing neutron energy through observation of their secondary scatters (as proposed in Ref.~\cite{Manly:2025pfm}). 

\begin{figure*}[htbp]
  \centering
   \begin{subfigure}[b]{0.48\textwidth}
      \centering
      \includegraphics[width=1.00\linewidth]{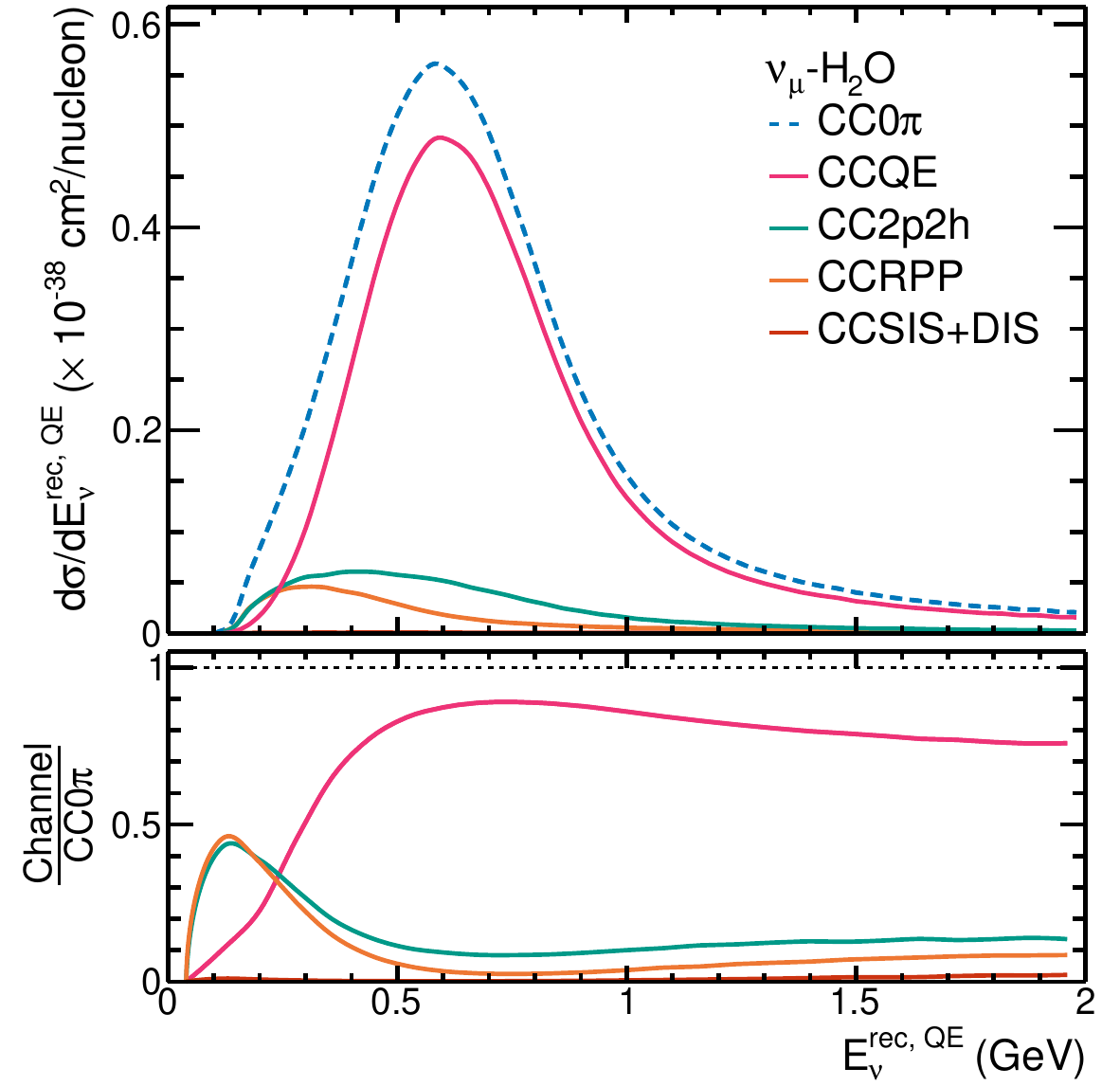}
      \caption{Hyper-K/T2K ND \numu--\water CC0$\pi$}
  \end{subfigure}
   \begin{subfigure}[b]{0.48\textwidth}
      \centering
      \includegraphics[width=1.00\linewidth]{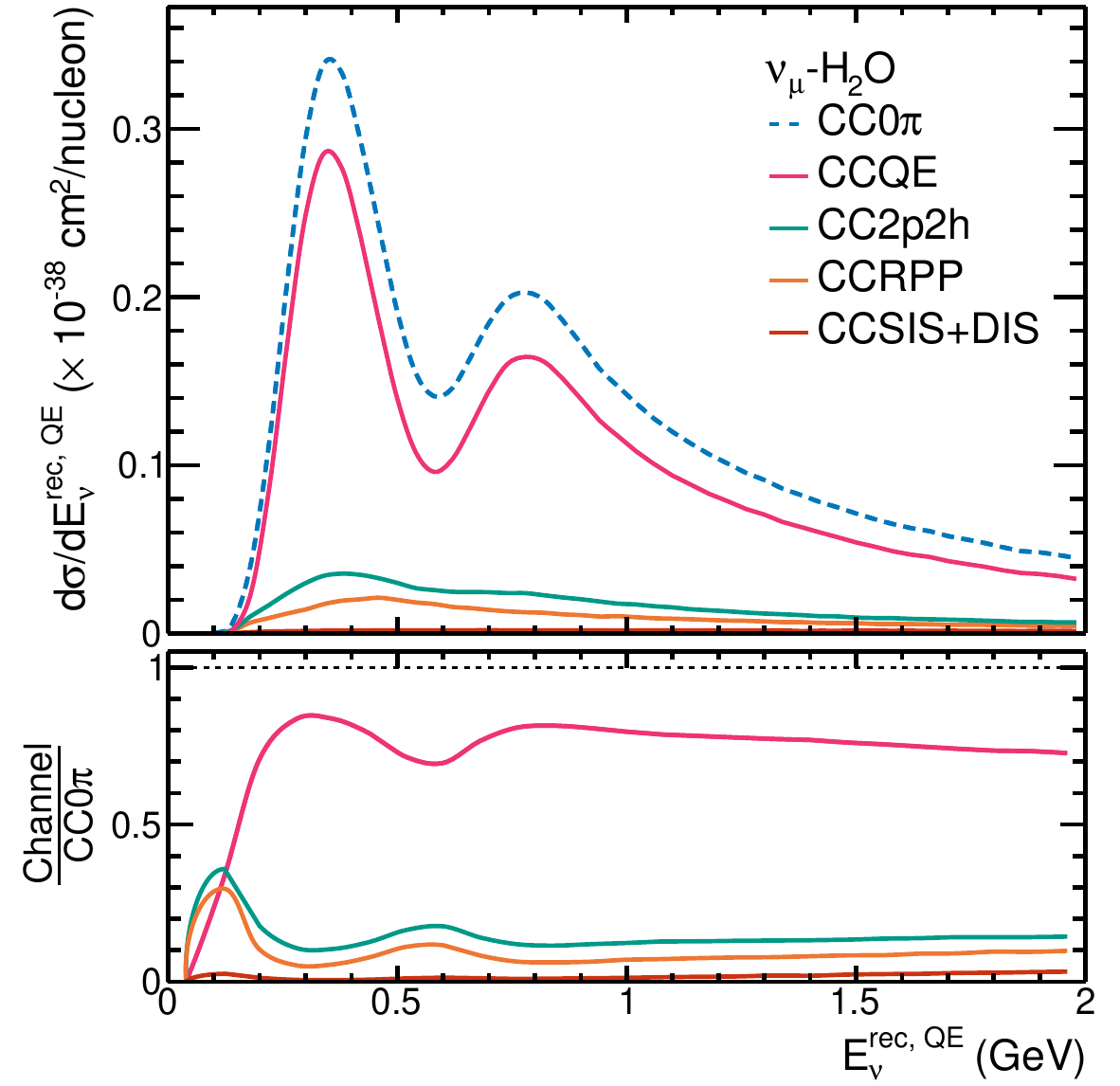}
      \caption{Hyper-K/T2K FD \numu--\water CC0$\pi$}
  \end{subfigure}
\\
  \begin{subfigure}[b]{0.48\textwidth}
      \centering
      \includegraphics[width=1.00\linewidth]{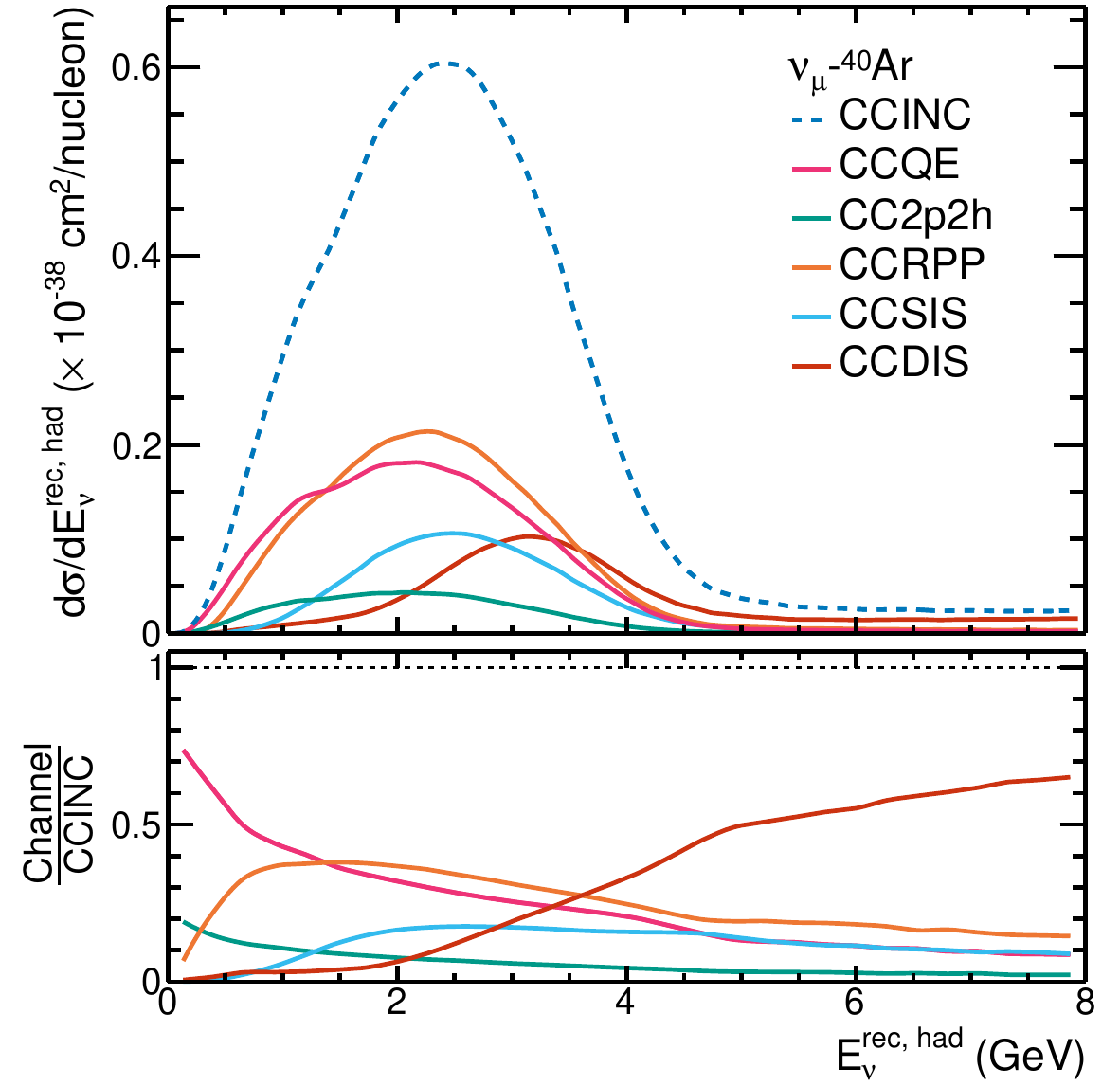}
      \caption{DUNE ND \numu--\argon CCINC}
  \end{subfigure}
  \begin{subfigure}[b]{0.48\textwidth}
      \centering
      \includegraphics[width=1.00\linewidth]{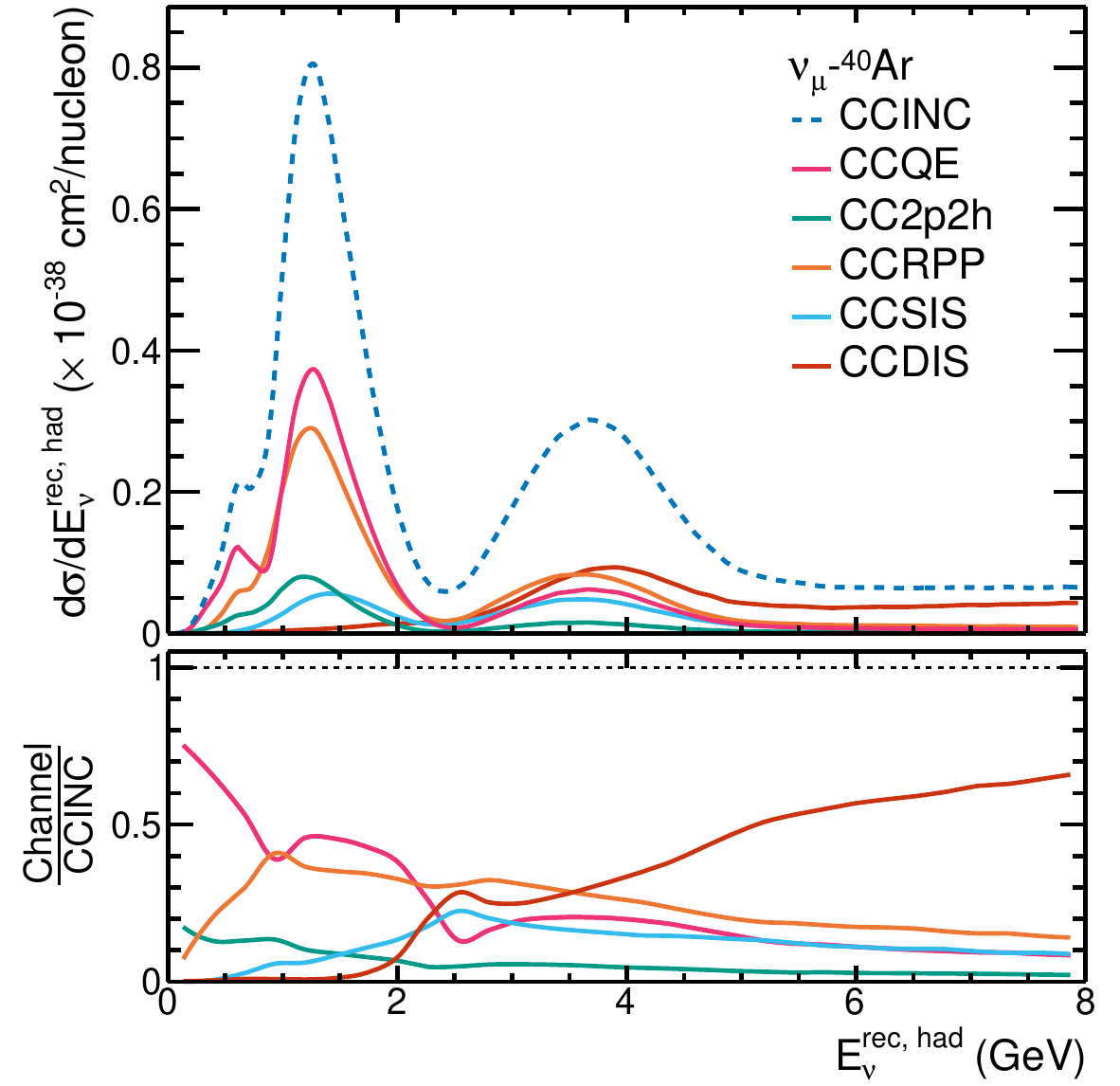}
      \caption{DUNE FD \numu--\argon CCINC}
  \end{subfigure}
  \caption{Comparison of the differential cross section as a function of reconstructed neutrino energy at the ND and FD for $\nu_\mu$--\argon CCINC events at DUNE and $\nu_\mu$--\water CC0$\pi$ events at Hyper-K/T2K, with contributions broken down by interaction channel, using the NuWro 25 generator. \autoref{eq:enuhad} (\autoref{eq:enuqe}) is used for the reconstructed neutrino energy at DUNE (Hyper-K/T2K).}
  \label{fig:enuspectcomp}
\end{figure*}

The cross section as a function of $E^{\mathrm{rec},\;\mathrm{had}}_{\nu}$ and $E^{\mathrm{QE},\;\mathrm{rec}}_{\nu}$ is shown in \autoref{fig:enuspectcomp} for CCINC interactions in DUNE, and CC0$\pi$ interactions for Hyper-K, before and after neutrino oscillations. 
As the relative contribution of interaction channels is different between the ND and FD (\autoref{tab:channelsoscnoosc}), the mapping between reconstructed and true neutrino energies is also different at each. In all cases, perfect detector reconstruction is assumed.

\subsubsection{Bias in kinematic neutrino energy reconstruction}

\begin{figure*}[htbp]
  \centering
  \begin{subfigure}[b]{0.48\textwidth}
      \centering
      \includegraphics[width=1.00\linewidth]{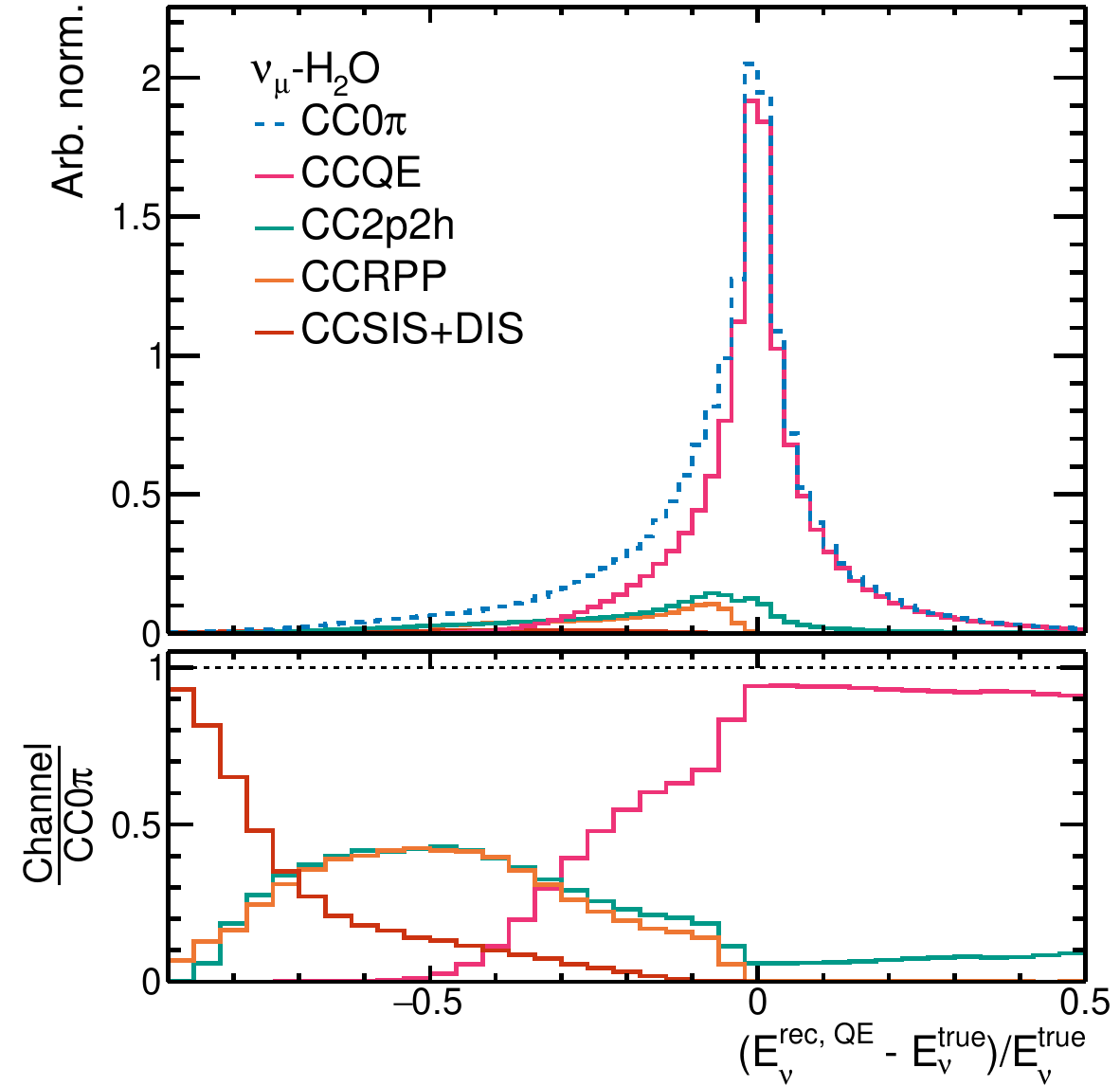}
      \caption{Hyper-K/T2K FD \numu--\water CC0$\pi$, rel. bias}
  \end{subfigure}
  \begin{subfigure}[b]{0.48\textwidth}
      \centering
      \includegraphics[width=1.00\linewidth]{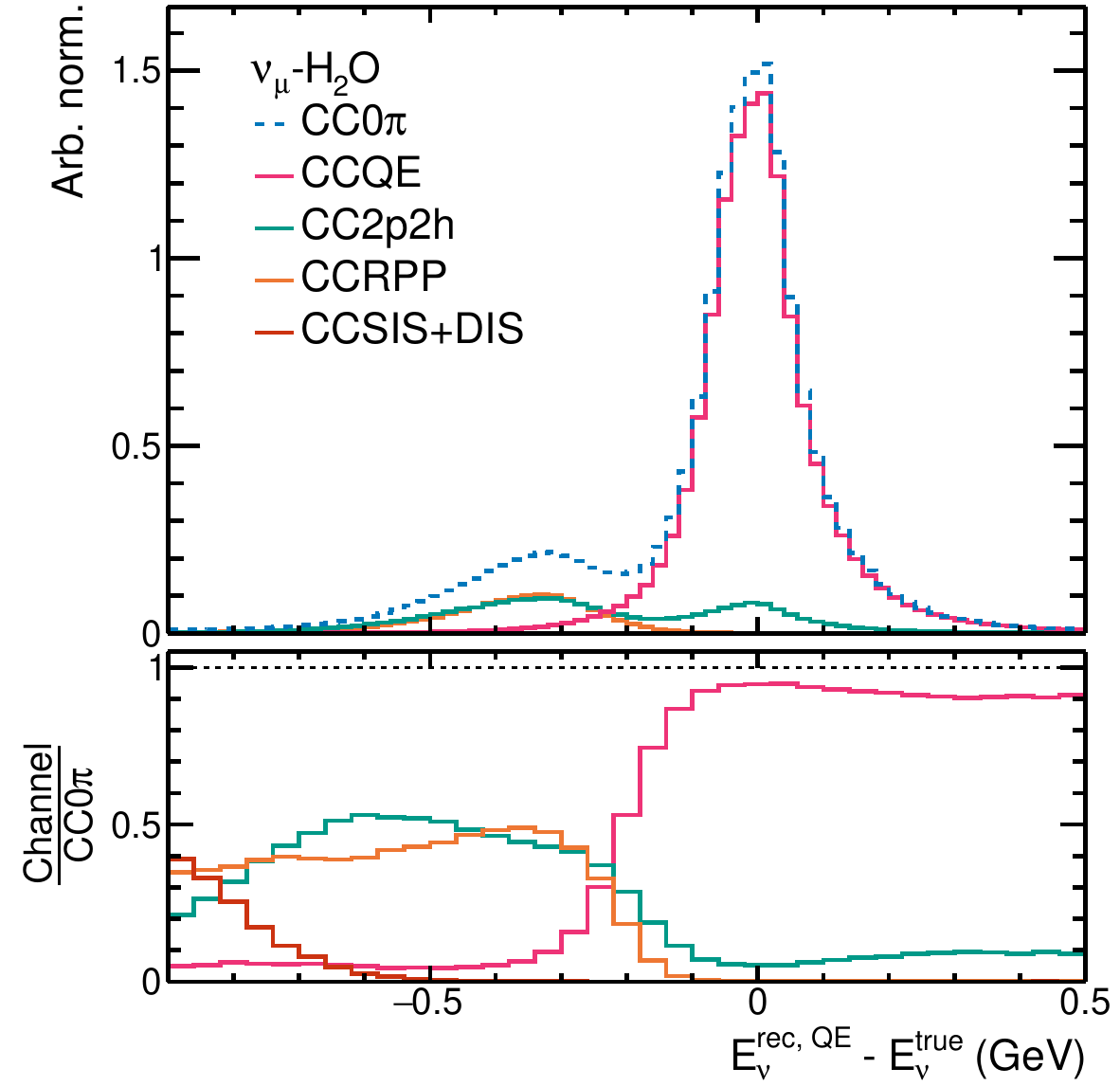}
      \caption{Hyper-K/T2K FD \numu--\water CC0$\pi$, abs. bias}
  \end{subfigure}
\\
   \begin{subfigure}[b]{0.48\textwidth}
      \centering
      \includegraphics[width=1.00\linewidth]{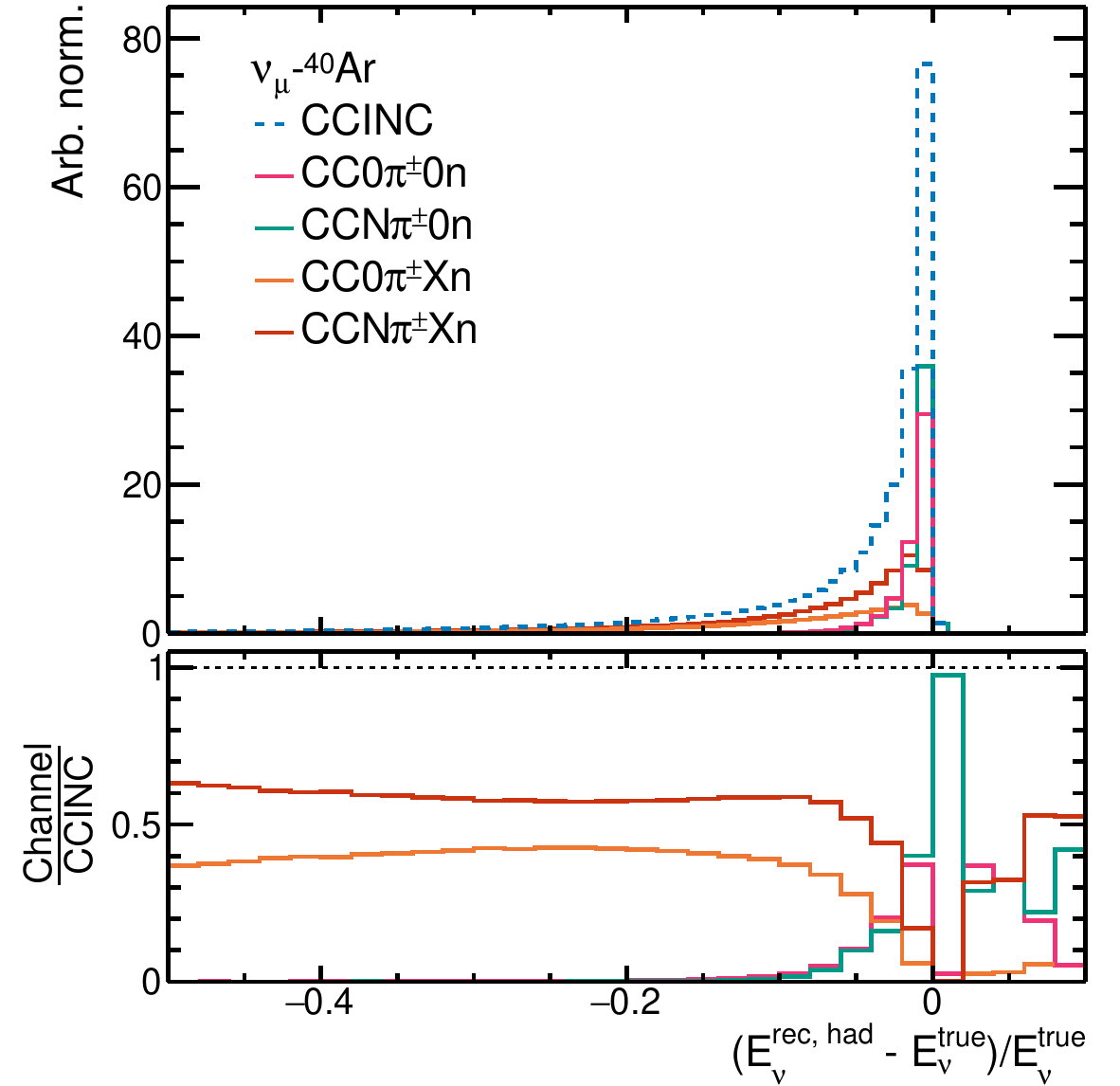}
      \caption{DUNE FD \numu--\argon CCINC, rel. bias}
  \end{subfigure}
   \begin{subfigure}[b]{0.48\textwidth}
      \centering
      \includegraphics[width=1.00\linewidth]{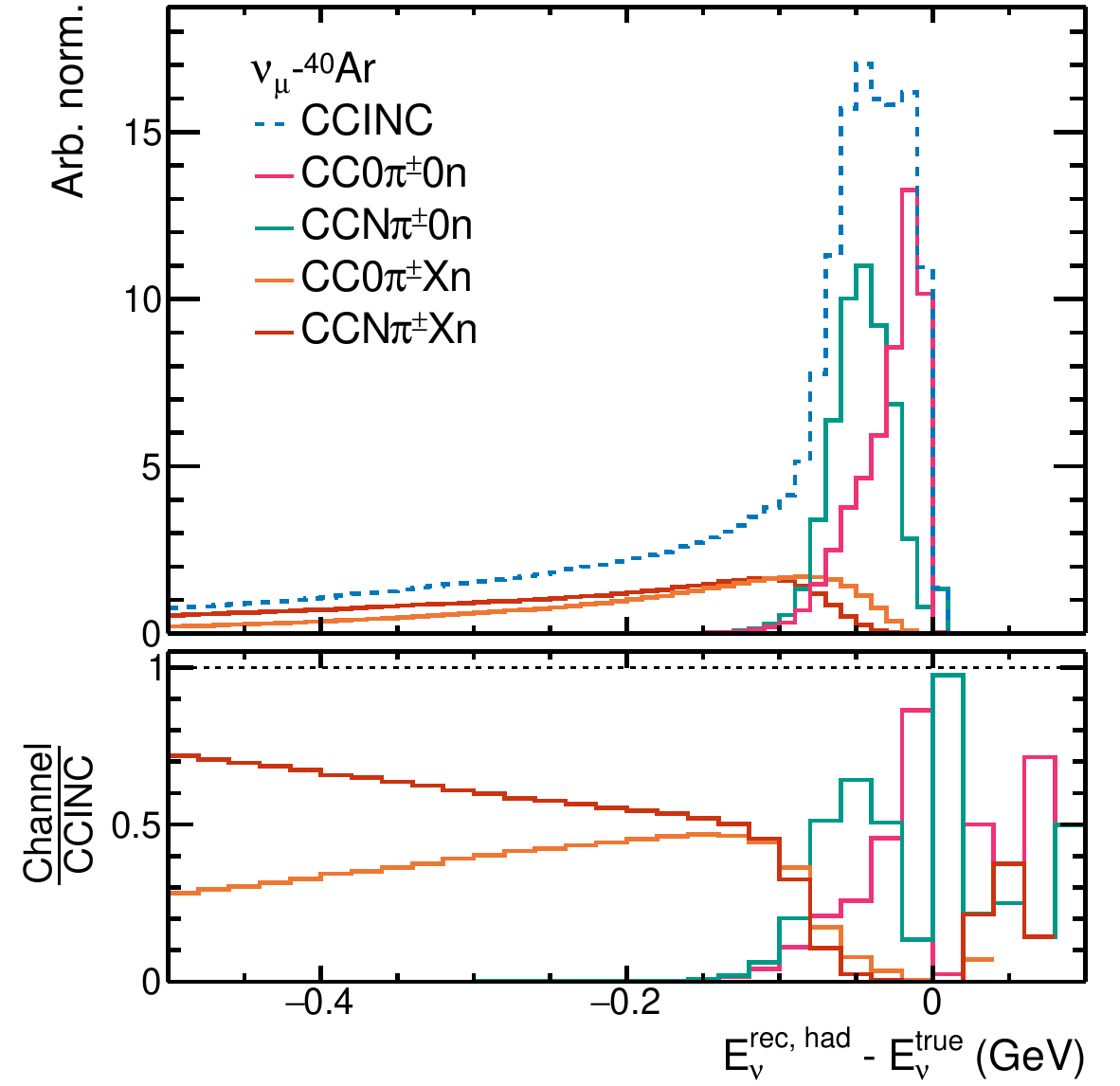}
      \caption{DUNE FD \numu--\argon CCINC, abs. bias}
  \end{subfigure}
  \caption{Comparison of the shapes of the relative and absolute bias in the reconstructed neutrino energy proxy with respect to the true neutrino energy at the FD of the DUNE and Hyper-K/T2K experiments from the NuWro 25 event generator. The different reconstructed neutrino energy proxies, nuclear targets and event topologies considered for DUNE and Hyper-K are the same as in \autoref{fig:enuspectcomp}. To best characterise sources of bias, the contributions to the cross section are broken down by topology for DUNE and interaction channel for Hyper-K/T2K.}
  \label{fig:enubiascomp}
\end{figure*}

Kinematic neutrino energy reconstruction, using \autoref{eq:enuqe}, is unbiased only in the case of CCQE neutrino interactions with a stationary nucleon bound with a fixed nuclear binding energy. 
The top panels of \autoref{fig:enubiascomp} show the bias in the reconstructed neutrino energy using \enuqe and illustrates the impact of deviations from these assumptions at Hyper-K/T2K FD. Even for CCQE interactions, there is significant spread in the distribution of bias from the Fermi motion of nucleons, whereas alterations in the binding energy from that assumed in \autoref{eq:enuqe} cause a shift in the mean of the distribution. 
Given that the $\sim$200 MeV/c scale of Fermi motion is comparable to the $\sim$600 MeV neutrino energies relevant to Hyper-K/T2K, the spread is significant. 
The slight asymmetry for CCQE interactions is due to a polarisation effect in which motion of the nucleon towards the incoming neutrino impacts the cross section for the interaction more than motion away from it due to the exponential rise in the CCQE cross section as a function of neutrino energy, as seen in \autoref{fig:oscxsecfluxcomp}.
For CCQE events, the binding energy shifts the peak with an absolute bias of a similar size to the differences between its assumed value in \autoref{eq:enuqe} and the real binding energy. 
Whilst this is only a few-MeV effect, Hyper-K sensitivity studies have shown that few-MeV energy-scale alterations can be sufficient to cause alterations to the inferred oscillated spectrum (the event rate spectra given by ~\autoref{eq:evtratenumu}~or \autoref{eq:evtratenue}) that have degeneracies with changes to the oscillation parameters~\cite{Munteanu:2022zla, Hyper-Kamiokande:2025fci}. 
Moreover, even with a perfect selection of CC0$\pi$ interactions (as used for \autoref{fig:enubiascomp}), non-QE interaction channels contribute. These are mainly 2p2h interactions and RPP interactions where the pion is absorbed through FSI. Applying \autoref{eq:enuqe} to these interactions does not give the correct neutrino energy, as illustrated in \autoref{fig:enubiascomp}.
The overall bias in the kinematic method is therefore subject to how robustly the non-QE contributions to CC0$\pi$ are modelled. 

\subsubsection{Bias in calorimetric neutrino energy reconstruction}
\label{subsec:biasinErechad}

Unlike the kinematic case, the top panels of \autoref{fig:enubiascomp} show that the calorimetric reconstructed neutrino energy using the proxy variable in \autoref{eq:enuhad} is almost always less than the true neutrino energy, as the visible energy should not be more than the incoming neutrino energy\footnote{The small number of events where the reconstructed energy is larger than true neutrino energy are due to occasional edge cases in the generators where energy is not properly conserved.}. Rather than separating these plots by interaction channel, they are split by interaction topologies to highlight which interactions have neutrons and pions in the final state. This demonstrates that the neutrino energy reconstruction bias is almost entirely driven by the energy carried away by neutrons, with any biases of $\gtrsim$100 MeV being almost exclusively driven by interactions containing neutrons. The majority of this bias corresponds to the total kinetic energy of neutrons produced in interactions, either directly from an interaction vertex (as can occur in, for example, RPP interaction with a $\Delta^+$) or ejected by nucleon or pion FSI. \autoref{fig:enubiascomp} shows that a large fraction of these neutrons are from pion-producing interactions. Additional small biases are introduced by the energy lost to overcoming the nuclear removal energy. This is responsible for almost all of the bias seen in the zero neutron topologies.  

\begin{figure*}[htbp]
  \centering
   \begin{subfigure}[b]{0.48\textwidth}
      \centering
      \includegraphics[width=1.00\linewidth]{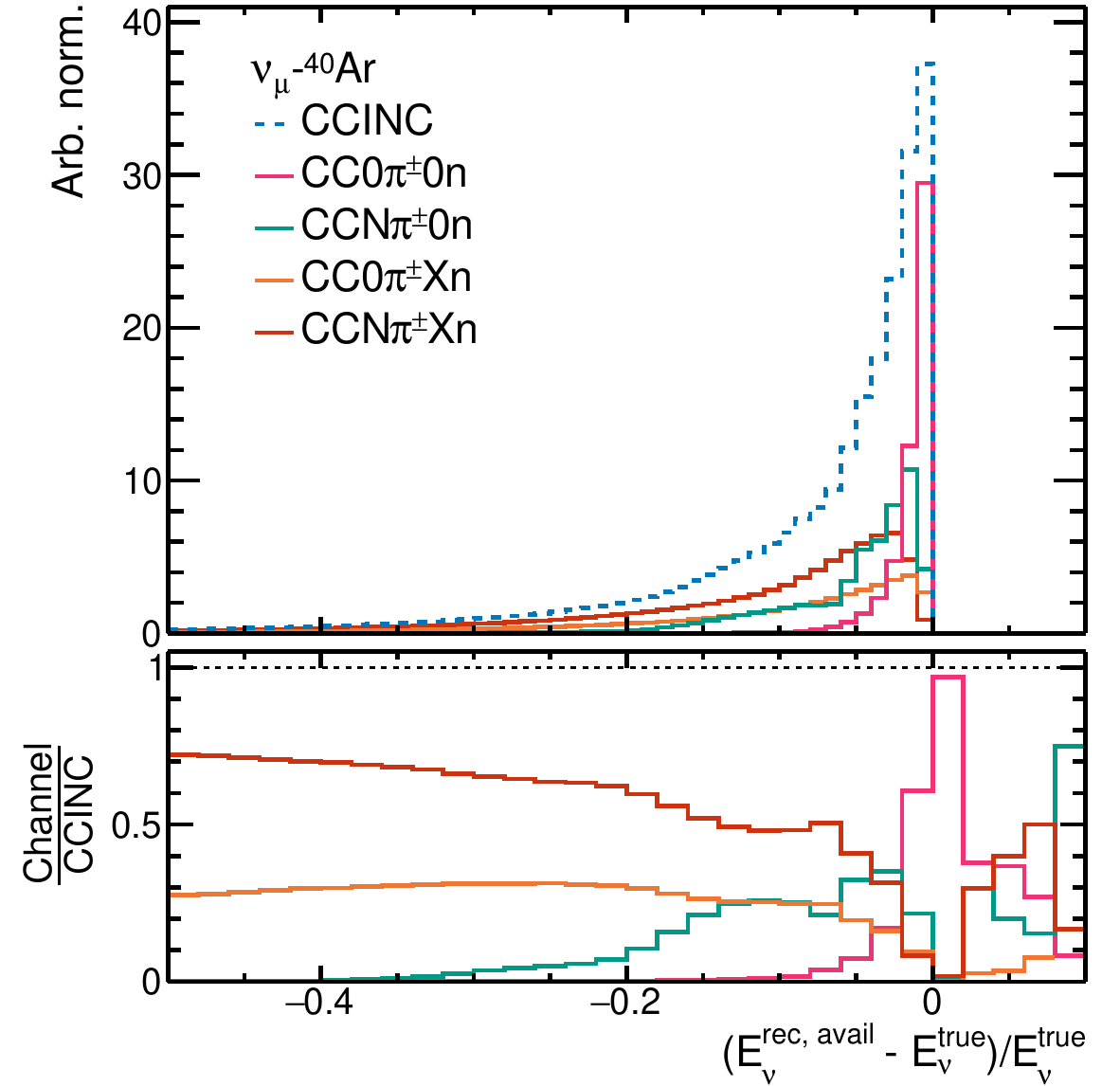}
      \caption{DUNE FD \numu--\argon CCINC, rel. bias}
  \end{subfigure}
   \begin{subfigure}[b]{0.48\textwidth}
      \centering
      \includegraphics[width=1.00\linewidth]{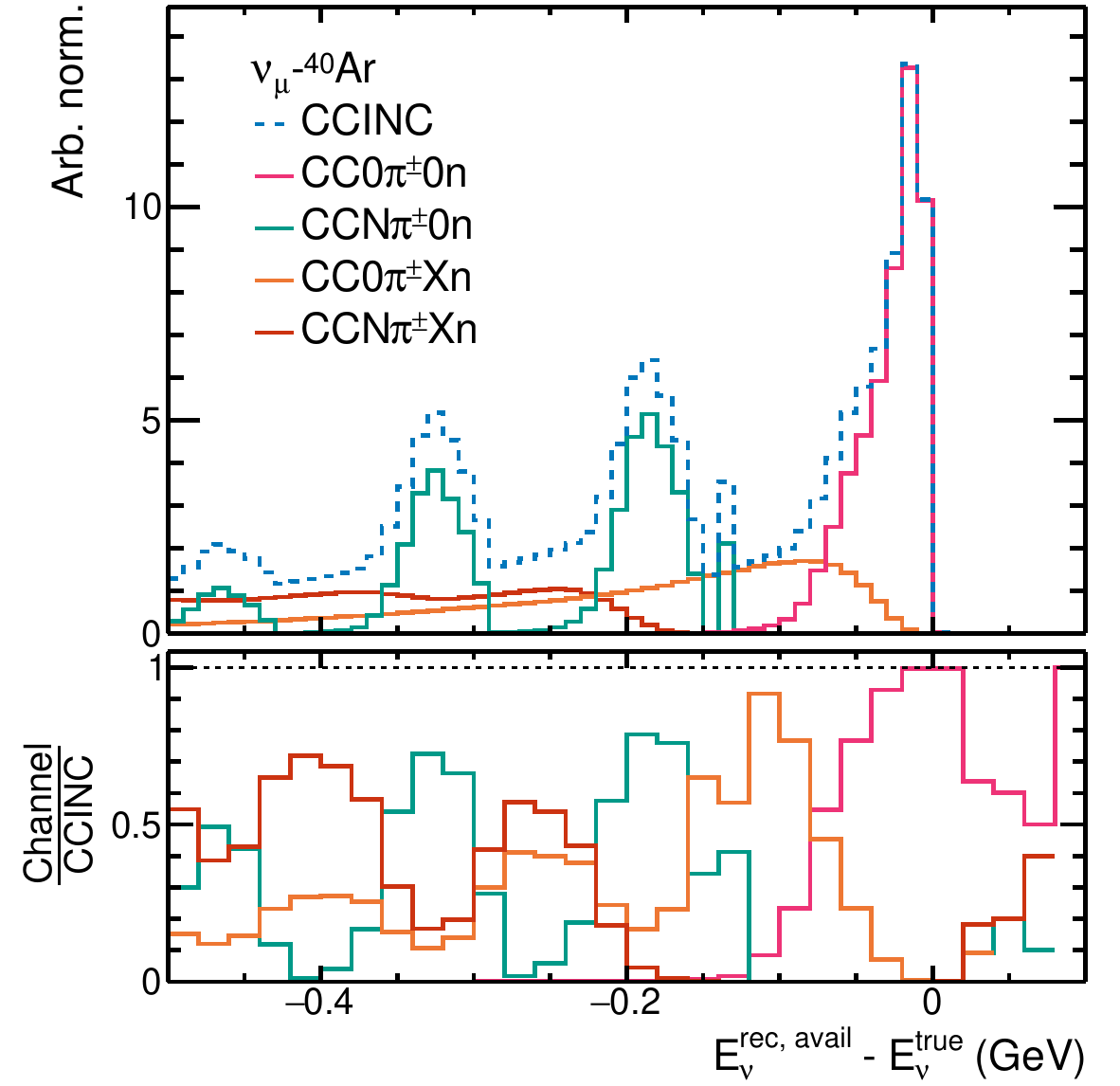}
      \caption{DUNE FD \numu--\argon CCINC, abs. bias}
  \end{subfigure}
  \caption{Comparison of the shape of the relative and absolute bias in the reconstructed neutrino energy proxy with respect to the true neutrino energy at the FD of the DUNE experiment from the NuWro 25 event generator as shown in \autoref{fig:enubiascomp}, but using $E^{\mathrm{rec},\;\mathrm{avail}}_{\nu}$ as defined in \autoref{eq:enuavail} rather than \autoref{eq:enuhad}.
  }
  \label{fig:enubiaswopions}
\end{figure*}

In the case of the proxy variable for neutrino energy given in \autoref{eq:enuavail}, the absolute and relative neutrino energy bias is shown in \autoref{fig:enubiaswopions}. This variable introduces an additional bias due to missing charged pion mass energy. This is responsible for the spikes seen in the absolute neutrino energy bias. These spikes are separated by the pion mass and correspond to interactions with different charged pion multiplicities. Their width is primarily due to binding energy in pion-producing interactions. Whilst this induces an additional bias with respect to the case of \autoref{fig:enubiascomp}, the overall bias is still dominated by energy carried away by neutrons.

Overall, the calorimetric neutrino energy reconstruction is very sensitive to the fraction of final-state hadronic energy carried by neutrons. This is particularly sensitive to the  details of FSI modelling, such as the probability for neutrons to be ejected at each step in an intranuclear cascade and the typical energy transferred to or from neutrons at those steps.

\subsubsection{Comparison of the bias across the two methods}

\begin{figure*}[htbp]
  \centering
   \begin{subfigure}[b]{0.48\textwidth}
      \centering
      \includegraphics[width=\linewidth]{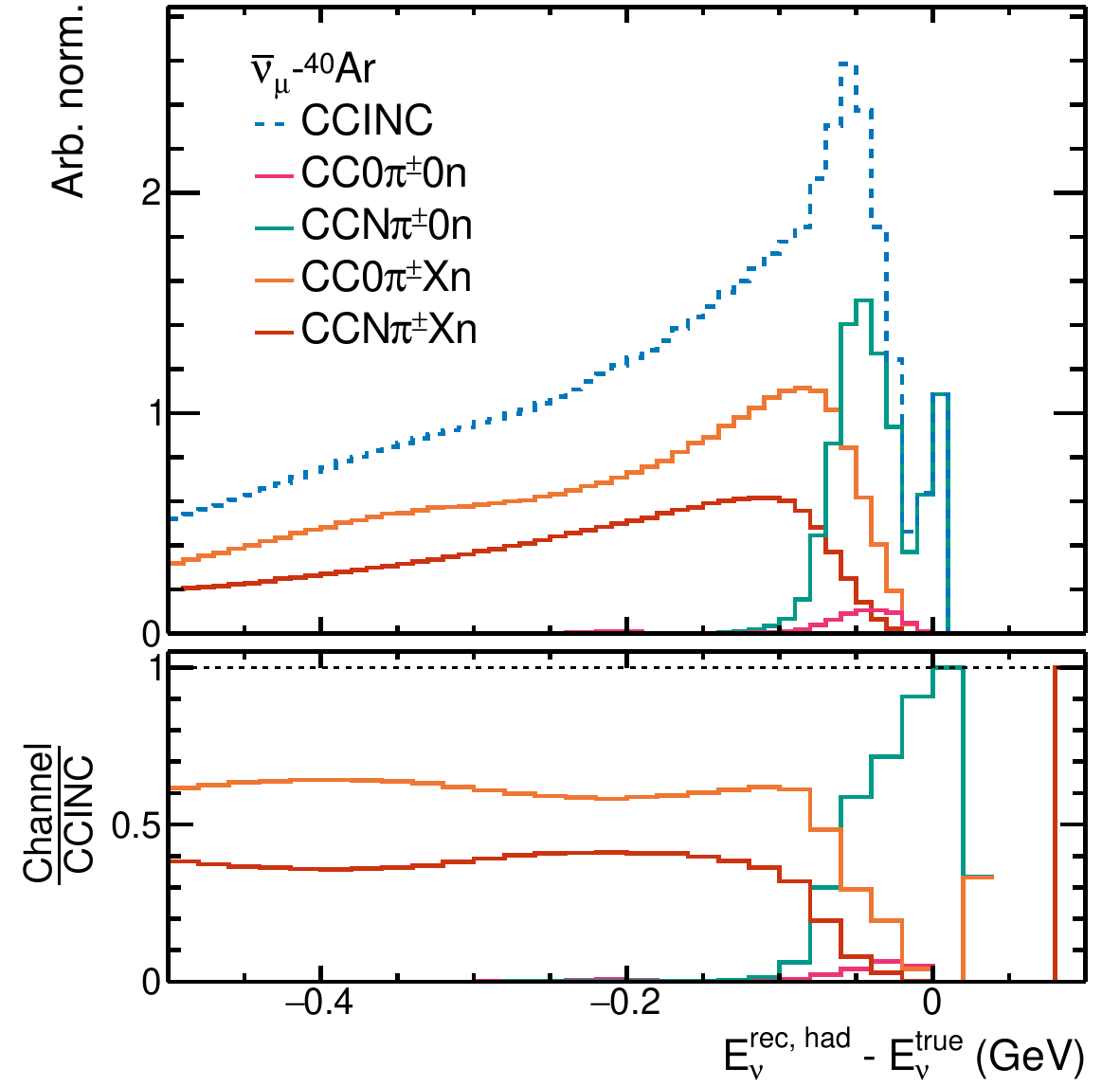}
      \caption{DUNE FD \numub--\argon CCINC abs. bias}
  \end{subfigure}
   \begin{subfigure}[b]{0.48\textwidth}
      \centering
      \includegraphics[width=\linewidth]{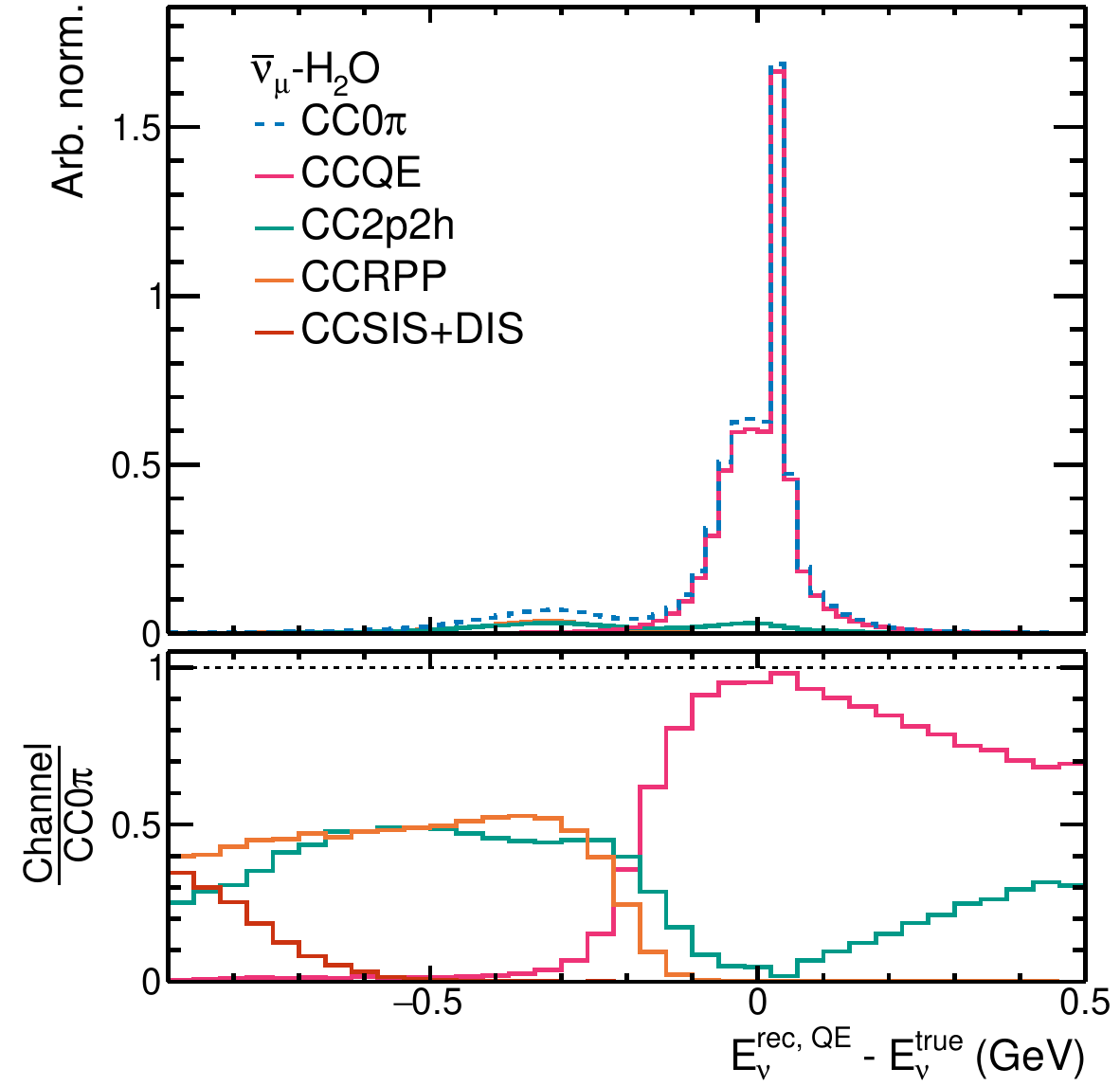}
      \caption{Hyper-K/T2K FD \numub--\water CC0$\pi$ abs. bias}
  \end{subfigure}
  \caption{Comparison of the shape of the absolute bias in the reconstructed neutrino energy proxy with respect to the true neutrino energy at the FD of the DUNE and Hyper-K/T2K experiments from the NuWro 25 event generator, as shown in \protect\autoref{fig:enubiascomp}, but for antineutrino interactions using antineutrino-enhanced beam running fluxes.}
  \label{fig:enubiasanucomp}
\end{figure*}

\autoref{fig:enubiascomp} compares the kinematic and calorimetric methods. It is clear that calorimetric neutrino energy reconstruction reduces the bias and spread through the use of more than just the outgoing lepton kinematics.
The calorimetric method is applicable to all CC interaction channels, whereas the kinematic method needs to be modified depending on the event topology, and is difficult to reliably apply for events with higher particle multiplicities.
The asymmetry between neutrino and antineutrino energy reconstruction is very different between the two methods. \autoref{fig:enubiasanucomp} shows that the performance of the calorimetric method is significantly degraded for antineutrino interactions, where the pre-FSI interaction topologies are much more likely to contain neutrons. Conversely, the kinematic method has similar performance for neutrino and antineutrino interactions, therefore leaving it  with better resolution in the latter case.
The kinematic method slightly improves for antineutrino interactions on targets containing hydrogen, such as C$_8$H$_8$ or H$_2$O. The hydrogen is responsible for the spike just above zero in \autoref{fig:enubiasanucomp}\footnote{It is not precisely at zero because \autoref{eq:enuqe} assumes a binding energy, whereas for hydrogen there is no binding energy.}.

Although a neutrino energy reconstruction method with small bias ultimately improves the resolution to neutrino oscillation parameters, it is at least as important that the size and shape of the bias under modifications to the neutrino interaction model is small and well understood. A reconstruction method that seems to provide better resolution and smaller bias may still result in large uncertainties for neutrino oscillation measurements if the bias is not well known. 
In general, the calorimetric method relies on the detailed predictions of hadron kinematics, which are typically poorly understood due to the complexity of the relevant nuclear effects, whereas the kinematic method is less sensitive to these issues. A detailed evaluation of the plausible alterations to the bias is discussed in \autoref{sec:impactonoa}.

\subsection{Sample migration and backgrounds}
\label{subsec:samplemigration}

Event selections used for neutrino oscillation analyses suffer from backgrounds whose size and shape are sensitive to neutrino interaction modelling. Backgrounds include the contribution of low momentum \textit{soft} pions to the CC$0\pi$ selections used by T2K and Hyper-K, \textit{wrong-sign} neutrinos (neutrinos entering antineutrino selections or vice-versa), wrong flavour neutrinos, NC interactions, radiative photons and intrinsic electron (anti)neutrino contamination to neutrino beams. Such backgrounds are projected to make up 10--65\% of the event selections at DUNE and Hyper-K~\cite{DUNE:2020jqi,Hyper-Kamiokande:2025fci}. A breakdown of these backgrounds from the experiments is shown in \autoref{tab:bkgs}, which includes detector effects.

\begin{table*}[hbtp]
\centering
\begin{threeparttable}
\begin{tabular}{l c|cccc|c}
\hline\hline
Experiment & Selection                             & NC    & Wrong sign    &  Wrong flavour   & Beam $\nueany$ & Total bkgd.\\ \hline
\multirow{4}{*}{DUNE} & $\nu_\mu\rightarrow\nu_\mu$              & 2.6\% & 6.7\% & 0.8\% & N/A          & 10.1\%   \\
     &$\bar{\nu}_\mu\rightarrow\bar{\nu}_\mu$  & 2.5\% & 36.2\%  & 0.8\% & N/A          & 39.5\%   \\
     & $\nu_\mu\rightarrow\nu_e$                & 4.7\% & 0.8\% & 2.8\% & 12.8\%         & 21.2\%   \\
     & $\bar{\nu}_\mu\rightarrow\bar{\nu}_e$    & 8.4\% & 20.0\%  & 5.9\% & 30.5\%         & 64.8\%   \\ \hline
\multirow{4}{*}{Hyper-K} & $\nu_\mu\rightarrow\nu_\mu$             & 3.5\% & 5.3\% & 0.03\% & N/A        & 8.8\%  \\
 &$\bar{\nu}_\mu\rightarrow\bar{\nu}_\mu$ & 2.6\% & 34.3\%& 0.01\% & N/A        & 36.9\% \\
 &$\nu_\mu\rightarrow\nu_e$\tnote{1}             & 5.9\% & 0.5\% & 0.3\% & 12.2\%     & 18.9\% \\
 &$\bar{\nu}_\mu\rightarrow\bar{\nu}_e$   & 9.8\% & 14.9\% & 0.6\% & 23.4\%   & 48.6\% \\\hline\hline
\end{tabular}

\begin{tablenotes}
\item[1]{For Hyper-K, $\mathit{\numu\rightarrow\nue}$ includes both the electron neutrino appearance event selections from Ref.~\cite{Hyper-Kamiokande:2025fci} (i.e. with and without a decay electron).} 
\end{tablenotes}
\end{threeparttable}
\caption{The breakdown of background contributions to the simulated DUNE and Hyper-K FD event selections used in oscillation analyses, as reported in Refs.~\cite{DUNE:2020jqi,Hyper-Kamiokande:2025fci}. 
Backgrounds are interactions that are not oscillated CC events of the right sign and flavour. 
NC includes all neutrino flavours.
Wrong sign refers to CC events with the right flavour but wrong sign. Wrong flavour refers to CC events with any sign that has the wrong flavour.
Beam \nueany refers to CC backgrounds from intrinsic electron neutrinos \textit{and} antineutrinos. The \textit{soft pion backgrounds} for Hyper-K discussed in the text are not included here.
The simulations assume normal neutrino mass ordering and $\deltacp=-\pi/2$, other than the DUNE $\nu_\mu$ and $\bar{\nu}_\mu$ samples which use $\deltacp$~=~0 (but these samples have minimal sensitivity to $\deltacp$). 
The DUNE selections only include interactions between 0.5 and 10 GeV reconstructed neutrino energy.
\label{tab:bkgs}}
\end{table*}

\subsubsection{Neutral current interactions}
One source of background is NC interactions that produce a particle which is misidentified as a charged lepton. For example, an NC$1\pi^{\pm}$ interaction in which the charged pion is misidentified as a charged muon might be classified as a $\nu^{\bracketbar}_\mu$ CC interaction, or an NC$1\pi^0$ interaction in which a $\pi^0\rightarrow\gamma\gamma$ decay photon may be mistaken for an electron causing the event to be classified as a $\nu^{\bracketbar}_e$ CC interaction. ND constraints on these contributions are challenging as ND RPP and SIS events are dominated by $\nu^{\bracketbar}_{\mu}$ CC1$\pi^{\pm}$ events, and the relationship between pion production topologies is difficult to model, as discussed in \autoref{sec:spp}.

The Hyper-K and DUNE experiments project NC backgrounds at the level of 3--10\% after selection cuts (see \autoref{tab:bkgs}). Since NC backgrounds are produced independently of the flavour and sign of the incoming neutrino, the largest relative contribution is in simulated event selections where the signal flux and cross section are lower, as is the case for selections of antielectron neutrino appearance events.

\subsubsection{Wrong-sign backgrounds}
\label{subsubsec:wsbkg}

The magnetic horns used in modern accelerator beams are capable of efficiently selecting neutrinos or antineutrinos depending on the chosen mode of experimental operation but some backgrounds from wrong-sign (anti)neutrinos remain. 
Facilities operating in neutrino-enhanced mode are typically able to produce neutrino beams with only 2--5\% contamination from wrong-sign antineutrinos. In antineutrino-enhanced mode the wrong sign contribution is significantly higher due to the production of more positive than negative secondaries. Coupled with the significantly higher cross section for neutrinos over antineutrinos, wrong-sign backgrounds contribute at the 20--40\% level for event selections using antineutrino-enhanced beams (see \autoref{tab:bkgs}). The exact size and shape of these contributions vary as a function of neutrino energy and depend on the details of the interaction model used to predict them, for which there are significant uncertainties, as well as the detector performance and design. For example, in the case of antineutrino-enhanced mode for Hyper-K/T2K, the range of event generators used throughout this work predict a wrong-sign contribution of the CC$0\pi$ event rate between $\sim$20--30\% of the right-sign CC$0\pi$ event rate in the energy region of the neutrino oscillation maximum ($\sim$0.6 GeV).

These wrong-sign contributions introduce a specific challenge for CP violation measurements, for which the sensitivity is related to the relative event rates between neutrino and antineutrino modes. However, the impact of this background can be reduced in a number of ways, including: improvements in beamline modelling; constraining the background \textit{in situ} using magnetised near detectors which separate right- from wrong-sign (anti)neutrino interactions by identifying the charge of the outgoing lepton; or, through novel techniques such as the recent addition of Gadolinium to the T2K FD, which has dramatically increased its ability to tag final-state neutrons in neutrino interactions.

\subsubsection{Wrong flavour backgrounds}

Imperfect event selections at experiments' FDs can lead to contamination from interactions of the wrong flavour. As shown in \autoref{tab:bkgs}, this is typically very small, occurring mostly in cases where a muon (anti)neutrino interaction produces a low momentum muon alongside a photon, which is reconstructed as an electron (for example, from $\pi^0\rightarrow\gamma\gamma$ decay). How often this occurs depends on the cross section for such interactions and on the proportion of the energy carried away by the outgoing photon. Such interactions represent a corner of neutrino--nucleus interaction kinematic phase space that is difficult to model. As was highlighted for the NC case, it is challenging to constrain the $\pi^0$-production cross section from RPP and SIS interactions with ND data due to different modelling approaches to relating $\pi^0$-production to the dominant $\nu^{\bracketbar}_{\mu}$ CC1$\pi^{\pm}$ topology.

For DUNE, $\nu^{\bracketbar}_{\tau}$ CC interactions cause an additional background in muon and electron (anti)neutrino selections, due to leptonic tau decays with electrons or muons in the final state~\cite{DUNE:2020jqi}. Tau (anti)neutrino cross sections are not well studied, however, this background is small at few-GeV energies due to the suppression of the $\nu^{\bracketbar}_{\tau}$ CC interactions from the large tau mass, with a threshold of $\sim$3.5 GeV.

\subsubsection{Beam $\nu^{\bracketbar}_e$ backgrounds}

Intrinsic electron (anti)neutrino contributions to the neutrino beam from the decay of muons (produced by pion decay) or kaons in the neutrino beam decay tunnel introduce a background to electron (anti)neutrino appearance event selections. Whilst this contamination is small for both DUNE and Hyper-K/T2K, at the percent level~\cite{T2K:2012bge, DUNE:2020ypp}, it is still significant compared to the small electron (anti)neutrino appearance samples from $\nu_{\mu}^{\bracketbar} \rightarrow \nu_{e}^{\bracketbar}$ oscillation. Indeed, \autoref{tab:bkgs} shows that this is the dominant background for electron (anti)neutrino appearance event selections. This background has a significantly different neutrino energy spectrum to the muon (anti)neutrino component of the beam, making constraints from ND muon (anti)neutrino measurements challenging (as described in \autoref{subsec:NDtoFD}). Direct measurements of this intrinsic background at a ND is possible, but low statistics and backgrounds from photon-producing muon (anti)neutrino interactions make this non-trivial (see, for example, Ref.~\cite{T2K:2020lrr}).

\subsubsection{Soft pion backgrounds in CC$0\pi$ selections}

For Hyper-K/T2K analyses in which kinematic neutrino energy reconstruction is used, a core strategy is to isolate event samples which are dominated by specific interaction channels. The most important of these are events with a single reconstructed Cherenkov ring in the FD. These single-ring samples are dominated by CC$0\pi$ events, but events with real soft pions enter the sample due to detector reconstruction thresholds or potentially through secondary interaction processes which make positive identification of the real pion challenging. Some of the below-threshold pion background can be removed through identification of delayed Michel electrons~\cite{MINERvA:2014ogb,T2K:2016cbz,T2K:2023smv,MiniBooNE:2010eis,Super-Kamiokande:2023ahc}, coming from $\pi^+\rightarrow\mu^+\rightarrow e^+$ decays inside the detector, but this is not 100\% efficient for $\pi^{+}$ and is unreliable for $\pi^-$ identification as the $\pi^-$ is frequently captured before decay~\cite{Ashery:1986nt}.

\begin{figure}[htbp]
  \centering
  \centering
  \includegraphics[width=0.98\linewidth]{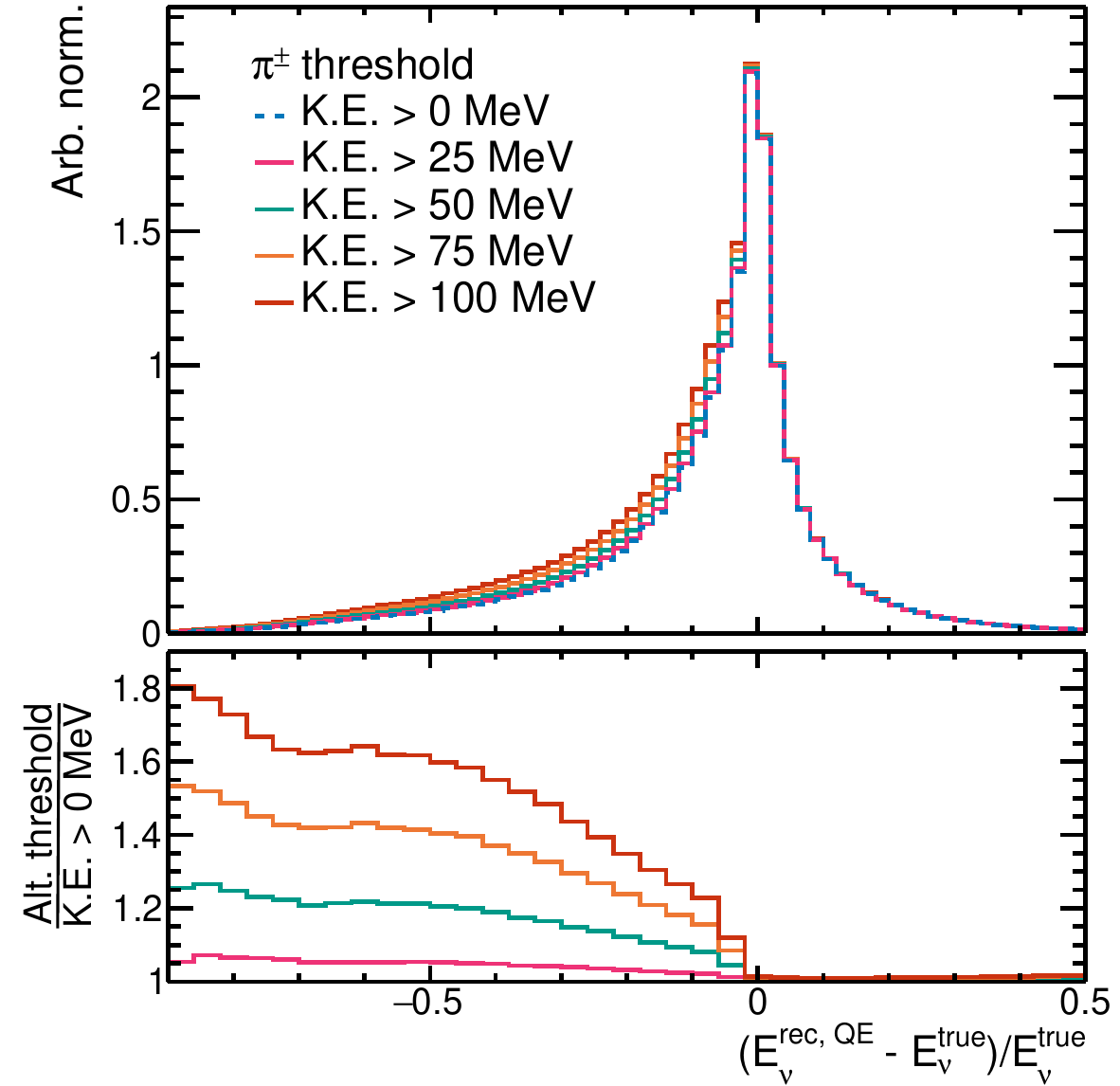}
  \caption{Neutrino energy reconstruction bias for \numu--\water CC0$\pi$ \textit{plus} a representative CC$N\pi^{\pm}$ background in which the final-state charged pions are below a given kinetic energy (K.E.) threshold, for a variety of thresholds. Interactions are simulated with the oscillated Hyper-K/T2K FD flux and the NEUT 580 event generator configuration.}
  \label{fig:cc0pimigration}
\end{figure}

Real pions which are produced but not identified in the detector bias the kinematic neutrino energy reconstruction in the same way as CCnon-QE events in which pions are absorbed through FSI before escaping the nucleus (discussed in \autoref{subsec:enurec}). \autoref{fig:cc0pimigration} shows how the neutrino energy bias for Hyper-K/T2K changes with different pion reconstruction thresholds. The higher the threshold, the more the high-bias tail of the distribution is populated by soft non-CC0$\pi$ events. For a detector sensitive to pions with kinetic energy above 100 MeV (approximately the Cherenkov threshold for pions in water) and operating in the Hyper-K/T2K neutrino beam, a CC0$\pi$ sample would have $\sim$12\% contributions from unseen soft pions which increase the high-bias tail by 30--50\%.

\subsubsection{Radiative photons}
As discussed in \autoref{sec:models}, radiative corrections of the tree-level neutrino interaction cross section can introduce real photon emission. Hard photons can lead to the identification of secondary electromagnetic signatures or electron-like Cherenkov rings in experimental event selections. This is a particular challenge for electron (anti)neutrino appearance searches where additional electromagnetic reconstructed objects can lead to the misclassification of true $\nu^{\bracketbar}_e$ interactions, or the mis-identification of $\nu^{\bracketbar}_\mu$ topologies as $\nu^{\bracketbar}_e$ events~\cite{Tomalak:2022xup, Tomalak:2021hec}.
Since the probability of event misidentification is expected to be dependent on the type of interaction ($\nu^{\bracketbar}_e$CC$0\pi1\gamma$ events are more likely to be misclassified as NC$1\pi^{0}$ interactions than $\nu^{\bracketbar}_\mu$CC$0\pi 1\gamma$ events), and the momentum transfer (higher $Q^{2}$ means increasing likelihood of RPP or SIS/DIS contributions), these effects could become a significant source of uncertainty for neutrino oscillation analyses in future experiments. The size of this background in realistic experimental event selections remains unquantified.

\section{Current constraints on neutrino interaction models}
\label{sec:constraints}

As introduced in \autoref{sec:models}, simulating the details of few-GeV neutrino--nucleus interactions requires a variety of modelling approximations. An open question is whether these approximations are {\it good enough} for accelerator neutrino oscillation experiments, without introducing limiting biases into an analysis. The overall validity of neutrino interaction models, including all of these necessary approximations, can be assessed by confronting the predictions of the models with measurements of few-GeV neutrino interactions made with beams of unoscillated neutrinos; either by an oscillation experiment's ND or in separate dedicated neutrino-scattering experiments. 
Beyond neutrino scattering, there is also data from electron- and hadron-scattering experiments, which can be used to measure the vector component of the weak interaction and provide additional handles on characterising a range of important nuclear effects, such as initial-state nuclear models and FSI.

Dedicated reviews of how recent neutrino interaction models compare to measurements~\cite{NuSTEC:2017hzk,Katori:2016yel,Avanzini:2021qlx,Betancourt:2018bpu,Mahn:2018mai} broadly conclude that:
\begin{itemize}
    \item no model has been shown to give a quantitatively good description of global neutrino scattering measurements; 
    \item models generally agree with measurements of low-multiplicity processes better than high-multiplicity processes;
    \item models agree with measurements of lepton kinematics better than hadron kinematics or correlations between lepton and hadron kinematics;
    \item models agree with measurements of interactions on carbon better than interactions on heavier targets.
\end{itemize}

In this section, we re-emphasise these conclusions through a broad overview of how model predictions compare to global measurements. We further compare a subset of the generator configurations from \autoref{sec:models} to a few select measurements to demonstrate their capacity to predict the cross sections of different interaction channels and observed topologies.

\subsection{Historical context}
\label{subsec:nuinthistory}

The field of GeV-scale neutrino--nucleus interaction cross-section measurements has grown rapidly over the last two decades.
A seminal review paper from 2013~\cite{Formaggio:2012cpf} reported that ``most of our knowledge of [GeV-scale] neutrino cross sections comes from measurements conducted in the 1970’s and 1980’s using either bubble chamber or spark chamber detectors'', which collected a few 10s, to a few 10,000s, of interactions in different channels. Through a broad community effort over the past decade, this situation has radically changed.

A plethora of new experiments and advances in neutrino beam technology has allowed for megawatt beams and neutrino detectors often exceeding 10 ton active mass, meaning that measurements are no longer necessarily dominated by statistical uncertainties. 
For example, in 2010, MiniBooNE measured the $\nu_\mu$ CC0$\pi$ and CCQE\footnote{The CCQE cross section relied on subtracting the non-QE background from a CC0$\pi$ event selection, which depends strongly on the input model assumption, so the ancillary CC0$\pi$ presented in an Appendix to Ref.~\cite{MiniBooNE:2010bsu} is now considered to be the key result of this work.} cross sections on mineral oil using 146,070 interactions~\cite{MiniBooNE:2010bsu}, increasing the statistical power compared to bubble chamber measurements by an order of magnitude.
Just over 10 years later, a MINERvA analysis of a similar interaction topology used more than one \textit{million} interactions~\cite{MINERvA:2019gsf}, and NOvA's recent cross-section measurements contain similarly-sized samples (for example, Ref.~\cite{NOvA:2021eqi}). 
The recently started SBND experiment will allow neutrino--argon measurements with similar statistics~\cite{SBND:2025lha}.

Beyond higher statistics, modern detectors have been designed to be sensitive to both the leptonic and hadronic final state of neutrino--nucleus interactions, allowing measurements that can more precisely characterise complex nuclear effects.
For example, measurements of imbalances between the outgoing lepton and hadron kinematics on the plane transverse to the incoming neutrino in CC$0\pi$ and CC1$\pi$ final states from T2K~\cite{T2K:2018rnz,T2K:2021naz}, MINERvA~\cite{MINERvA:2018hba,MINERvA:2019ope,MINERvA:2020anu}, MicroBooNE~\cite{MicroBooNE:2023krv,MicroBooNE:2023cmw,MicroBooNE:2023tzj} and ICARUS~\cite{ICARUS:2026nbg} have provided some separation of nuclear ground state effects from those related to FSI and multi-nucleon interactions~\cite{lu:2015hea,Furmanski:2016wqo,Baudis:2023tma, Dolan:2018zye,Filali:2024vpy,Dolan:2018sbb}. MINERvA has also used variables characterising imbalances between the leptonic and hadronic final state to isolate interactions on hydrogen in plastic scintillator, thereby reporting a cross section devoid of nuclear effects~\cite{MINERvA:2023avz}. 

Experiments have additionally made \textit{joint} and \textit{ratio} measurements in order to study the energy, flavour and nuclear-target dependence of neutrino interaction cross sections. 
T2K has made a number of correlated measurements of neutrino interactions in the same suite of detectors under different running conditions (such as muon neutrinos and antineutrinos~\cite{T2K:2020txr, T2K:2020sbd}), on different nuclear targets (for example, C$_{8}$H$_{8}$ and $\text{H}_2\text{O}$~\cite{T2K:2020jav}), as well as similar measurements using different detectors in the same neutrino beam (such as the INGRID and ND280 detectors~\cite{T2K:2023qjb}).
MINERvA's measurements of cross-section ratios between different nuclear targets for the same final state~\cite{MINERvA:2016oql,MINERvA:2022djk,MINERvA:2022esg,MINERvA:2023kuz,MINERvA:2025tem} share a similar goal, and measurements of the same topology (CC0$\pi$ and CCINC) but with different neutrino beam profiles (low energy \textit{LE}~\cite{MINERvA:2018vjb, MINERvA:2020zzv, MINERvA:2018hqn,MINERvA:2015ydy} and medium energy \textit{ME}~\cite{MINERvA:2022bno,MINERvA:2021owq, MINERvA:2019gsf,MINERvA:2021wjs, MINERvA:2025tem}) additionally probe the energy dependence of the interactions.

Along with a large influx of new measurements over the last decade, there has been an active push within the community to develop and apply new cross-section extraction methods which improve the power or robustness of results~\cite{Abe:2025yaa, Andreassen:2019cjw, Koch:2019jqr, Koch:2020oyf, Koch:2021yda, Koch:2022qsz, Koch:2024tit, Gardiner:2024gdy, Tang:2017rob, Huang:2025ziq, Licciardi:2021kzl}. A key issue is model dependence---the degree to which an extracted result may be biased by assumptions made in the analysis. Despite this ongoing work, at present there is no agreed upon approach for minimising the potential for model dependence, or for testing for its impact on published results. Interpreting the growing body of neutrino interaction measurements and validating or tuning cross-section model predictions with them are complicated by this situation. Attempts to compare models to measurements across multiple experiments and interaction topologies find significant tension between them~\cite{Betancourt:2018bpu, Avanzini:2021qlx, Mosel:2017nzk, Sobczyk:2014xza, Filali:2024vpy, Wolfs:2025ofb}. However, it is not obvious how much of the tension is from missing or inadequate physics in the interaction models, or deficiencies in the global dataset. This is a major challenge for extracting data-driven constraints on neutrino interaction uncertainties---incorrect models can lead to biased measurements which are then tuned to those models, a potentially degrading circularity.

Some conventions have been near-universally adopted by experiments producing neutrino cross-section measurements. One example is the use of signal definitions which are defined in terms of observable particles, such as CC0$\pi$, rather than primary processes which require explicitly model-dependent corrections, such as CCQE. Another example is only making measurements which are differential in observable particle kinematics\footnote{This refers to final-state particles observable in a detector, usually including limits on their kinematics to be within a detector's acceptance.}, rather than applying model-dependent corrections to measure unobservable variables, such as neutrino energy and four-momentum transfer. 
A modern cross-section measurement might be made in bins of \textit{reconstructed} four-momentum transfer, built using only observable lepton kinematics with the assumption of a CCQE interaction, $Q^2_{\mathrm{QE}}$ (see \autoref{eq:enuqe}), rather than relying on a particular model to report a measurement in \textit{true} $Q^2$. Other widely adopted conventions may need to be rethought to maximise the utility of future measurements. For example, almost all recent cross-section measurements report their uncertainties under a Gaussian approximation, which is not expected to hold for most neutrino cross-section measurements that are limited by systematic uncertainties~\cite{DAGOSTINI1994306, Chakrani:2023htw, Abe:2025yaa}.

The growing awareness of the challenges in extracting robust cross-section measurements has led to experiments releasing significantly more information about their analysis methods, which in turn has driven work towards developing methodological improvements. For example, recent measurements by T2K~\cite{T2K:2018rnz,T2K:2020sbd,T2K:2020txr,T2K:2023qjb} and MicroBooNE~\cite{MicroBooNE:2021sfa, MicroBooNE:2023krv, MicroBooNE:2024tmp} have provided \textit{unregularised} results, which removes bias and underestimation of uncertainties in the process of \textit{unfolding} the cross section (the transformation from reconstructed-level detector quantities to physics variables)~\cite{Gardiner:2024gdy,Koch:2022qsz}. Comprehensive data releases are important tools for future-proofing results. For example, the MINERvA open dataset~\cite{Fine:2020snd,osti_3022562} allows subsequent data reanalysis with new methods, potentially providing a blueprint for how the circular problem of cross-section modelling choices impacting cross-section results can be resolved.

\subsection{Alternative probe data}
\label{subsec:altprobes}
In addition to neutrino-scattering data, aspects of neutrino--nucleus interaction modelling can also be studied by benchmarking against measurements of electron, photon, and hadron scattering. 

\subsubsection{Electron scattering}
Electron and neutrino scattering experiments probe nucleons or nuclei in similar ways. Unlike current neutrino beam experiments, where the initial state neutrino energy is not known, electron scattering experiments measure both the initial- and final-state lepton, thereby allowing the momentum transfer of the interaction to be well constrained~\cite{Amaro:2019zos, Ankowski:2022thw}\footnote{Radiative corrections to the cross section~\cite{Mo:1968cg, Afanasev:2023gev}, which account for the radiation of real photons or exchange of virtual photons, must be considered to infer energy or momentum transfer from observed scattered electrons. Electron scattering results often come with a correction applied, which is model dependent, and is an important caveat to the often stated assertion that electron-scattering measurements are clean.}. 
This permits measurements targeted to precise regions of the kinematic phase space to study individual nuclear effects. 
However, electron scattering is only able to constrain the vector current component of weak interactions, leaving the axial and pseudoscalar components of neutrino interactions unconstrained. 
A successful description of precision electron scattering measurements therefore provides a powerful but incomplete benchmark for neutrino scattering models.

Electron-scattering data has been used to provide constraints on a variety of nuclear effects which have been used to build or tune neutrino-scattering models. For example, the knowledge of the incoming electron energy allows detailed inferences of nuclear spectral functions, multi-nucleon correlations, and their impact on overall leptonic interaction cross-sections~\cite{CLAS:2005ola,Hen:2014nza}, and nuclear transparency~\cite{Ershova:2023dbv,Niewczas:2019fro,CLAS:2025fqh,Mosel:2019vhx}. 
Recently, the GENIE, NEUT and NuWro event generators have been modified to simulate electron-scattering data and have been compared to such measurements, revealing modelling issues and motivating new treatments of systematic uncertainties for oscillation measurements~\cite{electronsforneutrinos:2020tbf,Abe:2024avs,Ankowski:2020qbe,Ershova:2023dbv,Niewczas:2019fro,CLAS:2025fqh}. GiBUU has been benchmarked against electron scattering measurements for much longer (see, for example, Refs.~\cite{Leitner:2007px, Leitner:2008fg, Leitner:2008ue, Leitner:2009ke, Lalakulich:2013tca, Mosel:2016inz, Mosel:2017anp}).
Electron (and photon) scattering data have also been used to probe resonance excitations in RPP interactions, for example constraining their interferences and relative phases~\cite{SajjadAthar:2022pjt, Nakamura:2015rta, Kabirnezhad:2020wtp, Kabirnezhad:2024cor, Sobczyk:2018ghy}, and to study nuclear PDFs in DIS interactions~\cite{Kovarik:2010uv,Muzakka:2022wey,Paukkunen:2013grz,Eskola:2016oht}. 
However, being largely motivated by nuclear physics rather than neutrino physics, the range of historical measurements do not cover the full phase space and range of nuclear targets which are of interest for current and future neutrino oscillation experiments~\cite{Ankowski:2022thw}. They also typically either do not measure the outgoing hadronic system or, when they do, are limited to studying very narrow regions of phase space to investigate specific effects. Moreover, many of these historical measurements provide incomplete data releases (for example they may be missing a full evaluation of systematic uncertainties) and/or apply model-dependent corrections for nuclear or radiative effects based on superseded models~\cite{Arrington:2011dn, Ye:2017gyb}.

A new generation of electron-scattering measurements is being conducted at Jefferson lab (JLab) and the Mainz Microtron (MAMI) with the explicit purpose of providing measurements to improve neutrino--nucleus interaction modelling. 
These experiments cover a wider range of kinematic phase space and nuclear targets than historical measurements.
Results from the CLAS experiments at JLab~\cite{Burkert:2020akg,Arrington:2021alx,CLAS:2003umf}, which can measure both the leptonic and hadronic final state of interactions, have significantly contributed to the understanding of neutrino energy reconstruction and the impacts of nuclear effects through measurements of few-GeV electron scattering~\cite{CLAS:2025fqh, CLAS:2022odn, CLAS:2021neh, CLAS:2018xsb}. 
Such measurements have provided a novel means to benchmark important components of neutrino interaction models. Current comparisons of event generator predictions to these measurements demonstrate unsatisfactory agreement, exposing clear deficiencies in their predictions~\cite{CLAS:2021neh,Isaacson:2025cnk}.

\subsubsection{Hadron scattering}
Hadron-scattering measurements play a major role in constraining the FSI models~\cite{Dytman:2021ohr,Nikolakopoulos:2022qkq,Niewczas:2019fro, PinzonGuerra:2018rju}.
Although FSI describes hadron scattering \textit{inside} the nucleus, they can be related to \textit{external} hadron--nucleus scattering~\cite{Salcedo:1987md}. For example, the DUET experiment~\cite{DUET:2015ybm, DUET:2016yrf} made measurements of several exclusive pion--nucleus interaction channels (such as pion--nucleus charge exchange or pion absorption), directly probing those considered in classical nuclear cascade models.
These complement a rich history of measurements on different target materials, incoming pion type, and reaction channel~\cite{Allardyce:1973ce,Binon:1970ye,Ingram:1982bn,Jones:1993ps, Ashery:1981tq, Nakai:1980cy,Bellotti:1973vn, Navon:1983xj, Cronin:1957zz, Wilkin:1973xd, Clough:1974qt, Carroll:1976hj, LADS:2001clz, LADS:2000pzh, LADS:1999dyv, LADS:1998pec, Kotlinski:1998vh}. 
Similar historical measurements are also available for proton--nucleus scattering~\cite{Wilkins:1963zza,Dicello:1970mx,MAKINO1965378,Slaus:1975zz,McGill:1974zz,Renberg:1972jf,PhysRev.95.1268,GOLOSKIE1962474,Chen:1955nkq}. 
Measurements from the LArIAT~\cite{LArIAT:2019kzd,LArIAT:2021yix} and ProtoDUNE~\cite{DUNE:2017pqt,DUNE:2020cqd,DUNE:2025pda,DUNE:2025zhx} experiments are providing new pion--argon and proton--argon scattering measurements. Additional measurements of proton- and pion- nucleon and nucleus scattering are planned by the HADES experiment~\cite{Messchendorp:2025men,Hojeij:2023rsq,Ramstein:2024uea}.

Hadron-scattering data have been used to benchmark and motivate uncertainties within FSI models specifically for use in oscillation measurements. The pion FSI model in NEUT, based on the Salcedo--Oset model~\cite{Salcedo:1987md,Oset:1987re}, has been tuned to give good agreement to DUET and other measurements~\cite{PinzonGuerra:2018rju}. 
Similar efforts have been pursued by the NOvA experiment, tuning the GENIE \textit{hN} model~\cite{Dolce_FSI}. 
Additional but less mature efforts are in progress for proton scattering~\cite{Ma:2017hkg,Nonnenmacher:2020bly}. Although useful, the measurements used for tuning do not cover the full range of phase kinematics relevant to FSI processes and models tend to diverge in regions not constrained by them~\cite{PinzonGuerra:2018rju}. 

Whilst hadron-scattering data provides an important tool for benchmarking FSI models, extant measurements generally only provide cross sections as a function of the \textit{incoming} hadron's energy. Few measurements are differential in the kinematics of the final-state particles, which limits their utility for constraining FSI models used for neutrino scattering. As with electron-scattering measurements, historical hadron-scattering measurements tend not to include complete systematic error estimates, making them challenging to use in quantitative analyses.  

Neutron interactions are also important for neutrino oscillation experiments. A wealth of differential measurements to constrain neutron scattering exists for proton beam experiments across various targets, which have been categorised through community efforts~\cite{Otuka:2014wzu}. 
Most of the available measurements exist for neutron kinetic energies between 0--100 MeV, as the cross section peaks in the 10--20 MeV region. However, the majority of neutrons emitted in few-GeV neutrino interactions have higher energies. There are ongoing efforts to extend evaluated nuclear data libraries to support benchmarking at relevant energies~\cite{JENDL4-HE,BROWN20181,Kohley:2012awa,Agarwal:2022kiv,MINERvA:2023ikp, MINERvA:2019wqe}.

\subsection{Tensions between models and neutrino cross-section measurements}
\label{subsec:nuintmodelcomp}

\subsubsection{Pionless interactions}
CC0$\pi$ is the dominant interaction topology at the predominantly sub-GeV energies of the T2K and Booster\footnote{BNB, the beamline which has provided the primary neutrino beam for SciBooNE~\cite{SciBooNE:2006asj}, MiniBooNE, SBND, MicroBooNE and ICARUS\cite{ICARUS:2023gpo} at Fermilab.} neutrino beamlines. Modern neutrino event generator predictions for CC0$\pi$ are dominated by CCQE interactions, with additional, energy-dependent, contributions from 2p2h and RPP in which outgoing pions are absorbed due to FSI (see~\autoref{tab:channelsoscnoosc}).

In Cherenkov detectors, such as MiniBooNE and the Hyper-K/T2K FD, protons produced by few-GeV (anti)neutrino interactions are almost always below Cherenkov threshold, so CC0$\pi$ events are identified by isolating events where only a single lepton-like ring is observed. Pions are identified either by the presence of a second ring, and/or through the observation of a decayed signal when below-threshold pions decay and produce a Michel electron. 
In tracking detectors such as MINERvA, the T2K off-axis ND (ND280), and LArTPC detectors, CC0$\pi$ interactions are selected by the presence of a single lepton-like track or shower from the interaction vertex, with no additional tracks identified as charged pions, or isolated photon-induced showers from the decay of neutral pions or other mesons\footnote{Typically CC0$\pi$ implicitly also includes restrictions on charged $K^{\pm}$-like tracks or displaced vertices from $\Lambda^{0}$ or $K^{0}$ decays, although these are only relevant for higher energy beams.}. However, additional tracks consistent with protons are allowed. As in the Cherenkov case, the identification of Michel electrons from charged pion decay are sometimes used to tag pions below detector tracking thresholds. 
Fine-grained tracking detectors have the significant advantage that outgoing proton kinematics can be reconstructed alongside the lepton kinematics. 

Early analyses of the muon (anti)neutrino CC$0\pi$ topology were made by MiniBooNE on a hydrocarbon target (mineral oil, $\text{CH}_{2.08}$) in the early 2010s~\cite{MiniBooNE:2007iti, MiniBooNE:2010bsu, MiniBooNE:2013qnd}. These were primarily framed as CCQE measurements, and included a model-dependent background subtraction of pion absorption FSI, but comparisons of \textit{CCQE} models showed poor agreement with them. Initially, large modifications to the axial form factor and Fermi surface momentum were considered as an effective remedy. However, such models were theoretically unsatisfactory and unable to describe similar measurements from the NOMAD experiment~\cite{NOMAD:2009qmu}. Collaborations with the nuclear theory community led to the inclusion of additional 2p2h~\cite{Martini:2009uj,Nieves:2011pp} components to the comparisons, as earlier explored by electron scattering experiments, enabling an improved description of the MiniBooNE measurements~\cite{Nieves:2011yp, Martini:2011wp, Nieves:2013fr, Martini:2013sha}.
More recent analyses of CC$0\pi$ topologies by T2K~\cite{T2K:2016jor, T2K:2020jav, T2K:2020txr, T2K:2020sbd, T2K:2017qxv, T2K:2018rnz}, MINERvA~\cite{MINERvA:2022bno, MINERvA:2022mnw, MINERvA:2017dzh, MINERvA:2019ope, MINERvA:2019gsf, MINERvA:2023kuz, MINERvA:2018hba, MINERvA:2015jih},  MicroBooNE~\cite{MicroBooNE:2023krv,MicroBooNE:2024zkh, MicroBooNE:2023tzj, MicroBooNE:2023cmw,MicroBooNE:2022tdd,MicroBooNE:2025ooi} and ICARUS~\cite{ICARUS:2026nbg} have shown that, although 2p2h interactions are undoubtedly needed, no model is able to provide a consistently good description of all the measurements across nuclear targets and neutrino energies.

\begin{figure}[hpbt]
	\centering
    \includegraphics[width=1\linewidth, trim=0 0 0 0, clip]{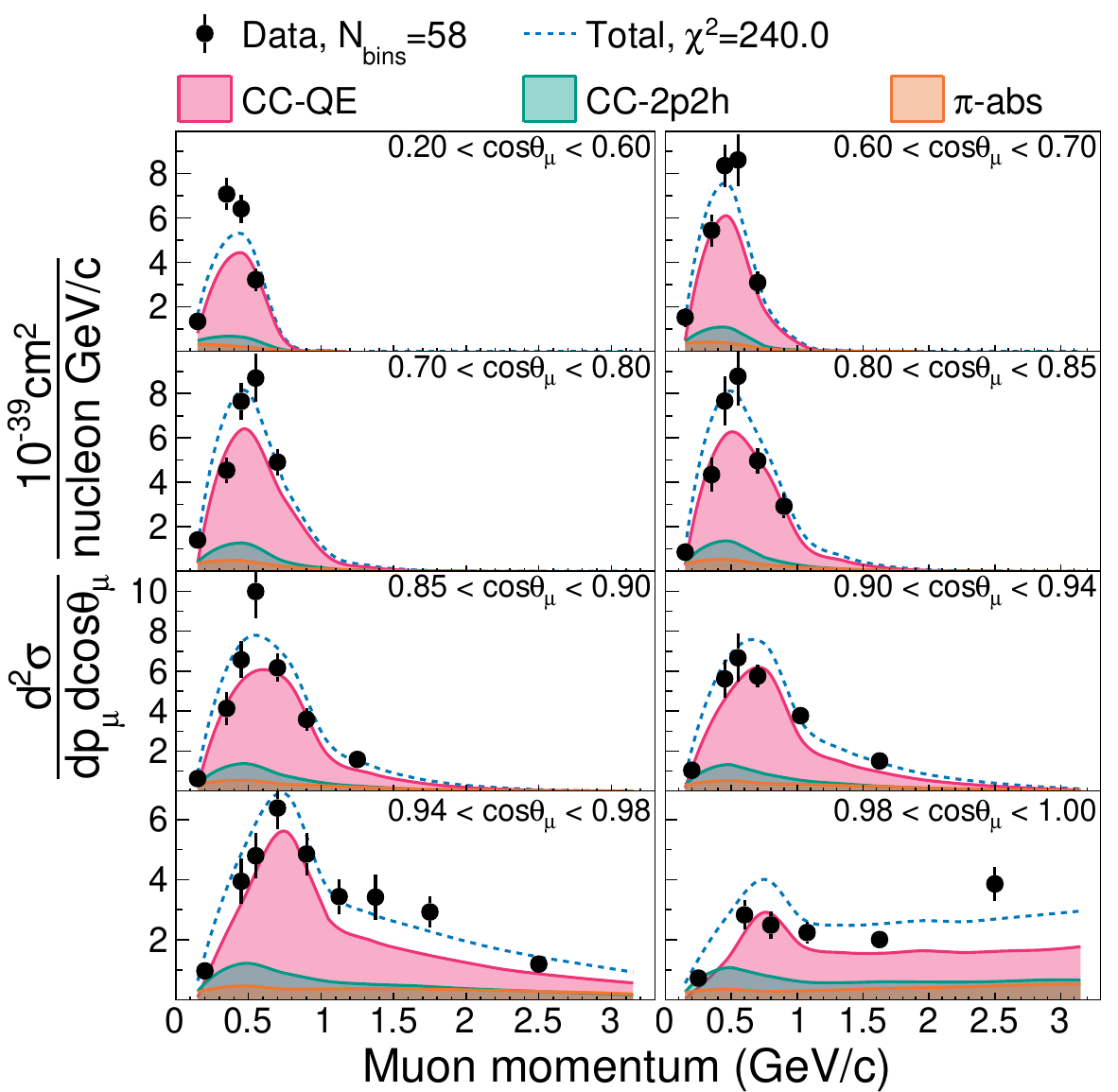}
	\caption{The T2K measurement of the  $\nu_\mu$CC0$\pi$ double-differential cross section on C$_8$H$_8$ as a function of outgoing muon kinematics~\cite{T2K:2020jav} is shown overlaid with predictions from the GENIE CRPA model, broken down by interaction channel. \textit{$\pi$-abs} refers to contributions involving pion-absorption FSI. Each row has a common y-axis limit, but they vary across rows. The $\chi^2$ value given in the legend indicates the level of agreement between the model and measurement and is calculated using the full covariance matrix provided in the experimental data release.}
	\label{fig:crpacc0picomp}
\end{figure}

Due to the broad-band nature of neutrino beams, multiple interaction processes always contribute to neutrino--nucleus cross section measurements, making it challenging to identify sources of mismodelling.
However, one region of data--model disagreement which has been observed by a number of experiments with various nuclear targets, and for both muon neutrinos and muon antineutrinos, is forward-angle, low-momentum transfer, CC0$\pi$ interactions, in which models typically overestimate the measured cross sections. This has been observed by MiniBooNE~\cite{MiniBooNE:2007iti},  T2K~\cite{T2K:2020jav, T2K:2023qjb, T2K:2017qxv, T2K:2016jor}, MINERvA~\cite{MINERvA:2025tem, MINERvA:2022bno, MINERvA:2022mnw, MINERvA:2019gsf, MINERvA:2018hqn, MINERvA:2018vjb, MINERvA:2013kdn, MINERvA:2013bcy} and MicroBooNE~\cite{MicroBooNE:2025ooi, MicroBooNE:2024yzp, MicroBooNE:2023cmw, MicroBooNE:2020akw}. 
~\autoref{fig:crpacc0picomp} shows a T2K $\nu_\mu$--C$_8$H$_8$ CC0$\pi$ cross-section measurement as a function of $\cos\theta_\mu$ and $p_\mu$~\cite{T2K:2020jav}, compared to the GENIE CRPA model, which demonstrates this issue. Poor overall agreement is indicated by the large $\chi^2$ (given in the legend), and the model predicts too large a cross section in the regions where the outgoing muon has momenta comparable to the peak of T2K's neutrino energy and is at a shallow angle with respect to the incoming neutrino (high $\cos\theta_\mu$). Similar observations have been made with other models~\cite{Avanzini:2021qlx}.
This region is predicted to contain CCQE interactions with low momentum transfer, which are notoriously difficult to model due to contributions from numerous nuclear effects (see \autoref{fig:q0GenComp} and the discussion in \autoref{sec:nuclear}), alongside significant contributions from non-CCQE processes. The disagreement between generator prediction and measurement could therefore be related to issues in modelling nuclear effects for CCQE interactions or the mismodelling of 2p2h and/or RPP with pion absorption~\cite{Dolan:2021rdd}.

In contrast to the apparent need for a low momentum transfer suppression, \autoref{fig:crpacc0picomp} also shows that the high-angle region is under-predicted for $0.2<\cos\theta_{\mu}<0.7$, which roughly corresponds to $Q^2\gtrsim0.25~\text{GeV}^2/c^4$ when accounting for the T2K neutrino energy distribution. Whilst nuclear effects are less impactful at this higher momentum transfer, it is unclear whether the discrepancy is due to mismodelling of the fundamental neutrino--nucleon CCQE interaction, non-QE contributions, or nuclear effects.

Because of the challenges in constraining complex neutrino interaction modelling issues with lepton kinematics alone, the community has responded over the last decade with a variety of new cross-section measurement approaches. Several of these follow the same core idea of reporting multiple correlated measurements and leveraging differences in how various nuclear effects affect each one to infer new information. These include making measurements on multiple interaction targets in the same beam; on two different detectors at different off-axis angles in the same beam; and of neutrino and antineutrino interactions in the same detector.

MINERvA and T2K have both made measurements of CC$0\pi$ interactions with the same beam but different nuclear targets. MINERvA has used a range of interaction targets from carbon to lead~\cite{MINERvA:2017dzh,MINERvA:2025tem,MINERvA:2023kuz}, which has shown that many generators struggle to correctly predict the $A$-scaling\footnote{$A$ is the mass number of the target nucleus, for example 12 for carbon-12.} of CC$0\pi$ cross sections. T2K has made CC$0\pi$ cross-section measurements on carbon and oxygen using its primary ND280 ND~\cite{T2K:2020jav} and the WAGASCI detector~\cite{t2k:2025kda}.
T2K has also made other correlated CC$0\pi$ measurements aimed at separating nuclear effects or interaction processes: correlated measurements of CC0$\pi$ muon neutrino and antineutrino cross sections~\cite{T2K:2020sbd} are potentially sensitive to the axial-vector interference term in the cross section~\cite{Ericson:2015cva}; and measurements taken using detectors at different off-axis angles from the incoming neutrino beam~\cite{T2K:2023qjb} sample a different (but correlated) flux distribution, making the combination weakly sensitive to the energy dependence of the CC$0\pi$ cross section. However, since none of the cross-section models considered in either analysis is able to provide a good description of the measurements, it is challenging to draw any strong conclusions.
Similarly, comparing measurements of equivalent observables from multiple experiments can potentially help identify the source of modelling issues. For instance, comparisons between MINERvA and T2K (both using a C$_8$H$_8$ target) can probe energy-dependence, whilst comparisons between T2K and MicroBooNE (which have similar neutrino fluxes) can probe the $A$-scaling of nuclear effects. 

Many modern experiments have high-resolution tracking abilities, making it possible to study the outgoing hadronic final state, which is a powerful new capability for accessing more information about nuclear modelling issues in CC0$\pi$ interactions. Both proton~\cite{T2K:2018rnz,MicroBooNE:2022emb,MicroBooNE:2024zkh} and neutron~\cite{MicroBooNE:2024hun, MINERvA:2023ikp,MINERvA:2019wqe,MINERvA:2023avz} multiplicities~\cite{Munteanu:2019llq,Baudis:2023tma,Dolan:2021hbw,Manly:2025pfm} have been measured by a variety of experiments. However, generators generally struggle to predict the relative contributions of different nucleon multiplicities to the cross sections, broadly indicating potential issues with FSI or 2p2h modelling, without providing a clear direction for model improvement. Additionally, MINERvA has made a measurement of the muon neutrino CC0$\pi$ cross section on hydrocarbon, differential in three dimensions, the muon kinematics and the total kinetic energy of all protons~\cite{MINERvA:2022mnw}. The latter acts as an approximation of the energy transfer and is accessed via calorimetry rather than by direct proton tracking, meaning no kinematic threshold is imposed\footnote{At the cost of adding weak dependence on the number of protons which are assumed to contribute (see the discussion of quenching in \autoref{subsec:enurec}).}. The ability to explore the cross section as a function of both lepton and hadron kinematics is a stringent probe of CC0$\pi$ modelling, possible thanks to MINERvA's high-statistics dataset. While every model tested by MINERvA in Ref.~\cite{MINERvA:2022mnw} showed significant tension with the measurement, a qualitative analysis suggested that discrepancies in the shallow-angle region (characterised by low transverse momentum) are generally at higher total proton kinetic energies, which is dominated by non-CCQE contributions. 

\begin{figure*}[hpbt]
	\centering
	\begin{subfigure}[b]{0.32\linewidth}
		\centering
		\includegraphics[width=\linewidth, trim=15 0 11 0, clip]{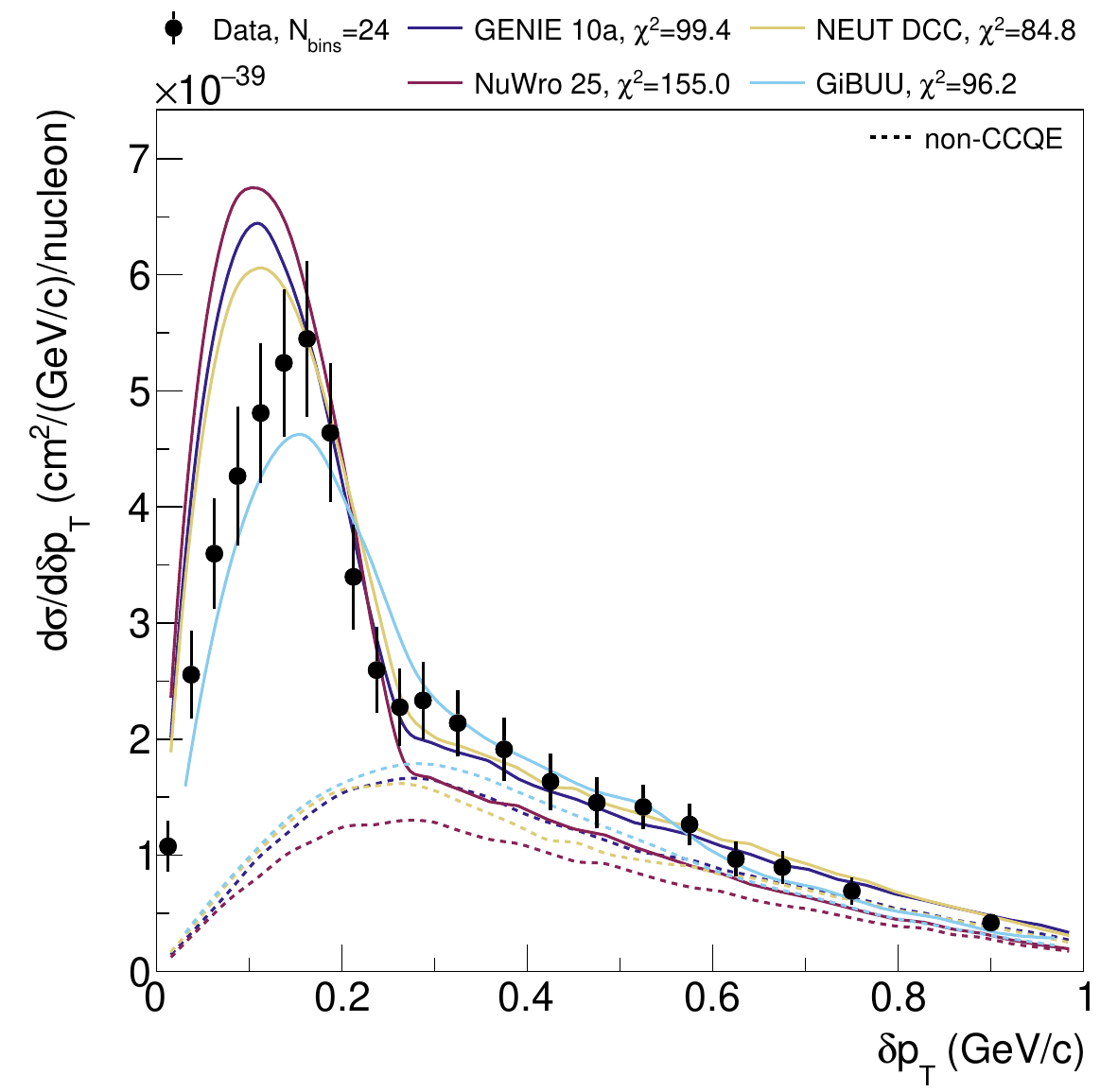}
		\caption{MINERvA}
	\end{subfigure}
	\begin{subfigure}[b]{0.32\linewidth}
		\centering   
		\includegraphics[width=\linewidth, trim=15 0 11 0, clip]{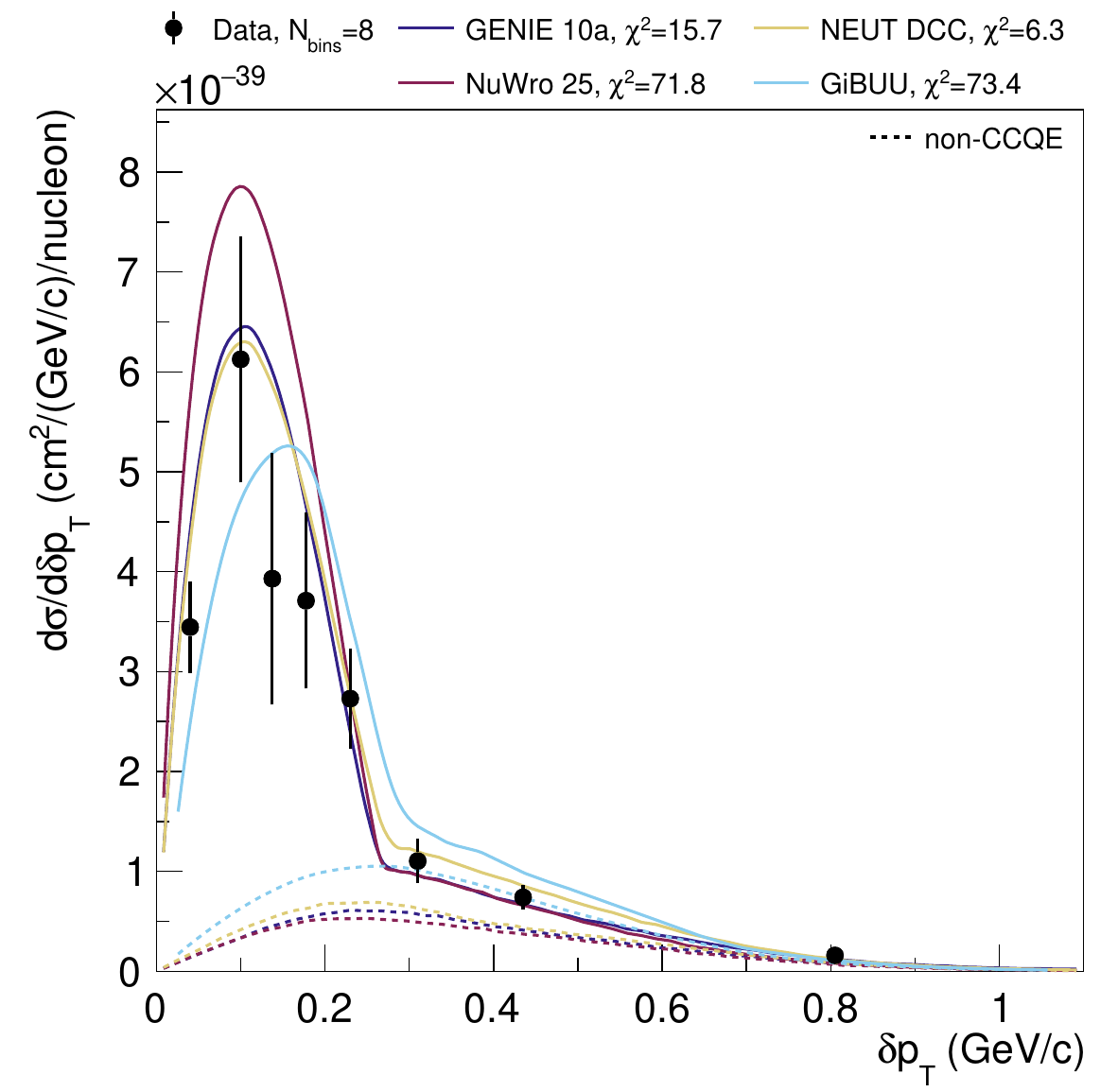}
		\caption{T2K}
	\end{subfigure}
\begin{subfigure}[b]{0.32\linewidth}
		\centering   
		\includegraphics[width=\linewidth, trim=15 0 11 0, clip]{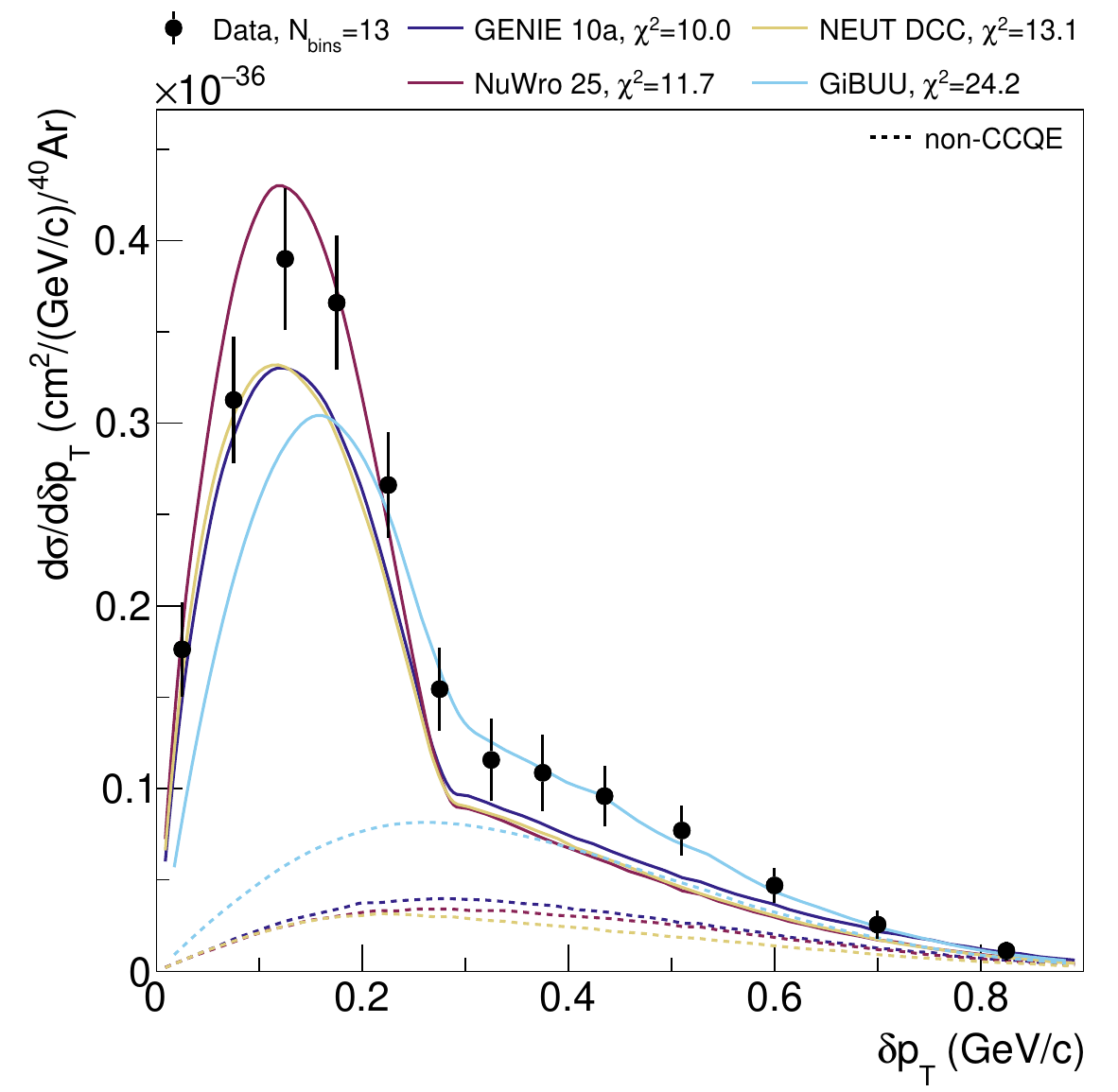}
		\caption{MicroBooNE}
	\end{subfigure}
\caption{Comparisons of event generator configurations to measurements of the transverse momentum imbalance in $\nu_{\mu}$CC$0\pi N$p events from MINERvA on C$_8$H$_8$~\cite{MINERvA:2018hba}, T2K on C$_8$H$_8$~\cite{T2K:2018rnz}, and MicroBooNE on $^{40}\text{Ar}$~\cite{MicroBooNE:2023tzj}. The $\chi^2$ shown in the legend indicate the level of agreement between the models and measurements and are calculated using the full covariance matrix provided in the experimental data releases.}
	\label{fig:cc0pinp_minerva_t2k_uboone}
\end{figure*}

Another approach to using hadronic information to help inform cross-section modelling issues has been to make measurements of observables that quantify imbalances between lepton and nucleon kinematics~\cite{lu:2015hea, Furmanski:2016wqo,MicroBooNE:2023krv, Baudis:2023tma, MINERvA:2019ope}. These complex \textit{derived} variables make it possible to \textit{partially} characterise nuclear effects and stress-test models, without requiring the very large statistics needed to make high-dimensional differential cross sections. Measurements of lepton--hadron imbalances in CC0$\pi$ interactions have been made by T2K~\cite{T2K:2018rnz}, MINERvA~\cite{MINERvA:2019ope,MINERvA:2018hba,MINERvA:2025tem}, MicroBooNE~\cite{MicroBooNE:2023tzj, MicroBooNE:2023cmw, MicroBooNE:2023krv,MicroBooNE:2024yzp} and ICARUS~\cite{ICARUS:2026nbg}. 
As an example, \autoref{fig:cc0pinp_minerva_t2k_uboone} shows comparisons of some of the generators considered in this work to measurements of missing momentum in the plane transverse to the incoming neutrino, $\delta p_{\mathrm{T}}$~\cite{lu:2015hea}\footnote{It is important to note that each experiment reports slightly different measurements. Each considers cross sections covering different ranges of muon and proton kinematics and uses slightly different definitions of $\delta p_{\mathrm{T}}$. T2K and MicroBooNE always consider the imbalance between the outgoing muon and highest momentum outgoing proton, whilst MINERvA consider the highest momentum proton below 1.2~GeV.} from all three experiments. The shape of the region $\delta p_{\mathrm{T}}\lesssim230\ \mathrm{MeV}/c$ offers enhanced sensitivity to the Fermi motion in CCQE interactions, whilst the normalisation and size of the tail are more sensitive to nucleon FSI, SRC contributions to CCQE, non-CCQE interactions and pion absorption~\cite{Lu:2015tcr,Dolan:2018zye, Filali:2024vpy}. 
No model is able to accurately predict all of the measurements, as indicated by the large $\chi^2$ for at least one of them. This suggests that there are important deficiencies in the modelling of nuclear effects, although it is not possible to unambiguously identify the source of the deficiency.

The range of recent, high-statistics, CC0$\pi$ measurements have the potential to provide a significant amount of information about a variety of nuclear effects which are highly relevant for oscillation experiments. However, as has been remarked several times, no model is able to provide a consistently good description of them. This reflects a fundamental challenge for neutrino interaction model development: measurements are \textit{weakly} sensitive to multiple contributing interaction processes and nuclear effects simultaneously. It is consequently difficult to unambiguously attribute discrepancies to specific modelling deficiencies and therefore to use neutrino--nucleus measurements to directly drive targeted model improvements.

\subsubsection{Pion production}
\label{sec:pionprodconstraint}

Pion production interaction channels are dominant for neutrinos with $\enutrue \gtrsim 1$ GeV, and contribute significantly to the event rates at all current and next-generation LBL experiments (\autoref{tab:channels}). As described in~\autoref{sec:models}, there are multiple interaction processes that contribute to the pion production cross section, including RPP, SIS, DIS and coherent interactions. 
The relative contribution from each process varies as a function of the four-momentum transfer, $Q^2$, and the invariant mass of the hadronic system, $W$. For lower energy neutrino beams (Hyper-K/T2K, BNB), RPP dominates which, at these energies, is driven mostly by the excitation of the $\Delta(1232)$ resonance. For higher-energy experiments like MINERvA and DUNE, there are significant RPP contributions from resonances heavier than the $\Delta(1232)$, from the non-perturbative SIS/DIS transition region for $W\gtrsim2~\text{GeV}/c^2$, and from pure DIS, which requires $Q^2\gtrsim3~\text{GeV}^2/c^4$. 
In historical bubble chamber measurements on hydrogen and deuterium targets, $W$ and $Q^2$ could be reliably reconstructed from measured particle kinematics.
However, for heavier nuclear targets, attempting to do the same is fraught with model dependence\footnote{In Ref.~\cite{MINERvA:2014ogb}, a MINERvA analysis attempted to isolate two CC pion-production samples with either $W<1.4$~GeV$/c^2$ or $W<1.8$~GeV$/c^2$. However, the analysis involved a correction from a reconstructed proxy for $W$ to true $W$. Modelling this mapping involves very similar challenges to modelling the mapping between $E^{\mathrm{rec},\;\mathrm{avail}}_{\nu}$ and true neutrino energy, opening the MINERvA correction up to all the challenges discussed in~\autoref{subsec:biasinErechad}.}. 
As a result, direct comparisons between pion-production cross-section measurements made in neutrino beams with different energies is even more challenging than for CC$0\pi$, as each experiment samples very different contributions from RPP, SIS and DIS depending on the incoming neutrino energy distribution. 

\begin{figure}[hptb]
  \centering
  \begin{subfigure}[b]{0.9\linewidth}
    \centering
    \includegraphics[width=1.00\linewidth]{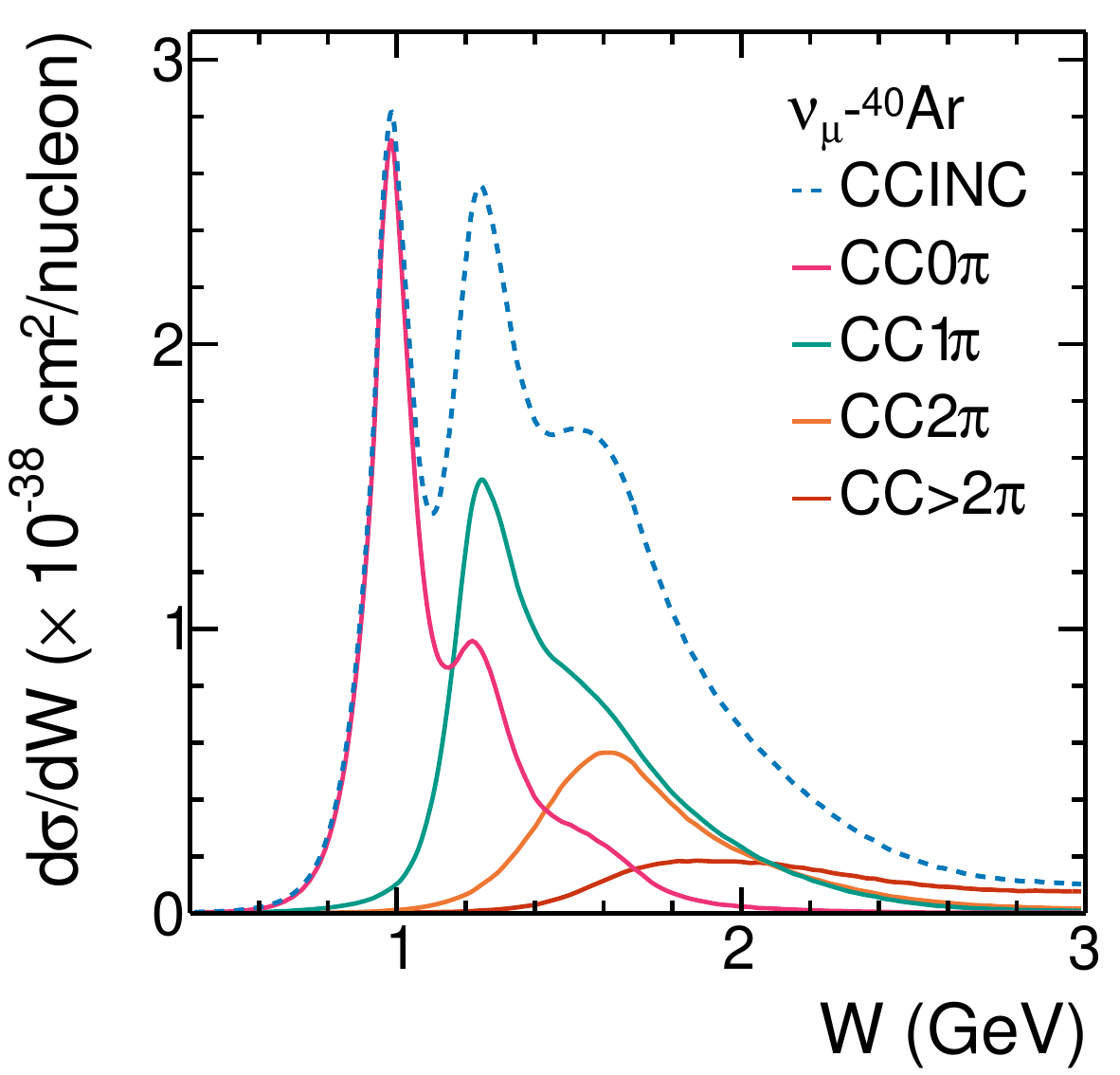}
    \caption{DUNE ND}
  \end{subfigure}
  \begin{subfigure}[b]{0.9\linewidth}
    \centering
    \includegraphics[width=1.00\linewidth]{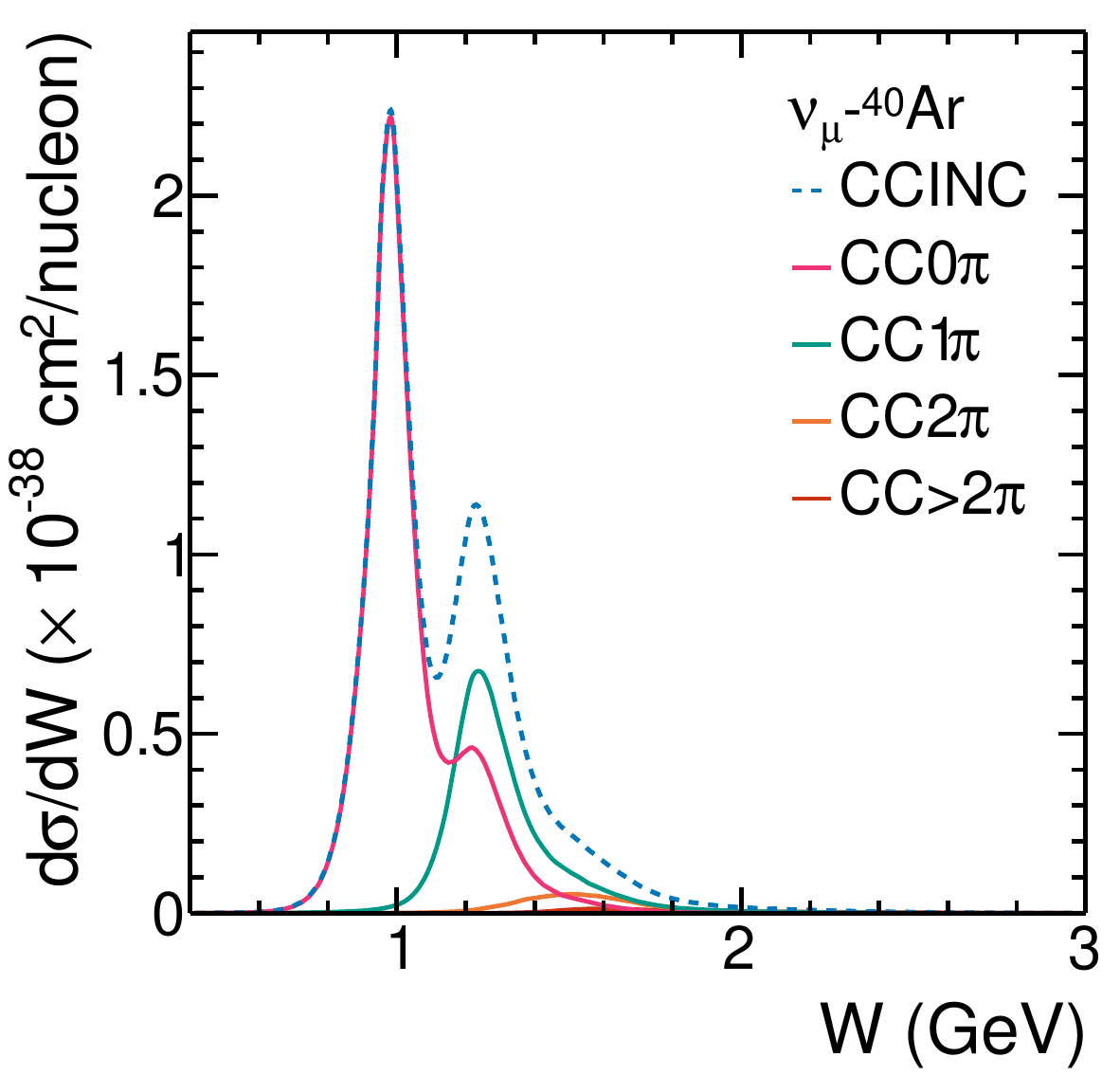}
    \caption{SBND}
  \end{subfigure}
\caption{The hadronic mass spectrum for \numu--\argon CCINC events in the DUNE ND and SBND, assuming the struck nucleon is at rest. As the SBND flux is not currently public, the SciBooNE flux was used~\cite{SciBooNE:2010slc} as a very close approximation. Each figure is broken down by the pion content in the final state.}
\label{fig:DUNE_SBND_HadMass}
\end{figure}

To illustrate how pion production measurements with different experiments probe different physics processes, \autoref{fig:DUNE_SBND_HadMass} shows the predicted hadronic mass spectra for \numu--\argon CCINC interactions at the DUNE ND and at SBND (which will be very similar to other BNB experiments including MicroBooNE, MiniBooNE and ICARUS) separated by the number of pions in the final state. 
The spectra are markedly different. Above the QE-peak around the nucleon mass, DUNE has contributions from multiple resonances, whereas the lower energy BNB beam mostly probes $\Delta(1232)$. DUNE also has significant contributions from SIS and DIS at higher $W$, for all pion multiplicities, whereas these contributions are very small at the BNB. Similarly, DUNE has a significant fraction of $\geq$2$\pi$ events, whereas these contributions are small at the BNB. Pion-production measurements at the BNB therefore only probe a subset of the physics processes that will contribute for DUNE, even for the same final-state topology.
Although the interpretation is complex, and experiments do not measure the same mix of processes or effects, \autoref{fig:DUNE_SBND_HadMass} shows that single- and multi-pion production topologies help to explore different regions of $W$.
Additionally, measurements of different CC and NC single-pion production interaction topologies are sensitive to different pion production model elements in complex ways, so the relationship between any pair of channels, for example $\bar{\nu}_\mu$ NC1$\pi^{0}$ and $\nu_{\mu}$ CC1$\pi^{+}$, is highly non-trivial. However, if these relationships can be understood, comparing measurements of two or more topologies may in principle allow constraints of particular interaction model elements.

A variety of neutrino and antineutrino CC and NC pion production cross sections on nuclear targets have been measured by a number of experiments including K2K~\cite{K2K:2004qpv, K2K:2008tus}, SciBooNE~\cite{SciBooNE:2009nlf,SciBooNE:2010lca},  MiniBooNE~\cite{MiniBooNE:2009dxl,MiniBooNE:2010cxl,MiniBooNE:2010eis}, MINOS~\cite{MINOS:2016yyz},  ArgoNeuT~\cite{ArgoNeuT:2018und}, T2K~\cite{T2K:2016cbz,T2K:2017epu,T2K:2019yqu,T2K:2021naz,T2K:2025kdk,T2K:2025wde,T2K:2025smz}, MINERvA~\cite{MINERvA:2014ogb,MINERvA:2015slz,MINERvA:2016sfc,MINERvA:2019rhx,MINERvA:2020anu,MINERvA:2022djk,MINERvA:2019kfr}, NOvA~\cite{NOvA:2023uxq,NOvA:2025vdg}, and MicroBooNE~\cite{MicroBooNE:2022zhr,MicroBooNE:2024bnl,MicroBooNE:2024sec,MicroBooNE:2025pvb}.
The general challenges and outlook associated with interpreting these measurements are similar to those highlighted for the CC0$\pi$ case. Firstly, no generator is able to fully describe all features of global measurements. Very broadly, they are able to qualitatively approximate the shape of cross-section measurements as a function of outgoing leptonic variables for CC interactions ($p_\mu$, $\theta_\mu$) but fail to capture the shape of the hadronic distributions (pion and nucleon kinematics)~\cite{Mahn:2018mai,Mosel:2017nzk,Avanzini:2021qlx,Betancourt:2018bpu}. Moreover, since historical bubble chamber experiments did not provide many measurements of the outgoing hadrons' kinematics, these are poorly benchmarked even for (anti)neutrino--\textit{nucleon} interactions. Secondly, since most pion production measurements are sensitive to a wide range of contributing interaction processes and nuclear effects, it is seldom straightforward to interpret how to use discrepancies to improve models. 
This challenge is amplified by the wider range of interaction channels, the multiple resonances that contribute and their interferences, as well as the non-resonant background. 

As for CC0$\pi$, the last decade has seen experiments adopt new approaches to use correlations between measurements to attempt to separate different contributing interaction processes or nuclear effects. These include correlated measurements on different nuclear targets, or of neutrino and antineutrino interactions, and  measurements of hadronic or derived kinematic variables, as well as approaches unique to pion production which exploit the differences between pion production topologies.

\begin{figure*}[hbtp]
	\centering
	\begin{subfigure}[b]{0.48\textwidth}
		\centering
		\includegraphics[width=\linewidth, trim=15 0 0 0, clip]{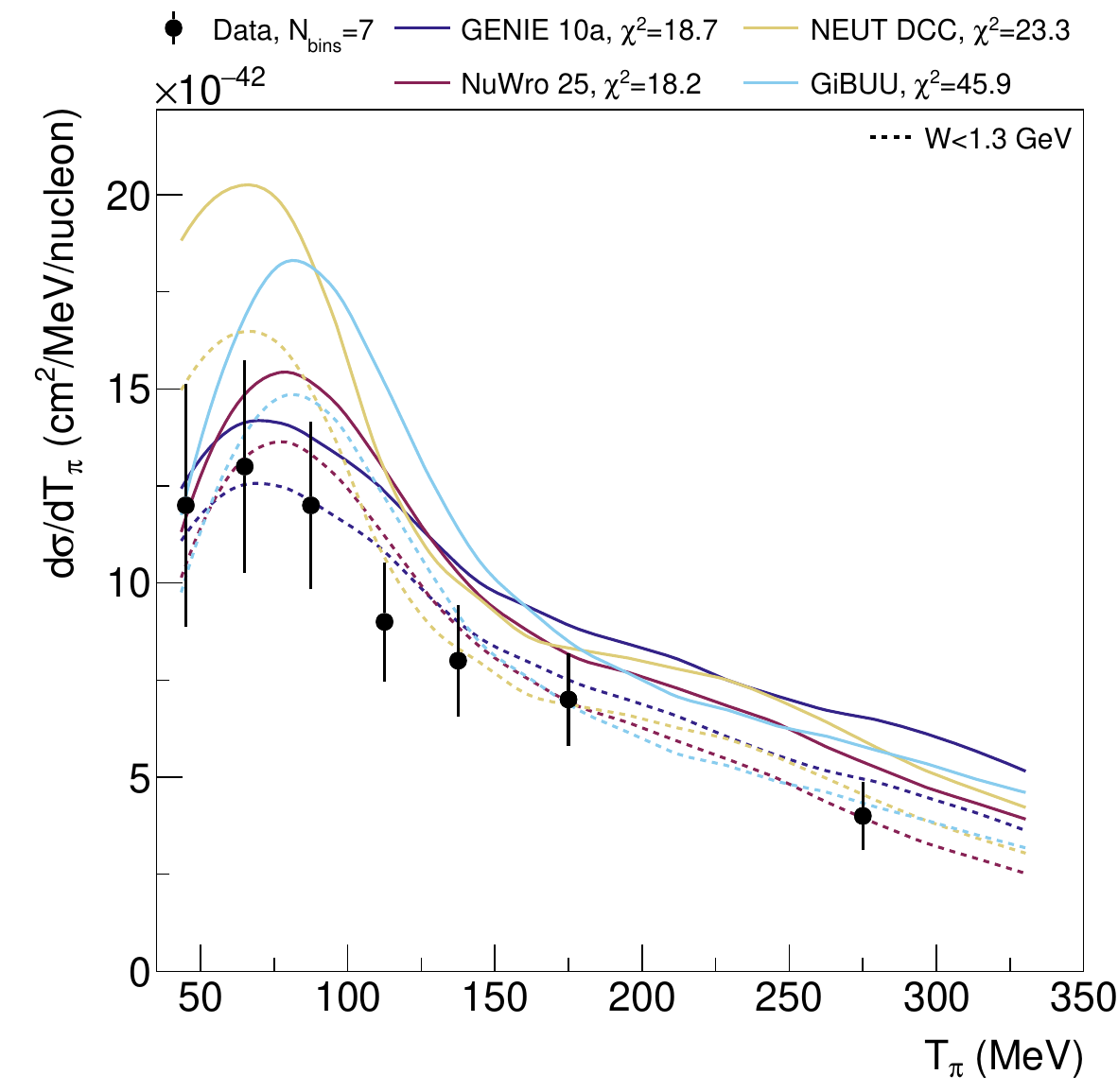}
		\caption{CC$1\pi^\pm$}
	\end{subfigure}
	\begin{subfigure}[b]{0.48\textwidth}
		\centering
		\includegraphics[width=\linewidth, trim=15 0 0 0, clip]{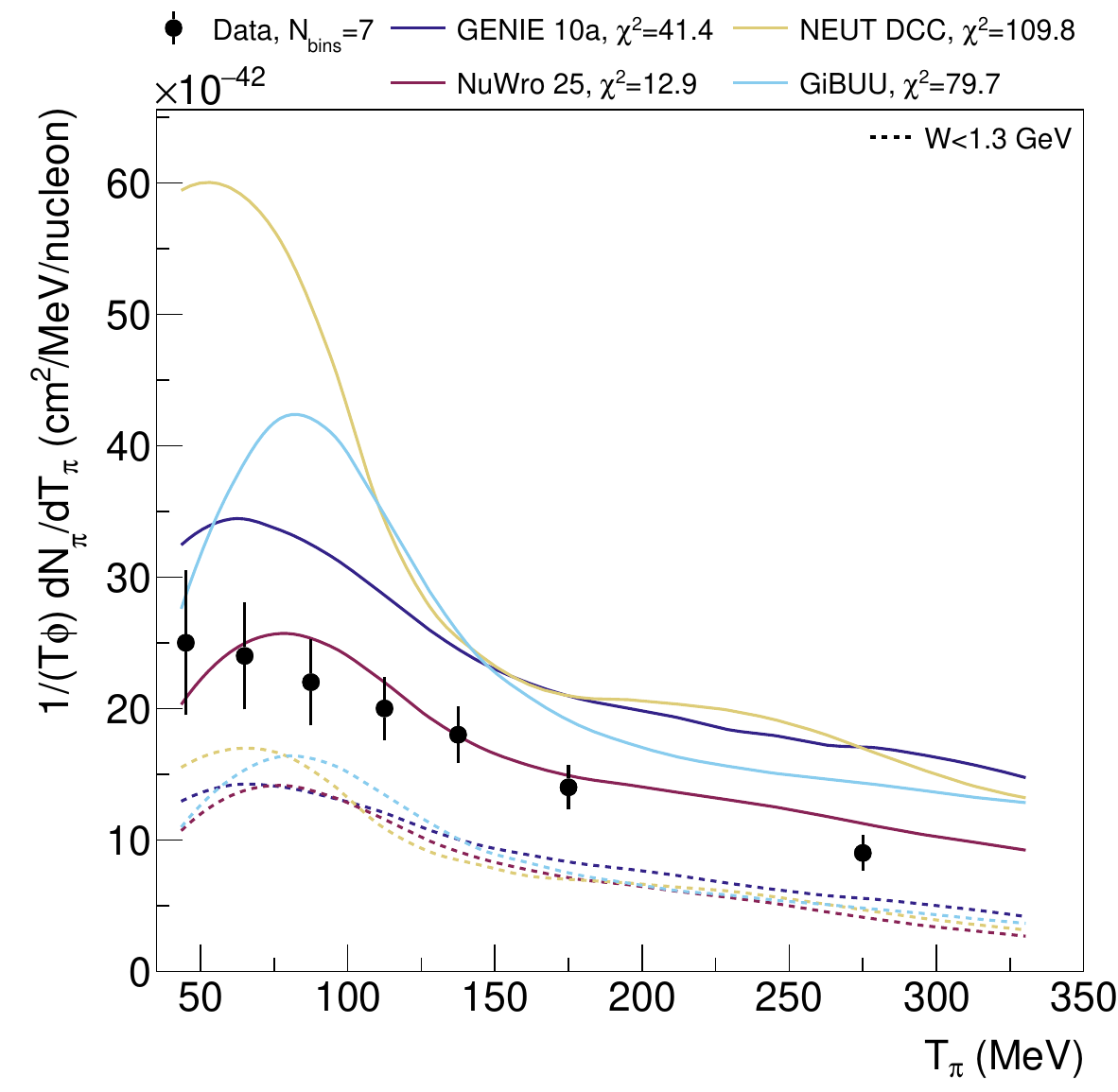}
		\caption{CC$N\pi^\pm$}
	\end{subfigure}
	\caption{CC$1\pi^\pm$~\cite{MINERvA:2014ogb} (left) and CC$N\pi^\pm$~\cite{MINERvA:2016sfc} (right) measurements made by MINERvA on a C$_8$H$_8$ target compared to current generator models. Dashed lines show the contributions that come primarily from the Delta resonance ($W<1.3~\mbox{GeV}$). The $\chi^{2}$ values are shown in the legend. The CC$N\pi^+$ measurement includes \textit{all} pions in an event, not just the highest momentum pion. The $\chi^2$ shown in the legend indicate the level of agreement between the models and measurements and are calculated using the full covariance matrix provided in the experimental data releases.}
	\label{fig:single_pi_minerva_1pi_npi}
\end{figure*}

Measurements of multi-pion production final states (for example, the CC$N\pi^\pm$ topology, where $N>1$) are broadly populated by interactions with higher hadronic mass, therefore including contributions from higher resonances, compared to single pion production (see \autoref{fig:DUNE_SBND_HadMass}). 
This is further illustrated in \autoref{fig:single_pi_minerva_1pi_npi}, which shows generator predictions of the CC$1\pi^\pm$~\cite{MINERvA:2014ogb} and CC$N\pi^\pm$~\cite{MINERvA:2016sfc} cross sections measured by MINERvA\footnote{The CC$1\pi^\pm$ and CC$N\pi^\pm$ measurement use signal definitions that require a proxy for $W$, $W_\text{exp}$, to be less than 1.4~GeV$/c^2$ and 1.8~GeV$/c^2$ respectively.}, with the predicted contributions from events with $W<1.3~\text{GeV}/c^2$ overlayed.
For both topologies, the $\chi^2$ values given in the legend show poor measurement--simulation agreement for all generators tested. For GENIE, NEUT and GiBUU, it is significantly worse for the CC$N\pi^\pm$ measurement. NuWro 25, which uses the Ghent single pion production model for RPP and SIS interactions, gives the best description of the CC$N\pi^\pm$ measurement, which appears to stem from the model's much lower prediction of the $W>1.3~\text{GeV}/c^2$ contribution to the cross section compared to other generators. However, the overall level of agreement for both samples remains poor.
The large model spread for the CC$N\pi^\pm$ measurement, which comes primarily from $W>1.3~\text{GeV}/c^2$ interactions, highlights the large variation in pion momenta predicted by SIS and DIS models.

Measurements of specific single-pion production topologies are also sensitive to different collections of resonances. For example, as discussed in \autoref{sec:models}, CCRPP neutrino--\textit{nucleon} interactions leaving a CC$1\pi^+1p$ final state have a larger proportion of contributions from the $\Delta(1232)$ resonance than the CC$1\pi^01p$ final state. Such measurements are also sensitive to FSI, as its differing effect on charged and neutral pions plays an important role in predicting the relative fractions of CC$1\pi^0$ and CC$1\pi^+$ interactions.
Generator--measurement comparisons show that simultaneously describing both charged and neutral pion production topologies with the same model is a challenge and that  
a good description of one topology does not guarantee good agreement for the other~\cite{MINERvA:2019kfr, T2K:2025yoy, Mosel:2017nzk,Yan:2024kkg,Avanzini:2021qlx,Betancourt:2018bpu,Nikolakopoulos:2022tut,Nikolakopoulos:2018gtf,Gonzalez-Jimenez:2017fea,Kabirnezhad:2017jmf}. 
NC topologies can also be sensitive to different resonances than their CC counterparts, but it is challenging to draw firm conclusions from current measurements. For example, a NC$1\pi^0$ measurement from MiniBooNE~\cite{MiniBooNE:2009dxl} suggests the need for a larger cross section than is implemented in many generators but a recent MicroBooNE~\cite{MicroBooNE:2024pdj} measurement agrees only when limited to an analysis without any visible protons. This could suggest an incorrect modelling of momentum transfer in the underlying neutrino--nucleon interaction model or it could be due to poorly modelled or missing nuclear effects. More data or theoretical work is likely to be needed to understand the observed differences.

\begin{figure}[hbtp]
	\centering
		\centering
		\includegraphics[width=\linewidth, trim=10 0 0 30, clip]{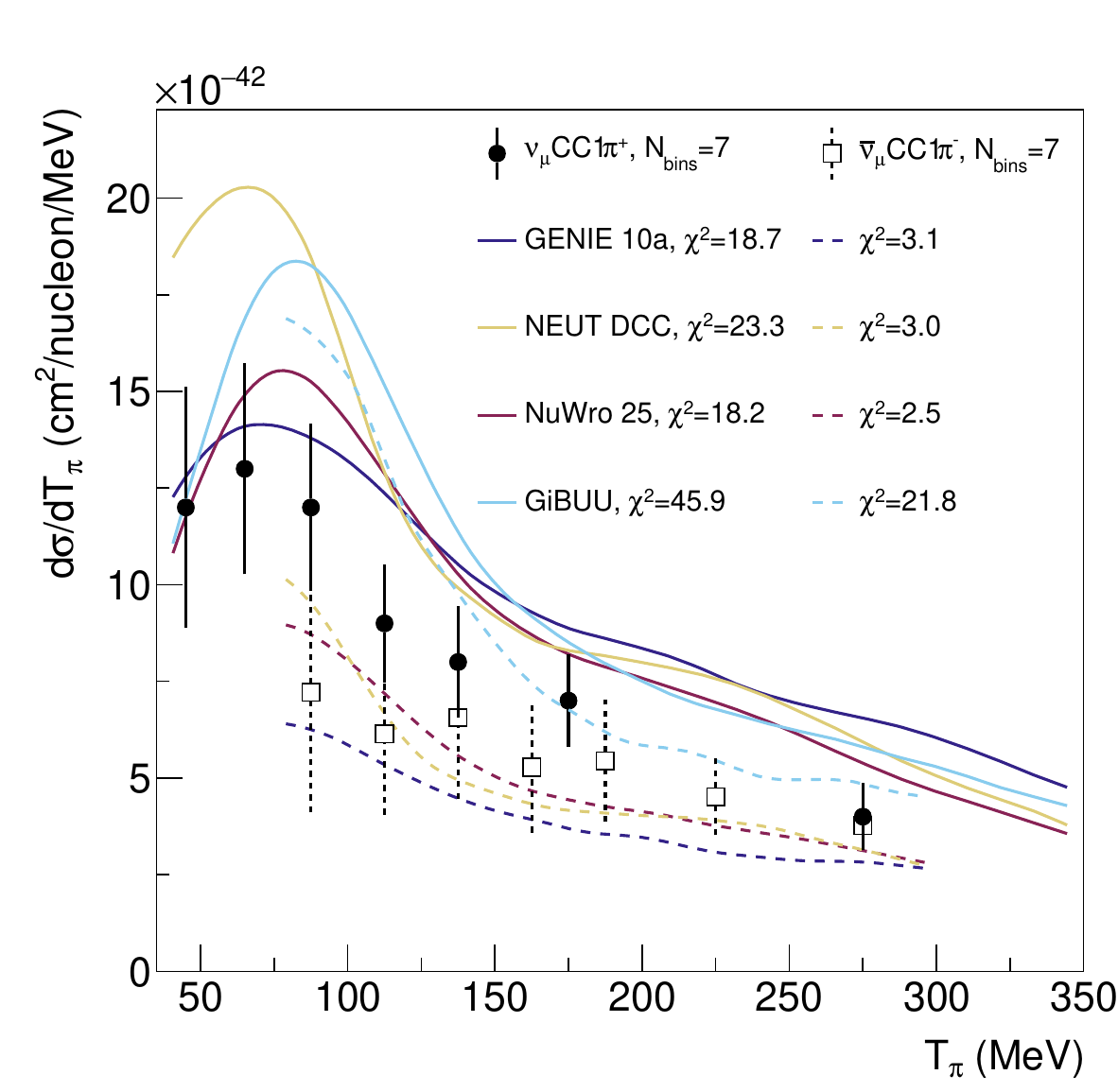}
	\caption{MINERvA's measurements of the differential cross sections on a C$_8$H$_8$ target in pion kinetic energy for $\nu_\mu$CC$1\pi^+$~\cite{MINERvA:2014ogb} (solid circle, solid lines) and $\bar{\nu}_\mu$CC$1\pi^-$~\cite{MINERvA:2019rhx} (open square, dashed lines) final states, compared to current generator models. The $\chi^2$ shown in the legend indicate the level of agreement between the models and measurements and are calculated using the full covariance matrix provided in the experimental data releases.}
	\label{fig:minerva_cc1pip_cc1pim}
\end{figure}

Similarly to comparing measurements of different interaction topologies, comparisons of neutrino and antineutrino pion production measurements may also help to characterise the source of model--measurement discrepancies, although measurements of the latter remain relatively rare.
MINERvA has made measurements of $\bar{\nu}_\mu$CC$1\pi^-$~\cite{MINERvA:2019rhx} and $\bar{\nu}_\mu$CC1$\pi^0$~\cite{MINERvA:2016sfc,MINERvA:2015slz} final states in their LE beam configuration. 
The $\bar{\nu}_\mu$CC$1\pi^-$ is overlaid with the aforementioned $\nu_\mu$CC1$\pi^+$ measurement in \autoref{fig:minerva_cc1pip_cc1pim} alongside generator predictions\footnote{Note that the two measurements impose a different hadronic mass cut: $W_\text{exp}<1.4$~GeV/c$^2$ and $W_\text{exp}<1.8$~GeV/c$^2$ for the neutrino and antineutrino measurements, respectively.}. 
The lower $\chi^2$ for the antineutrino measurement shows that, with the exception of GiBUU, the generator predictions are not in significant tension with the measurement, although this is likely to be due to the large uncertainties.
Importantly, the current BNB programme only includes neutrino-enhanced running. Whilst an additional antineutrino-enhanced running period has been discussed~\cite{SBND:2025lha}, there is currently no clear prospect for antineutrino measurements with the SBND or MicroBooNE.
This leaves limited prospects for antineutrino scattering measurements for single or multiple pion production on an argon target, which make up a large fraction of the interactions expected at DUNE (see \autoref{tab:channels}), before DUNE starts.

\begin{figure*}[hbtp]
	\centering
	\begin{subfigure}[b]{0.48\textwidth}
		\centering
		\includegraphics[width=\linewidth, trim=15 0 0 0, clip]{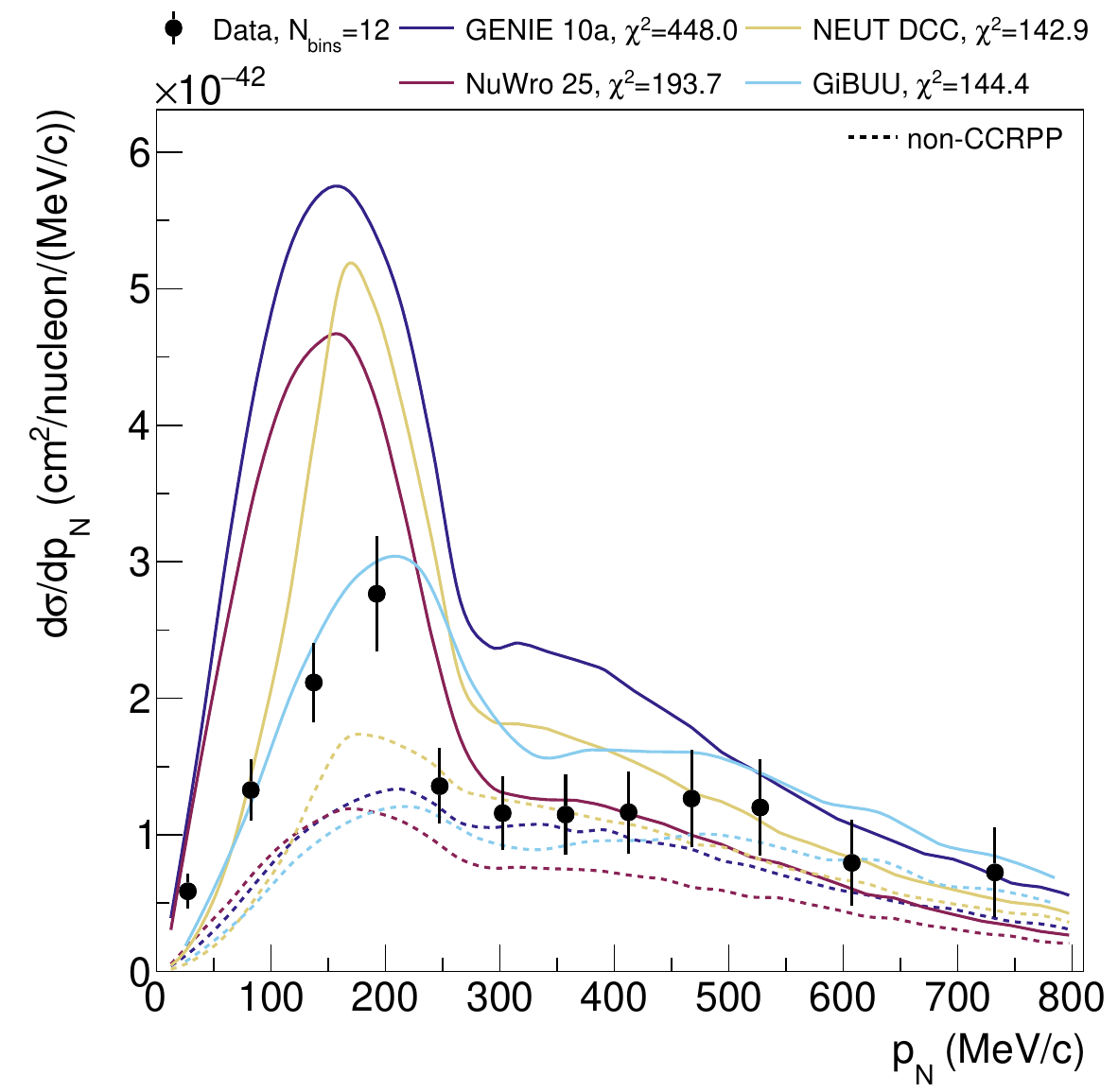}
		\caption{MINERvA CC$N\pi^0M$p}
	\end{subfigure}
	\begin{subfigure}[b]{0.48\textwidth}
		\centering
		\includegraphics[width=\linewidth, trim=15 0 0 0, clip]{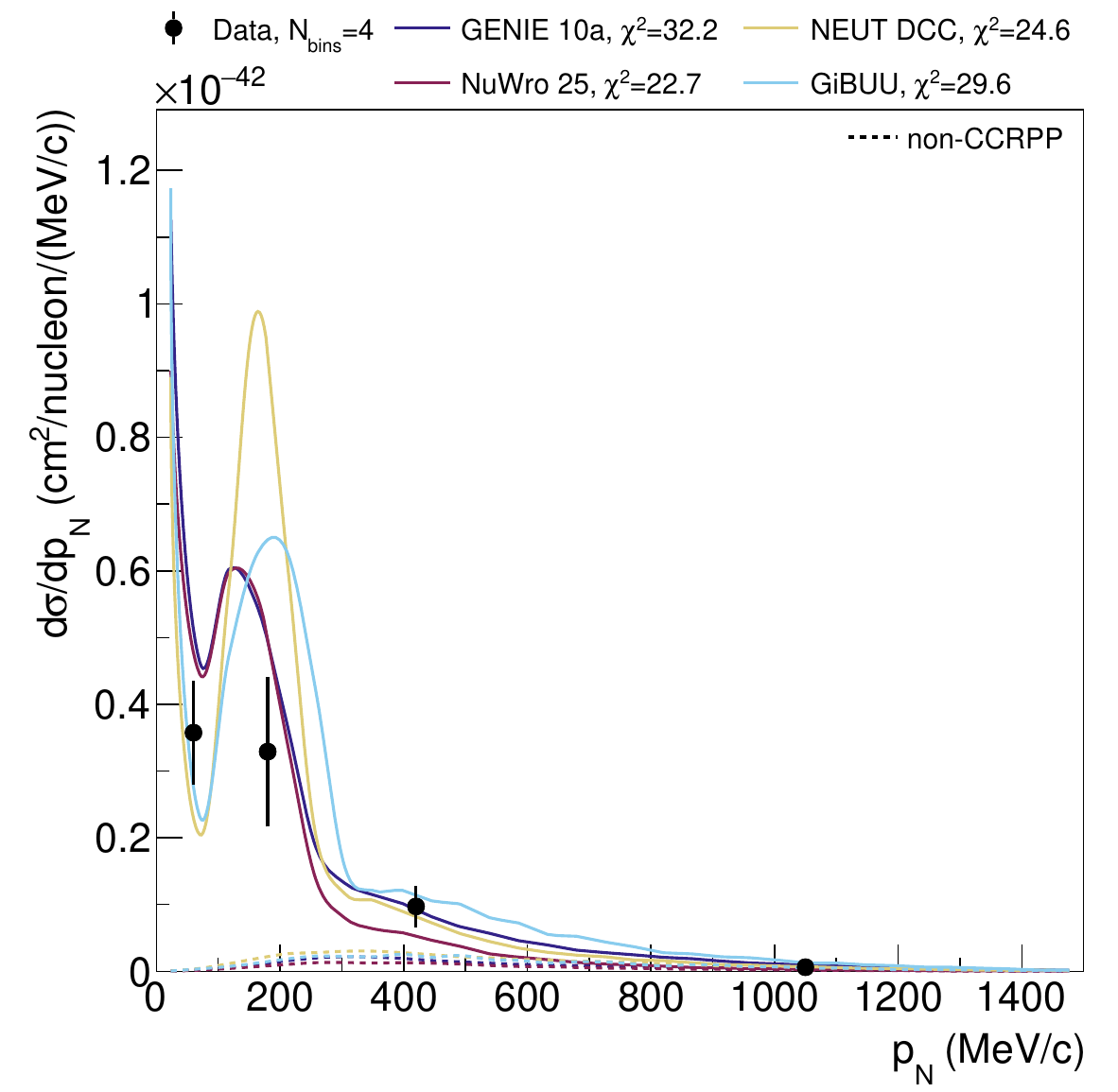}
		\caption{T2K CC$1\pi^+N$p}
	\end{subfigure}
	\caption{A measurement of the CC$N\pi^0M$p ($N>0, M>0$) (left) cross section from MINERvA on C$_8$H$_8$~\cite{MINERvA:2020anu} and the CC$1\pi^+N$p ($N>0$) cross section from T2K on C$_8$H$_8$~\cite{T2K:2021naz} (right) as a function of reconstructed initial state nucleon momentum, $d\sigma/dp_{\mathrm{N}}$, compared to current generator models. Dashed lines show the contributions that come from non-CCRPP processes, for example multi-pion and DIS interactions. The $\chi^2$ shown in the legend indicate the level of agreement between the models and measurements and are calculated using the full covariance matrix provided in the experimental data releases.}
	\label{fig:stv_pi_minerva_t2k}
\end{figure*}

As discussed for the CC0$\pi$ case, measurements of correlations between the leptonic and hadronic system can provide some information on the nuclear effects in pion production interactions. In particular, kinematic imbalances between the hadronic and leptonic system~\cite{lu:2015hea, Furmanski:2016wqo, Lu:2019nmf}, based on reconstructing both final-state pions and protons, allows for some separation of pion production events into regions where changes in FSI models and the nuclear ground state contribute differently. 
Such measurements have been made by MINERvA for the CC$N\pi^{0}M$p ($N>0,$ $M>0$) final state ~\cite{MINERvA:2020anu} and by T2K for the CC$1\pi^+ N$p final state~\cite{T2K:2021naz}.
The reconstructed initial state nucleon momentum~\cite{Furmanski:2016wqo} ($p_{\mathrm{N}}$) measured by the two experiments is shown in \autoref{fig:stv_pi_minerva_t2k}. Very broadly, the modelling of the RPP portion of the distribution below the Fermi momentum ($p_{\mathrm{N}}\lesssim$230~MeV/c) is more impacted by Fermi motion modelling whilst RPP interactions at higher $p_{\mathrm{N}}$ are more sensitive to FSI variations~\cite{Lu:2019nmf}. 
The peak at zero $p_{\mathrm{N}}$ in the T2K case is from CC$1\pi^+$ interactions on the hydrogen in the C$_8$H$_8$ target, which is not present for MINERvA due to $\numu$CC$1\pi^0$ resonant interactions happening only on neutrons. 
The large $\chi^2$ show that neither measurement is described well by any of the generators, suggesting that nuclear effects and hadron dynamics in pion production are poorly understood. 
It is interesting to note that the generator predictions for the RPP contribution to the MINERvA measurement differ much more than for MINERvA's CC$1\pi^+$ measurement, shown in \autoref{fig:single_pi_minerva_1pi_npi}. This demonstrates that, although the generators' RPP models may have similar predictions for the overall CC$1\pi^+$  RPP cross section, their treatment of outgoing protons and nuclear effects differ. 
The poor agreement between generators and both measurements suggests that the modelling of pion \textit{and} nucleon dynamics in pion production interactions is far from complete.

Also as in the CC0$\pi$ case, measurements of pion production have also been made on different nuclear targets as an additional means to characterise the role of nuclear effects. For example, MINERvA have studied the CC$1\pi^+$ cross section on carbon, hydrocarbon, water, iron and lead targets for interactions with low reconstructed hadronic mass in their ME beam configuration~\cite{MINERvA:2022djk}.
Comparisons of models to the measurement highlight large differences between generator predictions, even for the lighter targets, for instance, MINERvA's comparisons between GENIE v3 and MINERvA's tuned GENIE v2 show a 40\% difference with respect to the measured hydrocarbon cross section. As the size of the nucleus increases, measurement--model differences grow significantly, and no generator tested by MINERvA describes the measurement on all targets and in all variables.
A T2K CC$1\pi^+$ measurement at lower energies on water~\cite{T2K:2016cbz} shows satisfactory agreement with a NEUT simulation in lepton kinematics, and mild disagreements in pion kinematics, but has limited statistical power compared to the hydrocarbon analogue~\cite{T2K:2019yqu}.

Although the number of CC pion-production measurements has significantly increased over the last decade, there have been relatively few measurements on argon. ArgoNeuT~\cite{ArgoNeuT:2018und} made early measurements of the neutrino- and antineutrino--argon $\nu_\mu$CC$1\pi^\pm$ cross section with large ($\gtrsim$30\%) uncertainties. More recently, MicroBooNE has made $\nu_\mu$CC$1\pi^0$~\cite{MicroBooNE:2024bnl} and $\nu_\mu$CC$1\pi^+$~\cite{MicroBooNECollaboration:2025flv} cross-section measurements. In both cases, MicroBooNE report single-differential cross-section results as a function of lepton and pion kinematic variables, as well as the lepton--pion opening angle, and compare their results to a variety of neutrino generator predictions. In general, none of the tested generators are able to describe either dataset well across all variables.
MicroBooNE's $\nu_\mu$CC$1\pi^+$ analysis also demonstrates significant challenges posed by pion secondary interactions in argon, which complicates estimation of the pion's kinetic energy.
In order to make measurements as a function of pion kinematics, MicroBooNE developed a selection targeting pions without secondary interactions, which removed $\approx$50\% of interactions and limited the measurement of pion kinematics to $p_{\pi^+}\lesssim$300~MeV/c. The demonstrated difficulties in reliably studying the kinematics of charged pions in dense liquid argon detectors stands out as a potentially important challenge for DUNE.

In addition to RPP and SIS/DIS interactions, single pion topologies have important contributions from coherent neutrino--nucleus interactions at very low four-momentum transfer, often referred to as $t$ in the context of coherent measurements.
Since the interaction occurs off the entire nucleus, the pion does not undergo final-state interactions, and it is free of complications from initial-state motion. As such, it is possible to reconstruct the neutrino energy from coherent interactions in a way that is unperturbed by nuclear effects, although generally limited by the detector resolution.
The process is experimentally isolated by selecting events with low $t$ and low activity around the interaction vertex (to reject events with additional particles in the final state), which can remove much of the significant backgrounds from RPP interactions. 

MINERvA~\cite{MINERvA:2022esg,MINERvA:2017ipy,MINERvA:2014ani} has measured muon-neutrino CCCPP across a wide range of neutrino energies in the LE and ME beam configurations (1.5--20 GeV), on carbon, hydrocarbon, iron, and lead, targets, differential in both muon and pion kinematics. MINERvA has also measured the equivalent antineutrino cross section on a carbon target in its LE beam configuration~\cite{MINERvA:2017ipy,MINERvA:2014ani}.
T2K~\cite{T2K:2023xlh,T2K:2016soz} has measured the CCCPP neutrino and antineutrino cross sections on hydrocarbon for neutrino energies between 0.5--1.5 GeV, whilst ArgoNeuT~\cite{ArgoNeuT:2014uwh} has measured the same on an argon target for 3.5--16 GeV muon neutrinos and 2--5 GeV muon antineutrinos. 
T2K find that, whilst both the original Rein--Sehgal model~\cite{Rein:1982pf} and the alternative Berger--Sehgal model~\cite{Berger:2008xs}, as implemented in NEUT and GENIE, are compatible within measurement uncertainties, the former slightly overestimates their result whereas the latter slightly underestimates it. For the latest MINERvA result~\cite{MINERvA:2022esg}, the GENIE implementation of both models fail to describe all measurements, with the level of disagreement depending strongly on the nuclear target and interaction kinematics. 
The neutral current analogue, the NC$1\pi^0$ coherent interaction, has been measured across a wide range of neutrino energies (0.5--50 GeV) and targets for muon neutrinos and antineutrinos~\cite{Faissner:1983ng,Isiksal:1984vh,CHARM:1985bva,SKAT:1985uch,Baltay:1986cv,MiniBooNE:2008mmr, NOMAD:2009idt,SciBooNE:2010lca, MINOS:2016yyz,NOvA:2019bdw}.

\subsubsection{CC Inclusive (CCINC)}
Measurements that include all CC topologies provide an alternative benchmark to exclusive measurements of final-state topologies for assessing model performance. Measurements of CCINC interactions in the 1--4 GeV region are especially useful as they closely resemble the signal used for neutrino oscillation analyses with the DUNE or NOvA experiments. 
When made with higher energy ($\gtrsim3$~GeV) neutrino beams they provide samples which contain a large fraction of SIS/DIS interactions, without the need for model-dependent cuts on $W$. The impact of systematic uncertainties on such measurements can also be smaller than for more exclusive topologies, especially if experiments only attempt to reconstruct the outgoing charged lepton. Conversely, interpreting discrepancies between measurements and CCINC measurements can be even more challenging, since the source of the discrepancies can be from an even wider variety of contributing interaction channels.

Early few-GeV CCINC measurements on a range of nuclear targets came from NOMAD~\cite{NOMAD:2007krq}, MINOS~\cite{MINOS:2009ugl}, SciBooNE~\cite{SciBooNE:2010slc} and ArgoNeuT~\cite{ArgoNeuT:2014rlj}. These were followed by analyses from MINERvA in the LE beam configuration~\cite{MINERvA:2017ozn,MINERvA:2016ing} which reported results as a function of true neutrino energy, a choice of variable with the potential for model dependence.
T2K~\cite{T2K:2018lnf}, MicroBooNE~\cite{MicroBooNE:2019nio}, NOvA~\cite{NOvA:2021eqi} and MINERvA, in the LE~\cite{MINERvA:2020zzv} and ME~\cite{MINERvA:2021owq} beam configurations, have more recently followed up with CCINC measurements as a function of outgoing lepton kinematic variables. MINERvA report the outgoing muon's transverse and longitudinal momenta with respect to the incoming neutrino beam ($p_{\mathrm{T}}$ and $p_{||}$), whilst T2K and NOvA report muon momentum and angle with respect to the incoming neutrino. 
These analyses allow for some kinematic separation of the contributing interaction mechanisms. For example, in MINERvA the low $p_{\mathrm{T}}$ and low $p_{||}$ region is rich in resonant events with very little SIS or DIS contributions, whereas high $p_{\mathrm{T}}$ and high $p_{||}$ is dominated by SIS and DIS interactions.
Both of the MINERvA measurements presented as a function of lepton kinematics show poor agreement with the generators tested (by MINERvA). All generator configurations underestimate the cross section, but there appears to be a particular need for a suppression of predicted cross sections in the low transverse momentum region, where the RPP contribution is largest. At slightly lower energies, the NOvA measurement is dominated by RPP interactions and also finds that all of the generators tested (by NOvA) are strongly disfavoured by the data, with particularly poor agreement in the most forward regions. The T2K and MicroBooNE measurements are at lower energy where CCQE interactions dominate the cross section (see \autoref{tab:channels}). Generators consistently over-estimate the T2K measurement for forward-going muons, as was also found for CC0$\pi$ measurements (see \autoref{fig:crpacc0picomp}).

As was discussed in the context of exclusive topologies, over the last decade experiments have explored model--measurement discrepancies through simultaneous measurements of neutrino and antineutrino cross sections~\cite{MINERvA:2016ing,T2K:2017bvo} and of cross sections on different nuclear targets~\cite{MINERvA:2026apf}. Similarly, some CCINC measurements have reported the cross section as a function of outgoing lepton and hadron kinematics to better characterise the sources of model--measurement discrepancy. 

\begin{figure}[hbtp]
  \centering
  \begin{subfigure}[b]{\linewidth}
    \centering
    \includegraphics[page=9,trim={0mm 0mm 0mm 0mm}, clip, width=1.0\linewidth]{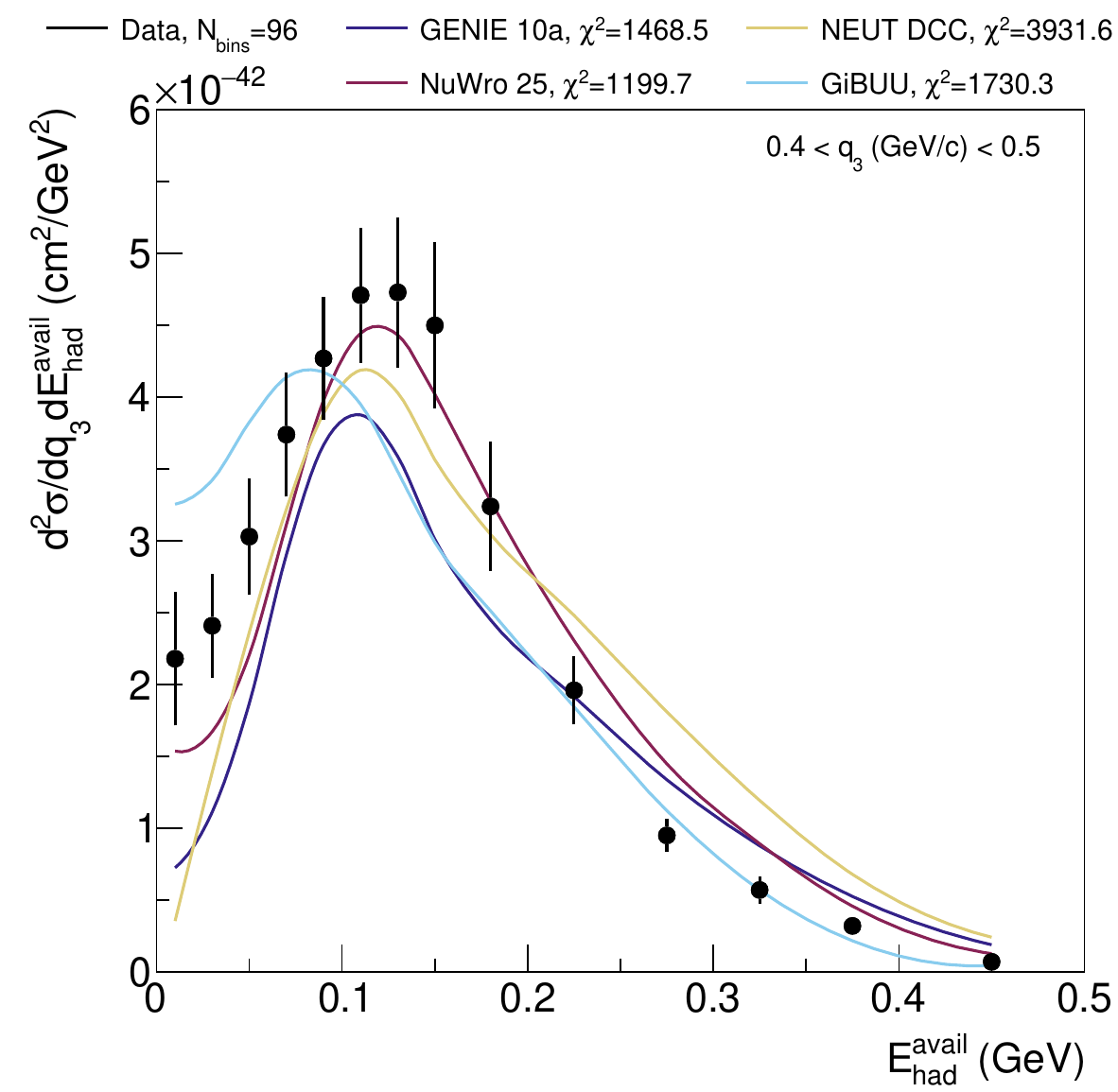}
    \caption{Generator comparison}
  \end{subfigure}
  \begin{subfigure}[b]{\linewidth}
    \centering
    \includegraphics[page=10,trim={0mm 0mm 0mm 0mm}, clip, width=1.0\linewidth]{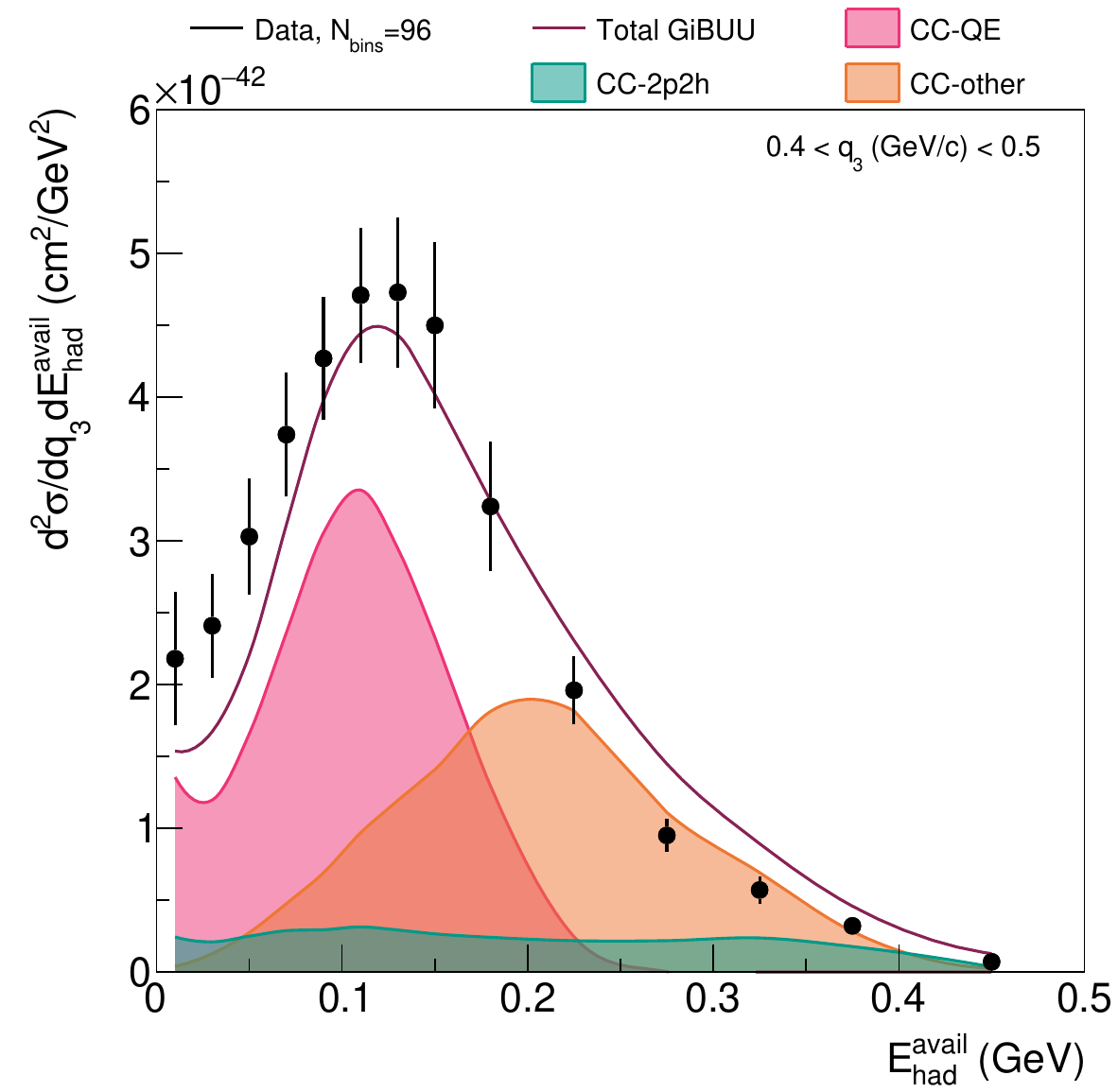}
    \caption{Channel breakdown for NuWro}
    \end{subfigure}
    \caption{Predictions of a single $|q_{3}|$ slice ($0.4<q_3~\text{GeV/c}<0.5$) of MINERvA's $\nu_\mu$--C$_8$H$_8$ CCINC low-recoil measurement shown for different generators (top) and the same slice broken down into interaction channels for NuWro 25, showing the different regions that CC-QE, CC-2p2h, and CC-other interactions occupy. The $\chi^2$ shown in the legend indicate the level of agreement between the models and measurements and are calculated using the full covariance matrix provided in the experimental data releases covering all of the bins of the measurement and not just the single $|q_3|$ slice shown here.}
    \label{fig:minerva-ccinc}
\end{figure}

Measurements from MINERvA~\cite{MINERvA:2015ydy,MINERvA:2018nab,MINERvA:2021wjs} and NOvA~\cite{NOvA:2024zmr} consider the total visible energy from the hadronic final state using $E_{\mathrm{had}}^{\mathrm{avail}}$ (introduced in \autoref{subsec:enurec}) alongside the absolute momentum transfer of the interaction, $|q_3|$. The former is correlated with energy transfer and so together they provide an approximation of a classic inclusive electron-scattering analysis (measuring $q_0$ for fixed $|q_3|$), at the cost of using a model-dependent mapping between $q_0$ and $E_{\mathrm{had}}^{\mathrm{avail}}$ to report $|q_3|$, which relies on inferring neutrino energy from visible energy deposits and so is subject to the issues described in \autoref{subsec:biasinErechad}. CCQE, 2p2h, RPP and SIS/DIS interactions occupy different regions in the $E_{\mathrm{had}}^{\mathrm{avail}}$--$|q_{3}|$ space and so such measurements can allow the partial separation of individual interaction channels despite the use of an inclusive signal definition. The first measurement of $E_{\mathrm{had}}^{\mathrm{avail}}$--$|q_3|$ came from MINERvA~\cite{MINERvA:2015ydy} in the LE beam configuration and was followed by measurements using antineutrinos~\cite{MINERvA:2018nab} and the ME beam~\cite{MINERvA:2021wjs}.
MINERvA used their LE neutrino analysis, in which they report discrepancies at low $E_{\mathrm{had}}^{\mathrm{avail}}$ and in the peak of the 2p2h distribution, to design an \textit{ad hoc} tune to their GENIEv2 model with an effective \textit{RPA correction} and large 2p2h enhancement~\cite{MINERvA:2015ydy}.
This \textit{MINERvA tune} appears to also bring their GENIEv2 model into better agreement with their LE antineutrino measurement~\cite{MINERvA:2018nab}, but not with their higher energy ME measurement~\cite{MINERvA:2021wjs}.\footnote{The MINERvA tuning effort described was achieved using data binned in reconstructed variables, before the unfolding or efficiency corrections made to extract a cross section. An important caveat is that model--data comparisons in the reconstructed space appear to show significantly less tension than in the unfolded space shown in~\autoref{fig:minerva-ccinc}.}
One slice of MINERvA's LE neutrino measurement ($0.4<q_3 <0.5$ GeV/$c$) is shown in \autoref{fig:minerva-ccinc}, overlaid with predictions from several generators and additionally broken down into true interaction channel contributions for NuWro 25. The latter shows that the measurement of $E_{\mathrm{had}}^{\mathrm{avail}}$ has \textit{some} ability to separate different interaction channels, although there is still substantial overlap between them. The very large $\chi^2$ values (for all bins in the measurement, not just the limited slice shown) show that the measurement strongly excludes all the generator predictions, with discrepancies prevalent across the full $E_{\mathrm{had}}^{\mathrm{avail}}$ range. Whilst the analysis provides some additional information about the source of modelling discrepancies compared to those that only report lepton kinematics, it remains difficult to draw firm conclusions. 

NOvA~\cite{NOvA:2026zup} and MicroBooNE~\cite{MicroBooNE:2023foc} have made complementary CCINC measurements that include hadron kinematic information, reporting triple-differential cross sections as a function of lepton kinematics and $E_{\mathrm{had}}^{\mathrm{avail}}$ or $E^{\mathrm{rec},\;\mathrm{had}}_{\nu}$ respectively. Another MicroBooNE analysis~\cite{MicroBooNE:2024zwf} measures the CCINC cross section with and without a final state proton above tracking threshold as a function of a variety of kinematic variables. None of these measurements are well described by any of the event generators tested by the respective experiments.

\subsubsection{Electron neutrino interactions}
\label{subsec:nuemeas}
As discussed in \autoref{subsec:samplemigration}, muon (anti)neutrino beams have an energy-dependent percent-level contamination of electron (anti)neutrinos. 
In principle, this could be leveraged to measure electron (anti)neutrino cross sections and directly probe $\nu^{\bracketbar}_{e}/\nu_{\mu}^{\bracketbar}$ cross section differences. However, this is challenging to do at a level of precision better than the few percent differences expected from theory for a number of reasons.
The electron (anti)neutrino contributions are more statistics limited and have challenging backgrounds from photon-producing muon-neutrino interactions. 
Additionally, as the wrong-sign electron neutrino contribution to the beam is significant compared to the right sign contribution, and because it is difficult to distinguish electrons from positrons, it is difficult to separate the electron neutrino and antineutrino contributions to any measurement. Consequently, experiments are forced to either measure the sum of the two or pick up potential model dependence by subtracting the wrong-sign contribution. Finally, even when made in the same beamline, measurements of electron and muon neutrino cross sections are not directly comparable as they are averaged over significantly different fluxes.

CCINC electron (anti)neutrino cross-section measurements on predominantly hydrocarbon targets have been made by T2K~\cite{T2K:2020lrr}, NOvA~\cite{NOvA:2022see} and MINERvA~\cite{MINERvA:2023ner}. T2K's ($\nu_e+\bar\nu_e$)--C$_8$H$_8$ measurement as a function of lepton kinematics shows some tension with the event generator predictions tested, but the model separation power of the analysis is limited by uncertainties from the subtraction of photon-related backgrounds, especially at low electron momentum, $p_e<300~\text{MeV/c}$. NOvA report a double-differential $\nu_e$ measurement as a function of outgoing lepton kinematics. The result shows reasonable agreement with a variety of generator predictions but also lacks constraining power due to large systematic uncertainties related to flux modelling and background subtraction. The MINERvA measurement provides $\nu_e$--C$_8$H$_8$ and $\bar\nu_e$--C$_8$H$_8$ cross sections using their ME beam configuration as a function of $E_{\mathrm{had}}^{\mathrm{avail}}$--$p_{\mathrm{T}}$ and $E_{\mathrm{had}}^{\mathrm{avail}}$--$|q_3|$. The measurement suggests that the baseline MINERvA generator model significantly underpredicts the cross section at low $p_{\mathrm{T}}$ and low $E_{\mathrm{had}}^{\mathrm{avail}}$ for both the $\nu_e$ and $\bar\nu_e$ measurements, as was also found for their analogous muon-neutrino measurement~\cite{MINERvA:2021wjs}. MINERvA have also made a $\nu_e$--C$_8$H$_8$ CC0$\pi$ measurement~\cite{MINERvA:2015jih} using the LE beam configuration. As with T2K, the MINERvA analysis struggles to reconstruct the lower-energy electrons ($p_e<2~\text{GeV/c}$) due to photon-related backgrounds.

CCINC ($\nu_e+\bar\nu_e$)--\argon cross sections have been measured by ArgoNeuT~\cite{ArgoNeuT:2020kir} and MicroBooNE~\cite{MicroBooNE:2021gfj,MicroBooNE:2021ppm}. The MicroBooNE measurements use the NuMI beam, as the detector's large off-axis angle from the beam centre amplifies the relative $\nu_e+\bar\nu_e$ contribution to the flux. MicroBooNE have further measured the $\nu_e$--\argon CC0$\pi$ cross section in both the NuMI~\cite{MicroBooNE:2026ifl} and BNB~\cite{MicroBooNE:2022tdd,MicroBooNE:2025aiw} neutrino beams. Similarly to the T2K C$_{8}$H$_{8}$ measurements, the results suffer from limited statistics and large backgrounds at low electron momenta. MicroBooNE find that the generators they compare against are broadly better at describing electron kinematics than hadron kinematics.

Additionally, T2K~\cite{T2K:2025smz} and MicroBooNE~\cite{MicroBooNE:2025pvb} have both reported \nue CC$1\pi^+$ cross-section measurements. These measurements are described well by generators within their currently large uncertainties.

\subsection{Cross-section measurements as inputs to neutrino oscillation analyses}
\label{subsec:xsecforoa}

Cross-section measurements have exposed serious deficiencies in various aspects of neutrino interaction modelling that oscillation experiments rely on (as introduced in \autoref{sec:howtooa}). For example, for experiments using kinematic reconstruction techniques, mismodelling of the angular distributions of pions or muons can be a major problem when considering the propagation between ND and FD with different angular acceptances. 
Similarly, experiments hoping to use calorimetric neutrino energy reconstruction for precision oscillation measurements must address the major discrepancies in cross-section measurements as a function of hadronic energy deposits, as this indicates deficiencies in modelling of the physics which determines the expected neutrino energy reconstruction bias. Unfortunately, it is unclear how to use cross-section measurements to overcome these challenges.

Ideally, generators would be accompanied by a parameterised uncertainty model that could be fitted to cross-section measurements in order to benchmark the parametrisation's suitability or to set prior parameter constraints for use in oscillation measurements. However, all efforts to tune theoretically justified model uncertainties have highlighted that the addition of {\it ad hoc} parameters to alter cross-section model predictions are required to achieve reasonable agreement, even to small subsets of the available neutrino--nucleus cross-section data~\cite{GENIE:2022qrc, GENIE:2021zuu,MINERvA:2019kfr,NOvA:2020rbg, Chakrani:2023htw, MicroBooNE:2021ccs, Wolfs:2025ofb}. The situation is further complicated by the possibility of uncertainty under-coverage or bias from model dependence in published measurements, making it unclear whether challenges in fitting experimental results stem from poor modelling or imperfect measurements.

Contrary to the case of neutrino--nucleus measurements, fits to measurements of neutrino--nucleon scattering~\cite{Kabirnezhad:2024cor, Meyer:2016oeg, Graczyk:2009qm} are far simpler and have provided useful constraints on parameterisations of the QE and RPP form factors~\cite{NOvA:2020rbg, MicroBooNE:2021ccs, T2K:2023smv}. However, the bubble chamber measurements used for these constraints, especially for pion production, may not be reliable and tensions between the available analyses have been identified~\cite{Wilkinson:2014yfa,Sobczyk:2014xza,Belusevic:1988ab,Nikolakopoulos:2022tut,Kabirnezhad:2017jmf}. Additionally, pion production bubble-chamber measurements do not typically include outgoing hadron kinematics, limiting their utility. New measurements of interactions on free nucleons, either directly with new hydrogen-target experiments~\cite{Alvarez-Ruso:2022ctb, Alvarez-Ruso:2022exy} or by using techniques to separate the hydrogen from composite targets~\cite{lu:2015hea, MINERvA:2023avz, Munteanu:2019llq, Baudis:2023tma, Duyang:2018xcc, Duyang:2019prb}, may therefore provide a useful means for cross-section measurements to directly constrain model parameterisations used for oscillation measurements. 

Whilst directly tuning models to neutrino--nucleus cross-section data is challenging, the disagreement between models and measurements can be used to motivate new model uncertainty treatments for use in oscillation measurements. 
For example, the T2K experiment introduced a low-$Q^2$ suppression as an alternative model to test the robustness of their oscillation measurements~\cite{T2K:2023smv}, which was motivated by an observed data--simulation discrepancy observed in MINERvA pion production measurements~\cite{MINERvA:2019kfr}.
However, without being able to isolate the cause of any discrepancy, this approach can only give a \textit{sense} of how important an observed discrepancy \textit{may} be, and is unlikely to be suitable for future precision oscillation measurements with DUNE and Hyper-K.

Future neutrino--nucleus cross-section measurements from high performance detectors remain a potentially useful tool. In particular, measurements focussed on better characterising the source of discrepancies seen today may be of value, for example using the higher statistics expected with SBND~\cite{SBND:2025lha} with respect to MicroBooNE or the higher precision offered by T2K's recent ND upgrade~\cite{T2K:2019bbb} with respect to its predecessor. 
Furthermore, the first cross-section measurements from the NuMI beam with ICARUS~\cite{ICARUS:2023gpo} may provide some sensitivity of neutrino--argon cross sections for neutrino energies in the 1--4 GeV range relevant for DUNE. Similarly, the \textit{2x2} DUNE ND prototype, positioned on-axis in the 2--8 GeV NuMI ME beam, will provide further complementary data, but will be limited by the small detector size and containment, particularly of the outgoing lepton~\cite{Russell:2024az}. 

Although there is a rich program of upcoming measurements planned, the limited prospects for precision cross-section measurements which explore the SIS and DIS regions relevant to DUNE, or antineutrino--argon interactions in general, remains problematic. 
Moreover, in spite of recent significant theory developments, current cross-section measurements cannot be described by any model and the exposed discrepancies have not enabled a breakthrough in interaction modelling. 
Exactly what future cross-section analyses can be made to confront the challenges facing DUNE and Hyper-K is therefore not clear. Conversely, it \textit{is} clear that the current cycle of benchmarking models against measurements---that may themselves depend on the models used to make them---is problematic and this issue would benefit from continued community attention.

\FloatBarrier
\section{Uncertainties in modelling crucial aspects of neutrino interactions for oscillation measurements}
\label{sec:issuesforoa}

The extrapolation of ND constraints to FD predictions in oscillation analyses requires the use of a neutrino interaction model (\autoref{sec:howtooa}). This section studies how that extrapolation depends on that model. As before in this work, we highlight the potential impact of making different interaction model choices using the spread in predictions from neutrino event generators described in \autoref{sec:evgen}, with the important caveat that there is no compelling reason to believe that this covers all plausible cases or fully describes the expected uncertainty on the neutrino interaction model. The discussion follows the critical modelling components described in \autoref{sec:howtooa}, and investigates the model spread for both a DUNE-like case, with CCINC interactions on an argon target, and a Hyper-K/T2K-like case, with CC0$\pi$ interactions on a water target. The generator predictions shown throughout the section depend on the observable under consideration. 
As GENIE 10b and 10c only differ from GENIE 10a in their treatment of hadron FSI, we do not include them for observables which are not {\it strongly} sensitive to hadron kinematic variations\footnote{Strictly speaking CC0$\pi$ cross sections always depend on FSI due to migrations of non-QE events into CC0$\pi$ samples, but GENIE 10b and 10c are neglected for legibility.}.

\subsection{Cross-section energy dependence}
\label{subsec:issuesEdep}

\begin{figure*}[htbp]
  \centering
  \captionsetup[subfigure]{aboveskip=0pt,belowskip=3pt}       
  \begin{subfigure}[b]{0.48\textwidth}
      \centering
      \includegraphics[width=1.00\linewidth]{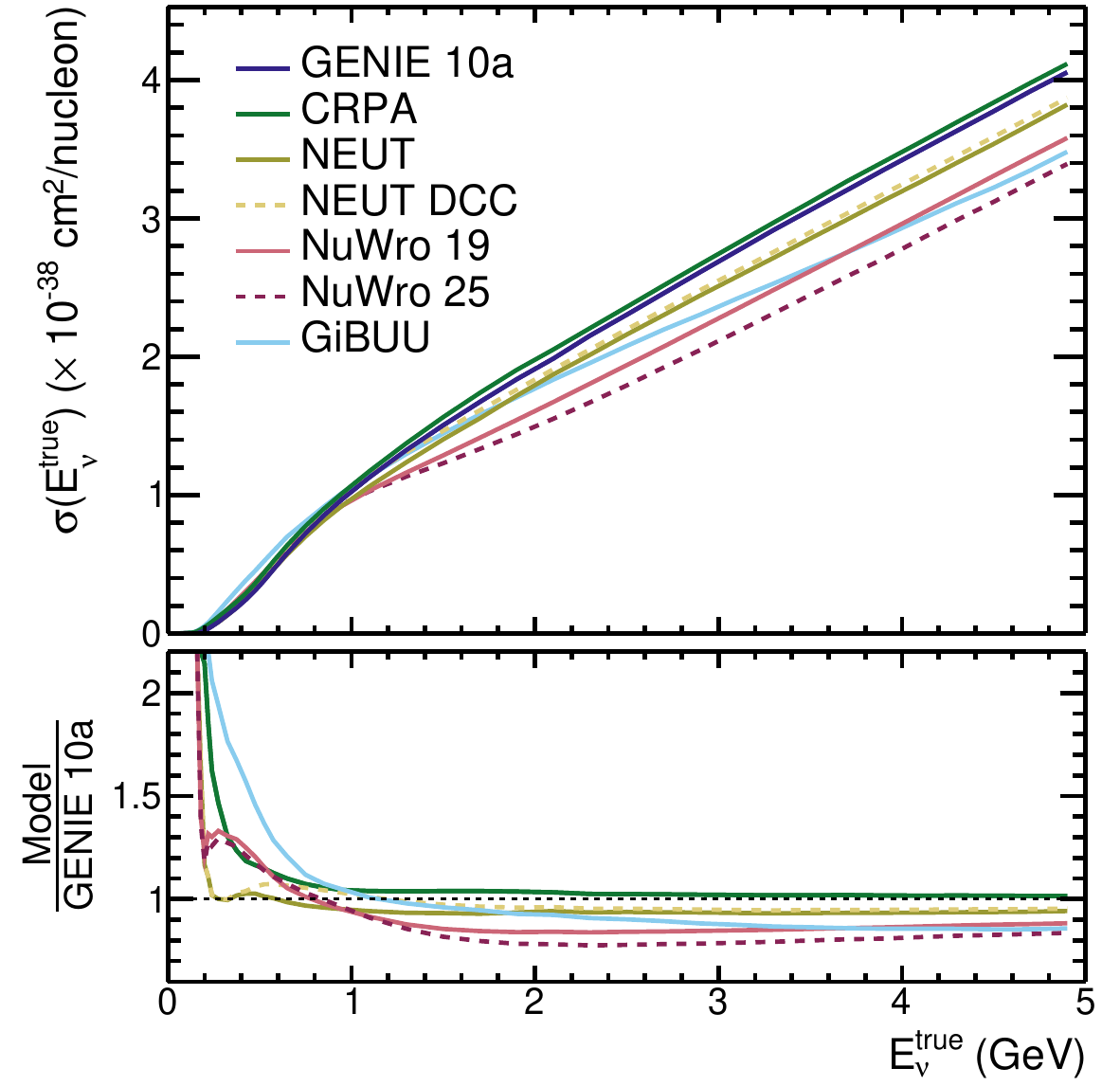}
      \caption{\numu--\argon CCINC}
  \end{subfigure}
  \begin{subfigure}[b]{0.48\textwidth}
      \centering
      \includegraphics[width=1.00\linewidth]{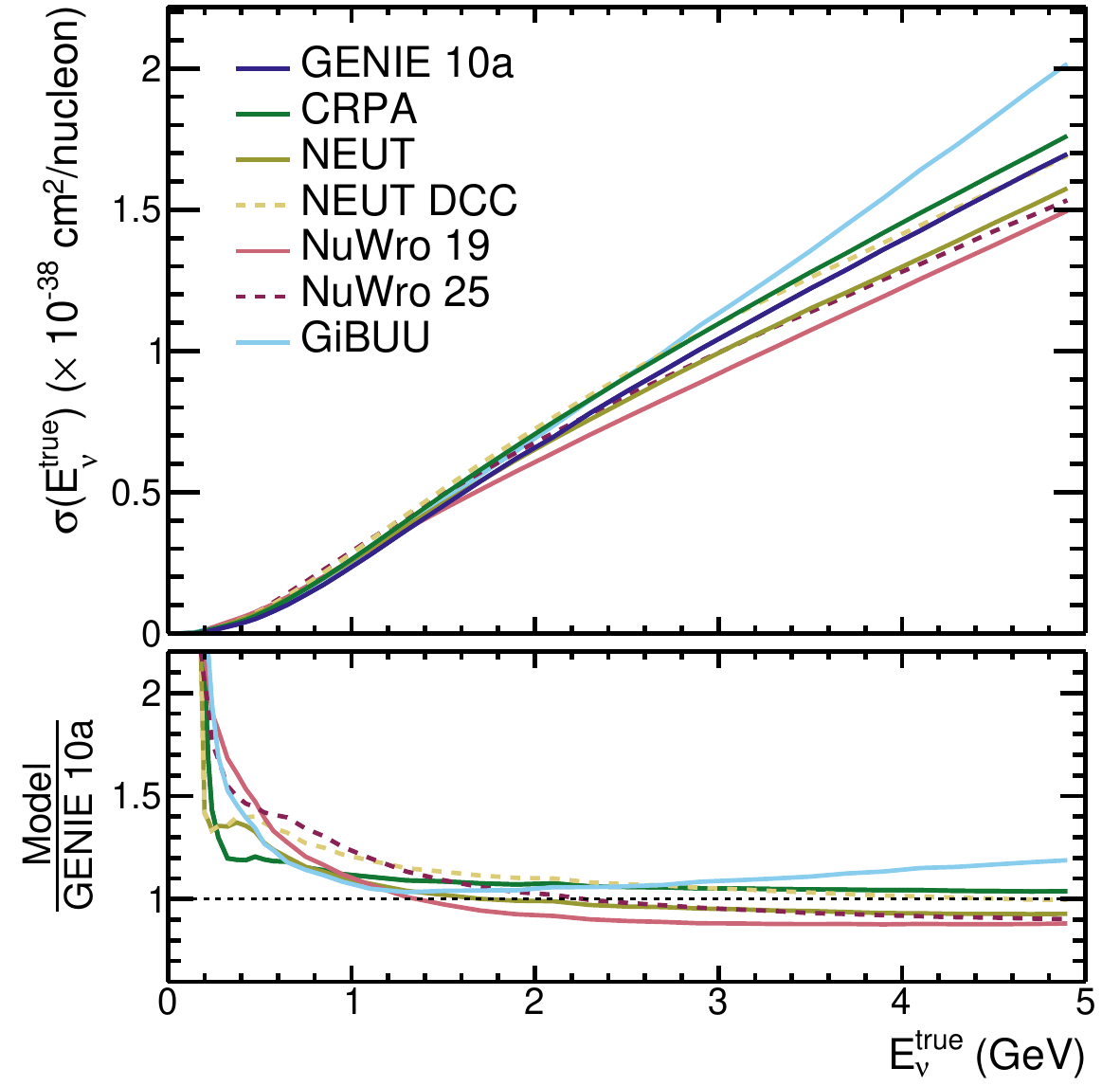}
      \caption{\numub--\argon CCINC}
  \end{subfigure}
  \\
   \begin{subfigure}[b]{0.48\textwidth}
      \centering
      \includegraphics[width=1.00\linewidth]{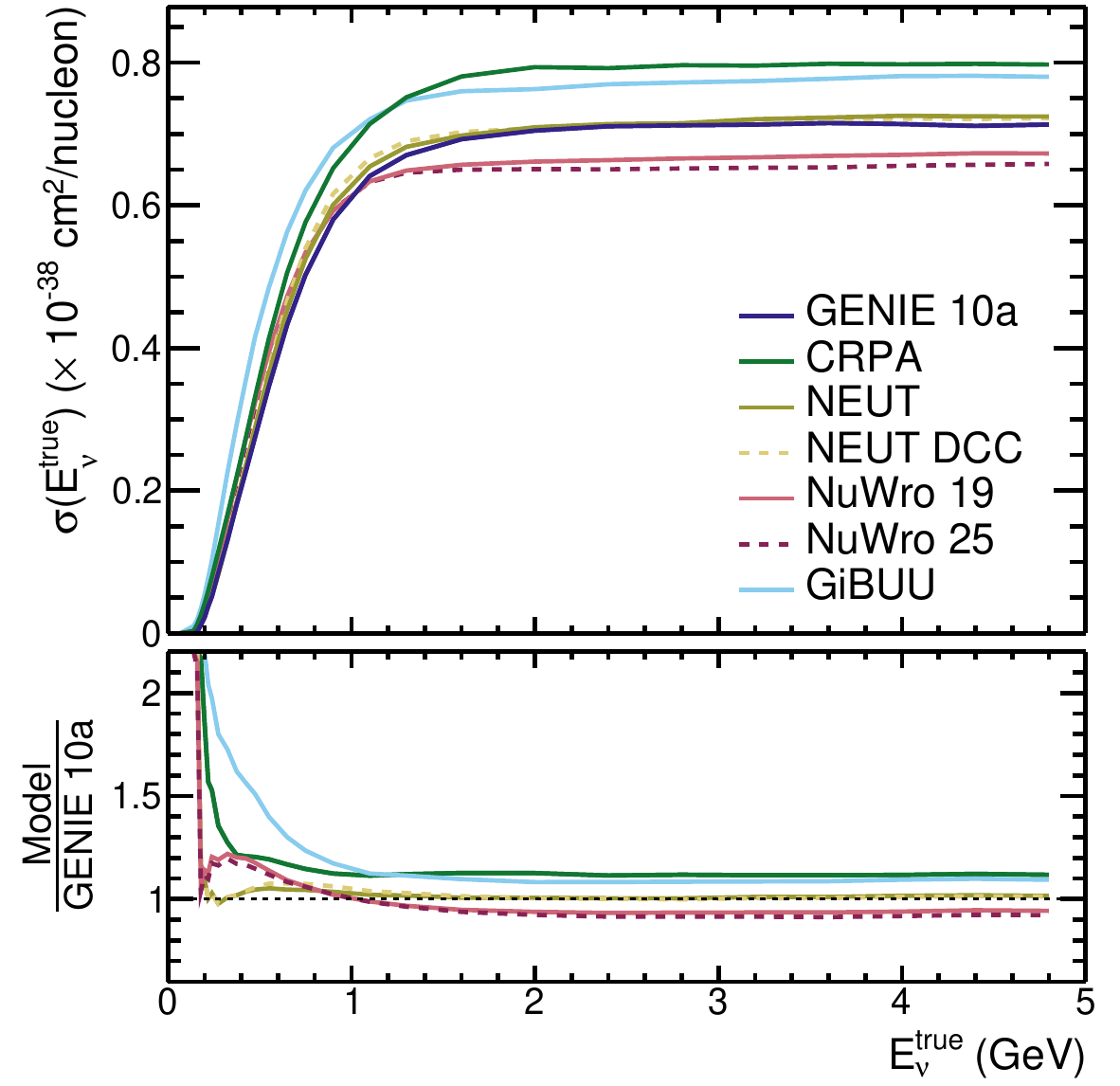}
      \caption{\numu--$^{16}$O CC0$\pi$}
  \end{subfigure}
   \begin{subfigure}[b]{0.48\textwidth}
      \centering
      \includegraphics[width=1.00\linewidth]{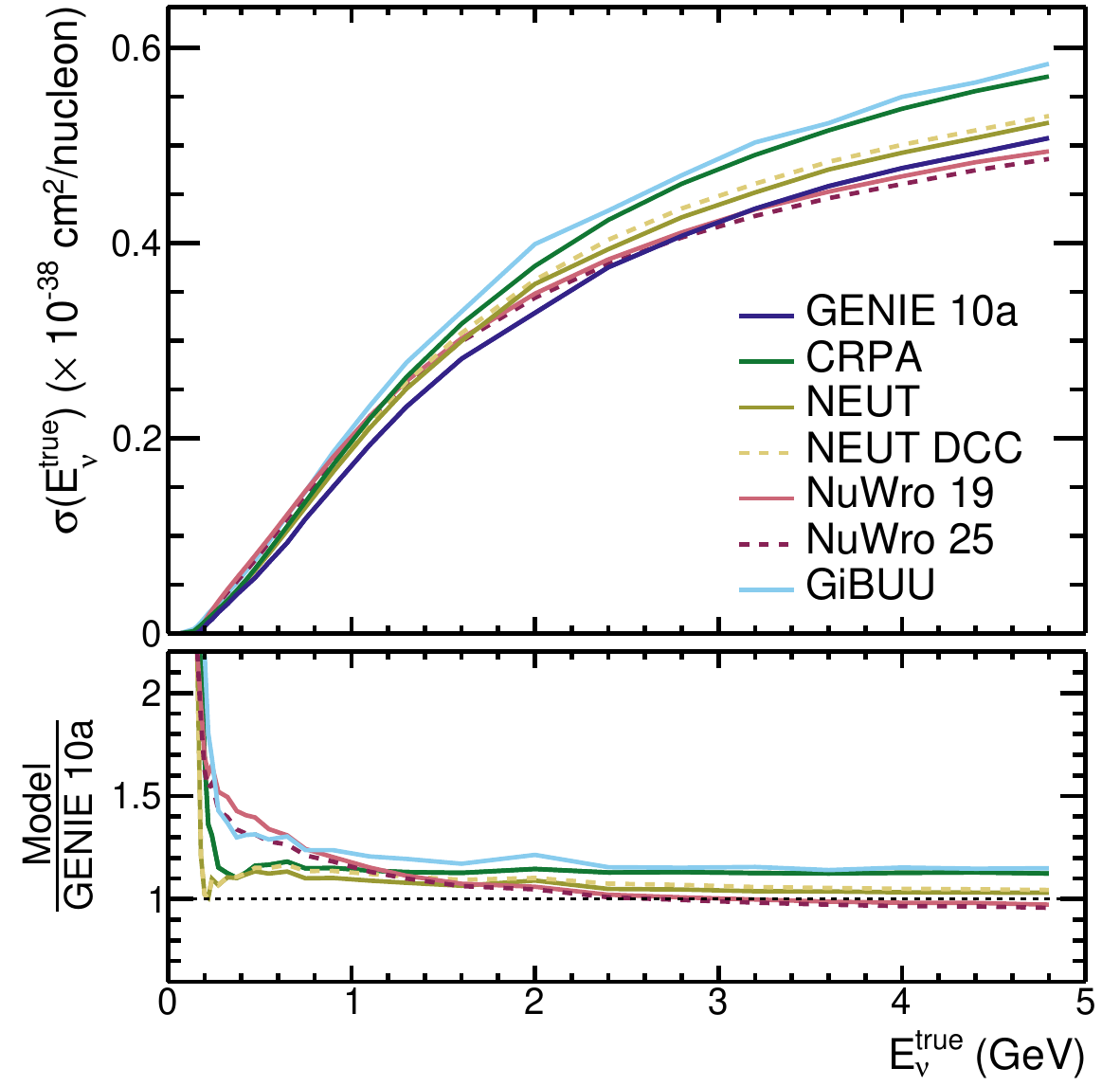}
      \caption{\numub--$^{16}$O CC0$\pi$}
  \end{subfigure}
  \caption{A comparison of the CCINC (CC0$\pi$) cross section as a function of \enutrue for \numu--\argon and \numub--\argon (\numu--$^{16}$O and \numub--$^{16}$O) interactions, for a variety of neutrino generators.
  }
  \label{fig:enu_xsec_gencomp}
\end{figure*}

Because of the radically different neutrino flavour and energy profiles at the ND and FD, the energy dependence of the cross section must be well understood to extrapolate ND constraints to FD predictions. \autoref{fig:enu_xsec_gencomp} shows the muon neutrino and antineutrino cross section as a function of neutrino energy for the DUNE and Hyper-K-like scenarios.
In both cases, there are greater than 20\% differences between generator predictions across the full neutrino energy range. Below $\sim$1 GeV the differences are even larger, approaching 30\% at the Hyper-K/T2K neutrino oscillation maxima ($\sim$0.6~GeV).
The pronounced differences at low energies are likely to be the result of threshold effects and different interaction mechanisms becoming accessible. However, it is difficult to assert that the higher-energy region is better under control, as the models for DIS interactions are very similar between neutrino event generators simply because they have received less community attention.

\begin{figure*}[htbp]
  \centering
  \begin{subfigure}[b]{0.48\textwidth}
      \centering
      \includegraphics[width=1.00\linewidth]{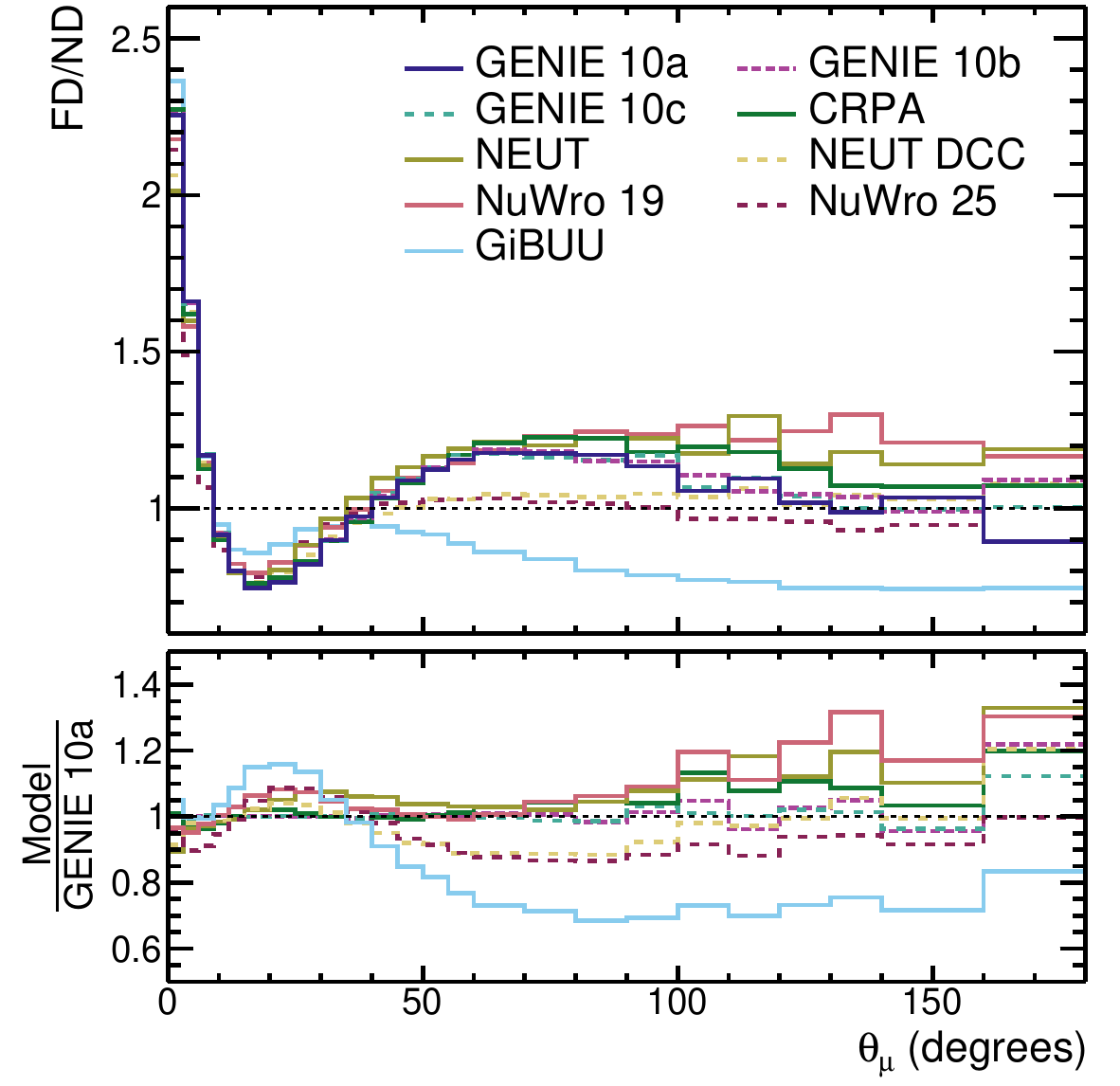}
      \caption{DUNE FD/ND \numub--\argon}
  \end{subfigure}
  \begin{subfigure}[b]{0.48\textwidth}
      \centering
      \includegraphics[width=1.00\linewidth]{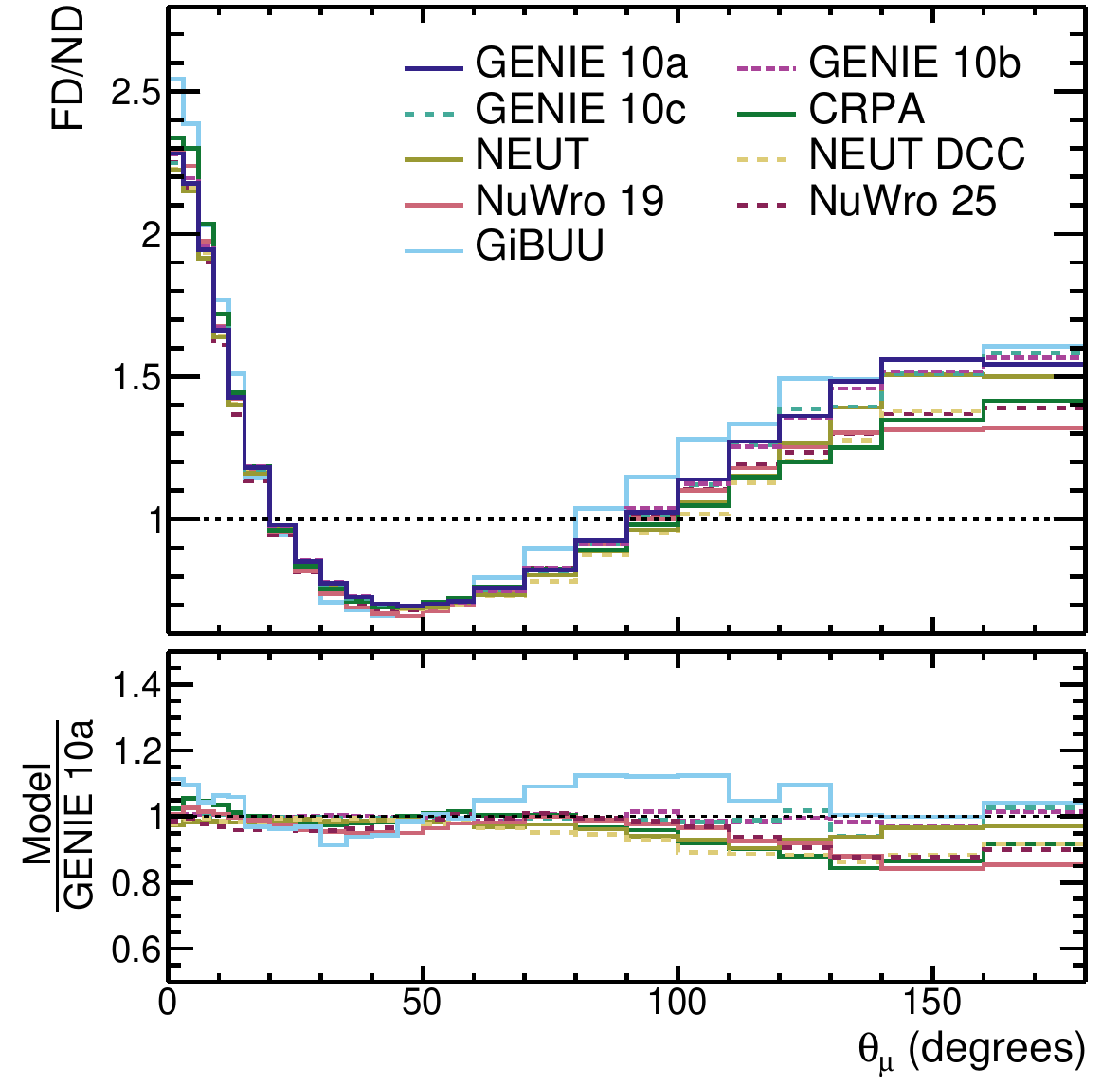}
      \caption{Hyper-K/T2K FD/ND \numub--\water}
  \end{subfigure}
  \caption{The FD/ND flux-averaged cross section ratio as a function of the outgoing muon angle for DUNE \numub--\argon CCINC interactions and Hyper-K/T2K \numub--\water CC0$\pi$ interactions, shown for a variety of neutrino generators.}
  \label{fig:angFDND_gencomp}
\end{figure*}

To extrapolate between the ND and FDs, models are relied upon not only to predict the evolution of the total cross section, but also how differential cross sections evolve as a function of neutrino energy.
\autoref{fig:angFDND_gencomp} shows the ratio of the differential cross section as a function of outgoing muon angle, averaged over the ND and FD antineutrino-enhanced fluxes, for the DUNE and Hyper-K-like scenarios. Whilst the generator models predict similar broad trends, they differ by at least 10\% across most of the angular phase space and 20--40\% for angles above 40(60)$^\circ$ for the DUNE (Hyper-K) case. Although not shown, differences between generators are slightly smaller (5--10\%) for the neutrino-enhanced fluxes.

\vspace{-3mm}
\subsection{Neutrino energy reconstruction}
\label{subsec:issuesEnurec}

\begin{figure*}[htbp]
  \centering
  \begin{subfigure}[b]{0.48\textwidth}
      \centering
      \includegraphics[width=1.00\linewidth]{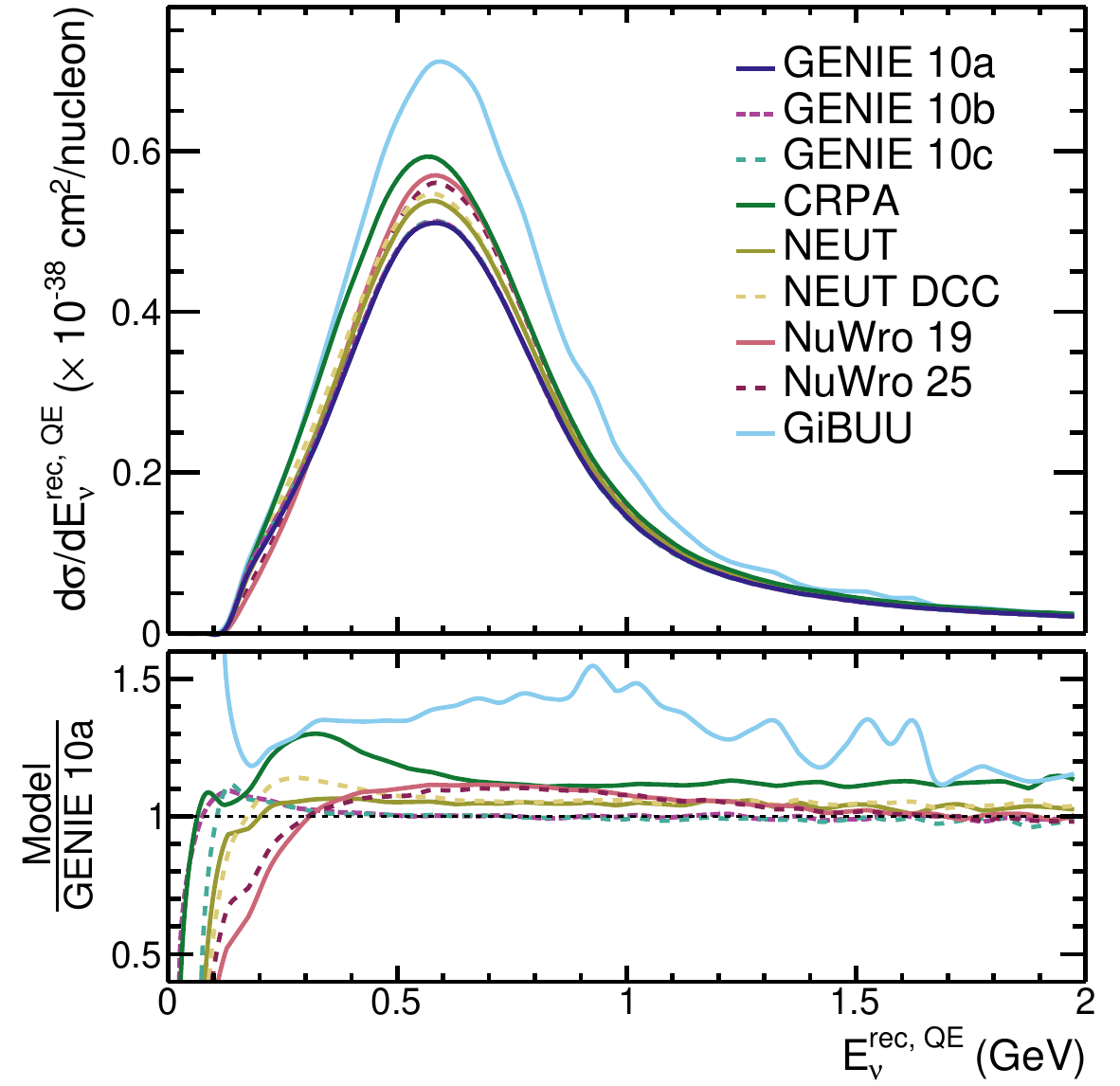}
      \caption{Hyper-K/T2K ND \numu--\water \enuqe spectra}
  \end{subfigure}
  \begin{subfigure}[b]{0.48\textwidth}
      \centering
      \includegraphics[width=1.00\linewidth]{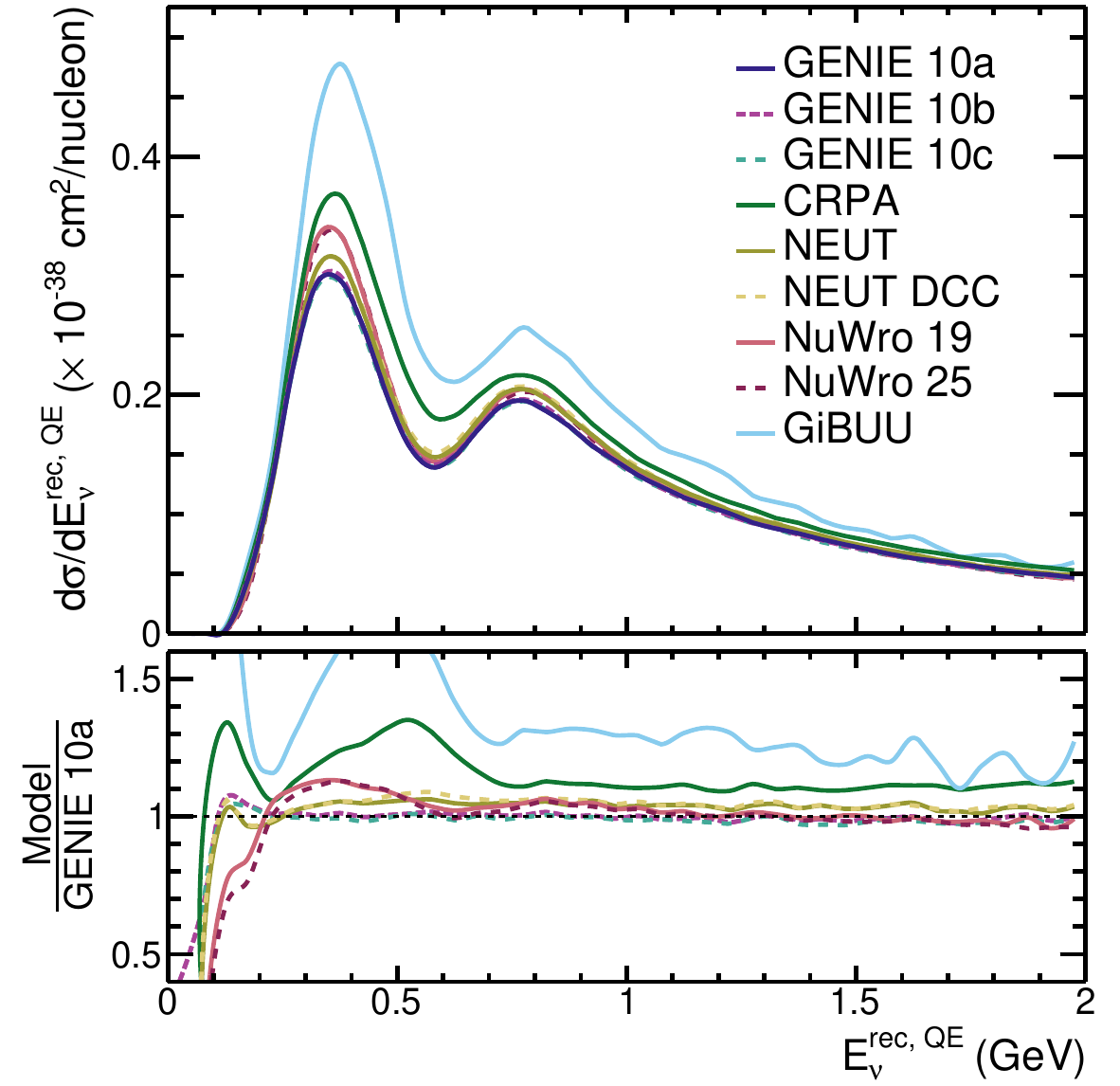}
      \caption{Hyper-K/T2K FD \numu--\water \enuqe spectra}
  \end{subfigure}
  \\
   \begin{subfigure}[b]{0.48\textwidth}
      \centering
      \includegraphics[width=1.00\linewidth]{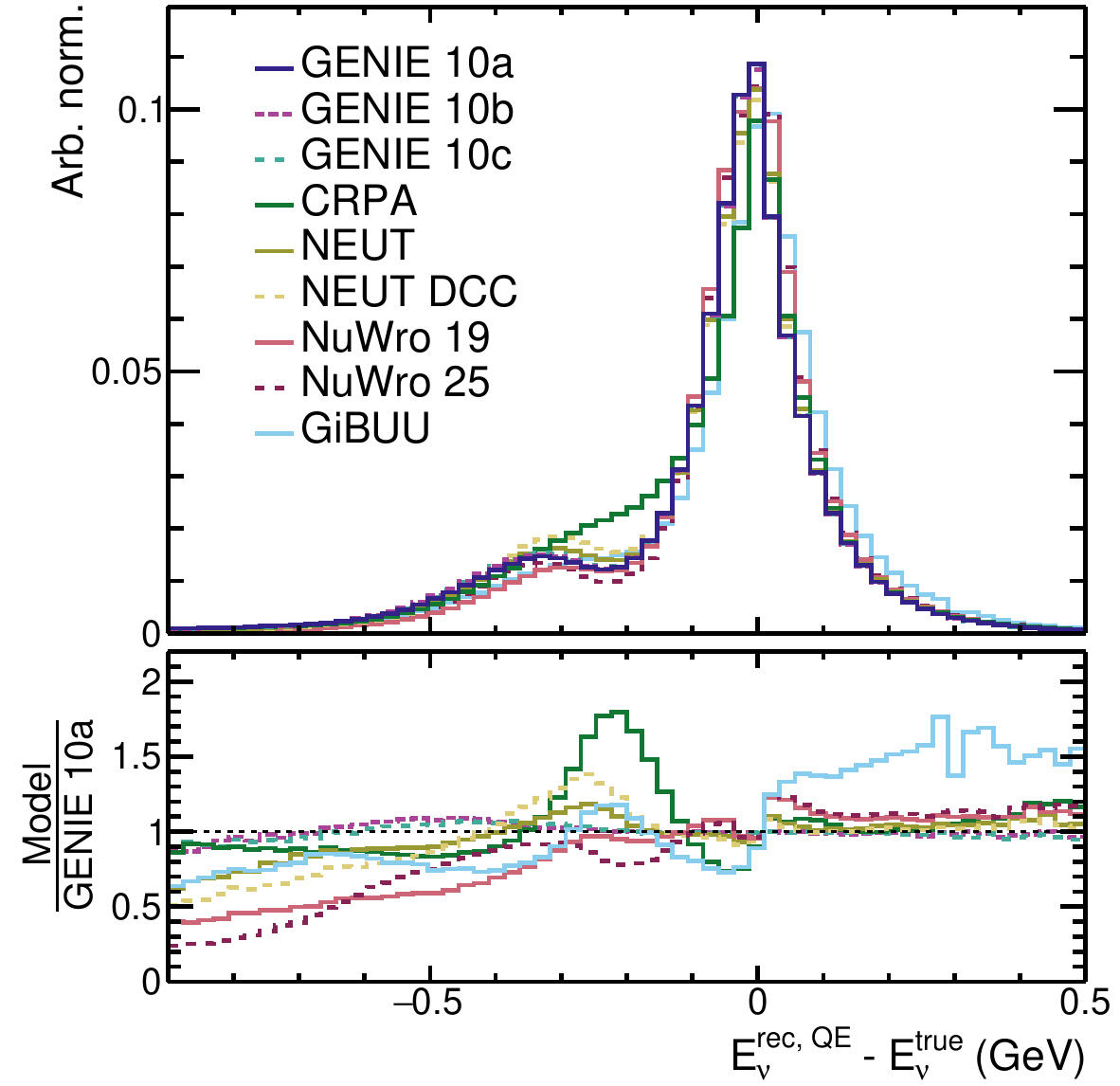}
      \caption{Hyper-K/T2K ND \numu--\water energy bias}
      \label{fig:T2K_erec_gencomp_noosc}
  \end{subfigure}
   \begin{subfigure}[b]{0.48\textwidth}
      \centering
      \includegraphics[width=1.00\linewidth]{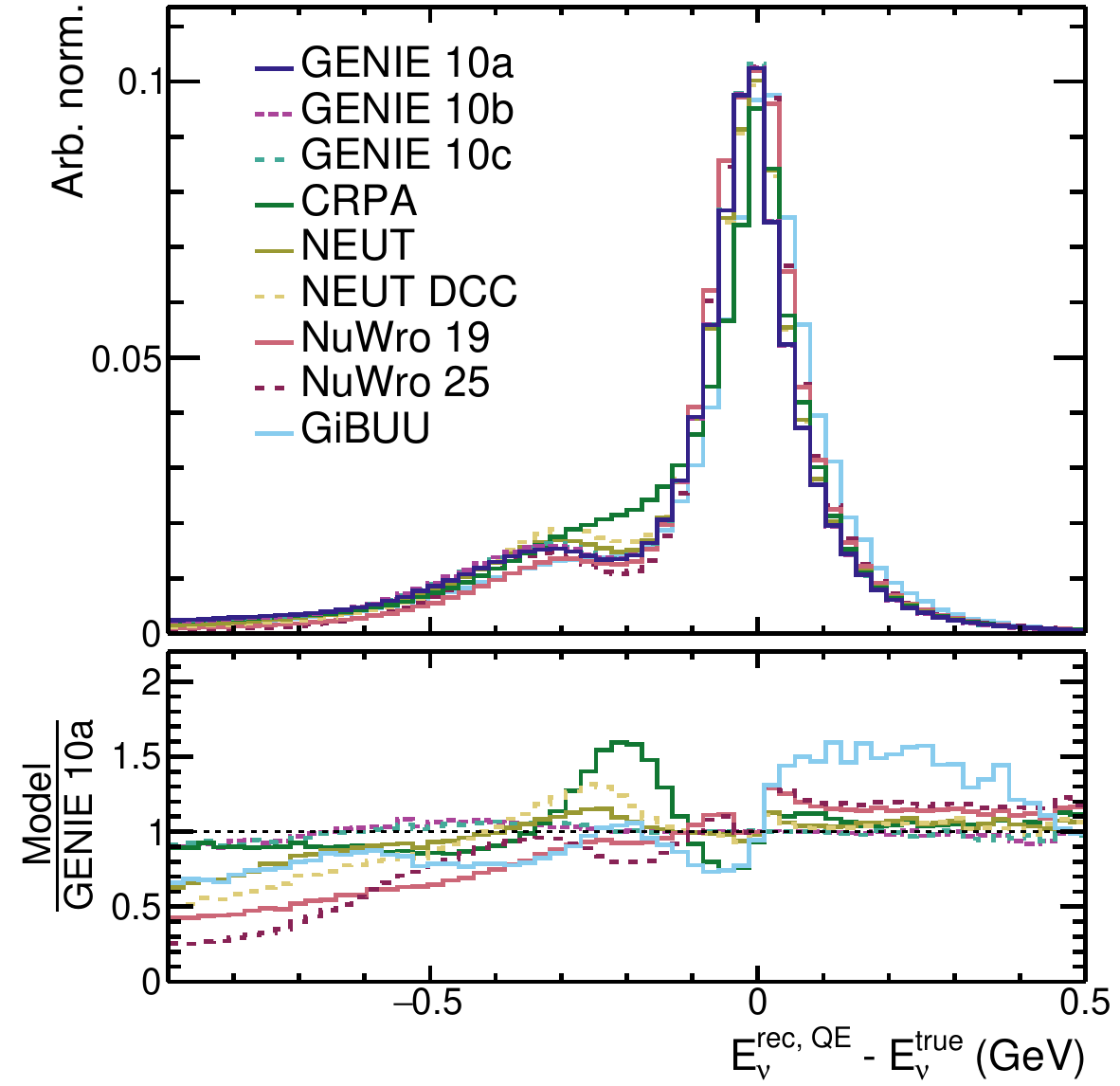}
      \caption{Hyper-K/T2K FD \numu--\water energy bias}
      \label{fig:T2K_erec_gencomp_osc}
  \end{subfigure}
  \caption{A comparison of the reconstructed neutrino energy spectra and absolute neutrino energy bias for \numu--\water CC0$\pi$ interactions using the Hyper-K/T2K ND unoscillated and FD oscillated fluxes, for a variety of neutrino generators.}
  \label{fig:T2K_erec_gencomp}
\end{figure*}

\begin{figure*}[htb]
  \begin{minipage}{\textwidth}
  \centering
  \captionsetup[subfigure]{aboveskip=0pt,belowskip=3pt}       
  \begin{subfigure}[b]{0.48\textwidth}
      \centering
      \includegraphics[width=1.00\linewidth]{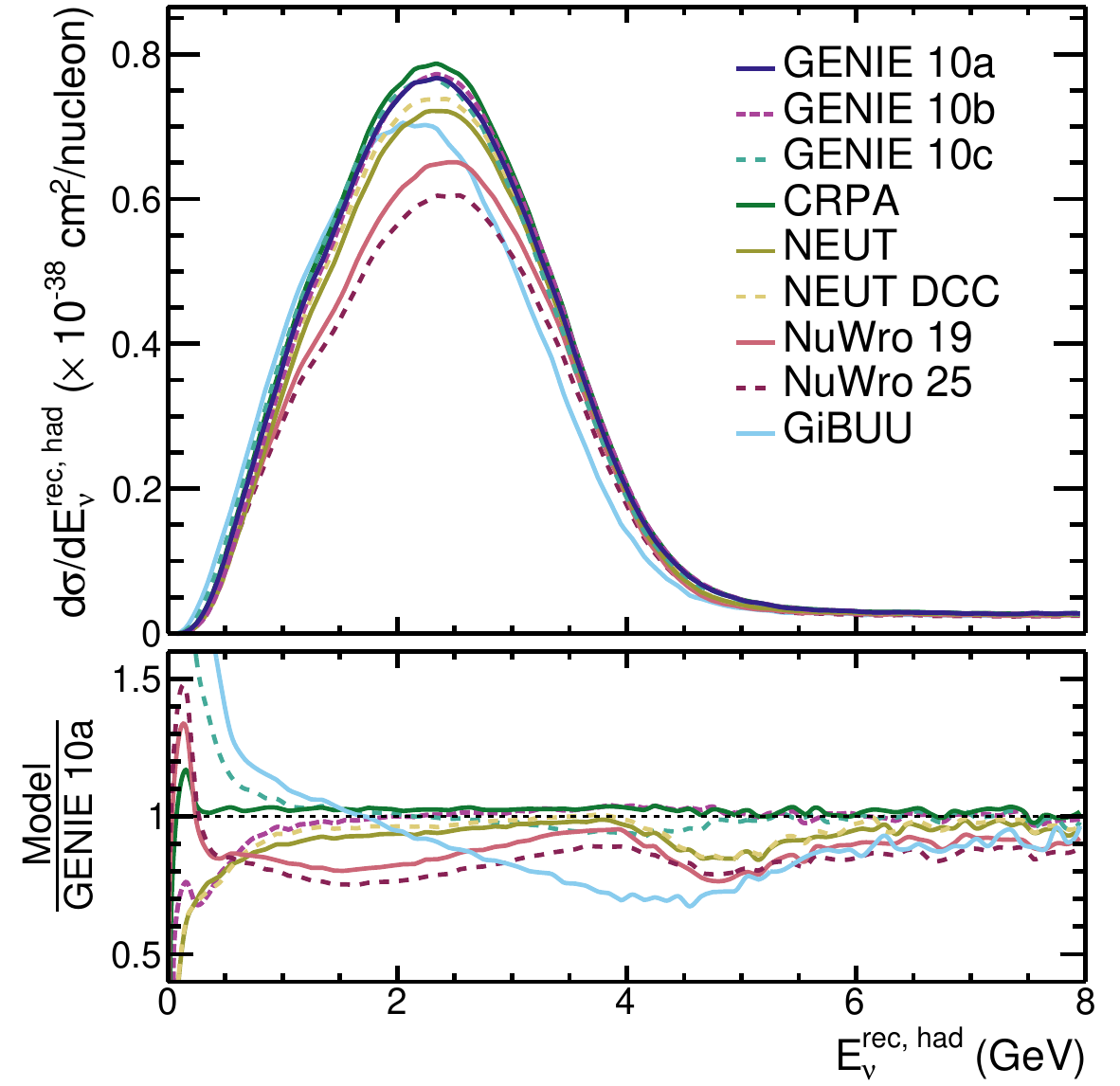}
      \caption{DUNE ND FHC \numu \enureco spectra}
  \end{subfigure}
  \begin{subfigure}[b]{0.48\textwidth}
      \centering
      \includegraphics[width=1.00\linewidth]{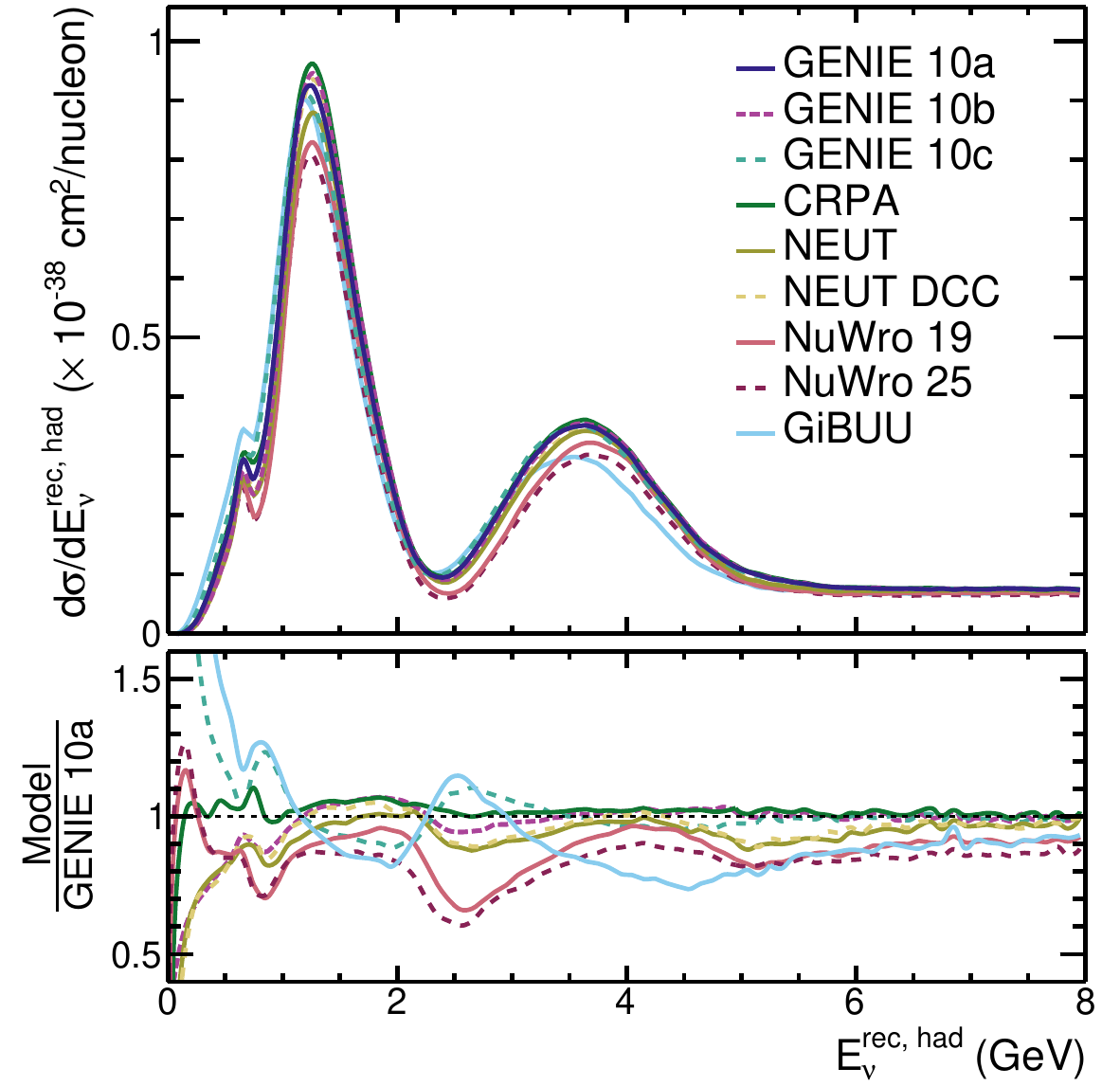}
      \caption{DUNE FD FHC \numu \enureco spectra}
  \end{subfigure}
  \\
   \begin{subfigure}[b]{0.48\textwidth}
      \centering
      \includegraphics[width=1.00\linewidth]{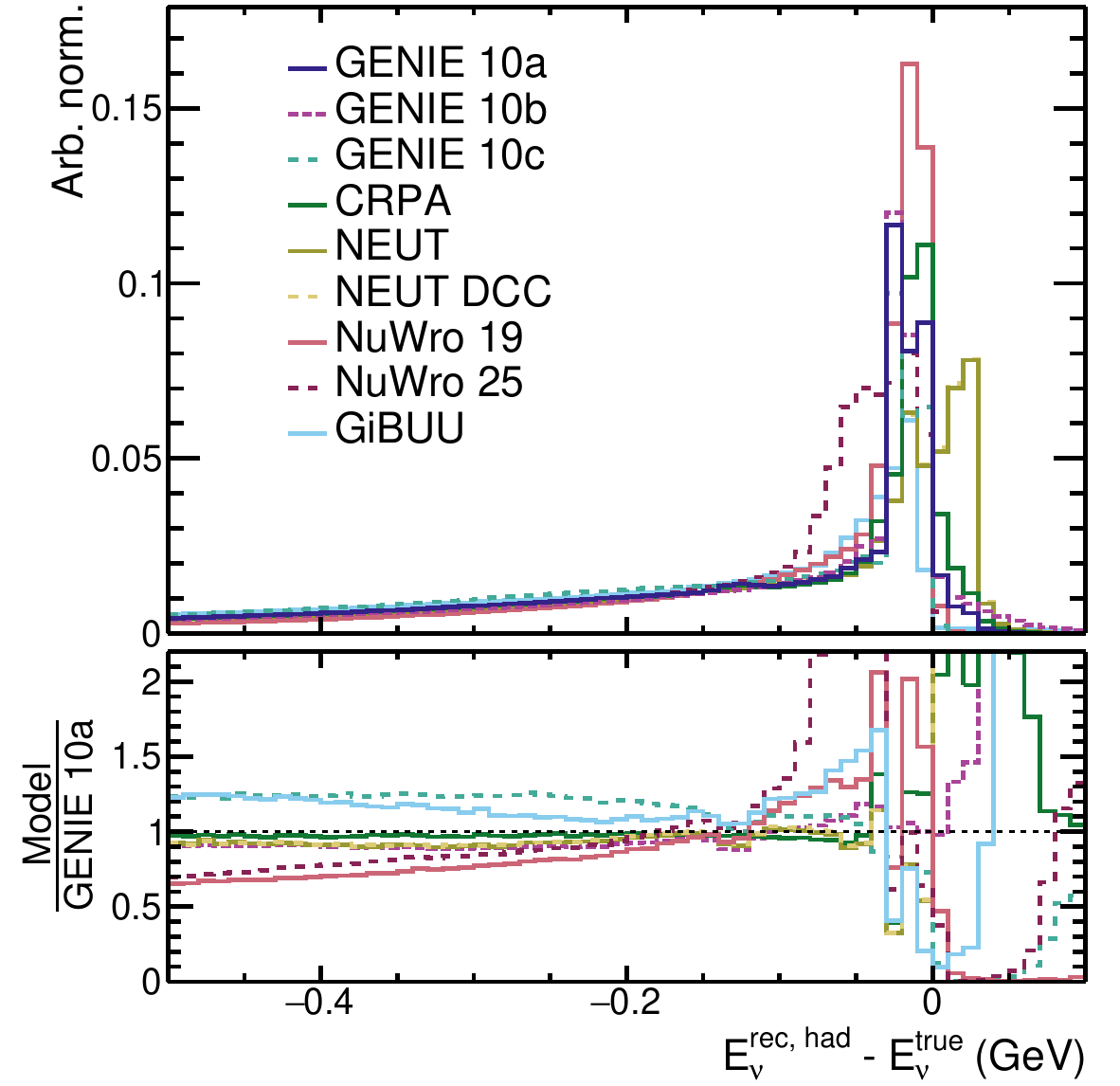}
      \caption{DUNE ND FHC \numu energy bias}
  \end{subfigure}
   \begin{subfigure}[b]{0.48\textwidth}
      \centering
      \includegraphics[width=1.00\linewidth]{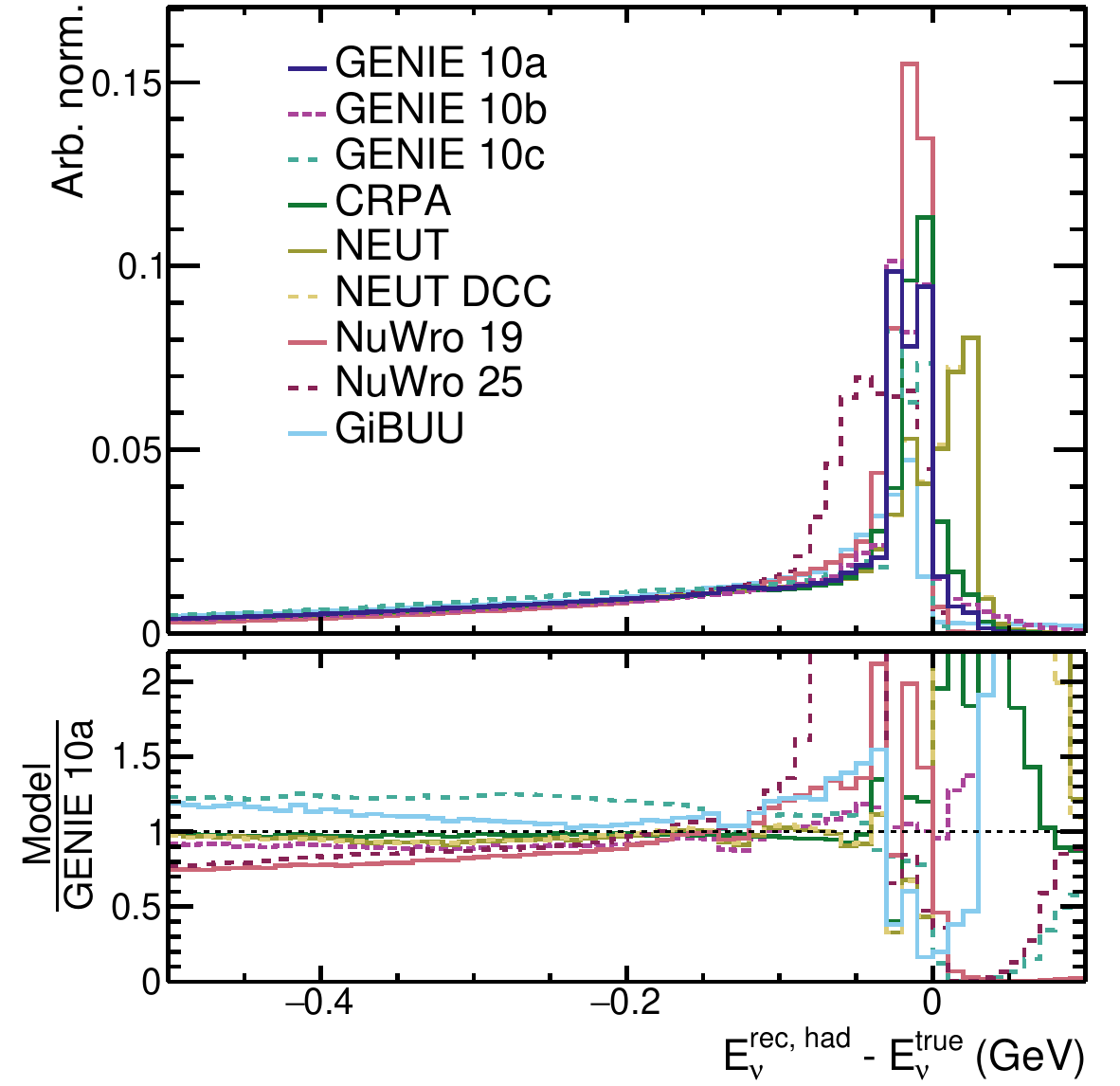}
      \caption{DUNE FD FHC \numu energy bias}
  \end{subfigure}
  \caption[]{A comparison of the reconstructed neutrino energy spectra and absolute neutrino energy bias for \numu--\argon CCINC interactions using the DUNE ND unoscillated and FD oscillated fluxes, for a variety of neutrino generators\footnote{Some NEUT- and GENIE-based models predict interactions in which the calorimetrically reconstructed neutrino energy is larger than the true neutrino energy, implying that there is more energy in the final state than the initial state. These non-energy conserving interactions are due to approximations made in the event generation: for NEUT it is due to the lack of binding energy for non-QE interactions; for GENIE it is related to the potential for acceleration of hadrons via FSI.}.}
  \label{fig:DUNE_erec_gencomp}
  \end{minipage}
\end{figure*}

A well understood, although not necessarily precise, metric for neutrino energy reconstruction is essential for making precision neutrino oscillation measurements. \autoref{fig:T2K_erec_gencomp} shows the spread of generator predictions for the Hyper-K/T2K oscillated and unoscillated \numu--\water CC0$\pi$ neutrino energy spectra, reconstructed using the kinematic method (\autoref{eq:enuqe}). \autoref{fig:DUNE_erec_gencomp} shows the equivalent for DUNE, using the calorimetric method (\autoref{eq:enuhad}) and \numu--\argon CCINC interactions. Both figures also show the neutrino energy bias.

It is clear that, for both Hyper-K/T2K and DUNE, the reconstructed energy spectra are very sensitive to model variations. For example, the relative proportion of events reconstructed within $\sim$10\% of the true neutrino energy can differ by more than 20\% and 50\% for the kinematic and calorimetric methods, respectively. Furthermore, the differences between generator models vary for unoscillated and oscillated fluxes. The fact that both neutrino energy reconstruction methods are themselves energy dependent underlines the challenge of extrapolating ND constraints to FD predictions. Further comparisons of the sensitivity of neutrino energy reconstruction to cross-section modelling can be found in Refs.~\cite{Leitner:2010kp,Martini:2012fa,Nieves:2012yz,Martini:2012uc,Ankowski:2015jya,Ankowski:2015kya,Nikolakopoulos:2018sbo,Ankowski:2014yfa,Friedland:2018vry,Coyle:2025xjk,Gallmeister:2025jwe,Dolan:2026dfu,Liu:2025hpl}.

\FloatBarrier

\subsection{Kinematic acceptance differences}
\label{subsec:issuesAccept}
\begin{figure}[H]
    \centering    \includegraphics[width=0.45\textwidth,clip,trim=6mm 2mm 16mm 15mm]{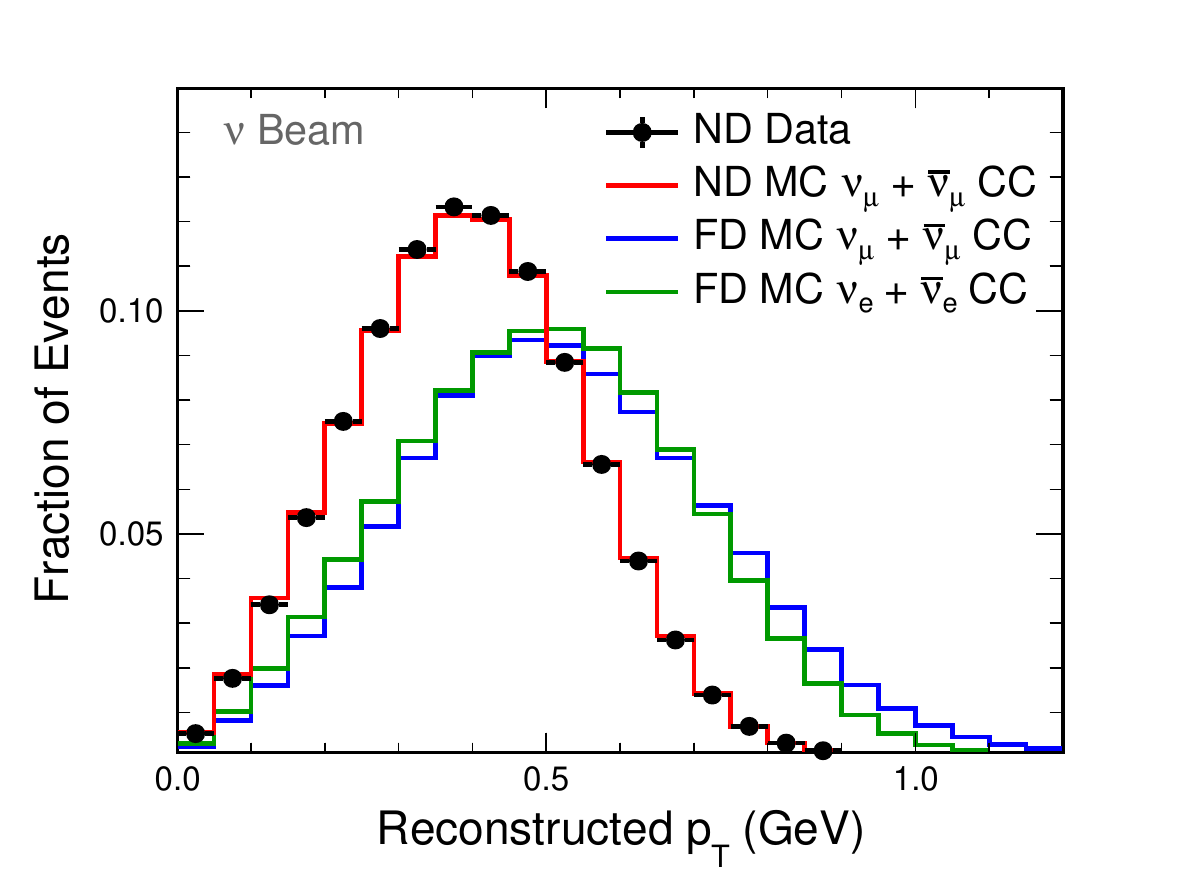}
    \caption{The fraction of events in reconstructed transverse lepton momentum for the NOvA ND and FDs, showing a mismatch in phase space due to detector geometry. Reproduced from Ref.~\cite{NOvA:2021nfi}.}
    \label{fig:nova_pt}
\end{figure}
Extrapolation of ND constraints to FD predictions must account for differences in their kinematic acceptances. In general, the FDs have a broader acceptance than the smaller NDs. For example, in the NOvA experiment, leptons must be well contained within the detector for accurate momentum reconstruction and those with high transverse momentum, with respect to the incoming neutrino, are more likely to be contained in the FD than the ND. \autoref{fig:nova_pt} illustrates the considerable acceptance mismatch this introduces. The situation is similar for T2K, whose primary ND (ND280) also has better acceptance for forward-going particles whilst its FD has uniform angular acceptance\footnote{Although recent and planned upgrades of Hyper-K/T2K NDs aim to address this~\cite{T2K:2019bbb,Scott:2016kdg}.}. DUNE's primary LArTPC ND will also be affected by a similar issue~\cite{DUNE:2021tad}, in which most high-angle muons will exit the LArTPC but not be matched to the downstream muon spectrometer needed for accurate momentum measurements.

\begin{figure*}[htbp]
  \centering
  \captionsetup[subfigure]{aboveskip=0pt,belowskip=3pt}       
  \begin{subfigure}[b]{0.48\textwidth}
      \centering
      \includegraphics[width=1.00\linewidth]{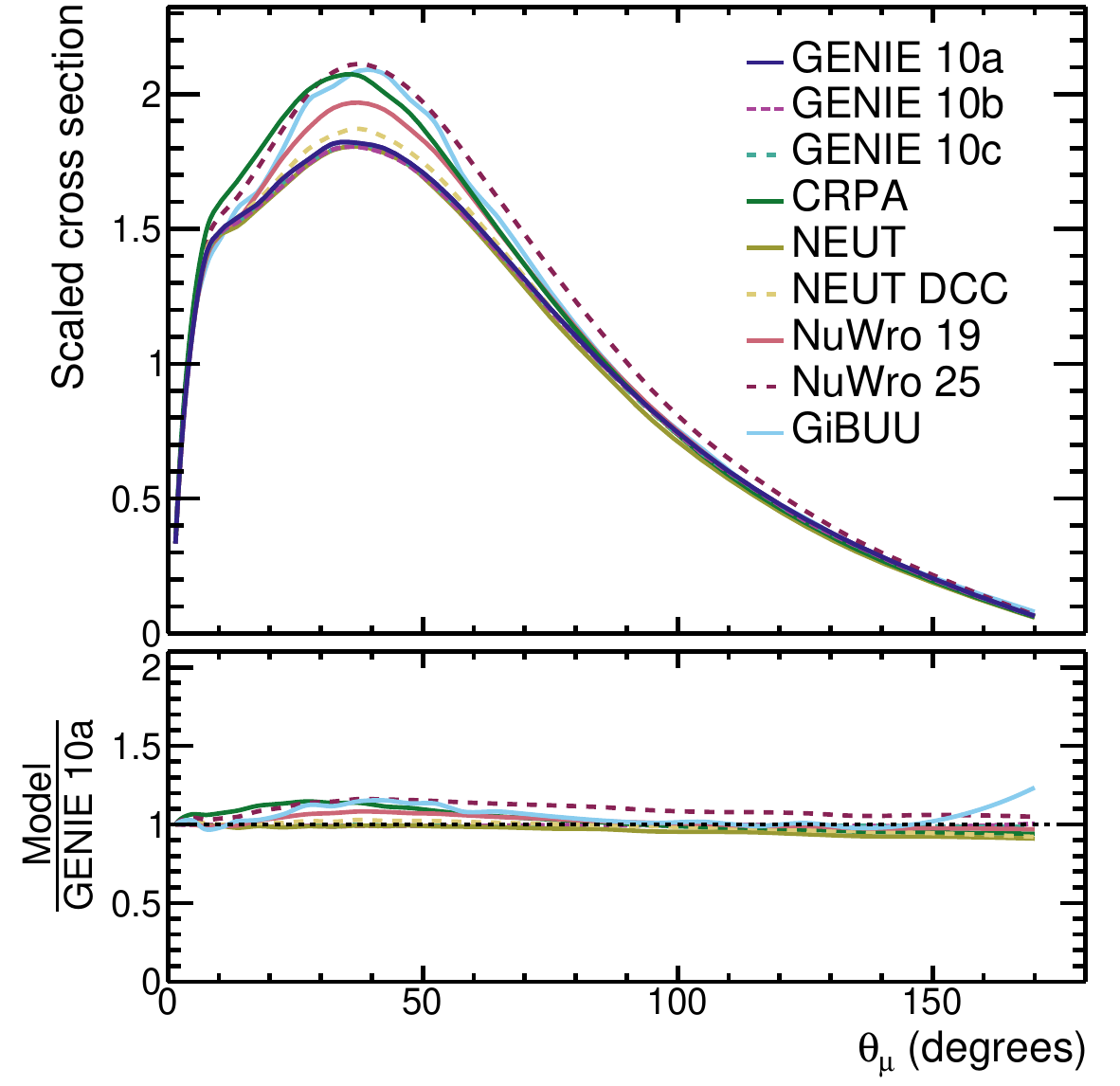}
      \caption{Hyper-K/T2K ND \numu--\water CC0$\pi$}
  \end{subfigure}
  \begin{subfigure}[b]{0.48\textwidth}
      \centering
      \includegraphics[width=1.00\linewidth]{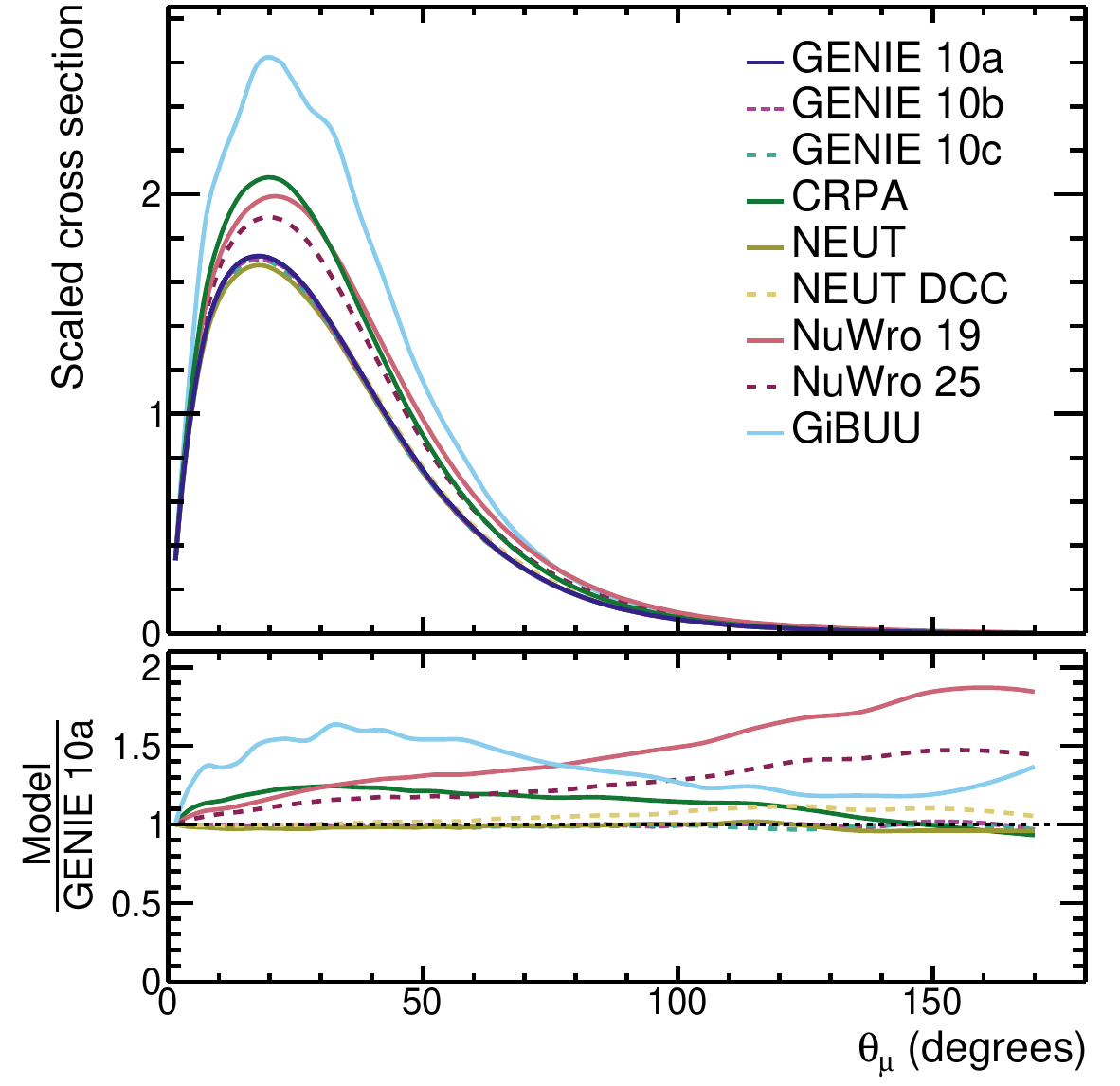}
      \caption{Hyper-K/T2K ND \numub--\water CC0$\pi$}
  \end{subfigure}
  \\
   \begin{subfigure}[b]{0.48\textwidth}
      \centering
      \includegraphics[width=1.00\linewidth]{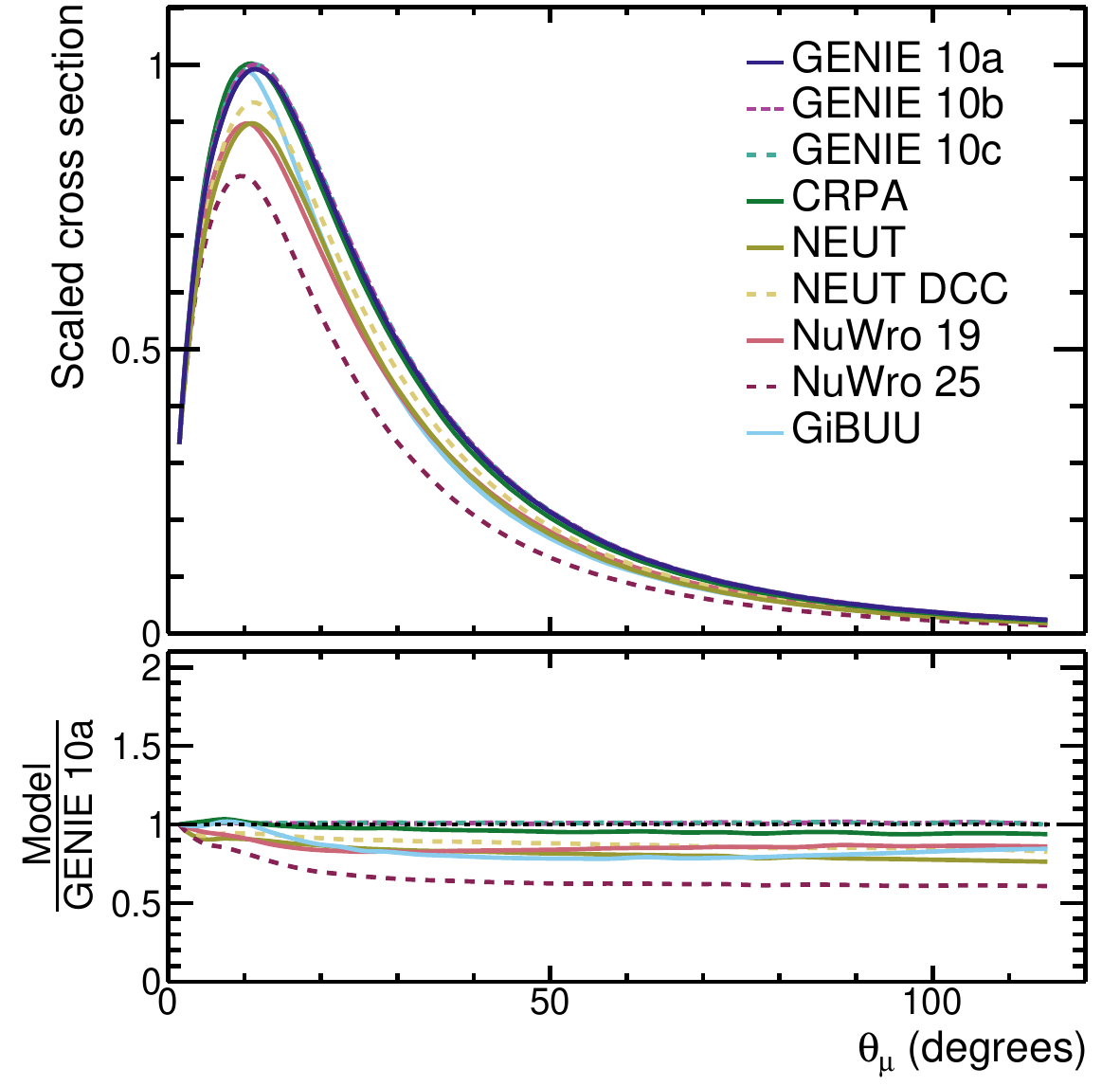}
      \caption{DUNE ND \numu--\argon CCINC}
  \end{subfigure}
   \begin{subfigure}[b]{0.48\textwidth}
      \centering
      \includegraphics[width=1.00\linewidth]{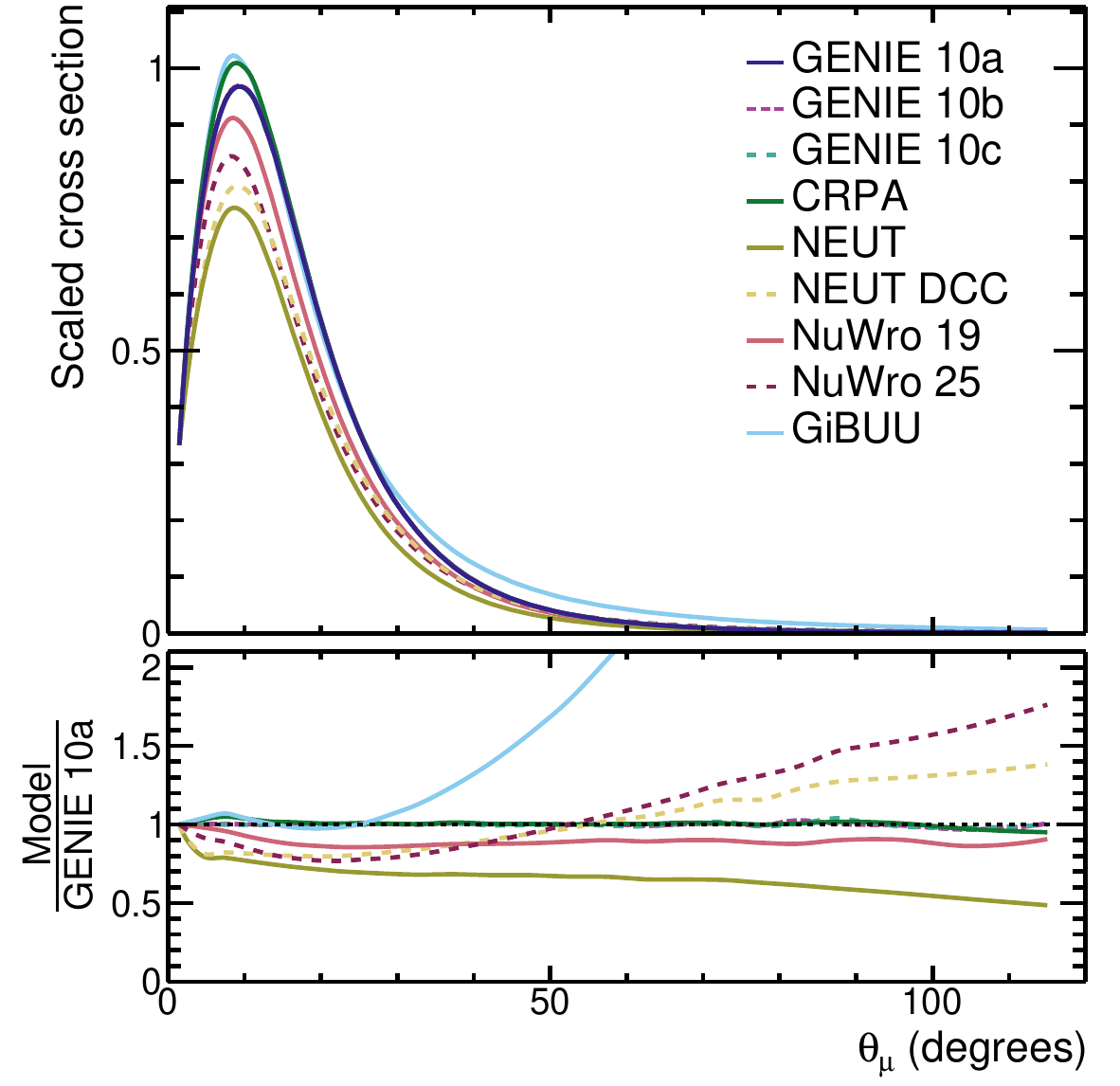}
      \caption{DUNE ND \numub--\argon CCINC}
  \end{subfigure}
  \caption{A comparison of the shape of the cross section as a function of the outgoing lepton angle with respect to the incoming neutrino (such that the value in the first bin, $0^{\circ} \leq \theta_{\mu} \leq 3^{\circ}$ is the same for all models), for CC0$\pi$ $\numu^{\bracketbar}$--\water interactions with the Hyper-K/T2K ND fluxes and CCINC $\numu^{\bracketbar}$--\argon interactions with the DUNE ND fluxes.
  }
  \label{fig:costheta_accept_gencomp}
\end{figure*}

Neutrino interaction models are needed to correct for these ND--FD acceptance differences. To illustrate the potential impact neutrino interaction mismodelling can have on this correction, \autoref{fig:costheta_accept_gencomp} shows a comparison of neutrino event generator predictions as a function of the angle between the outgoing lepton and the incoming neutrino. 
The distributions are scaled to give the same prediction in the most forward bin (0--3$^{\circ}$) to roughly demonstrate the impact that model differences could have when a forward ND constraint is extrapolated to a larger acceptance FD. The model differences at higher angles emulate the impact of using different models to extrapolate between them. 
The interaction models behave very differently in this toy extrapolation procedure, particularly for antineutrino interactions, exhibiting differences of over 20\% for most of the angular phase space. Whilst \autoref{fig:angFDND_gencomp} and \autoref{fig:costheta_accept_gencomp} both show the angular dependence of interaction model predictions, they demonstrate two separate challenges that oscillation experiments face. The former highlights the challenge in extrapolating an ND measurement to the FD without any acceptance restrictions, whilst the latter highlights challenges extrapolating from a limited acceptance ND to a full acceptance FD, independent of ND--FD flux differences.

Whilst a real measurement at a ND would utilise a broader range of angles, and partially mitigate this effect, \autoref{fig:costheta_accept_gencomp} qualitatively demonstrates the importance of a neutrino interaction model to extrapolate between different regions of kinematic acceptance. Additionally, although this example has focused on muon kinematic acceptances, outgoing hadron kinematic reconstruction performance and accessible kinematic phase spaces also generally differ between the ND and FD, leading to a significantly more complex picture.
Although difficult to assess without detailed information on detector acceptances, one hadronic acceptance example is the different proton tracking thresholds in T2K's ND280 and SK. At the ND280, the proton tracking threshold is $\sim$300 MeV/c due to the geometry of its plastic scintillator cubes~\cite{T2K:2019bbb}, whereas the Cherenkov threshold for protons in SK is around 1.4 GeV/c and so the vast majority are not observed. 

\subsection{Electron- and muon- neutrino and antineutrino cross section differences}
\label{subsec:issuesNueNumu}

Measurements of CP violation, the neutrino mass ordering, and the $\theta_{23}$ octant all rely on inferring oscillation probabilities from observed electron-neutrino appearance events in experiments' FDs.
Conversely, the NDs primarily measure muon (anti)neutrino interactions.
Precision oscillation measurements therefore require a detailed understanding of the relationship between electron (anti)neutrino and muon (anti)neutrino interactions.
Additionally, as CP violation and mass ordering measurements rely on differences between electron neutrino and antineutrino appearance, neutrino oscillation analyses further rely on the double cross section ratio between the four flavours, $(\bar{\nu}_e/\bar{\nu}_\mu)/(\nu_e/\nu_\mu)$.

\begin{figure*}[htbp]
  \centering
    \begin{subfigure}[b]{0.48\textwidth}
      \centering
      \includegraphics[width=0.90\linewidth]{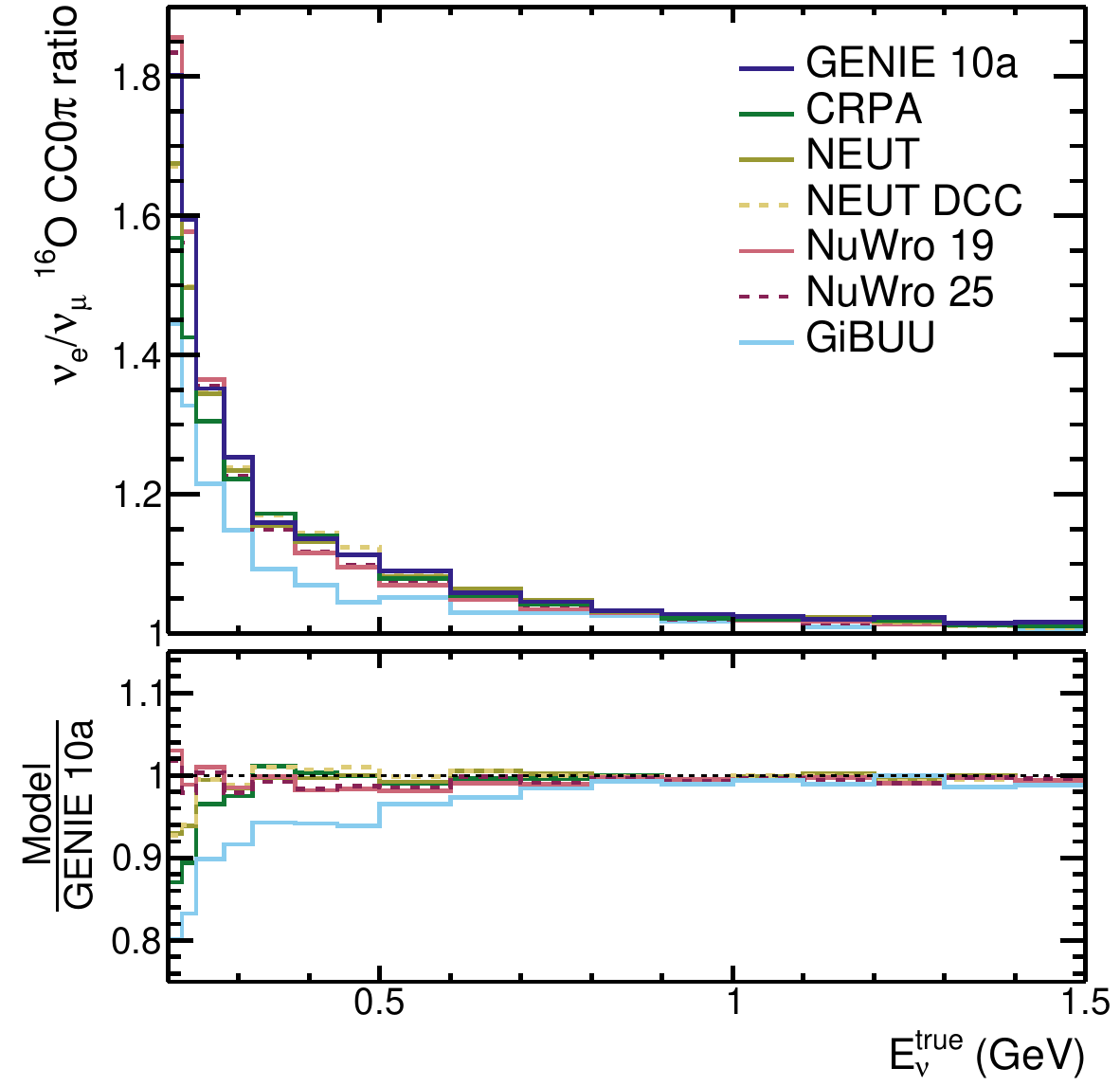}
      \caption{$\nu_e/\nu_\mu$, \water CC0$\pi$}
  \end{subfigure}
   \begin{subfigure}[b]{0.48\textwidth}
      \centering
      \includegraphics[width=0.90\linewidth]{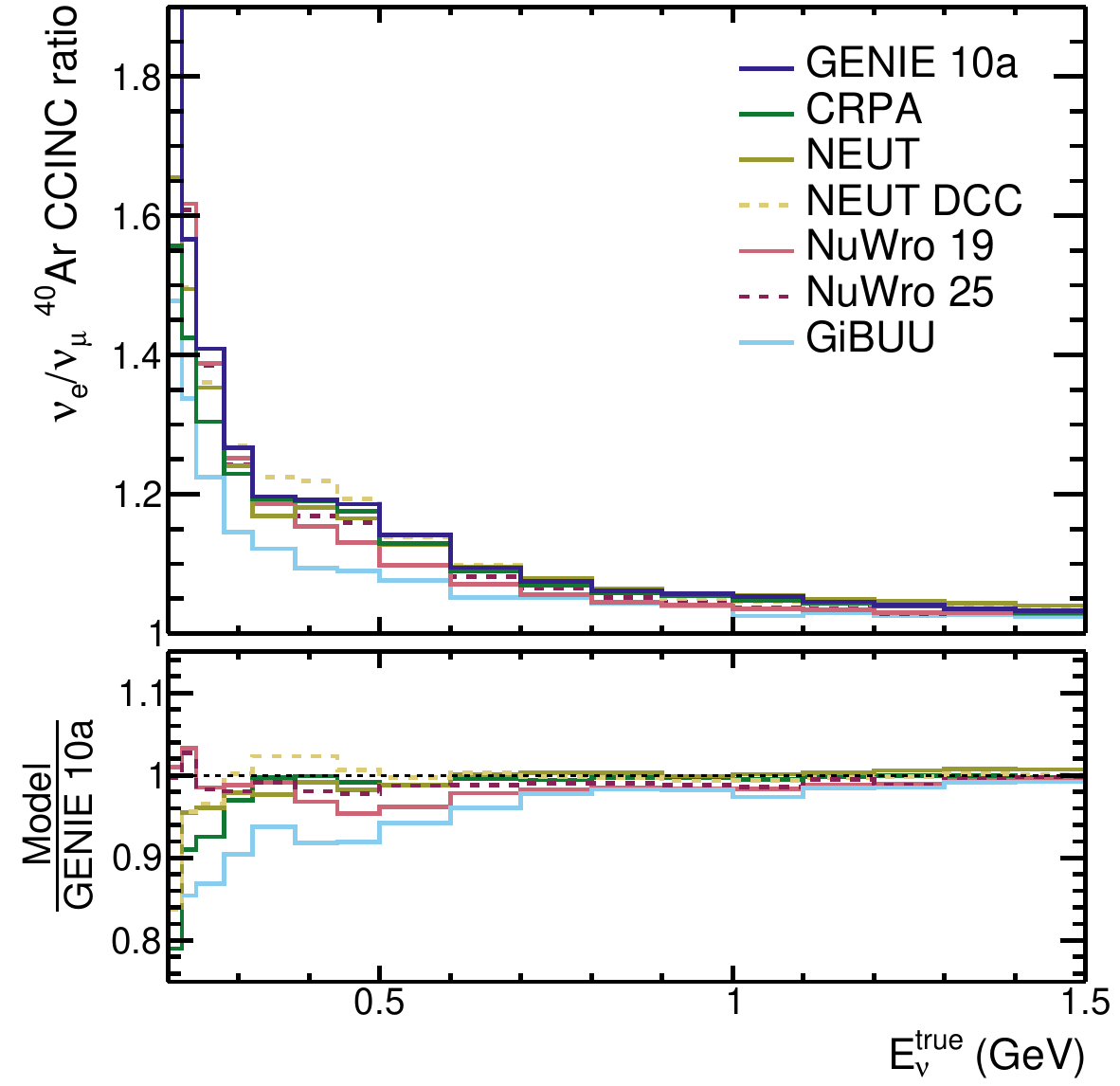}
      \caption{$\nu_e/\nu_\mu$, \argon CCINC}
  \end{subfigure}
      \begin{subfigure}[b]{0.48\textwidth}
      \centering
      \includegraphics[width=0.90\linewidth]{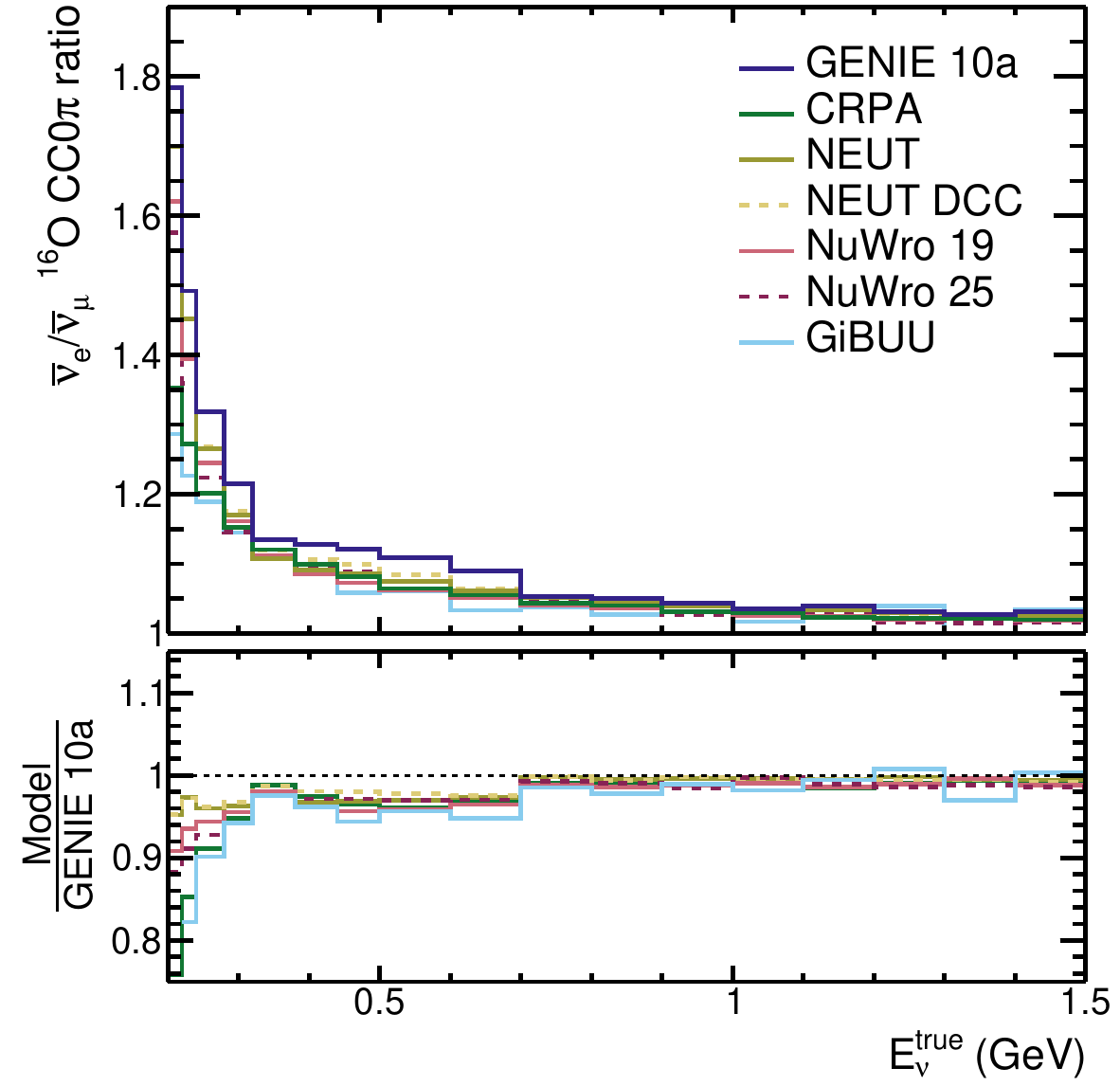}
      \caption{$\bar{\nu}_e/\bar{\nu}_\mu$, \water CC0$\pi$}
  \end{subfigure}
   \begin{subfigure}[b]{0.48\textwidth}
      \centering
      \includegraphics[width=0.90\linewidth]{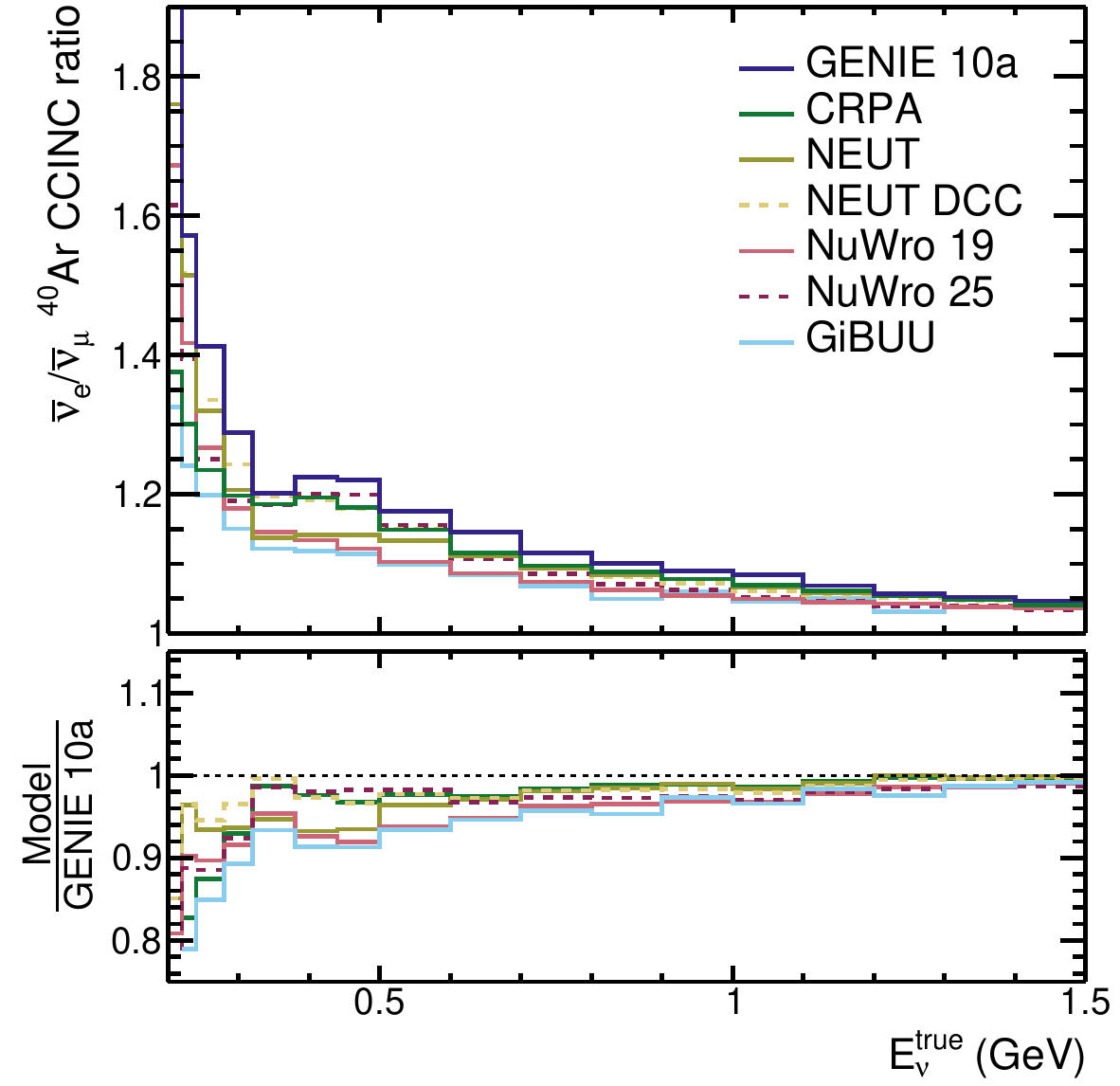}
      \caption{$\bar{\nu}_e/\bar{\nu}_\mu$, \argon CCINC}
  \end{subfigure}
  \begin{subfigure}[b]{0.48\textwidth}
      \centering
      \includegraphics[width=0.90\linewidth]{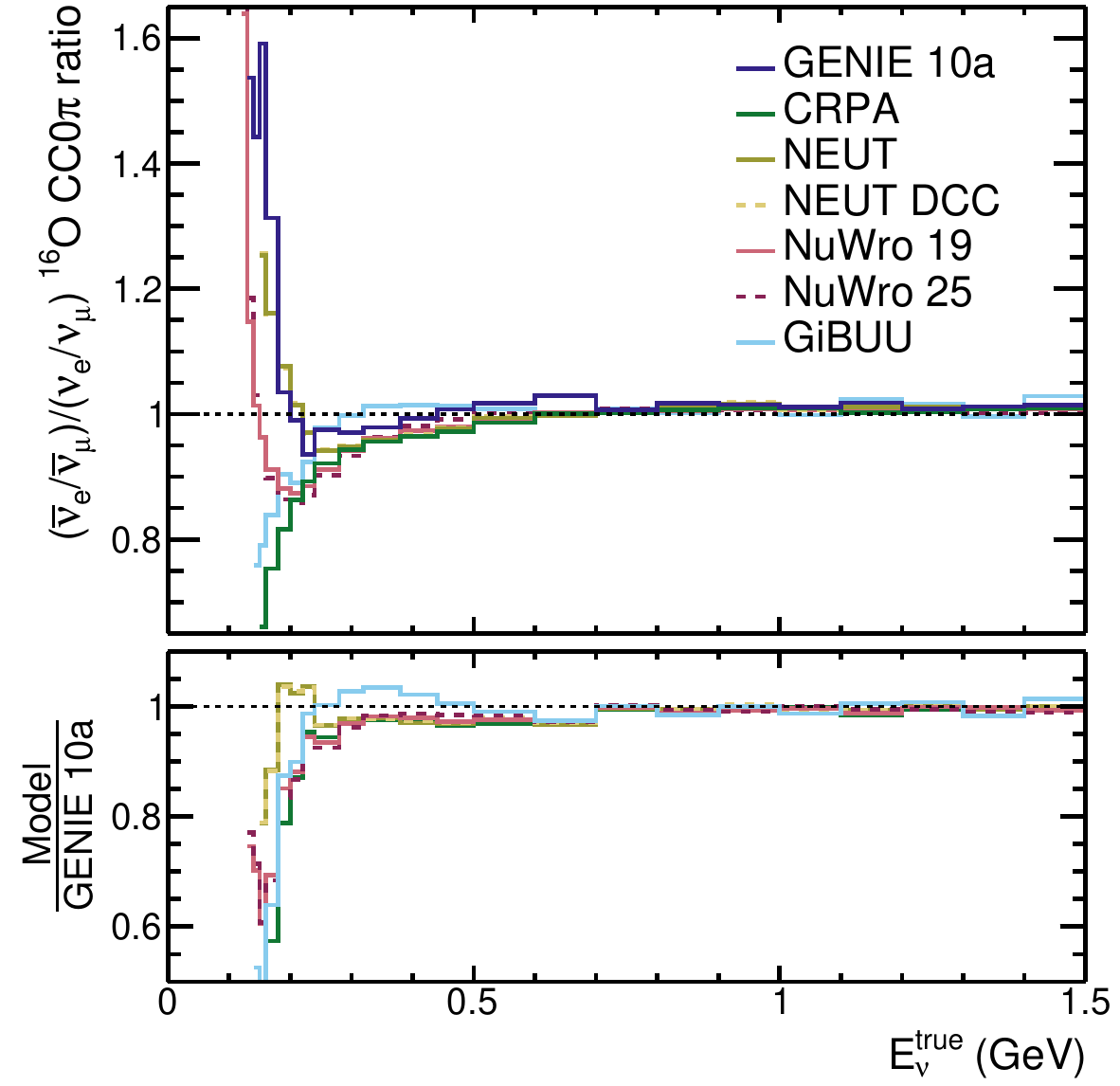}
      \caption{$(\bar{\nu}_e/\bar{\nu}_\mu)/(\nu_e/\nu_\mu)$, \water CC0$\pi$}
  \end{subfigure}
   \begin{subfigure}[b]{0.48\textwidth}
      \centering
      \includegraphics[width=0.90\linewidth]{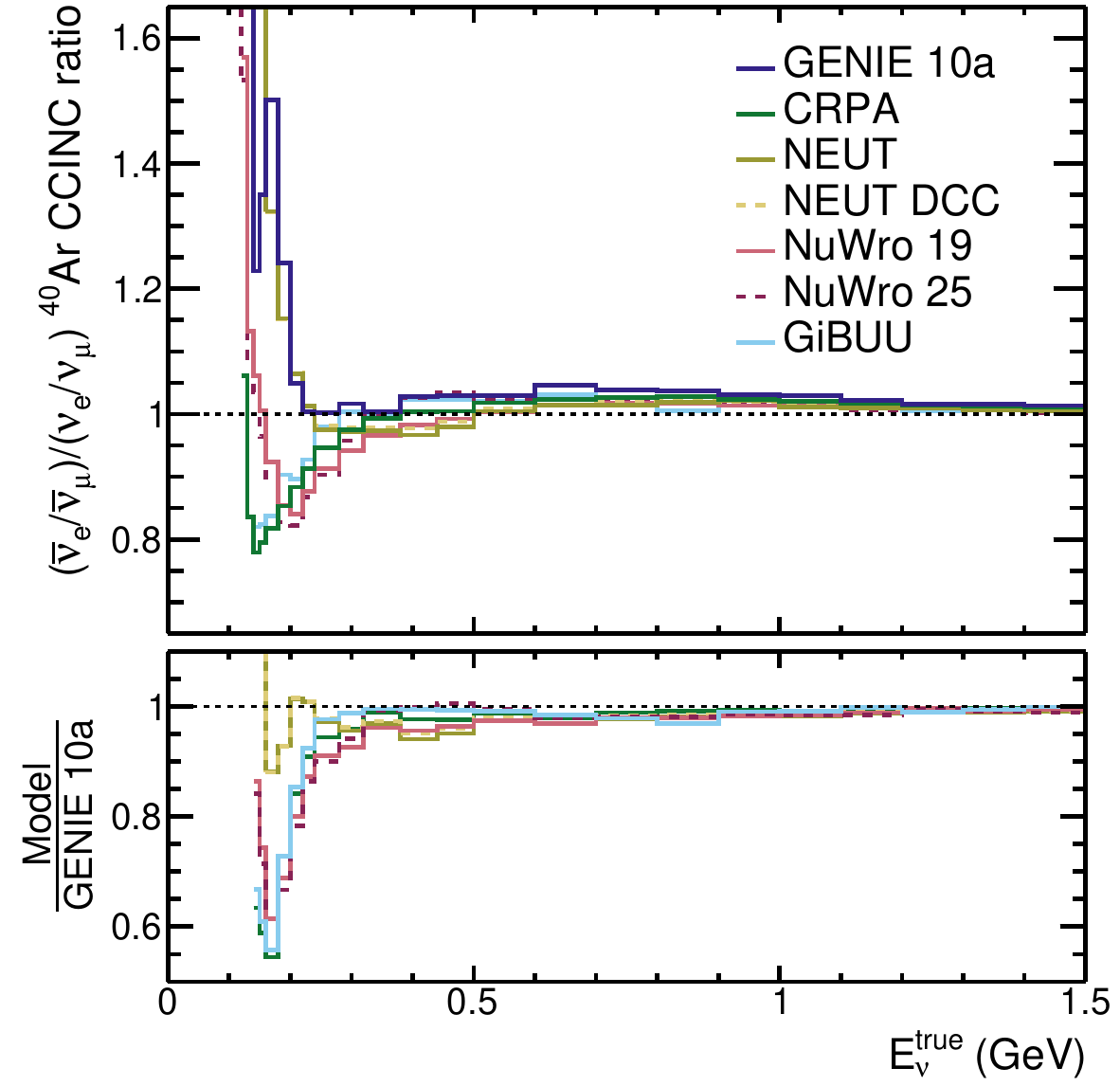}
      \caption{$(\bar{\nu}_e/\bar{\nu}_\mu)/(\nu_e/\nu_\mu)$, \argon CCINC}
  \end{subfigure}
  \caption{A comparison of the single $\nue^{\bracketbar}/\numu^{\bracketbar}$ and double $(\nueb/\numub)/(\nue/\numu)$ flavour ratios as a function of \enutrue for CC0$\pi$ interactions on \water and CCINC interactions on \argon, for a variety of neutrino generators.}
  \label{fig:double_flav_ratio_gencomp}
\end{figure*}

Event generator predictions of the single and double cross-section ratios are shown for the Hyper-K/T2K and DUNE-like scenarios in \autoref{fig:double_flav_ratio_gencomp}. Whilst there are very large differences at low neutrino energies ($\enu \lesssim$ 0.2~GeV for the double ratio and 0.5~GeV for the single ratio), the differences between models around the Hyper-K/T2K oscillation maximum ($\sim$0.6 GeV) are $\sim$5\%. Around the first oscillation maxima for the DUNE experiment ($\sim$2.5 GeV) differences are much smaller, although in the region of the second oscillation maxima ($\sim$0.8 GeV), which is also the region of interest for the SBN program, they also reach $\sim$5\%. Some of the differences between generator predictions have been explored in Refs.~\cite{Nikolakopoulos:2019qcr, Dieminger:2023oin}. For all models, both single ratios are greater than one ($\nu^{\bracketbar}_e/\nu^{\bracketbar}_{\mu} > 1$) up to much larger energies than the double ratio, indicating that the former are generally more likely to be sensitive to neutrino interaction modelling differences at energies relevant to oscillation experiments. Note that the GiBUU predictions have non-negligible statistical fluctuations, discussed in \ref{appen:gen}. 

\begin{figure*}[htbp]
  \centering
  \begin{subfigure}[b]{0.48\textwidth}
      \centering
      \includegraphics[width=1.00\linewidth]{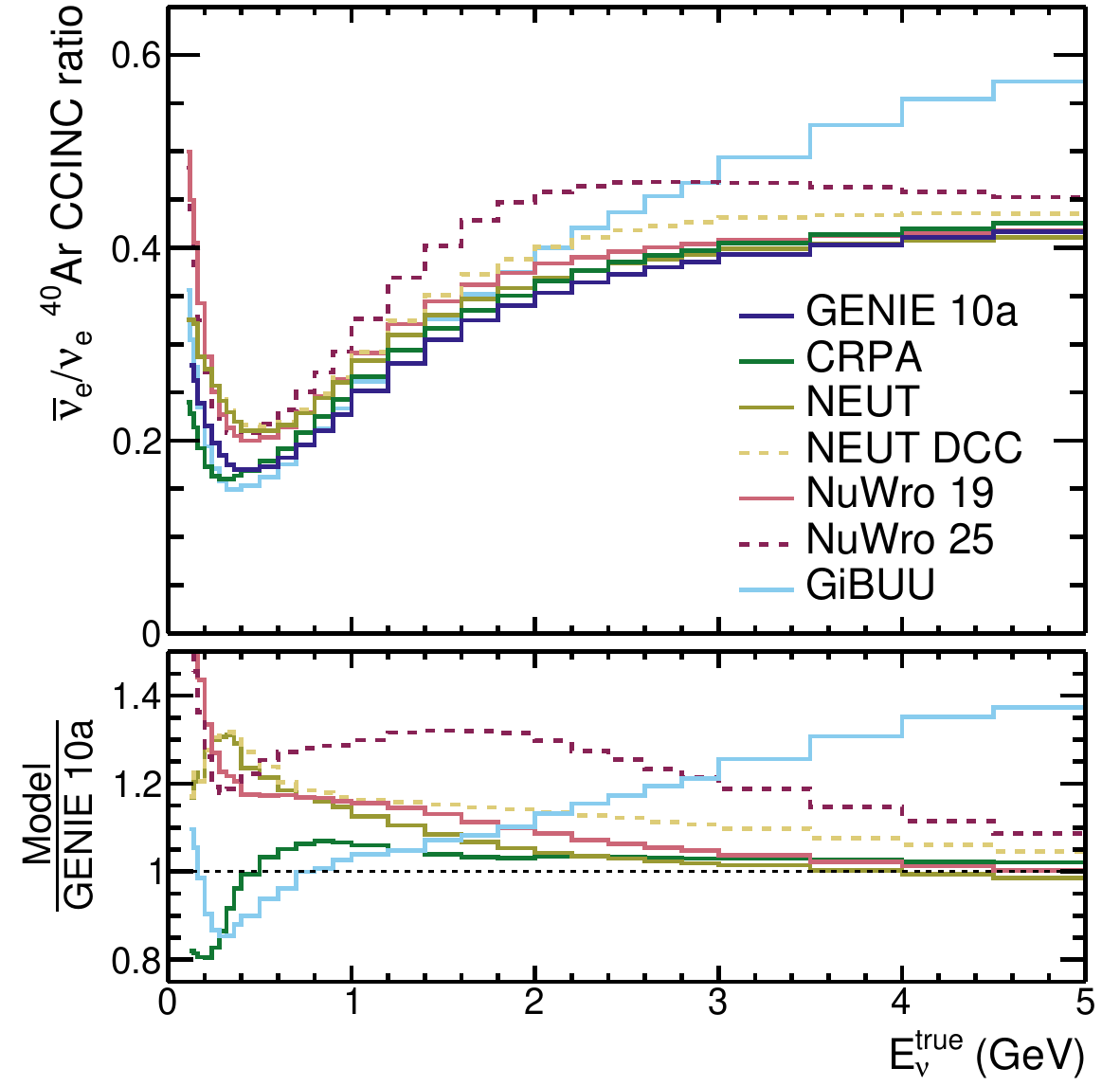}
      \caption{\nueb/\nue \argon CCINC}
  \end{subfigure}
  \begin{subfigure}[b]{0.48\textwidth}
      \centering
      \includegraphics[width=1.00\linewidth]{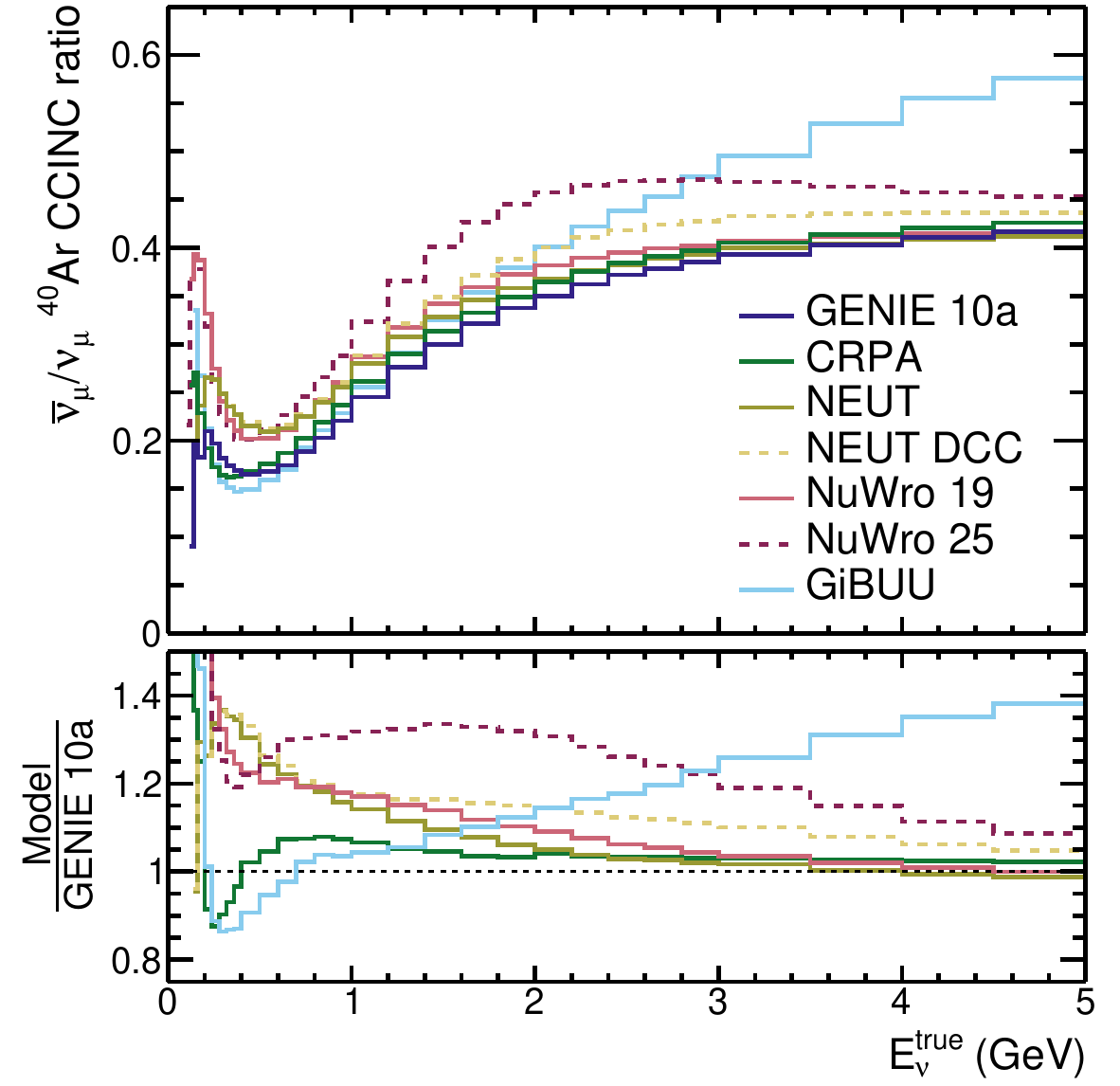}
      \caption{\numub/\numu \argon CCINC}
  \end{subfigure}
  \\
  \begin{subfigure}[b]{0.48\textwidth}
      \centering
      \includegraphics[width=1.00\linewidth]{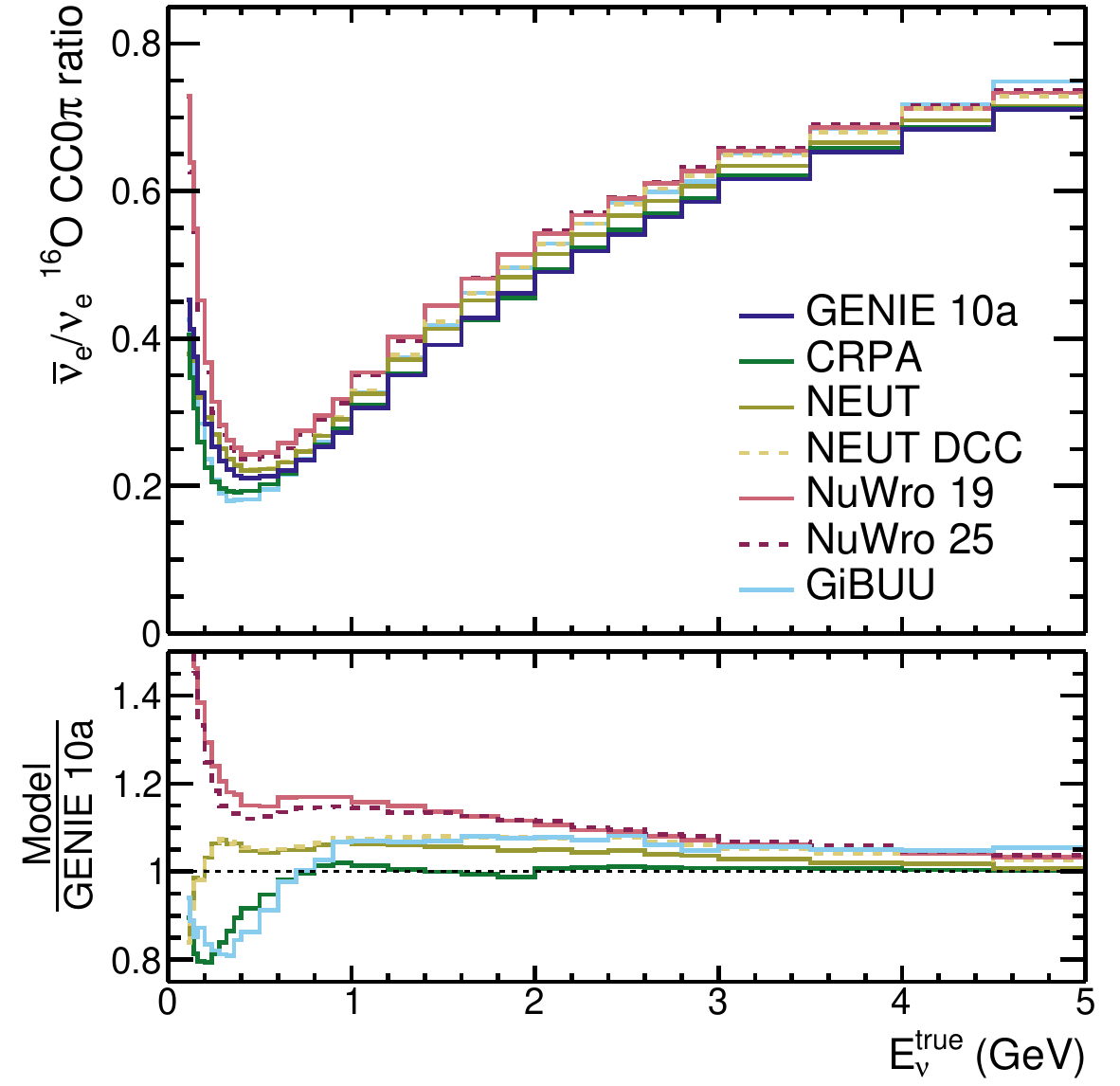}
      \caption{\nueb/\nue $^{16}$O CC0$\pi$}
  \end{subfigure}
  \begin{subfigure}[b]{0.48\textwidth}
      \centering
      \includegraphics[width=1.00\linewidth]{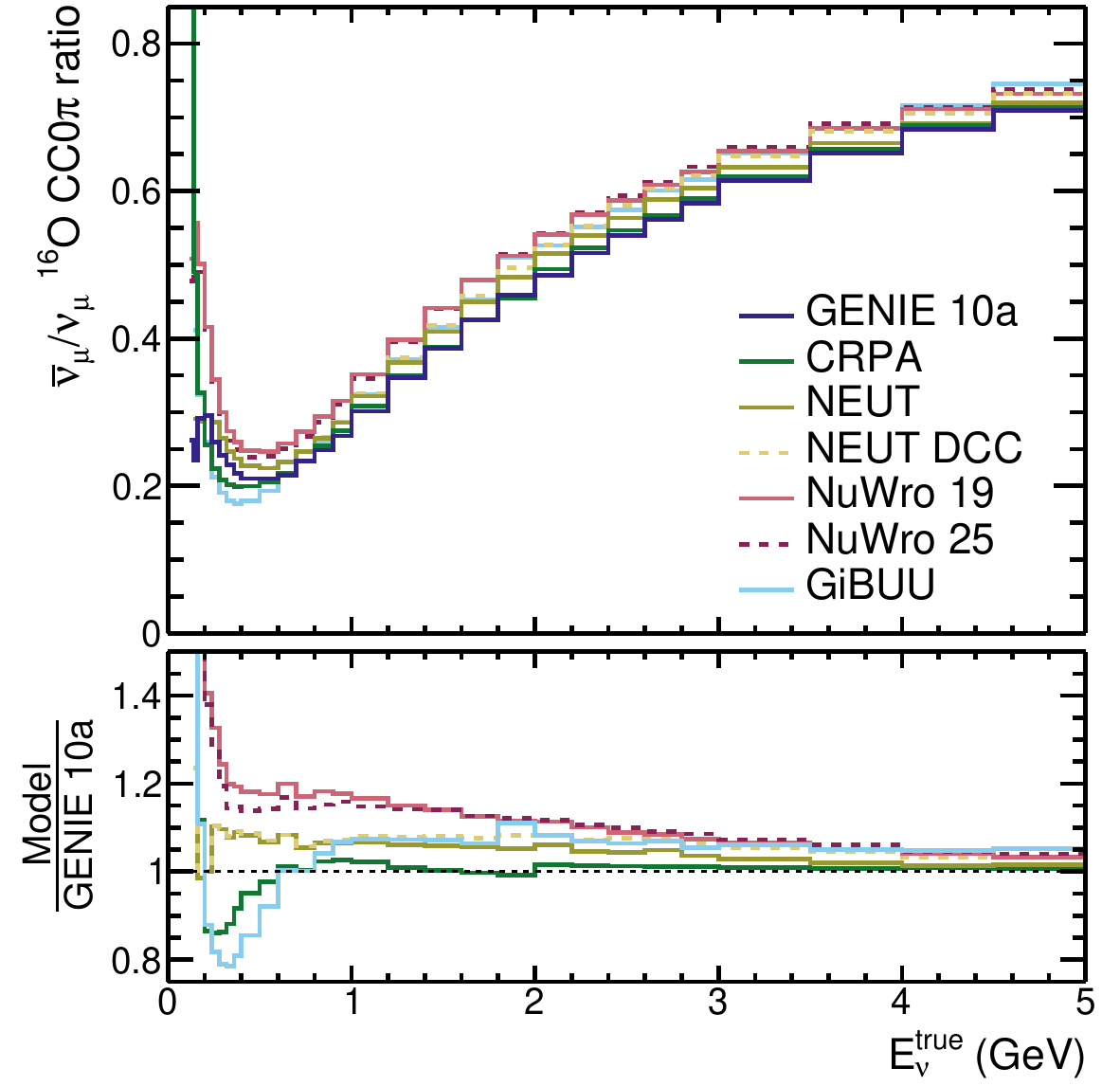}
      \caption{\numub/\numu $^{16}$O CC0$\pi$}
  \end{subfigure}
  \caption{A comparison of the \nueb/\nue and \numub/\numu cross-section ratio as a function of \enutrue for CC0$\pi$ interactions on \water and CCINC interactions on \argon, for a variety of neutrino generators.}
  \label{fig:single_flav_ratio_gencomp}
\end{figure*}

The model spread in the neutrino/antineutrino cross section ratios ($\nu_e/\bar{\nu}_e$ or $\nu_\mu/\bar{\nu}_\mu$) are not confined to low energies. This is illustrated in \autoref{fig:single_flav_ratio_gencomp}, which shows that  models vary by more than 20\% in their prediction of the ratio across the energy range of interest for Hyper-K and DUNE.

\begin{figure}[htbp]
  \centering
  \begin{subfigure}[b]{0.48\textwidth}
      \centering
      \includegraphics[width=1.00\linewidth]{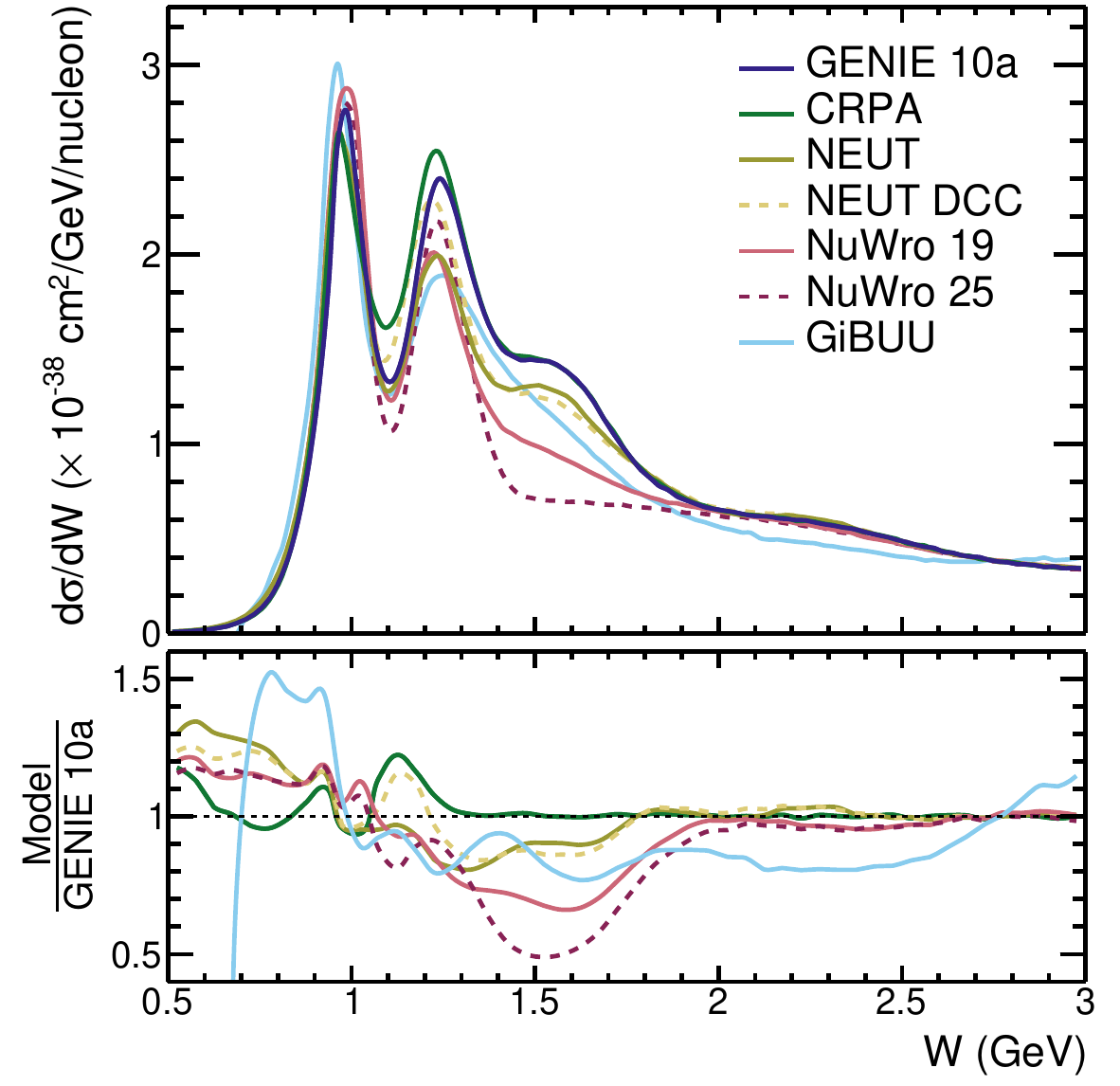}
      \caption{DUNE FD \numu--\argon}
  \end{subfigure}
  \begin{subfigure}[b]{0.48\textwidth}
      \centering
      \includegraphics[width=1.00\linewidth]{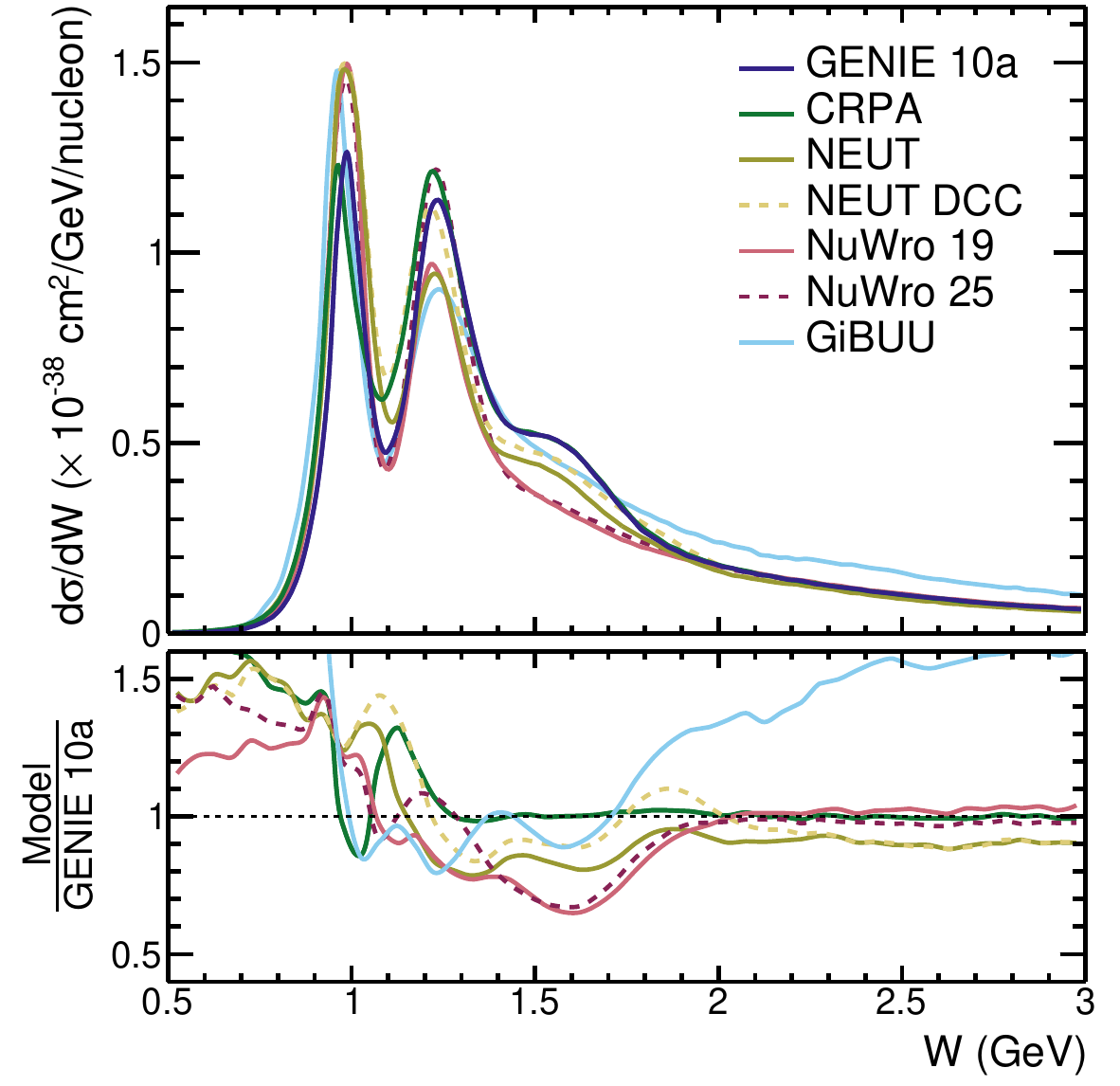}
      \caption{DUNE FD \numub--\argon}
  \end{subfigure}
  \caption{A comparison of the flux-averaged cross sections as a function of the invariant hadronic mass, assuming the struck nucleon is at rest, for CCINC \numu--\argon interactions with the DUNE neutrino-enhanced and antineutrino-enhanced flux at the FD, shown for a variety of neutrino generators.
  }
  \label{fig:W_nuanucomp}
\end{figure}

Beyond predicting the inclusive neutrino/antineutrino or flavour cross section ratios, experiments also rely on generator predictions of differential cross section ratios when extrapolating constraints from ND to FD. As an example, \autoref{fig:W_nuanucomp} shows the differential cross section as a function of the hadronic invariant mass for the neutrino- and antineutrino-enhanced fluxes for the DUNE-like scenario. The figure demonstrates not only that the predicted differential cross section is significantly different between generators but also that the relative behaviour of the neutrino and antineutrino cross sections differs between generators as a function of the invariant mass. 

\subsection{Nuclear target scaling}
\label{subsec:issuesAScale}

Significant proportions of the Hyper-K/T2K and DUNE NDs are composed of a different neutrino interaction target to the FD. In the case of Hyper-K/T2K, although there is water in the ND complex, much of the available precision ND neutrino interaction data are on hydrocarbon targets, whilst the FD is composed of water~\cite{Hyper-Kamiokande:2018ofw, T2K:2011qtm}\footnote{In the Hyper-K era, the IWCD will further mitigate this challenge~\cite{Hyper-Kamiokande:2025asb}. However, detailed hydrocarbon ND data will remain important for oscillation measurements.}.
For DUNE, although the primary LArTPC ND detector matches the FD, some ND subdetectors will use hydrocarbon targets~\cite{DUNE:2021tad}. For either Hyper-K or DUNE to utilise hydrocarbon ND samples as a constraint on neutrino interactions on water or argon, a robust model of the relationship between the cross-sections of these nuclear targets is required (\autoref{subsec:NDtoFD}). 

\begin{figure*}[htbp]
  \centering
  \captionsetup[subfigure]{aboveskip=0pt,belowskip=3pt}       
   \begin{subfigure}[b]{0.48\textwidth}
      \centering
      \includegraphics[width=1.00\linewidth]{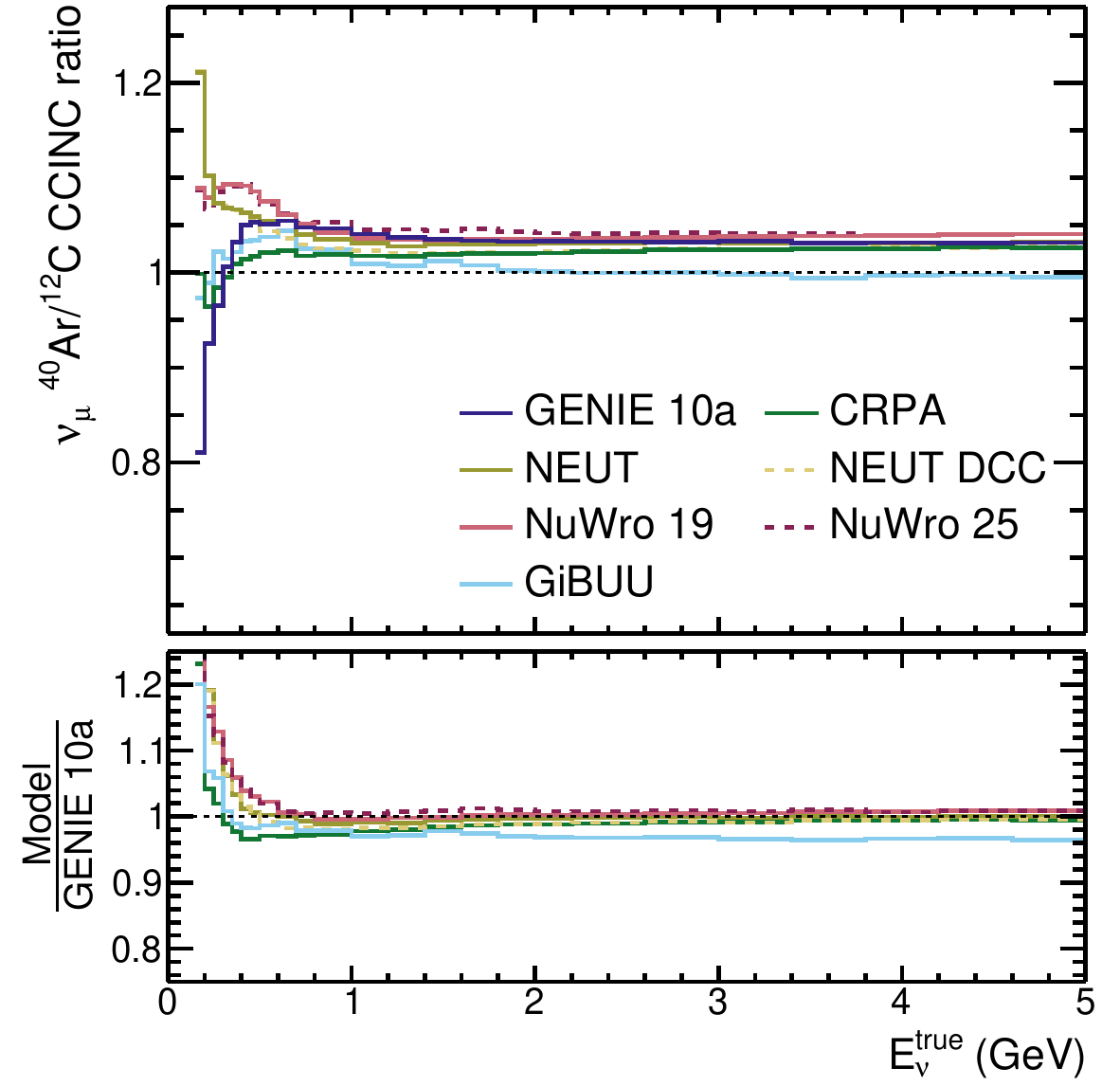}
      \caption{\numu--\argon/\numu--$^{12}$C CCINC}
  \end{subfigure}
   \begin{subfigure}[b]{0.48\textwidth}
      \centering
      \includegraphics[width=1.00\linewidth]{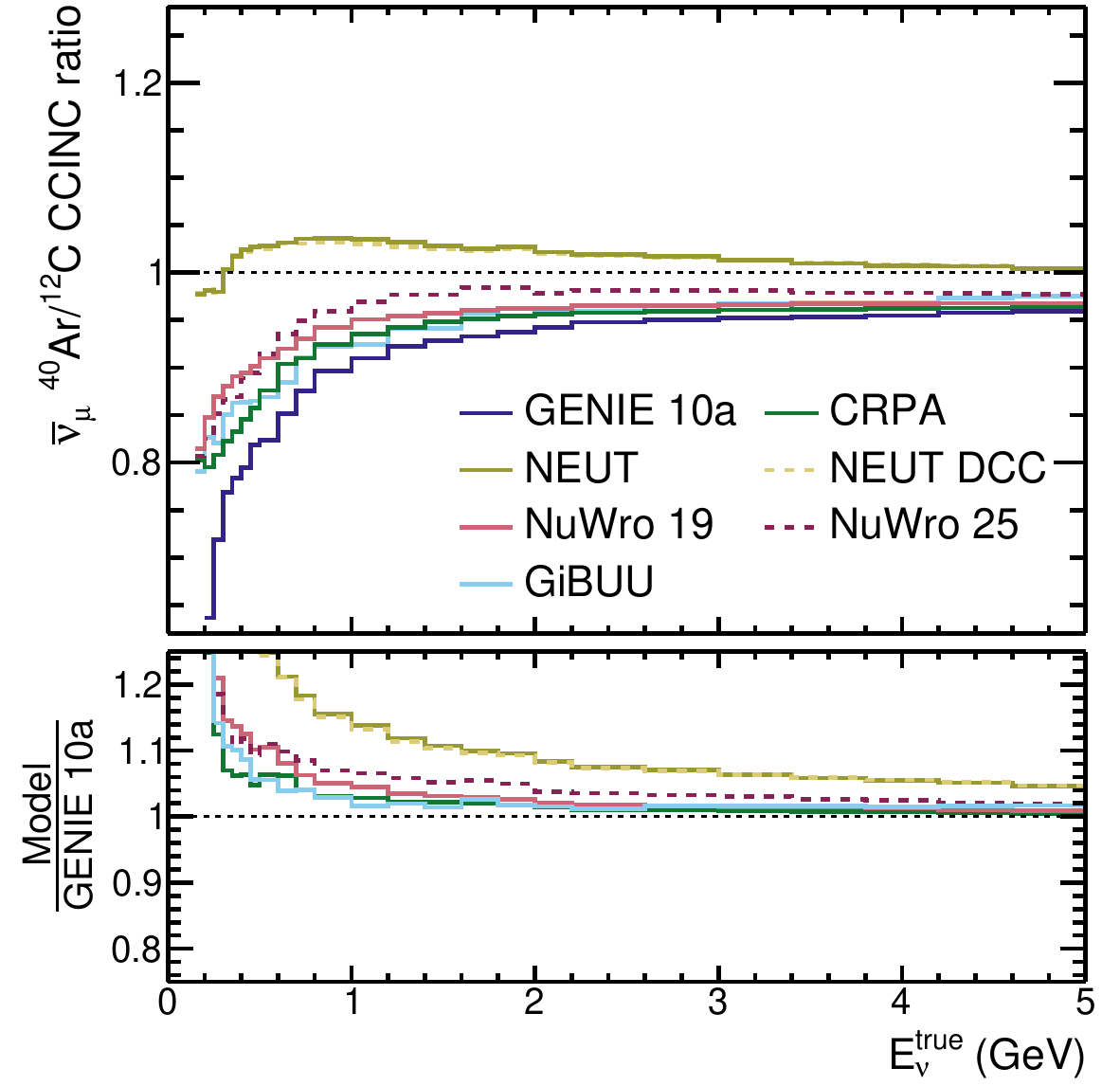}
      \caption{\numub--\argon/\numub--$^{12}$C CCINC}
  \end{subfigure}
\\
  \begin{subfigure}[b]{0.48\textwidth}
      \centering
      \includegraphics[width=1.00\linewidth]{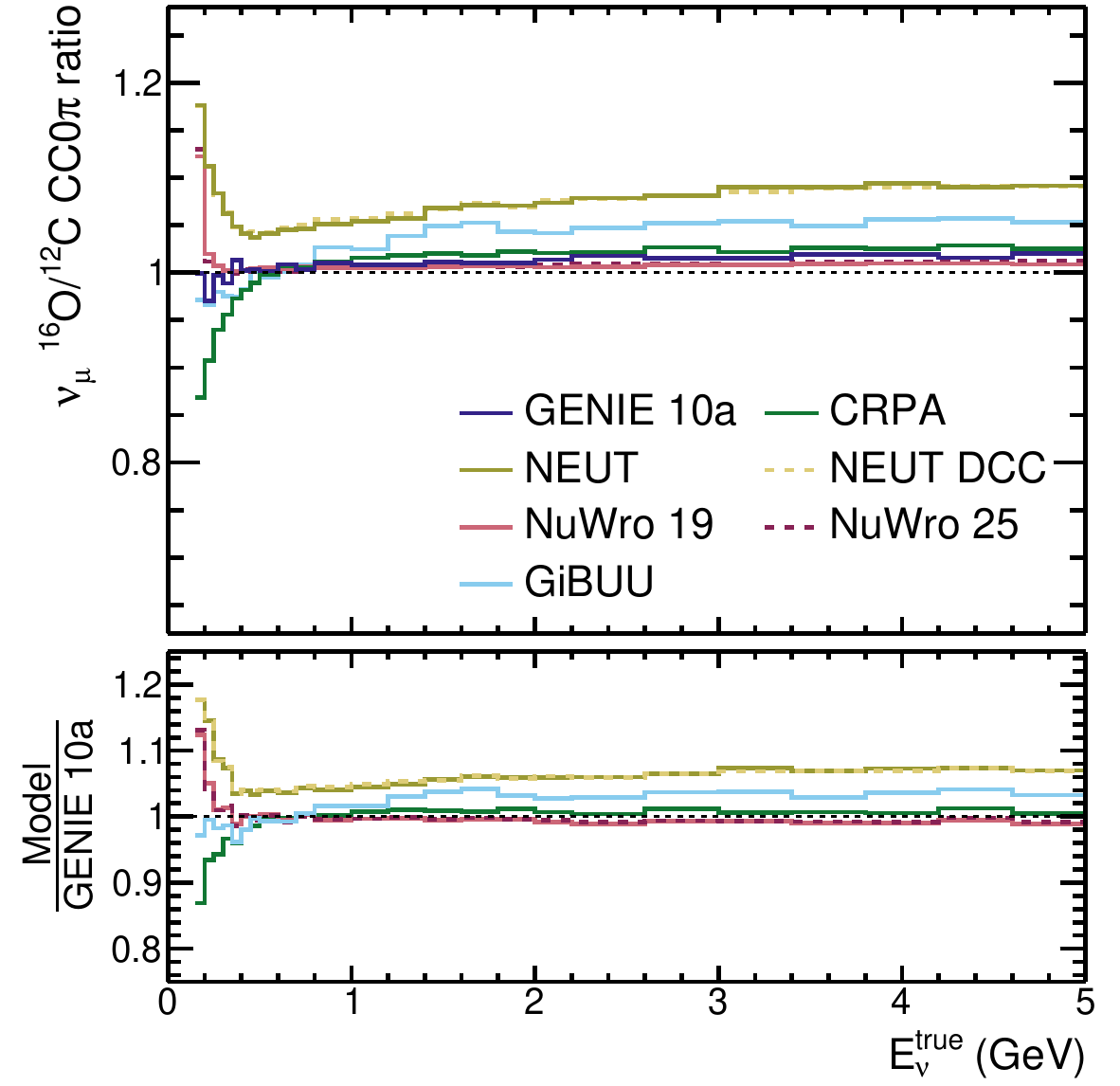}
      \caption{\numu--$^{16}$O/\numu--$^{12}$C CC0$\pi$}
  \end{subfigure}
  \begin{subfigure}[b]{0.48\textwidth}
      \centering
      \includegraphics[width=1.00\linewidth]{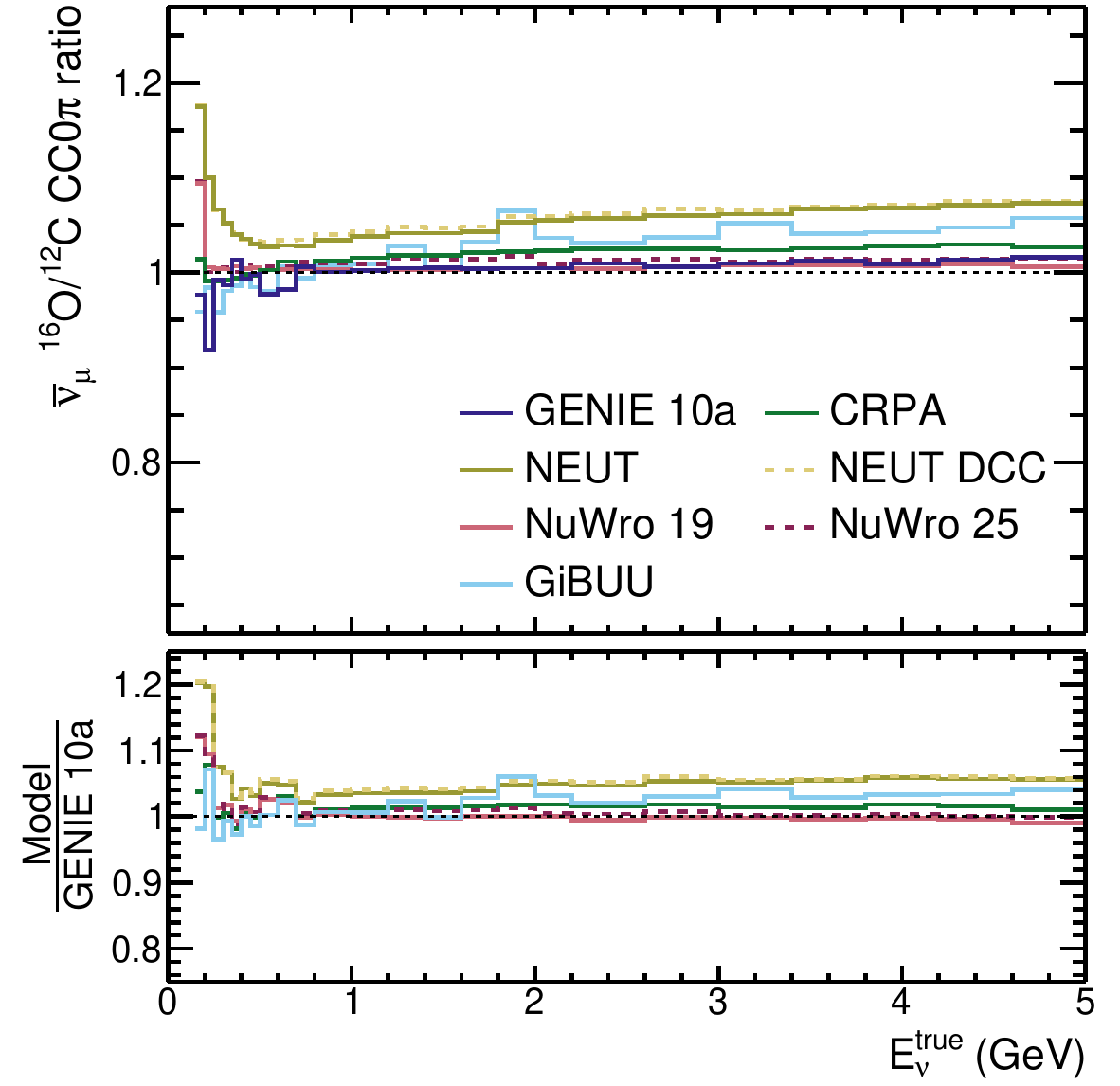}
      \caption{\numub--$^{16}$O/\numub--$^{12}$C CC0$\pi$}
  \end{subfigure}
  \caption{A comparison of the muon neutrino and antineutrino cross section (per nucleon) ratios as a function of \enutrue, for \numu--\argon/\numu--$^{12}$C CC0$\pi$ interactions and \numu--$^{16}$O/\numu--$^{12}$C CCINC interactions, for a variety of neutrino generators. 
  }
  \label{fig:ascale_gencomp}
\end{figure*}

\autoref{fig:ascale_gencomp} shows the ratio of cross sections on different nuclear targets for neutrinos and antineutrinos in the DUNE- and Hyper-K/T2K-like scenarios\footnote{The ratios of cross sections \textit{per nucleon} are shown. Since argon contains more neutrons than protons, the argon/carbon cross section ratio for (anti)neutrino interactions is (below) above one.}. Differences between models are at the level of $\sim$5\% for $\enu \gtrsim 0.4$ GeV and are relatively flat as a function of energy. NEUT is an outlier for both scenarios. For the Hyper-K/T2K-scenario this is due to different pion-absorption cross sections leading to more migration from pion production events into the CC0$\pi$ cross section. For the DUNE-scenario this is due to the issues with NEUT's treatment of cross-section scaling described in \ref{appen:gen}. Only the CRPA model predicts a cross section for interactions on oxygen that is significantly lower than that of carbon at low energies (expected since oxygen's doubly-magic structure implies larger binding energies). As for the flavour ratio predictions, GiBUU simulation contains non-negligible statistical fluctuations, as discussed in \ref{appen:gen}. 

\subsection{Backgrounds and sample migration}
\label{subsec:issuesbkg}

Backgrounds in the FD event selections used in neutrino oscillation analyses can contribute up to 50\% of the rate in some regions of phase space. In general, sources of background are tightly connected to the details of the event selection and are difficult to assess outside an experiment. Here, two examples of how backgrounds in event selections are susceptible to neutrino interaction modelling choices are given, one considering the impact of unreconstructed pions for Hyper-K/T2K and one considering wrong-sign neutrinos.

\begin{figure}[htbp]
  \centering
  \includegraphics[width=0.98\linewidth]{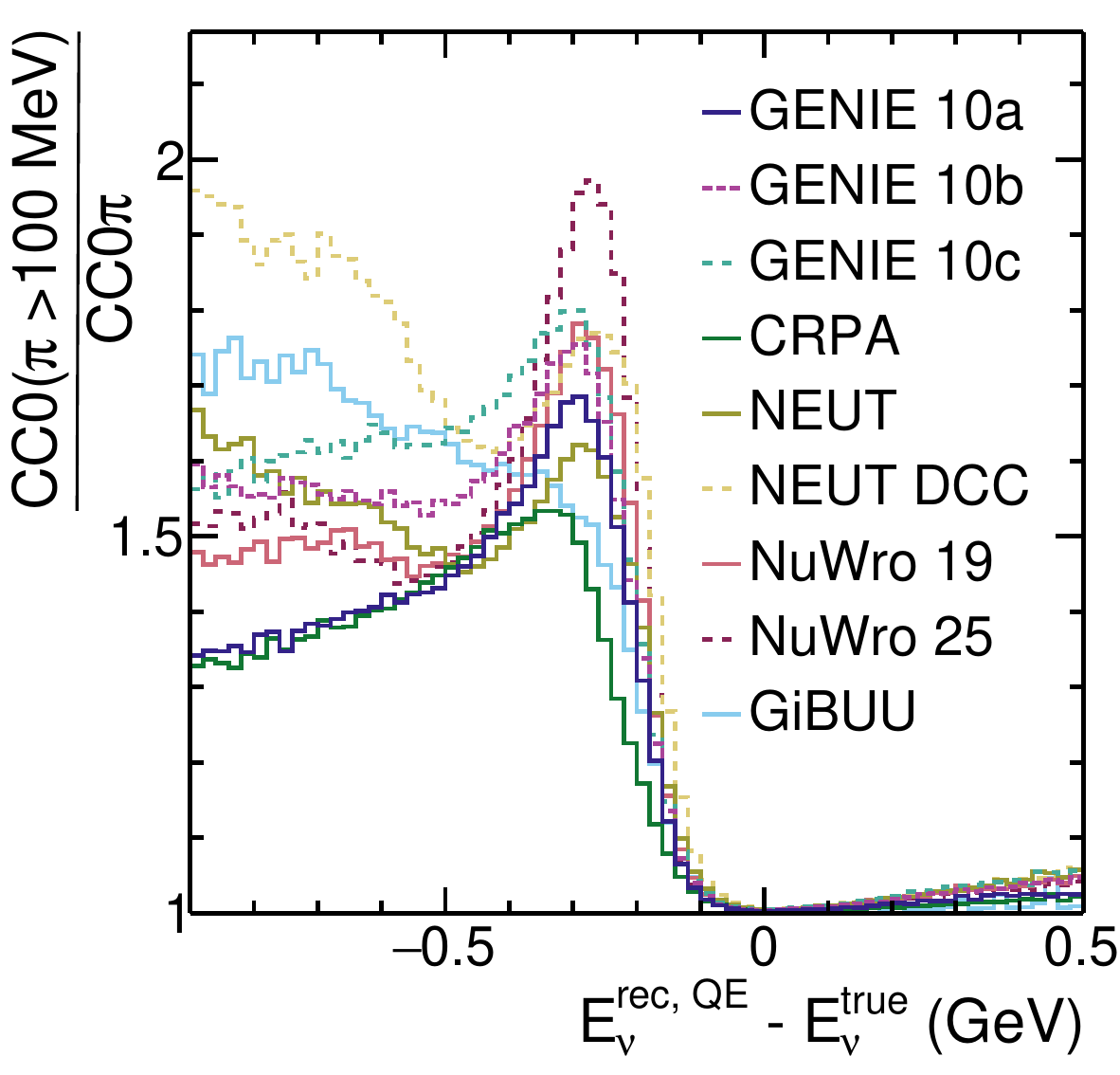}
  \caption{The ratio in the neutrino energy reconstruction bias for \numu--\water CC0$\pi$ events allowing charged pions with kinetic energies less than 100 MeV, with respect to the true CC0$\pi$ case, shown for a variety of neutrino generators.}
  \label{fig:cc0pimigrationgencomp}
\end{figure}

\autoref{fig:cc0pimigrationgencomp} shows a comparison of the neutrino energy reconstruction bias predicted between generators for CC$0\pi$-like topologies for two different cases: one in which only true CC0$\pi$ events are included; the other including pion-production events in which the pion kinetic energy is less than 100 MeV, close to the Cherenkov threshold for pions in water, and broadly illustrative of what can be expected in a detector like the Hyper-K FD. This provides a crude but qualitatively useful emulation of the impact of the pion background to the Hyper-K FD event selection and how sensitive this is to neutrino interaction model variations. It is immediately clear that the presence of the pion tracking threshold produces an increase in the high-bias tail region (\enuqe - \enutrue $\lesssim$ $-$0.1 GeV). This is caused by a higher proportion of RPP/DIS events entering the selection, for which neutrino energy is poorly reconstructed (see \autoref{subsec:enurec}). The impact of the pion threshold on the energy reconstruction bias varies very significantly between models. In a real analysis, some portion of the soft pion background could be eliminated by identifying the positron from the decay products of a $\pi^+$; for Super-K this is $>70\%$~\cite{Super-Kamiokande:2016exg}.  Conversely, $\pi^-$ are typically captured on nuclei before they can decay. This charge asymmetry makes dealing with this background more challenging for antineutrino than neutrino interactions, leading to complications in understanding the flavour ratios.

\begin{figure}[htbp]
  \centering
      \begin{subfigure}[b]{0.48\textwidth}
      \centering
      \includegraphics[width=1.0\linewidth]{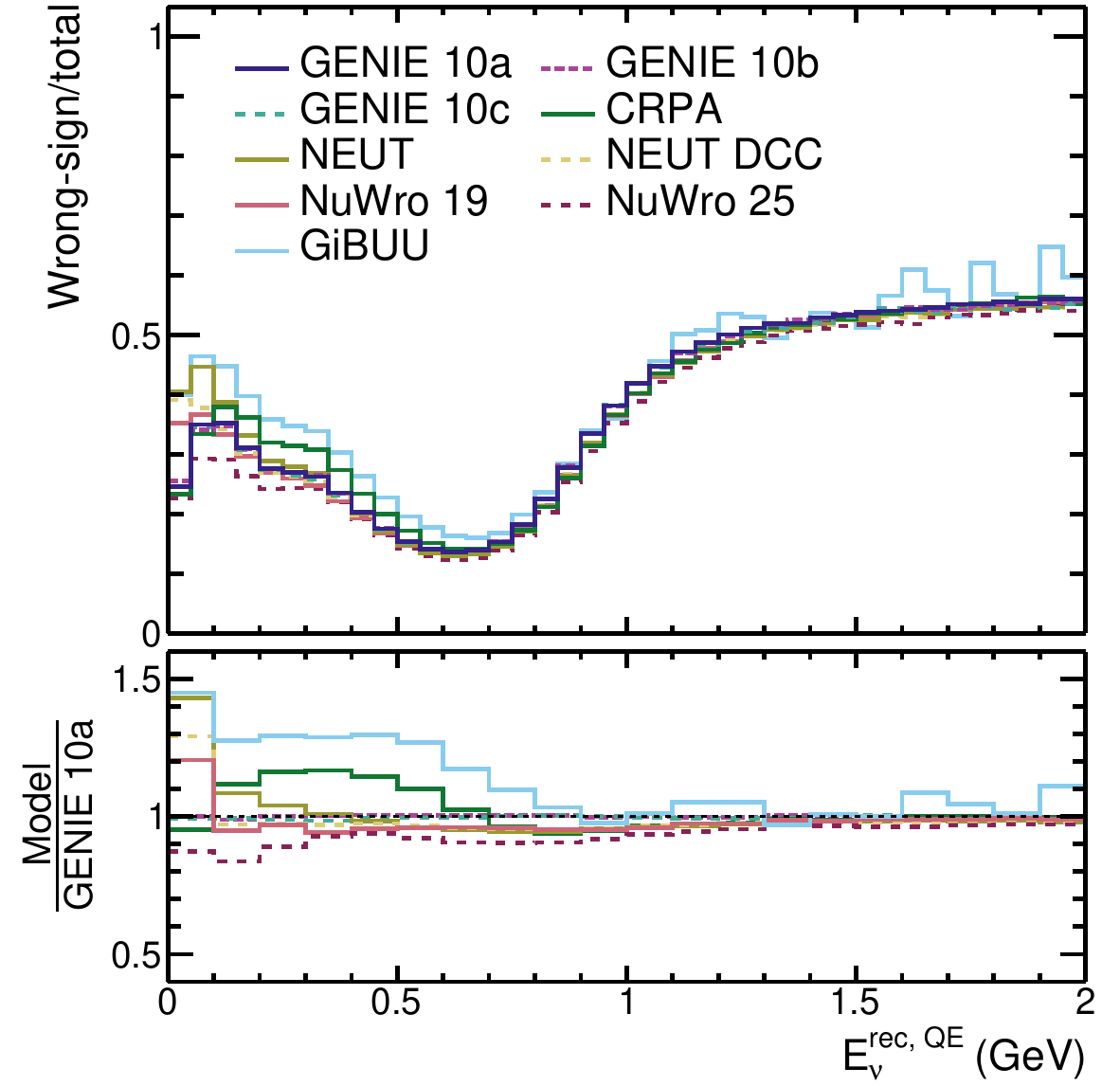}
      \caption{Hyper-K/T2K FD RHC \nueb--\water CC$0\pi$}
      \label{fig:skwrongsign_a}
  \end{subfigure}
     \begin{subfigure}[b]{0.48\textwidth}
      \centering
      \includegraphics[width=1.0\linewidth]{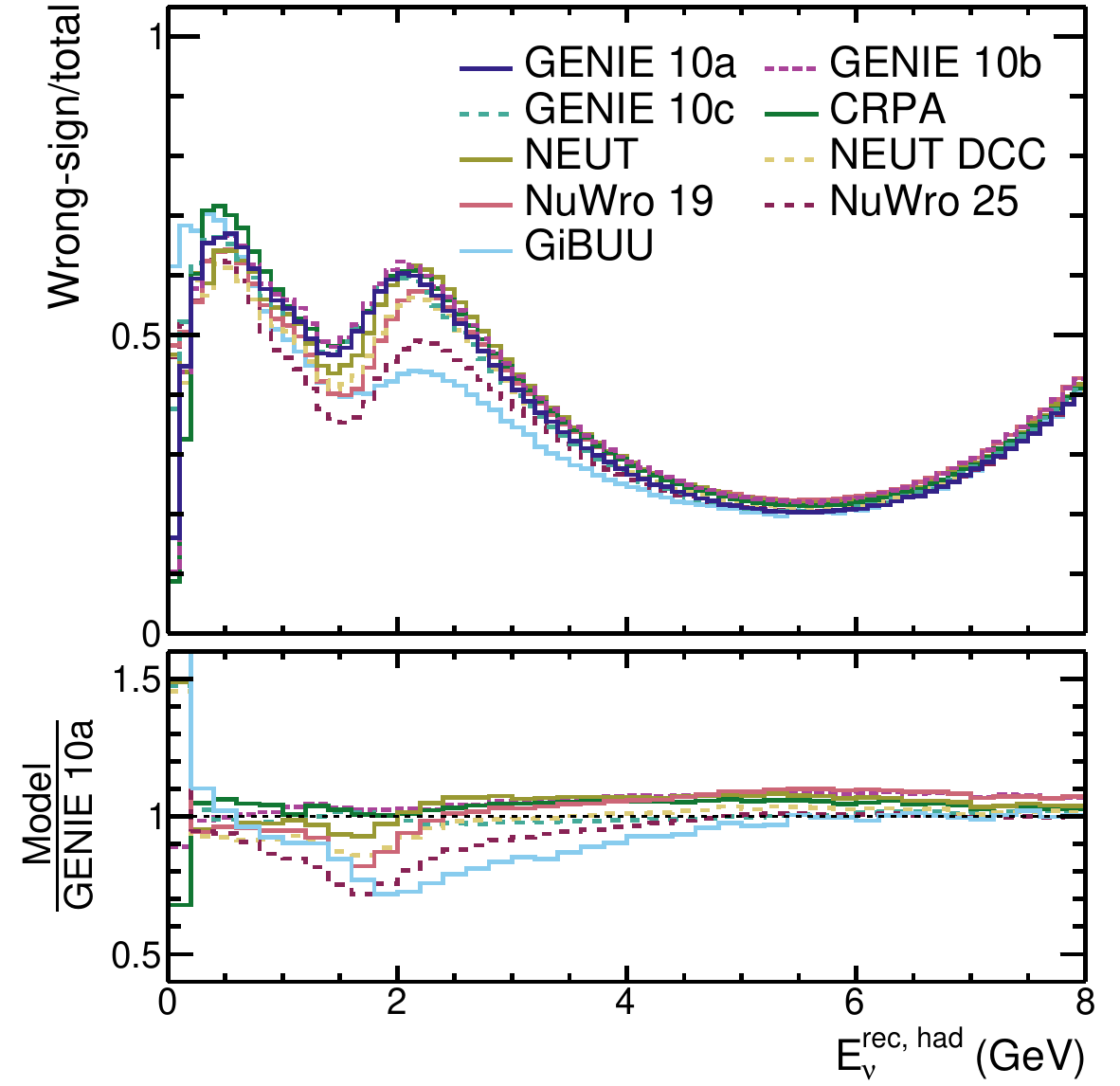}
      \caption{DUNE FD RHC \nueb--\argon CCINC}
      \label{fig:skwrongsign_b}
  \end{subfigure}  
     \caption{The estimated relative wrong-sign neutrino contribution to electron neutrino appearance event selections in the Hyper-K/T2K and DUNE FD for antineutrino mode operation as predicted by a variety of event generators, shown as a function of reconstructed neutrino energy. This is produced by comparing simulated interactions with the right- and wrong-sign flux predictions, accounting for the cross section and flux normalisation differences.}
     \label{fig:wrongsigngencomp}
\end{figure}

\autoref{fig:wrongsigngencomp} shows the estimated relative contribution of \nue (wrong-sign) interactions to \nueb selections at the DUNE and Hyper-K FDs as a function of reconstructed neutrino energy. This is produced by comparing simulated interactions with the right- and wrong-sign
flux predictions, accounting for the cross section and flux normalisation differences. The wrong-sign background is predicted to be significantly lower around the oscillation maximum for Hyper-K ($\sim$20\%) than for DUNE ($\sim$50\%), but both exhibit a large model spread of $\approx$30\% between generators. Whilst this background can be constrained by inferring the \numu contribution to the \numub flux with dedicated ND antineutrino mode selections, the small relative fraction of \numu interactions compared to \numub interactions and large wrong-sign flux uncertainties render this challenging. Some statistical separation of \nue from \nueb interactions may also be possible at FDs (see \autoref{subsubsec:wsbkg}), but the efficiency of any separation technique would also be sensitive to neutrino interaction modelling.

\subsection{Relationship between pion production topologies}

\label{subsec:issuespions}
\begin{figure*}[htbp]
  \centering
  \captionsetup[subfigure]{aboveskip=0pt,belowskip=3pt}       
   \begin{subfigure}[b]{0.48\textwidth}
      \centering
      \includegraphics[width=1.00\linewidth]{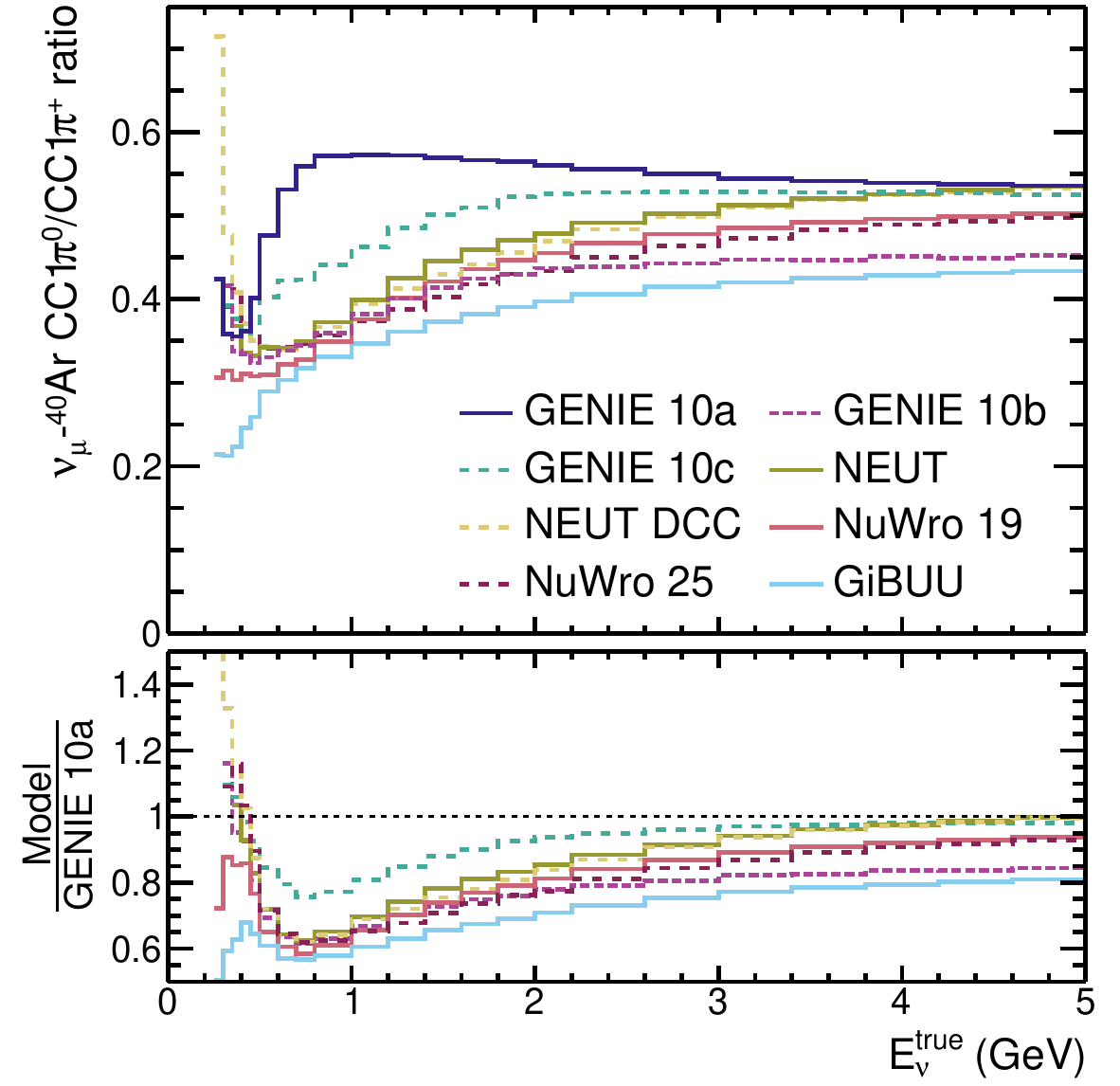}
      \caption{\numu--\argon CC1$\pi^{0}$/CC1$\pi^{+}$}
  \end{subfigure}
   \begin{subfigure}[b]{0.48\textwidth}
      \centering
      \includegraphics[width=1.00\linewidth]{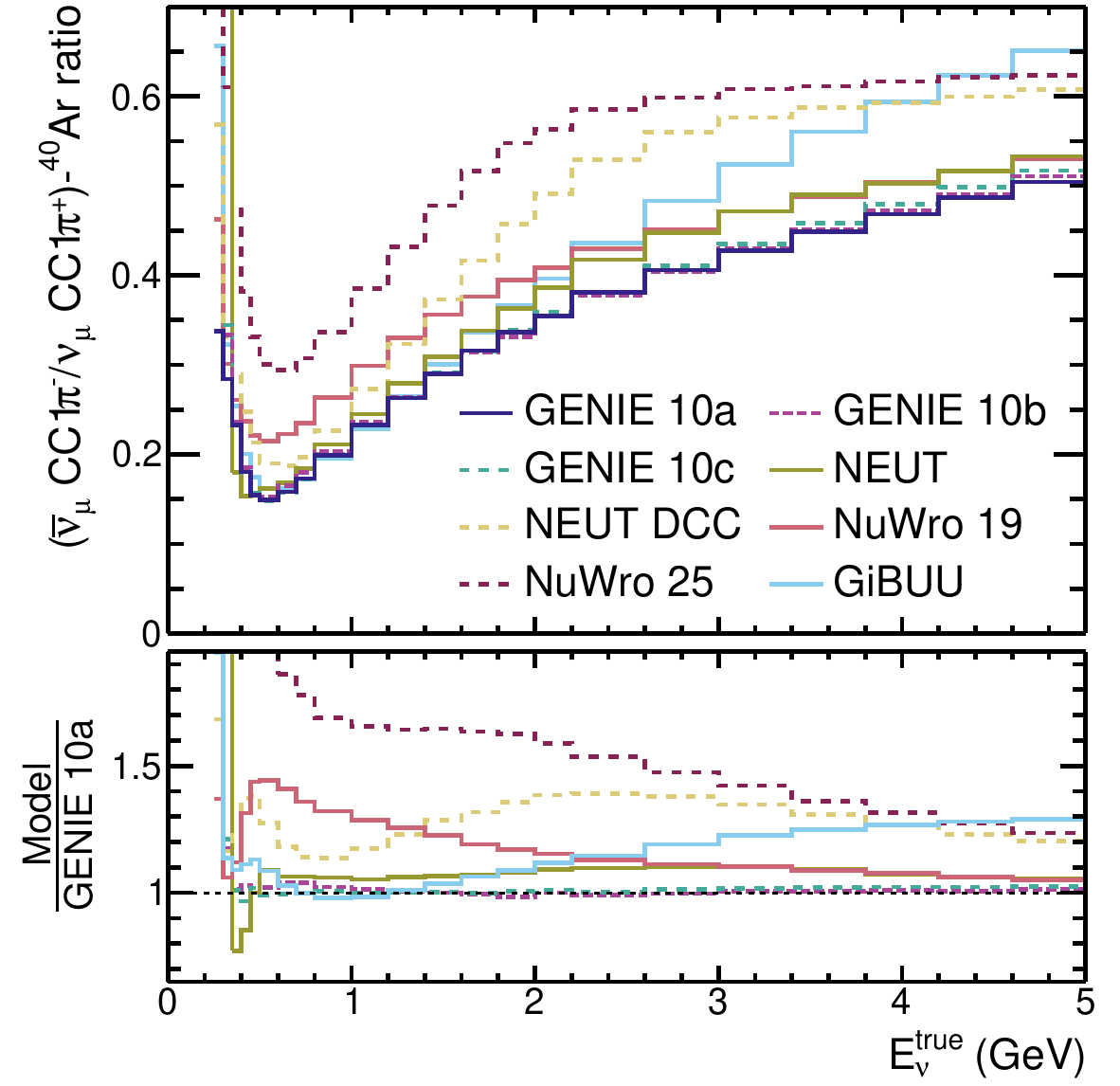}
      \caption{(\numub CC1$\pi^{-}$/\numu CC1$\pi^{+}$)--\argon}
  \end{subfigure}
  \begin{subfigure}[b]{0.48\textwidth}
      \centering
      \includegraphics[width=1.00\linewidth]{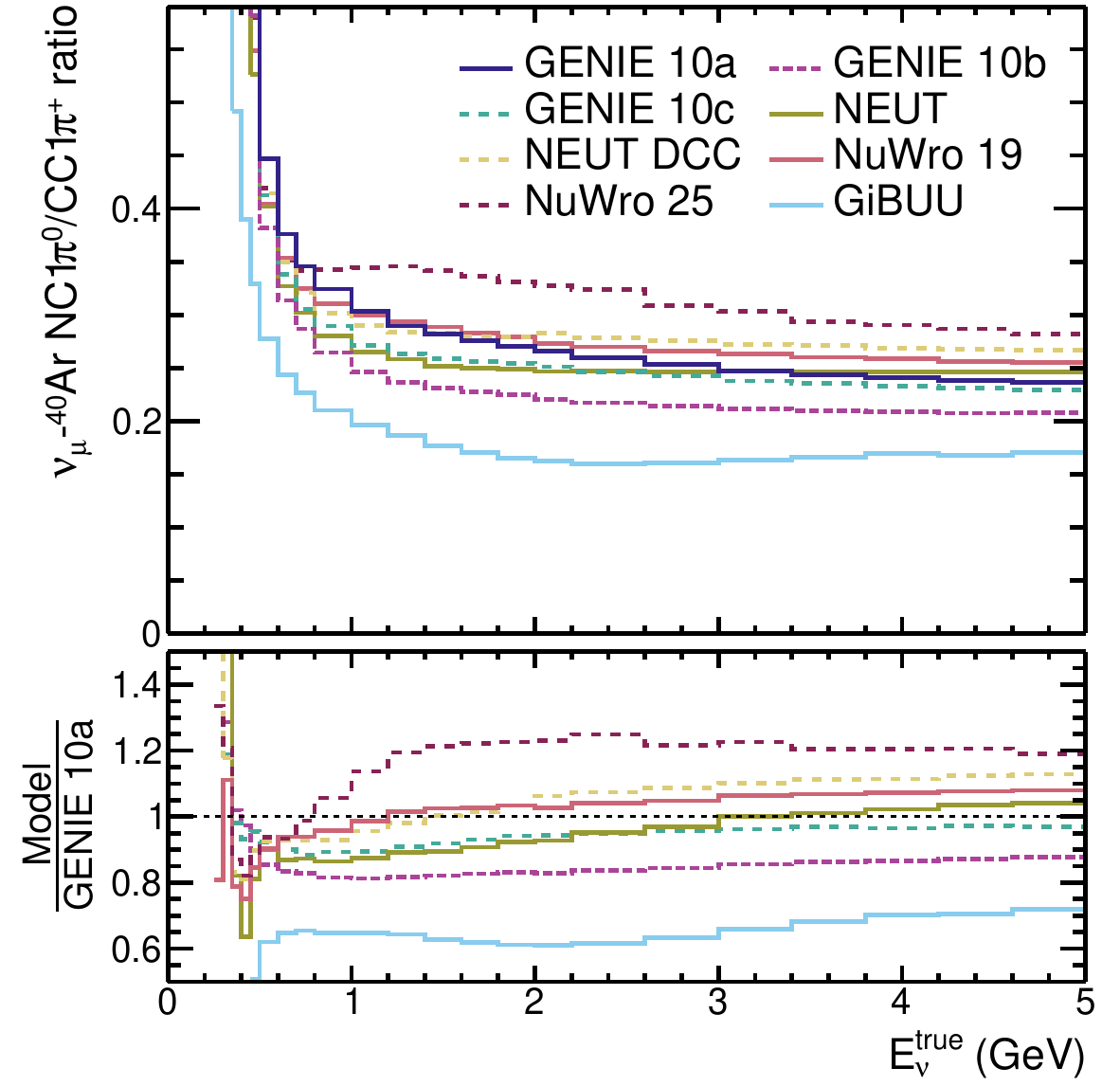}
      \caption{\numu--\argon NC1$\pi^{0}$/CC1$\pi^{+}$}
  \end{subfigure}
  \begin{subfigure}[b]{0.48\textwidth}
      \centering
      \includegraphics[width=1.00\linewidth]{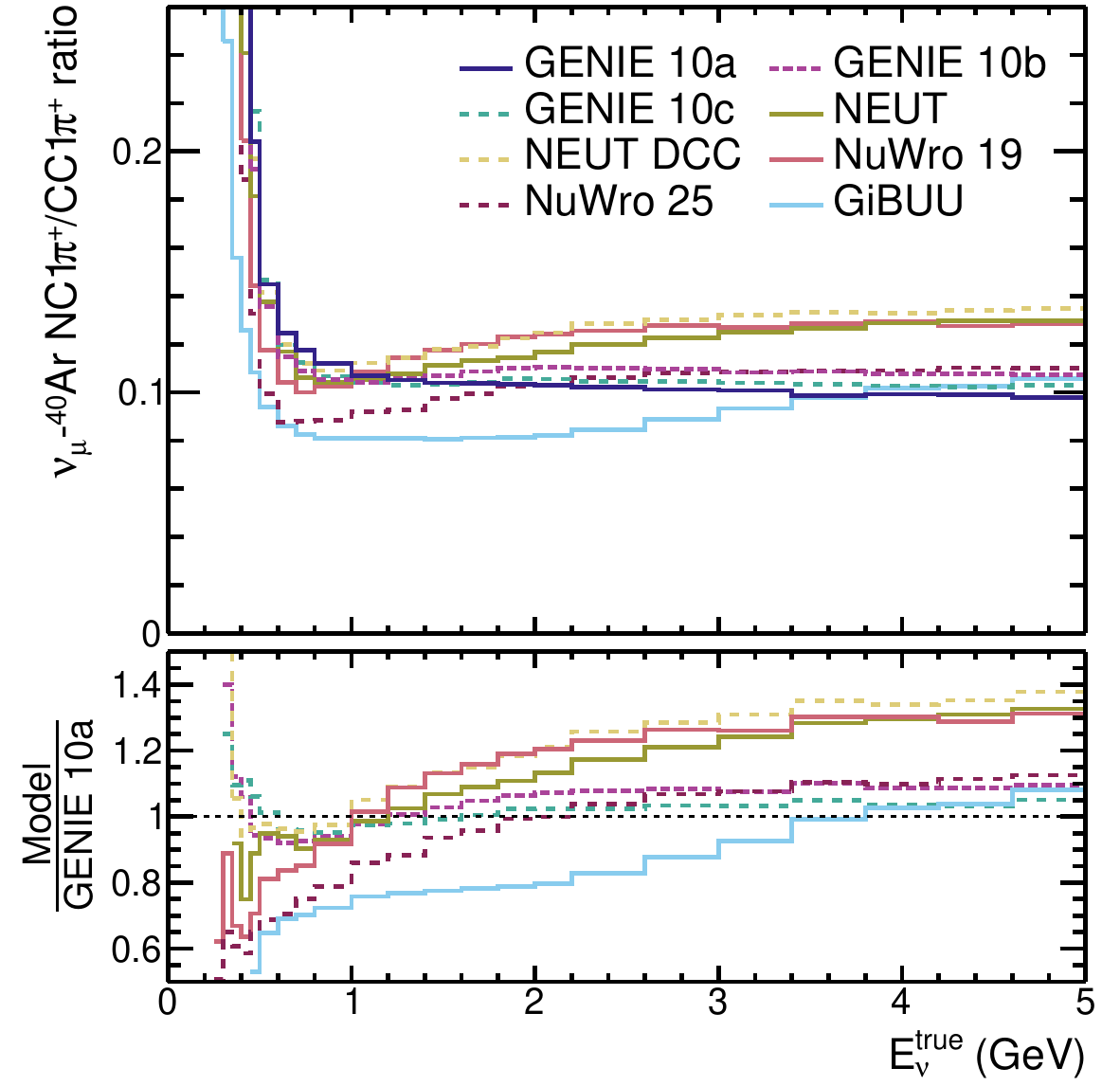}
      \caption{\numu--\argon NC1$\pi^{+}$/CC1$\pi^{+}$}
  \end{subfigure}
  \caption{Comparisons of the ratio between the cross sections (per nucleon) of a variety of $\numu^{\bracketbar}$--\argon pion production topologies with respect to the dominant \numu--\argon CC1$\pi^{+}$ topology as a function of \enutrue, for a variety of neutrino generators.}
  \label{fig:pion_ratios_gencomp}
\end{figure*}

\begin{figure*}[htbp]
  \centering
  \captionsetup[subfigure]{aboveskip=0pt,belowskip=3pt}       
   \begin{subfigure}[b]{0.48\textwidth}
      \centering
      \includegraphics[width=1.00\linewidth]{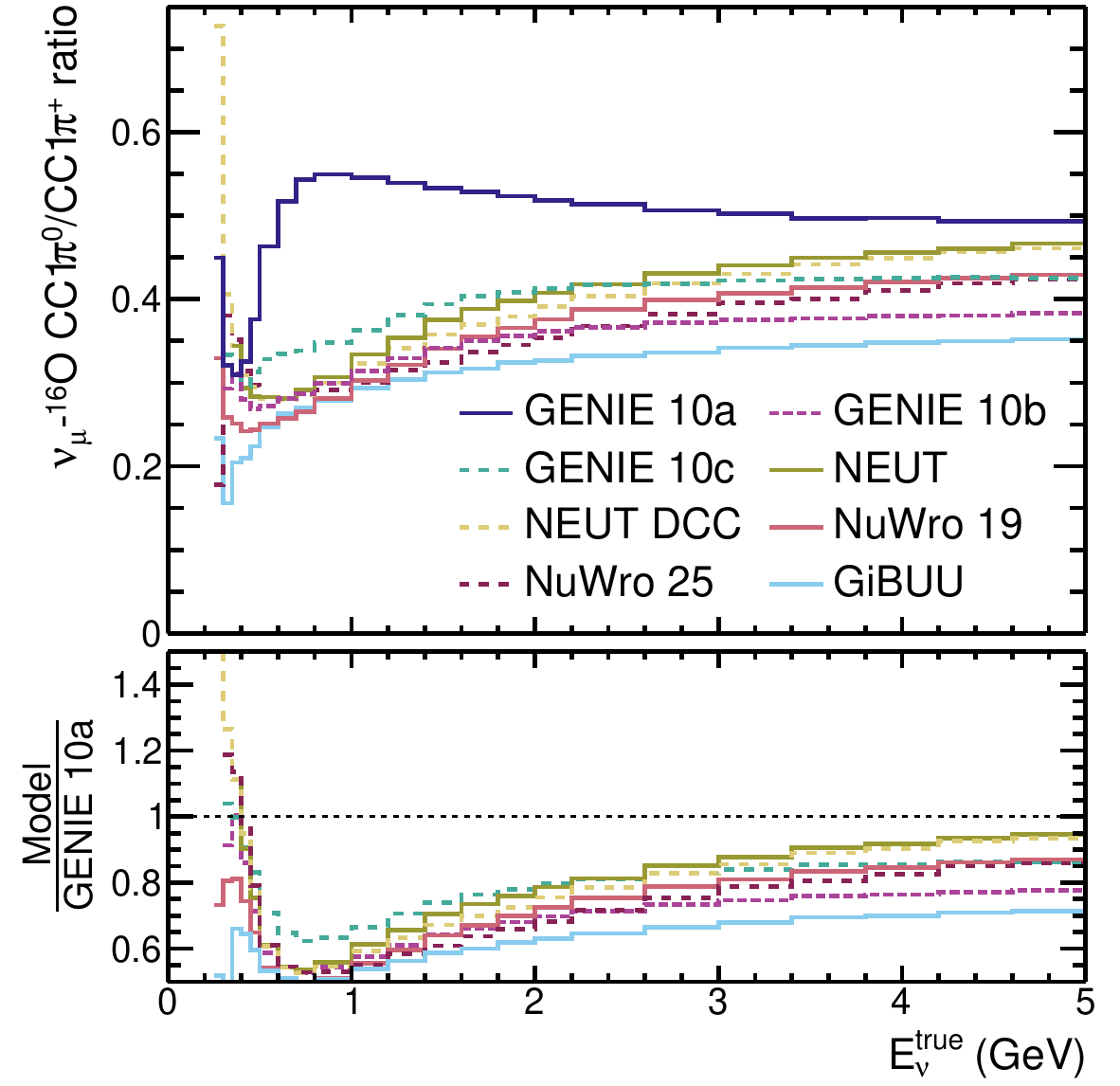}
      \caption{\numu--$^{16}$O CC1$\pi^{0}$/CC1$\pi^{+}$}
  \end{subfigure}
   \begin{subfigure}[b]{0.48\textwidth}
      \centering
      \includegraphics[width=1.00\linewidth]{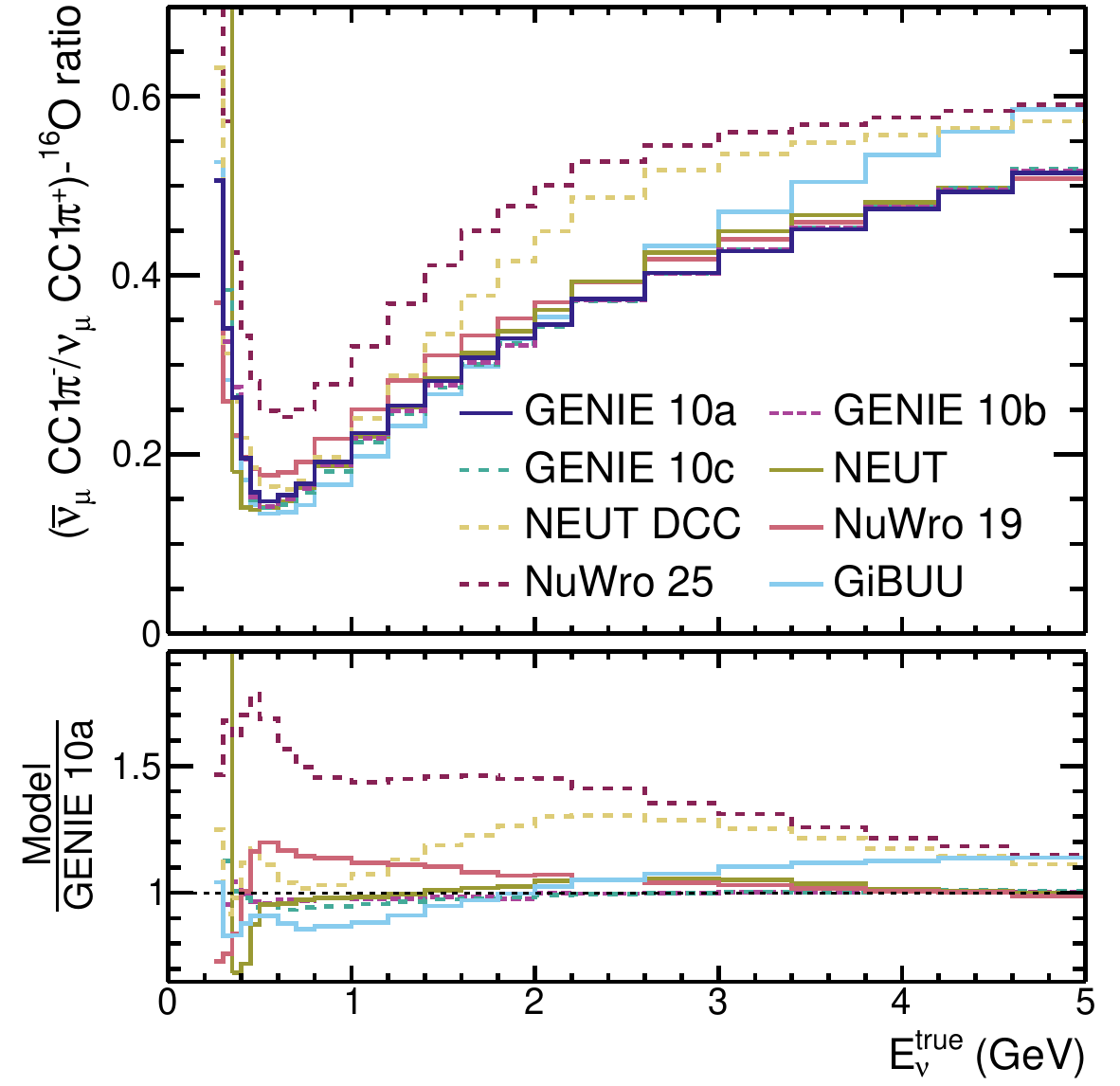}
      \caption{(\numub CC1$\pi^{-}$/\numu CC1$\pi^{+}$)--$^{16}$O}
  \end{subfigure}
  \begin{subfigure}[b]{0.48\textwidth}
      \centering
      \includegraphics[width=1.00\linewidth]{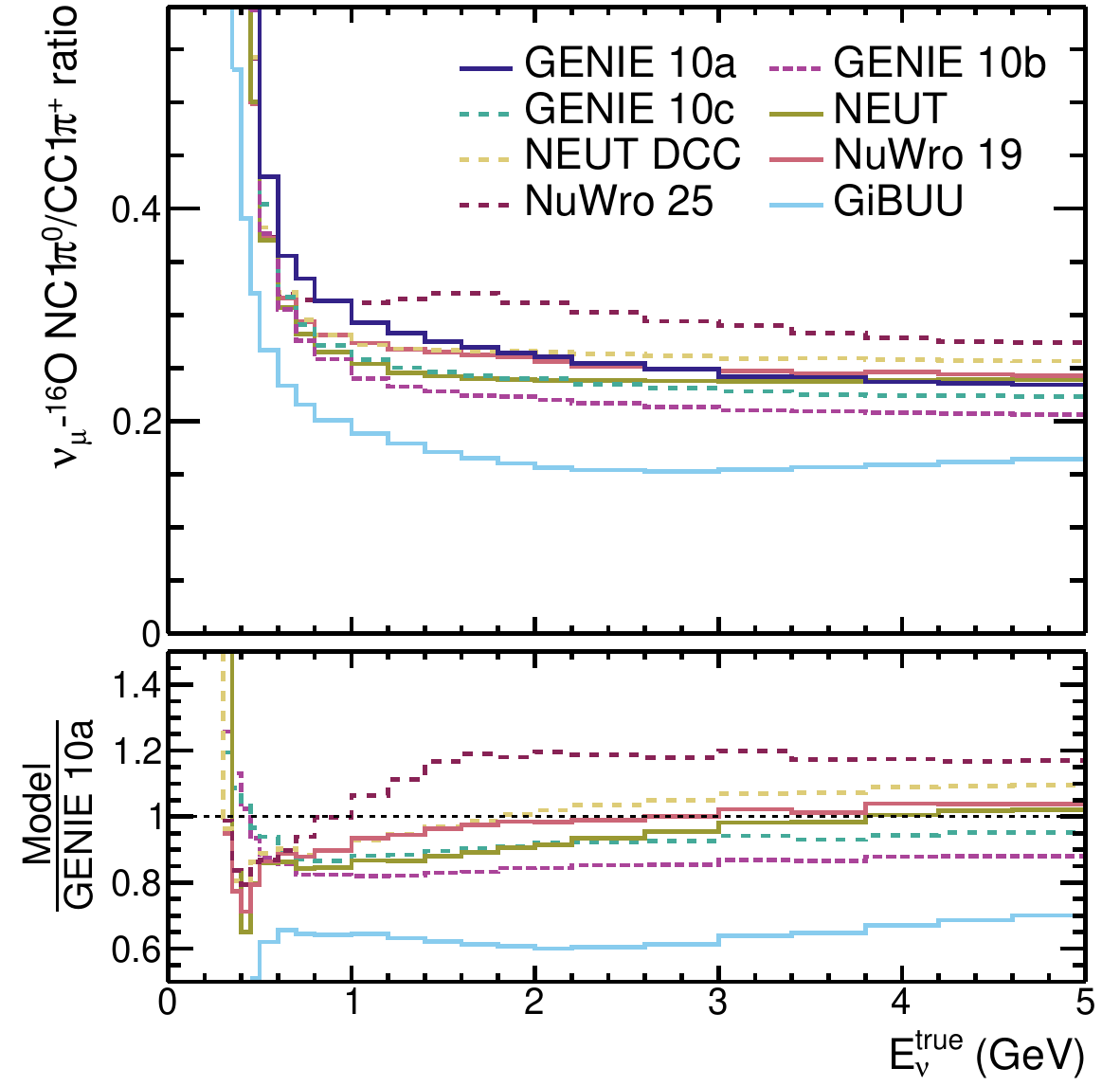}
      \caption{\numu--$^{16}$O NC1$\pi^{0}$/CC1$\pi^{+}$}
  \end{subfigure}
  \begin{subfigure}[b]{0.48\textwidth}
      \centering
      \includegraphics[width=1.00\linewidth]{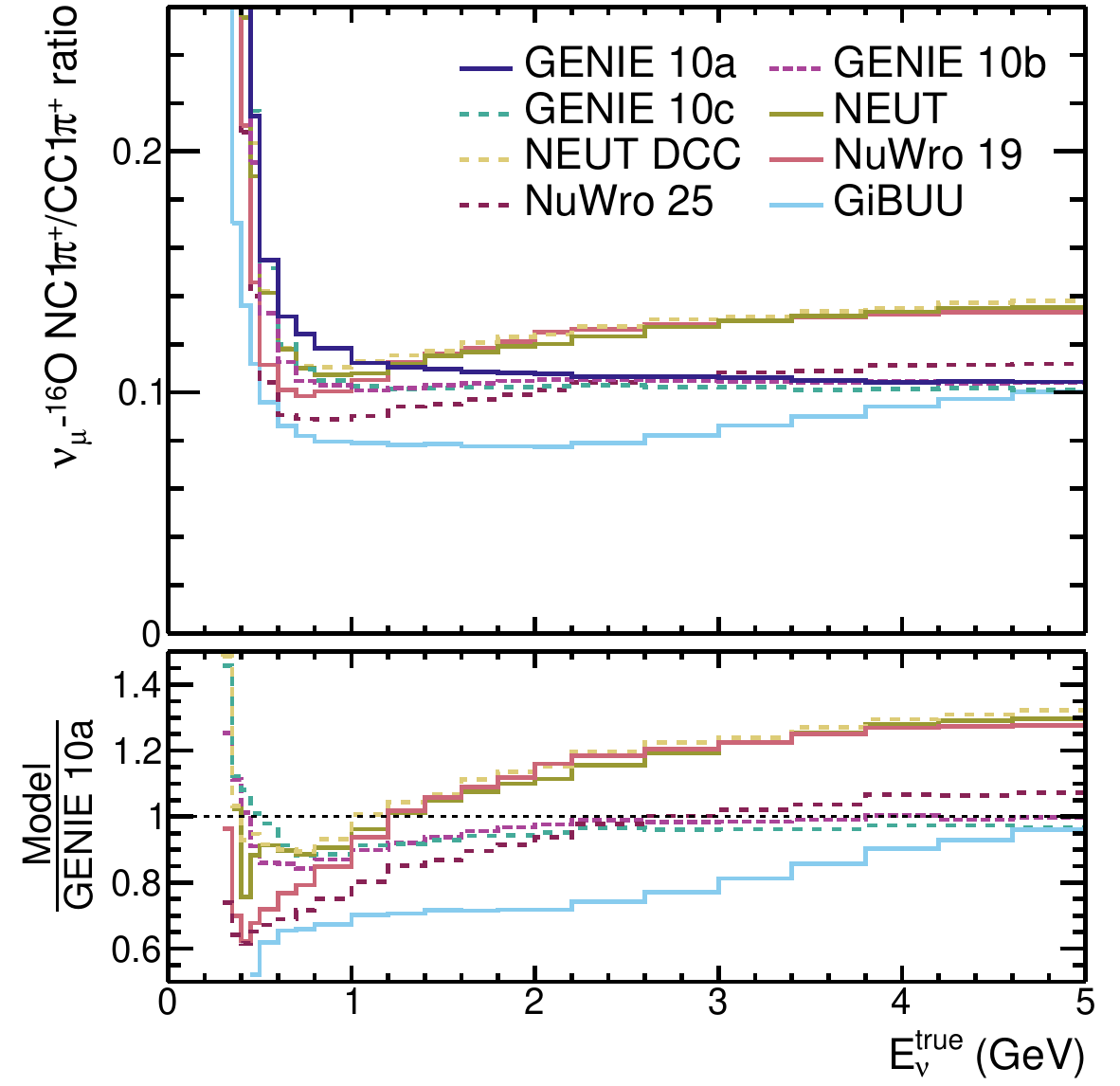}
      \caption{\numu--$^{16}$O NC1$\pi^{+}$/CC1$\pi^{+}$}
  \end{subfigure}
  \caption{Comparisons of the ratio between the cross sections (per nucleon) of a variety of $\numu^{\bracketbar}$--$^{16}$O pion production topologies with respect to the dominant \numu--$^{16}$O CC1$\pi^{+}$ topology as a function of \enutrue, for a variety of neutrino generators.}
  \label{fig:pion_ratios_gencomp_oxygen}
\end{figure*}

A variety of different pion production topologies contribute to FD event selections. These include the dominant $\nu$~CC1$\pi^{+}$ and $\bar{\nu}$~CC1$\pi^{-}$ topologies, which are a major contribution to the CCINC event rate used as the signal process for NOvA/DUNE, and are an important---often irreducible---background to the CC0$\pi$ event selections for Hyper-K/T2K (see \autoref{subsec:issuesbkg}). They also include CC1$\pi^{0}$ and multi-pion topologies, as well as additional contributions which come from misreconstruction of interactions that can cause misalignment between the true final-state topology and the reconstructed particle content. Three common examples are: 1) NC1$\pi^{\pm}$ interactions that can be misidentified as \numuany CC interactions since muons and charged pions can both produce long tracks with similar energy deposition; 2) NC1$\pi^{0}$ interactions that can be misidentified as $\nue^{\bracketbar}$ CC interactions if a photon from the $\pi^{0} \rightarrow \gamma\gamma$ decay is reconstructed as an electron; and 3) \numuany CC1$\pi^{0}$ interactions in which the muon is not identified, typically when the muon has low energy, can be misidentified as $\nue^{\bracketbar}$ CC interactions. Misidentification of pions, or their decay products, as outgoing leptons is also possible in multi-pion topologies, where the higher multiplicity gives more possibilities for particle misidentification.

ND events, which can be used to constrain pion production in an analysis, are dominated by \numuany CC interactions and ND event selections generally have to rely on identifying $\pi^\pm$, which only represent a subset of the pion production topologies relevant at the FD. The relationship between pion production at the ND and FD is challenging to model and constrain because each measured topology receives energy- and kinematic-dependent contributions from a variety of resonant and non-resonant pion production processes. For example, the \numu CC1$\pi^{+}$ topology that usually makes up a significant fraction of ND selections is dominated by the $\numu p \rightarrow \mu^{-} p \pi^{+}$ interaction, which proceeds purely through the $I_{3/2}$ channel (almost entirely via the $\Delta^{++}(1232)$), whereas the neutron-target channel, as well as the NC and CC1$\pi^{0}$ topologies that contribute as backgrounds at the FD, contain mixed-isospin contributions from multiple interfering resonances, which produce different characteristics in the hadronic system compared to the $\Delta(1232)$. Extrapolating an ND constraint on this mix of contributions to the FD implicitly relies on the relationship between these contributions within the model. FSI and other nuclear effects (see \autoref{sec:models}) further complicates the relationship between observable topology and the contributing processes: absorption removes pions from the final state and charge exchange migrates interactions directly between the $\pi^\pm$ and $\pi^{0}$ topologies such that propagating a constraint from one topology to another directly depends on the FSI model.

\autoref{fig:pion_ratios_gencomp} and \autoref{fig:pion_ratios_gencomp_oxygen} show how neutrino event generators differ in their prediction of the relationship between different pion production topologies as a function of \enutrue, for an argon and an oxygen target respectively\footnote{CRPA is not included in \autoref{fig:pion_ratios_gencomp} or \autoref{fig:pion_ratios_gencomp_oxygen}, as it only differs from GENIE 10a in its CCQE and 2p2h treatment.}. The two targets show broadly similar features. However, predictions for the \numu CC1$\pi^{0}$/CC1$\pi^{+}$, \numu NC1$\pi^{0}$/CC1$\pi^{+}$ and \numu NC1$\pi^{+}$/CC1$\pi^{+}$ ratios differ substantially among the generators, and the models disagree in different ways for each ratio, reflecting the complex relationships between the predictions of all of the contributing processes. The GENIE 10a, 10b and 10c models, which share the same hard-scattering physics but differ in their FSI treatment, show much smaller differences between the \numub CC1$\pi^{-}$/\numu CC1$\pi^{+}$ and \numu NC1$\pi^{+}$/CC1$\pi^{+}$ ratios, where the effect of FSI largely cancels between numerator and denominator, but differ significantly for the \numu CC1$\pi^{0}$/CC1$\pi^{+}$ and \numu NC1$\pi^{0}$/CC1$\pi^{+}$ ratios, where charge exchange feeds the $\pi^{0}$ channels from the much larger $\pi^{+}$ production channel. The generators also differ significantly in their predictions of the two dominant production channels, \numu CC1$\pi^{+}$ and \numub CC1$\pi^{-}$, which partly determines the wrong-sign contribution to antineutrino-enhanced selections at the unmagnetised FDs (see \autoref{subsec:issuesbkg}). The models vary considerably in their predictions of the ratios of the cross section between the considered pion production topologies, showing energy-dependent differences of $\approx$40\% or more across the majority of the 0--5 GeV energy range.
The differences grow very large near the interaction thresholds, but this is a region where the cross section is comparably small (see \autoref{fig:sigenu} and \autoref{fig:oscxsecfluxcomp}).

Overall, \autoref{fig:pion_ratios_gencomp} and \autoref{fig:pion_ratios_gencomp_oxygen} demonstrate that there is considerable model dependence when extrapolating predictions or constraints between pion production topologies. This has direct implications for oscillation experiments, where high-statistics CC samples at the ND are used to constrain a variety of CC and NC pion production topologies at the FD.

\subsection{Summary}
\label{subsec:issuessummary}

This section has demonstrated that many of the aspects of neutrino interaction modelling relied upon in neutrino oscillation analyses are subject to significant uncertainty, particularly those which affect the extrapolation from an ND constrained model to the FD. Although they have been presented in a factorised way, many of these issues are non-trivially interrelated. The examples shown are representative of the difficulties that will be faced by DUNE and Hyper-K, although the full experimental analyses have the ability to make much more sophisticated ND constraints than simple generator-level studies can capture. That said, many of the challenges discussed---detector acceptances, background rejection issues, and neutrino energy reconstruction---are significantly more complex in a real detector than these simplified studies. Although not strictly quantitative, the range and size of the effects underscore how difficult it is to reach the few-percent level uncertainties required in the ND to FD extrapolation for both experiments.

{\section{Uncertainties on neutrino oscillation parameters}
\label{sec:impactonoa}
Demonstrating the exact impact of cross-section model variations on measurements of oscillation parameters is challenging without access to each experiment's simulation and analysis frameworks. 
This is due to the complex interplay between cross-section, flux and detector-response models, particularly in understanding the impact of ND constraints on the systematic uncertainties, and extrapolating those constraints to the FD.
Current and future neutrino oscillation experiments often quantitatively assess the role of neutrino interaction uncertainties on their analyses.
In this section we review some of the key conclusions of these studies, organised by experiment. 
Some independent groups have also conducted studies that assess the impact of targeted cross-section model variations on FD observables and then estimate the impact on oscillation parameters measurements, although these studies are necessarily approximate. A review of some of this work~\cite{Coloma:2013rqa,Coloma:2013tba,Ankowski:2015kya,Jen:2014aja,Ankowski:2015jya,Ankowski:2016bji} can be found in Ref.~\cite{Ankowski:2016jdd} with further contributions in Refs.~\cite{Meloni:2012fq,Nagu:2019uco,Nagu:2019fvi,Devi:2022zfh,Coyle:2025xjk}.

\subsection{T2K and Hyper-K}
Recent T2K neutrino oscillation analyses can be found in Refs.~\cite{T2K:2023smv, T2K:2024wfn, T2K:2025yoy}, with additional work to improve and benchmark its treatment of CC$0\pi$ neutrino interactions and associated uncertainties in Ref.~\cite{Chakrani:2023htw}. 
The latest Hyper-K sensitivity studies~\cite{Hyper-Kamiokande:2025fci} use a model developed by T2K~\cite{T2K:2023smv} for neutrino interactions and their uncertainties, reduced by varying extents to account for projected improvements in modelling.

The dominant uncertainty on the electron-neutrino appearance channels for T2K and Hyper-K is related to the modelling of the double cross section ratio between the four flavours, $(\bar{\nu}_e/\bar{\nu}_\mu)/(\nu_e/\nu_\mu)$, (introduced in \autoref{subsec:issuesNueNumu})~\cite{Zhu:2023nsv, Munteanu:2022zla, Hyper-Kamiokande:2025fci}, directly impacting measurements of \deltacp, the mass ordering, and the octant of $\theta_{23}$.
The analyses include a partially correlated uncertainty on the normalisation of both the $\nue/\numu$ and $\nueb/\numub$ \textit{single} cross-section ratios. 
These are set to approximately account for a component due to radiative corrections to the cross section based on Ref.~\cite{Day:2012gb} and another due to nuclear effects~\cite{T2K:2023smv}. 
Small reductions of the resultant 4.9\% uncertainty on the double ratio significantly changes the proportion of values of \deltacp which Hyper-K can exclude at 5$\sigma$ significance, as demonstrated in \autoref{fig:hyperknuenumu}\footnote{The uncertainties shown in the plot body are the impact of the double cross-section ratio $(\bar{\nu}_e/\bar{\nu}_\mu)/(\nu_e/\nu_\mu)$ on Hyper-K sample event rates, which is smaller than the aforementioned 4.9\% errors as the samples have background components.}. 
Recent work~\cite{Tomalak:2021hec, Tomalak:2022xup} suggests that, provided appropriate corrections are included in simulations, the radiative correction uncertainty on the cross section can be significantly reduced to an almost negligible level at the energies and four-momentum transfer of Hyper-K\footnote{Whilst the uncertainty on the cross section is reduced, this work also stresses the importance of evaluating the impact of associated radiative hard photon emission on event selections employed at the FD.}.
\begin{figure}[hptb]
  \centering
  \includegraphics[width=\linewidth, clip, trim=11mm 5mm 18mm 0mm]{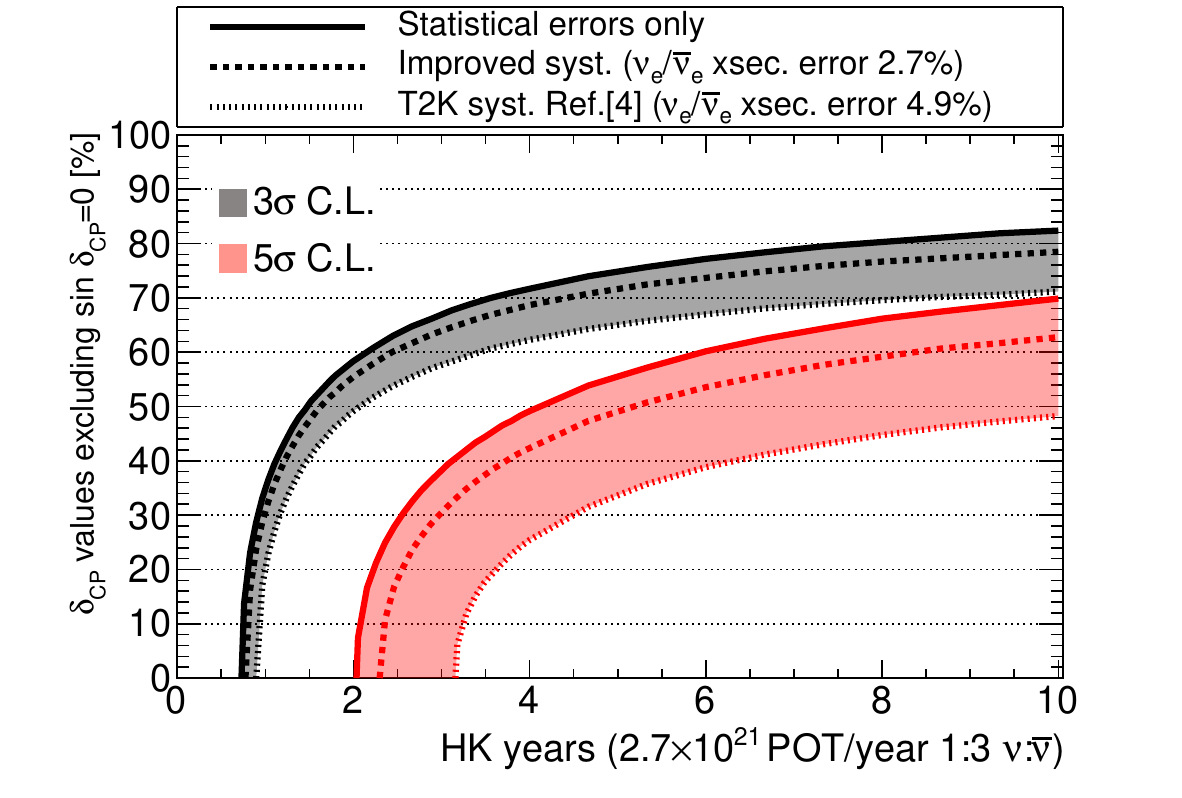}
  \caption{The sensitivity to exclude CP conservation ($\sin\deltacp=0$) at 3$\sigma$ (black) and 5$\sigma$ (red) significance, as a function of projected Hyper-K exposure. The width of the bands show the different assumptions for the systematics: statistics-only case (no systematics), T2K's uncertainty from a reference paper (Ref.~\cite{T2K:2023smv} in this article), and an improved uncertainty on the double ratio of the neutrino to antineutrino electron to muon neutrino cross-section ratio, amongst others. Reproduced from Ref.~\cite{Hyper-Kamiokande:2025fci}.}
  \label{fig:hyperknuenumu}
\end{figure}

An independent study~\cite{Dieminger:2023oin} of CC0$\pi$ interactions suggests that the uncertainty due to nuclear-effects on the $\nue/\numu$ and $\nueb/\numub$ single ratios used by T2K and Hyper-K may be underestimated. However, the study also finds that alterations to the interaction model shift the neutrino and antineutrino $\nue/\numu$ ratios in the same direction, such that the resulting uncertainty on the $(\bar{\nu}_e/\bar{\nu}_\mu)/(\nu_e/\nu_\mu)$ double ratio is smaller than the aforementioned 4.9\%. This highlights that, for determining the sensitivity of Hyper-K to exclude CP-conserving values of \deltacp, the correlation between the two single ratio uncertainties can be as important as their individual sizes.  

Beyond muon and electron neutrino differences, T2K performs \textit{simulated data studies} as part of its oscillation analyses to explore the impact of mis-modelling of neutrino interactions that may not be contained within the uncertainties it considers~\cite{T2K:2023smv}. 
In these studies, a neutrino interaction simulation that is different than T2K's base model is treated as if it were real data and the extracted oscillation parameters are compared to their true values to assess biases. 
In T2K's analysis published in Ref.~\cite{T2K:2023smv}, all such studies induce biases on oscillation parameters that are much smaller than the systematic uncertainties on them, which are in turn smaller than the statistical uncertainty. 
However, T2K's most recent neutrino oscillation analysis~\cite{T2K:2025yoy} found that simulated data in which CCQE interactions were modelled using CRPA (compared to T2K's baseline spectral function model) caused biases on measurements of \dm{32} that are larger than all other systematic uncertainties combined, but still significantly less than the sum of the statistical and systematic uncertainty~\cite{T2KDolanNuInt2022}. A similarly sized bias was found when the fraction of 2p2h and CCQE was modified based on the maximal variation allowed by T2K ND data. Comparable potential for biases was also reported in the joint T2K--SK oscillation analysis~\cite{T2K:2024wfn}.

\subsection{NOvA}

The most recent neutrino oscillation analysis from the NOvA experiment is presented in Ref.~\cite{NOvA:2025tmb}, which builds on previous analyses~\cite{NOvA:2021nfi}, with further details of the ND analysis, neutrino interaction model and the uncertainties described in Ref.~\cite{NOvA:2020rbg}. 
NOvA's most recent analysis included changes to the RPP interaction model, a new pion FSI tune, and a neutron model based on MENATE~\cite{Kohley:2012awa}.
NOvA provided an estimated breakdown of the sources of uncertainties on the measured neutrino oscillation parameters, evaluated at the best fit value, for earlier analyses~\cite{NOvA:2021nfi,miranda_elkins_2020_4253085},  which has been reproduced in \autoref{fig:novaOAParams}.
For these analyses, detector calibration uncertainties were dominant, although uncertainties related to neutrino interaction modelling play an important role for the three oscillation parameters given, particularly \deltacp. 
In the most recent analysis, neutrino interaction systematics are the most important systematic for measurements of \deltacp, followed by detector calibrations, and for $\sin^2\theta_{23}$ it is the second-most important systematic following the detector calibrations. The neutron modelling uncertainty was a large contribution to the uncertainty in earlier analyses (as shown in~\autoref{fig:novaOAParams}), but its contribution was reduced through detector response studies.
\begin{figure*}[htbp]
  \centering
  \includegraphics[width=0.85\linewidth]{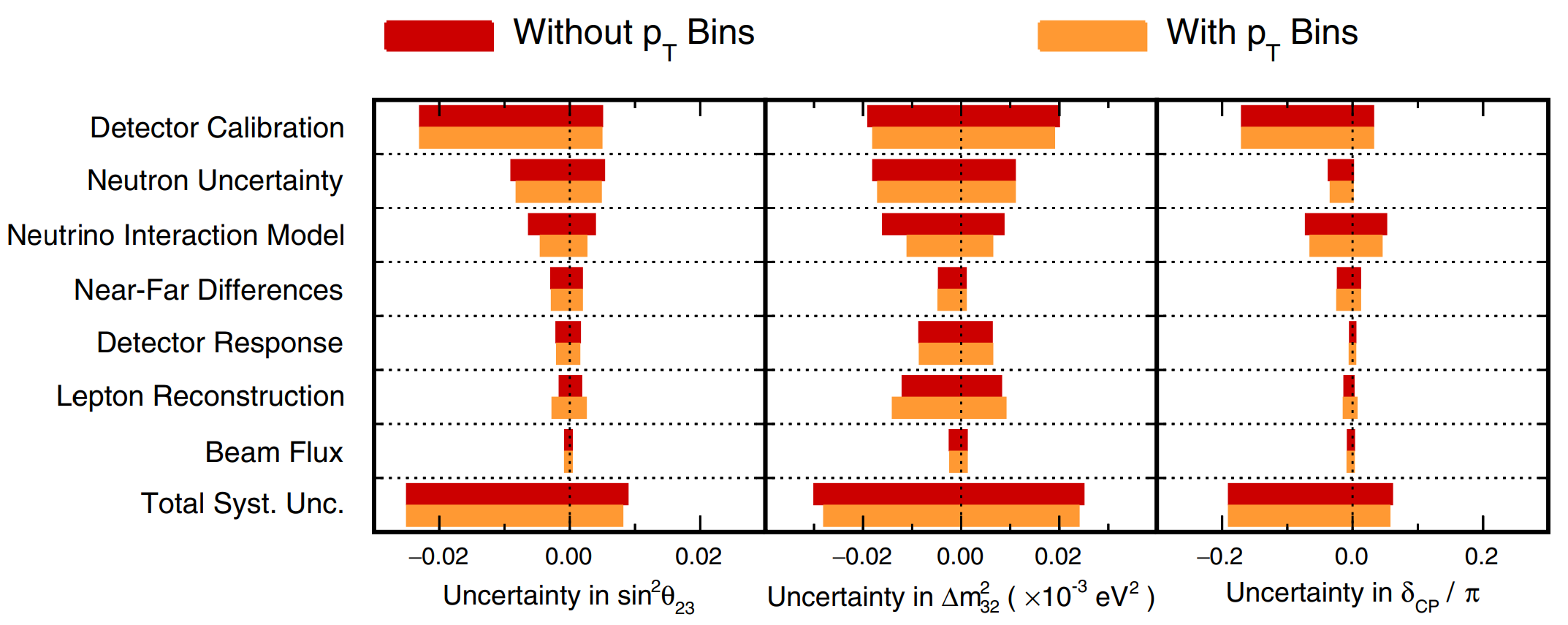}
  \caption{The impact of systematic uncertainties reported by the NOvA experiment on extracted neutrino oscillation parameters. It shows the impact with and without the additional constraints provided by binning the analysis in the transverse projection of the outgoing lepton momentum. Reproduced from Ref.~\cite{NOvA:2021nfi}. 
  }
  \label{fig:novaOAParams}
\end{figure*}

\subsection{DUNE}

With a wide-band on-axis beam and calorimetric energy reconstruction using LArTPC FDs, DUNE gains unique sensitivity to the evolution of the neutrino oscillation probability over a wide neutrino energy range, but also faces unique neutrino interaction modelling challenges. The projected DUNE sensitivity to three-flavour oscillation parameters has been described in detail in Refs.~\cite{DUNE:2020jqi, DUNE:2021mtg}, which include a bespoke treatment of cross-section uncertainties, built on top of a GENIEv2 baseline model and inspired by those used in currently running experiments. The collaboration has demonstrated that a high-performance ND is essential to sufficiently constrain the uncertainty model at the FD to deliver its physics goals~\cite{DUNE:2022aul}, and that upgrades beyond the initial ND complex~\cite{DUNE:2021tad} are likely to be needed to reach its ultimate precision goals~\cite{DUNE:2022aul, DUNE:2022yni}. Beyond uncertainties propagated through the oscillation analysis, both the DUNE collaboration and studies by independent groups have shown that mismodelling the performance of calorimetric neutrino energy reconstruction (for example by incorrectly simulating the fraction of hadronic energy carried away by neutrons) may bias the extracted oscillation parameters~\cite{Ankowski:2015kya, Friedland:2018vry, Nagu:2019uco, DUNE:2021tad, Coyle:2025xjk, Liu:2025hpl}.
\begin{figure}[hptb]
  \centering
  \includegraphics[width=0.98\linewidth, trim={0 0 1.5cm 1.15cm}, clip]{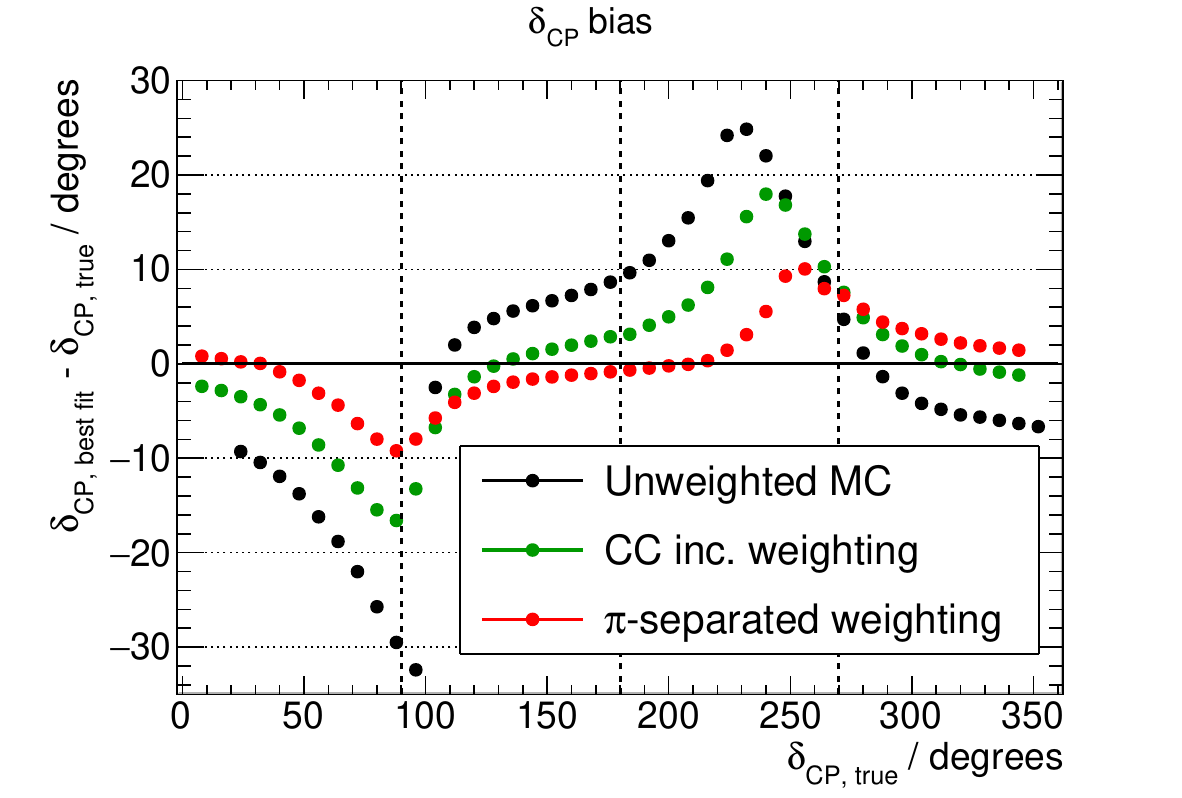}
  \caption{The absolute bias DUNE finds on the fitted value of \deltacp when analysing simulated data from NuWro. Black shows the bias without any ND, green shows the ability of a ND to partially mitigate the bias using a CCINC ND sample and red shows the improved ability to mitigate the bias by splitting the CCINC sample according to the number of final-state pions. Reproduced from Ref.~\cite{DUNE:2021tad}.}
  \label{fig:dune_dcp_bias}
\end{figure}

DUNE has directly studied the impact that mismodelling the neutrino cross section may have on its \deltacp resolution and CP-violation sensitivity. An example is given in \autoref{fig:dune_dcp_bias}, reproduced from Ref.~\cite{DUNE:2021tad}, in which the size of the bias on the measured value of \deltacp is shown as a function of the true value of \deltacp, when simulated data from one generator (NuWro) is fitted with another (GENIE).
The size of the bias is shown without an ND constraint (\textit{Unweighted MC}) and in two scenarios where an ND is used. \textit{CC inc. weighting} uses a constraint derived from a CCINC ND selection in which events are binned by reconstructed energy and three-momentum transfer, whereas \textit{$\pi$-separated weighting} uses a constraint in which the ND is additionally able to subdivide the events by the number of final-state pions. Both the no ND and CCINC ND constraint cases show biases in measurements of \deltacp that are comparable with the ultimate target precision for DUNE ($\sim$15--17\degree for $\deltacp=90\degree,270\degree$\cite{DUNE:2021mtg}). This is mitigated by the more exclusive ND constraint, demonstrating the importance of a capable ND to ensure DUNE can realise its ultimate physics reach. 

A study of specific FSI models within GENIE~\cite{Liu:2025hpl} found that their impact on a proxy for the reconstructed neutrino energy spectra in the DUNE FD is comparable to that of neutrino oscillation parameters at the precision that the next-generation experiments target. 
For example, \autoref{fig:FSIonDUNEOAdm32} shows that varying FSI models has a much greater and similarly-shaped effect on the approximated FD spectrum to variations of \dm{32} at the $\pm 0.4$\% level, corresponding roughly to both DUNE and Hyper-K's target precision~\cite{DUNE:2020ypp,Hyper-Kamiokande:2025fci}. 
Importantly, the study does not account for having an ND constraint on the FSI models, and so is representative of their importance if no robust parameterisation of FSI uncertainties can be developed and constrained at the DUNE ND. The study also found that the GENIE FSI variations considered were comparable to small variations of \deltacp at the level of the ultimate sensitivity projected by DUNE, but that they are not degenerate with variations of $\theta_{23}$.

\begin{figure}[htb]
\centering
\includegraphics[width=0.48\textwidth, clip,trim=0mm 5mm 10mm 22.5mm]{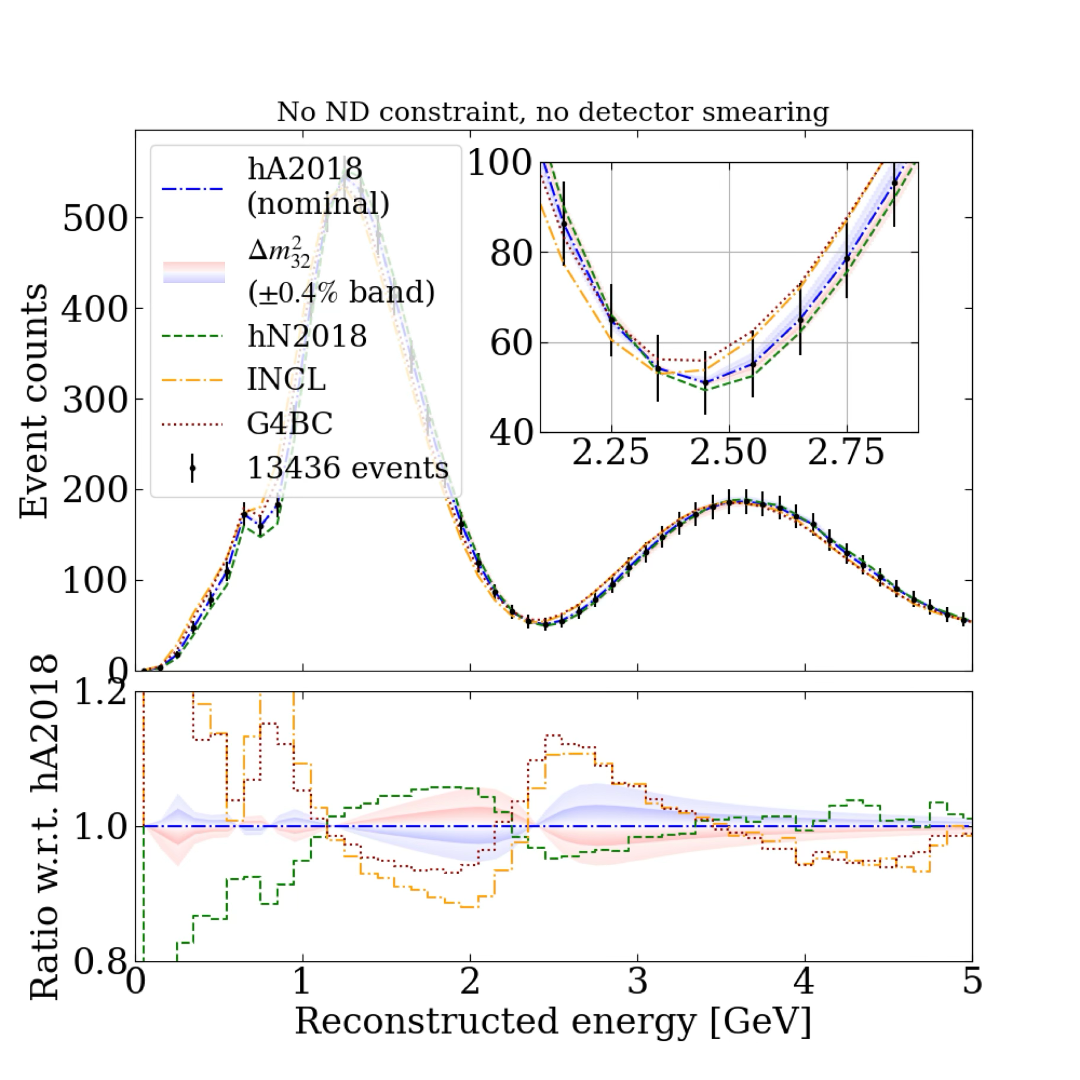}
\caption{Comparison of a simulated muon neutrino reconstructed energy spectra at the DUNE FD using different FSI models, to variations of \dm{32}. The reconstructed energy spectra under different FSI models are shown as histograms and variations of \dm{32} $\pm 0.4$\% are shown with shaded red (+) and blue (-) bands centred on the base model (\textit{hA2018}). The uncertainty bars are indicative of the DUNE experiment's statistical uncertainty after 10 years of operation. The inset provides a zoomed-in view of the first oscillation maximum. The lower panel displays ratios of FSI models and \dm{32} variations relative to base model. In the lower panel, the bands show 1$\sigma$ (inner) and 2$\sigma$ (outer) variations to \dm{32}. Note that this result does not consider detector smearing or potential ND constraints. Reproduced from Ref.~\cite{Liu:2025hpl}.} 
\label{fig:FSIonDUNEOAdm32}
\end{figure}

\section{Alternative approaches for systematic uncertainty reduction}
\label{sec:novelapproach}

The neutrino flux prediction is an important source of uncertainty in accelerator neutrino experiments which affects both the measurements of neutrino cross-sections and the extraction of neutrino oscillation parameters.
Although this review focuses on neutrino cross-section modelling and uncertainties, it is worth discussing methods by which accelerator neutrino flux uncertainties are constrained.
Improvements to flux modelling would directly improve the measurements of neutrino cross-sections, which benefits model developers. 
It can also help reduce the degeneracies between uncertainties in the neutrino flux and cross-section model in analyses of ND data in oscillation experiments.

\subsection{Hadron production uncertainties}
As introduced in \autoref{sec:intro}, accelerator muon neutrino and muon antineutrino beams are produced by impinging primary proton beams on a target and then using electromagnetic \textit{horns}~\cite{vanderMeer:1961sk} to focus secondary pions of a desired polarity and energy range which subsequently decay to produce (anti)neutrinos. The polarity of the horns can be changed to focus either positive pions, which predominantly decay to produce muon neutrinos, or negative pions, which predominantly decay to produce muon antineutrinos. Charged and neutral kaons, as well as sigma and lambda baryons, are also produced and decay to produce neutrinos.

Detailed simulations of the production and transport of particles through accelerator neutrino beamlines produce the neutrino flux prediction and associated uncertainties used for analysis~\cite{Fields:2025fhk}. A key source of uncertainty comes from the hadron production cross section for protons interacting in the target material, which is complicated by the finite extent of the targets, in which a secondary particle may interact before leaving. Subdominant effects come from interactions of secondary particles in other materials in the beamline, such as the electromagnetic horns, and unfocussed particles.

Data from a suite of hadron production experiments is used to constrain the neutrino flux predictions used by accelerator neutrino oscillation experiments~\cite{Kopp:2006ky, T2K:2012bge, MINERvA:2016iqn}. Additionally, dedicated hadron production experiments, or dedicated runs of existing experimental setups with tailored configurations, make measurements with the intent to reduce the flux uncertainties for specific neutrino experiments. These collide a proton beam of the same or similar energy either with the relevant target material, or even with a replica target of the neutrino experiment. 
Examples include HARP~\cite{HARP:2009pkw}, MIPP~\cite{Raja:2006pv}, NA61/SHINE~\cite{NA61:2014lfx} and EMPHATIC~\cite{EMPHATIC:2019xmc}. 
NA61/SHINE's measurements of proton scattering on a thin target with the same material and beam energy as T2K reduced the flux uncertainty to 10--15\%~\cite{NA61SHINE:2011dsu, NA61SHINE:2011tlp, T2K:2012bge, NA61SHINE:2013utd, NA61SHINE:2015bad}. 
When using a replica of the T2K target, this decreases to 5--10\%~\cite{NA61SHINE:2012vyd, NA61SHINE:2018rhe, NA61SHINE:2020iqu, T2K:2023smv}. NA61/SHINE have performed further analyses, targetting the reduction of flux uncertainties for Fermilab experiments~\cite{NA61SHINE:2018hif, NA61SHINE:2019aip, NA61SHINE:2019nzr, NA61SHINE:2022uxp, NA61SHINE:2023bqo} and a proposed lower-energy proton beam extension may offer additional improvements~\cite{Nagai:2783037}. These analyses demonstrate how hadron scattering data is used to reduce large \textit{a priori} flux uncertainties to tolerable levels for neutrino cross-section and oscillation experiments. It is hard to overstate the importance of these programs for current and future precision oscillation measurements.

\subsection{Direct flux constraints}

Although, in general, neutrino interaction cross sections are not well known---the subject of much of this review---a general approach to constraining the neutrino flux {\it in situ} involves isolating a subsample of neutrino interactions that have a \textit{relatively} well known cross section. 
When using this approach, special attention must be given to background modelling as processes or regions of phase space that contribute minimally to the total neutrino--nucleus event rate may be dominant backgrounds for these precisely selected subsamples. The uncertainties for these backgrounds may be very poorly understood, and may be misleadingly small, so a careful assessment needs to be made to confidently use any of these methods.

\subsubsection{Neutrino--electron elastic scattering and inverse muon decay (IMD)} are both lepton--lepton scattering processes, and are calculable with uncertainties at the 0.1\% level~\cite{Bardin:1986dk,Marciano:2003eq,Erler:2013xha,Tomalak:2019, Tomalak:2022uwv, Tomalak:2025tls}. The cross section for both processes is approximately three orders of magnitude smaller than the neutrino--nucleus cross section, but with modern high-intensity neutrino beams they offer a valuable handle on the flux uncertainty if they can be isolated experimentally.

\paragraph{Neutrino--electron elastic scattering, $\nu^{\bracketbar}_{l} + e^{-} \rightarrow \nu^{\bracketbar}_{l} + e^{-}$,} has a relatively simple experimental signature of a single, very forward, electron with no additional particles produced at the vertex. In principle, the neutrino energy can be reconstructed using the electron energy, $E_{e}$, and angle with respect to the incoming neutrino direction, $\theta_{e\nu}$, as
\begin{equation}
  E_{\nu} \approx \frac{E_{e}}{1-\frac{E_{e}(1-\cos\theta_{e\nu})}{m_{e}}},
  \label{eq:nue_scattering}
\end{equation}
where $m_{e}$ is the electron mass. However, this relationship is very sensitive to $\theta_{e\nu}$, which is limited by both detector reconstruction and the intrinsic divergence of accelerator neutrino beams, which is a few mrad at detectors close to the neutrino production point~\cite{Marshall:2019vdy}.
There are two key background processes that can mimic this channel: NC coherent $\pi^{0}$, in which only one of the photons from the $\pi^{0} \rightarrow \gamma\gamma$ decay is reconstructed; and $\nu^{\bracketbar}_{e}$--nucleus CC scattering, in which the energy transfer to the nucleus is low enough that no hadronic activity is reconstructed in the detector. Although the contribution from the latter is small, low energy transfer processes are not well modelled by current neutrino interaction models, as described in \autoref{sec:nuclear}. 

Neutrino--electron elastic scattering events have been selected and utilised by the MINERvA experiment in LE neutrino-enhanced mode and both neutrino- and antineutrino-enhanced ME modes~\cite{MINERvA:2022vmb,MINERvA:2019hhc,MINERvA:2015nqi}. By cutting on $E_{e}\theta_{e\nu}^2$, which is constrained to be $\leq 2m_{e}$ for neutrino--electron elastic scattering events, the background is greatly reduced, and the high $E_{e}\theta_{e\nu}$ sideband can be used to control residual background modelling uncertainties. 
MINERvA find $\mathcal{O}$(1000) events for each ME dataset and $\mathcal{O}$(100) events in their LE dataset after cuts and background subtraction. A key challenge in using this approach is that all neutrino species contribute, albeit with different cross sections. The MINERvA analyses incorporate a method that incorporates the constraint provided by this approach with the predicted correlations between neutrino species in the flux model, which are motivated by detailed tuning of the simulation to hadron production data.
Using this method, MINERvA were able to reduce their LE, neutrino-enhanced $\nu_{\mu}$ flux uncertainty from 9\% to 6\%~\cite{MINERvA:2015nqi}, and their ME, neutrino-enhanced $\nu_{\mu}$ (antineutrino-enhanced $\bar{\nu}_{\mu}$) flux uncertainty from 8\% to 4\%~\cite{MINERvA:2019hhc} (8\% to 5\%~\cite{MINERvA:2022vmb}).

Next generation accelerator neutrino beams will see orders of magnitude increases in observed neutrino scattering event rates at their near detectors. 
Following the work carried out by MINERvA, Ref.~\cite{Marshall:2019vdy} explored the potential to use the $\mathcal{O}$(5000/year) neutrino--electron elastic interactions in DUNE's LArTPC ND~\cite{DUNE:2021tad} to constrain the DUNE neutrino flux. The rate is sufficiently high that DUNE will be able to extract a percent-level flux {\it normalisation} constraint. However, the ability to constrain the flux {\it shape} is severely limited due to the aforementioned divergence of the neutrino beam. 
As the fundamental limitation is the beam and not the detector, neutrino--electron elastic scattering samples from smaller but higher-resolution detectors will not be able to provide a stronger constraint. Despite this promising result, no detailed estimation of the systematic uncertainty on the background rate was performed. Such an effort will be required before a robust constraint can be made by DUNE.

\paragraph{Inverse muon decay (IMD), $\nu_{\mu} + e^{-} \rightarrow \mu^{-} + \nu_{e}$,} has a similar signature to neutrino--electron elastic scattering with a single $\mu^{-}$ in the final state, and no hadronic activity at the vertex. It can only help to constrain the $\nu_\mu$ flux, as a $\bar\nu_\mu$ IMD would require a target positron. However, IMD has a relatively high threshold of $E_{\nu} \gtrsim 10.6$ GeV and so can only provide information about the high energy tail of the muon neutrino flux distribution for current and planned accelerator neutrino oscillation experiments. The largest expected backgrounds are from $\nu^{\bracketbar}_{\mu}$--nucleus scattering with low energy transfer to the hadronic final state. As in the case of neutrino--electron elastic scattering, this is poorly modelled by current neutrino interaction generators.

MINERvA have selected IMD samples containing $\mathcal{O}$(100) candidate events in both neutrino- and antineutrino-enhanced modes of their ME beam~\cite{MINERvA:2021dhf}. These samples were used to constrain the high-energy tail of the neutrino flux predictions using a similar method to their neutrino--electron elastic scattering analyses. 
Additionally, MINERvA have combined neutrino--electron elastic scattering and IMD information~\cite{MINERvA:2022vmb} to constrain the neutrino- and antineutrino-enhanced fluxes in their ME beam, which provides a blueprint for how future experiments can also incorporate this information into their analyses.

Although DUNE's neutrino beam will be higher in intensity than the MINERvA ME beam, it is peaked at a lower neutrino energy. 
As a result, DUNE's LArTPC ND will only sample $\mathcal{O}$(1000) IMD events per year~\cite{Marshall:2019vdy}. This may still provide a useful handle on the high-energy flux tail, in combination with the neutrino--electron elastic scattering.

\subsubsection{The low-$\nu$ method} was first developed and used in the context of the CCFR experiment~\cite{Mishra:1990ax, Auchincloss:1990tu,Seligman:1997fe} and is closely related to the low-$y$ method~\cite{Belusevic:1987rn,Belusevic:1988ab}. It comes from the observation that in the expression of the CCINC scattering cross section written in terms of nucleon structure functions, there is a term that is constant for small values of the energy transfer, $\nu = E_{\nu} - E_{\mu}$. The CCINC cross section, differential in $\nu$, is found by integrating $d^2\sigma/d\nu dx$ over $x = Q^2/(2M \nu)$,
\begin{align}
  \frac{\mathrm{d}\sigma}{\mathrm{d}\nu} &= \frac{G^2_{\mathrm{F}} M}{\pi} \int_0^1 \Big(F_2 - \frac{\nu}{E_\nu}  \left[F_2 \varmp xF_3\right]
      + \frac{\nu}{2E_\nu^2}  \left[\frac{Mx(1-R_{\mathrm{L}})}{1+R_{\mathrm{L}}}F_2\right]  \nonumber \\
      &+ \frac{\nu^2}{2E_\nu^2}  \left[\frac{F_2}{1+R_{\mathrm{L}}} \varmp xF_3\right]  \Big)\,\mathrm{d}x,
\label{eq:low-nu}
\end{align}
\noindent where $M$ is the struck nucleon mass, $F_2$ and $xF_3$ are structure functions, $R_{\mathrm{L}}$ is the structure function ratio $F_2/(2xF_1)$, $G_{\mathrm{F}}$ is Fermi's constant, and the $(+)-$ is used for (anti)neutrinos~\cite{Mishra:1990ax}. For interactions with low relative $\nu$, the $\nu/E_{\nu}$ terms are suppressed, leaving only the constant term proportional to $F_2$. If a sample of low-$\nu$ interactions can be selected, the constant cross section may be able to be leveraged to extract a flux {\it shape} constraint.

The low-$\nu$ method has been used for, or discussed in the context of, a variety of experiments, including (in chronological order): CCFR ($30 \leq E_{\nu} \leq 360$ GeV)~\cite{Seligman:1997fe}; NuTeV ($30 \leq E_{\nu} \leq 360$ GeV)~\cite{NuTeV:2005wsg}; NOMAD ($3 \leq E_{\nu} \leq 100$ GeV)~\cite{Petti:2006tu}\footnote{Although the flux constraint described in Ref.~\cite{Petti:2006tu} does not appear to have been used for subsequent analysis.}; MINOS ($2 \leq E_{\nu} \leq 10$ GeV)~\cite{MINOS:2009ugl, Bhattacharya:2009zza} and MINERvA ($2 \leq E_{\nu} \leq 10$ GeV)~\cite{MINERvA:2016ing,MINERvA:2021mpk}. Over time, the method has been applied to experiments with lower average neutrino energies, moving away from a regime in which the event rate is strongly dominated by DIS interactions, and into a regime where the event rate is dominated by other interaction processes where low-energy nuclear effects are important, which may not scale with neutrino energy in a way that is compatible with the low-$\nu$ method. Moreover, at lower energies a sufficiently accurate reconstruction of energy transfer is very challenging.

Ref.~\cite{Wilkinson:2022dyx}\footnote{Written by the authors of this review.} 
explores the potential for using the low-$\nu$ method in DUNE. It ultimately concludes that given the energy range important for DUNE ($1 \leq E_{\nu} \leq 6$ GeV), and the precision to which modern accelerator neutrino fluxes are constrained through hadron production experiments, the cross-section modelling uncertainties in any low-$\nu$ sample are not as well controlled as the flux prediction. Indeed, a low-$\nu$ region at DUNE energies is shown to be a region where the cross section shape is particularly poorly known. Overall, the low-$\nu$ method is unlikely to provide a useful \textit{in situ} constraint on the flux in the precision era and the same conclusion certainly applies in the lower energy Hyper-K beam. However, for higher-energy applications, or applications in which less precision is required, the low-$\nu$ method could be useful~\cite{Wilkinson:2023vvu}.

\subsubsection{Hydrogen extraction}
Another proposed method for constraining the flux {\it in situ} is to select a sample of events which are from interactions on hydrogen. Such interactions are free from nuclear effects, meaning the neutrino energy reconstruction is largely unbiased, opening the possibility of constraining both the flux shape and normalisation. Neutrino--deuterium scattering measurements may also be valuable, but there are suggestions that non-trivial nuclear effects in deuterium could be a barrier (see \autoref{sec:nucleon}). A recent review of the utility of neutrino--hydrogen and neutrino--deuterium data can be found in Ref.~\cite{Alvarez-Ruso:2022ctb}. 

Although there is a compelling physics case for making neutrino--hydrogen measurements, high-density hydrogen (or deuterium) targets bring significant safety concerns. However, ideas are still being explored, both to use liquid targets in dedicated experiments~\cite{Alvarez-Ruso:2022ctb} or to use high-density gases in multi-purpose detectors~\cite{Hamacher-Baumann:2020ogq}. Alternatively, compound targets that include hydrogen are safer and typically easier to scale up, and interactions on hydrogen may be extracted through statistical subtraction or kinematic separation. Although easier from a detector design standpoint, compound target analyses are vulnerable to mismodelling of backgrounds, and require high-resolution, low-threshold, detectors.

Separating the hydrogen component through statistical subtraction involves having a detector system with both hydrogen-enriched and unenriched components, and then subtracting the rates, or carrying out a combined fit to extract the hydrogen component. An example is using both polystyrene-based plastic scintillator (C$_8$H$_8$), and graphite (pure carbon)~\cite{Duyang:2018xcc, DUNE:2021tad}. 
The challenge for these subtraction methods is understanding the difference in detector performance between the two components. For example, plastic scintillator is an active detector material, whereas graphite is passive, making it likely that reconstruction performance for interactions that begin on each will not be identical. Another challenge is gaining sufficient statistics to make a precision measurement on hydrogen. Although the target materials are different, this technique is similar in principle to the extraction of neutrino--water interactions in the T2K ND, which has plastic scintillator, and mixed plastic scintillator and passive water targets~\cite{T2KND280FGD:2012umz}. 

Kinematic separation techniques rely on the observation of the complete final state in an interaction, and correlations between multiple particles. Essentially, events that have no momentum imbalance in the plane transverse to the incoming neutrino direction are a signature of the initial target being at rest (hydrogen) rather than bound in a nucleus (other elements in the compound). 
Interaction channels for which this technique is discussed in the context of future precision experiments~\cite{lu:2015hea, Duyang:2018xcc, Munteanu:2019llq, Hamacher-Baumann:2020ogq, DUNE:2021tad, Baudis:2023tma} include antineutrino CCQE $\bar{\nu}_{\mu} + p \rightarrow \mu^{-} + n$ interactions, and CCRPP $\nu_{\mu} + p \rightarrow \mu^{-} + \pi^{+} + p$ and $\bar{\nu}_{\mu} + p \rightarrow \mu^{+} + \pi^{-} + p$ interactions. The CCQE case is only applicable to antineutrino interactions and introduces  significant challenges associated with the identifying and reliably reconstructing the neutron from the primary interaction vertex. 
The CCRPP case, whilst applicable to neutrino and antineutrino interaction channels, requires detectors to accurately track at least three outgoing particles. 
Typically this is only possible for a subset of interactions, which limits the available statistics. Moreover, the detector resolution smears the reconstructed energy and direction of all outgoing particles, distorting the reconstructed transverse momentum imbalance for interactions on hydrogen, making it harder to separate them from the non-hydrogen component. 
MINERvA published a measurement of the CCQE case~\cite{MINERvA:2023avz}, extracting the \numub--hydrogen cross section from a selection of interactions on a C$_8$H$_8$ target through measurements of kinematic imbalances between the outgoing muon and neutron. 
However, the measurement faces large uncertainties from carbon backgrounds, as only $\approx$30\% of selected interactions are on hydrogen. The recently installed upgrade to T2K's ND (which will go on to serve as Hyper-K's ND) may be able to substantially improve on this, with significantly higher projected hydrogen purity~\cite{T2K:2019bbb,  Munteanu:2019llq, Baudis:2023tma}.

If future measurements enable the isolation of a high-purity sample of neutrino--hydrogen interactions, extracting a flux constraint still requires knowledge of the neutrino--hydrogen cross section. As discussed in \autoref{sec:nucleon}, continuing advances in LQCD calculations make precision knowledge of the CCQE cross section plausible but the pathway to precision constraints on CCRPP is less clear, especially for resonances beyond the $\Delta(1232)$. Conversely, if the neutrino flux can be well constrained, such samples can instead be used to constrain the neutrino--hydrogen cross section.
A combination of the low-$\nu$ method and hydrogen extraction has been proposed~\cite{Petti:2019asx,Duyang:2018xcc} as a way to mitigate some of the aforementioned issues with the low-$\nu$ method, and extract a robust flux constraint from a hydrogen-enriched sample of interactions.

Overall, whilst there are questions over whether the neutrino--hydrogen cross section is well enough understood theoretically to unambiguously {\it measure} the flux directly, it is clear that clean measurements of neutrino scattering on hydrogen breaks many of the degeneracies that make constraining cross-section models so challenging. 

\subsubsection{Charged-current coherent pion production} (CCCPP), introduced in \autoref{sec:evgen}, represents a small subset of single pion production interactions in which a neutrino transfers a small amount of momentum to a target nucleus, producing a charged lepton and pion whilst leaving the nucleus in its initial state. Since only a tiny fraction of the incoming neutrino energy is taken by nuclear recoil, the neutrino energy can be reliably inferred from the final state interaction products.
Moreover, the Adler relation~\cite{Adler:1964yx,ADLER1968189} relates the CCCPP cross section to the differential cross-section for elastic pion--nucleus scattering, for interactions where an incoming pion interacts with the nucleus as a whole. Therefore, if tight constraints on the differential pion--nucleus scattering cross section can be achieved, then a measurement of an event rate of CCCPP interactions can be leveraged to provide a constraint on the incoming neutrino flux normalisation and shape~\cite{Jung:2025bqh}. However, challenges in using a CCCPP-driven flux-constraint remain. Recent measurements of neutrino-induced CCCPP from T2K~\cite{T2K:2023xlh} and MINERvA~\cite{MINERvA:2022esg} have sizeable backgrounds which must be either reduced or modelled sufficiently well so as not to undermine a flux constraint. Moreover, the identification of pion--nucleus elastic scattering is challenging and the process has never been measured on an Argon nucleus. Additionally, the validity of the Adler relation is not guaranteed. Ref.~\cite{Jung:2025bqh} simulates a flux constraint for DUNE using a CCCPP measurement at its ND, accounting for a predicted background, and incorporating an \textit{in situ} measurement of elastic pion--nucleus scattering from measurements of secondary interactions of neutrino-produced pions. Neglecting uncertainties related to detector performance, the analysis projects a flux constraint at the $\approx$1.5\% level. 

\subsection{Novel neutrino beams}
\label{sec:altexpdes}
Rather than using dedicated measurements to constrain the flux \textit{in situ}, an alternative approach is to construct dedicated facilities which use novel beam designs that produce better known neutrino fluxes to make precision cross-section measurements.

ENUBET~\cite{Longhin:2014yta, Torti:2020yzn, ENUBET:2023hgu} is a proposal for a monitored neutrino beam, in which the decay pipe is instrumented in order to tag positrons from $K^+ \rightarrow \pi^{0}e^{+}\nu_{e}$ decays, and muons from two-body kaon and pion decays. By measuring the charged leptons produced in association with neutrinos, ENUBET would be able to make a precise determination of the flux normalisation to $\mathcal{O}(1\%)$ for $\nu_{e}$ and $\mathcal{O}(0.1\%)$ for $\nu_{\mu}$. Measurements in the resulting beam would not be subject to the usual large flux uncertainties. 

The proposed \textit{neutrino tagging} technique~\cite{Baratto-Roldan:2024bxk, Perrin-Terrin:2021jtl} takes the monitored beam concept further, using a specialised decay pipe instrumented with fast ($\sim$GHz) detectors to measure the momentum of decaying mesons and their subsequent decay products. Neutrinos measured in a downstream detector can then be individually associated with a specific meson decay, constraining their energy and direction precisely. To avoid excessive pile-up of mesons, this requires a significantly lower power proton beam than is used in conventional neutrino experiments ($<1~\textrm{kW}$), which is slowly extracted from an accelerator over many seconds. 

The recently formed nuSCOPE collaboration has proposed a neutrino beam that combines monitoring and tagging using the SPS accelerator at CERN~\cite{Acerbi:2025wzo}. 
Using a 500~ton detector, around one million CC muon neutrino interactions can be observed in 5--7 years of operation, where the neutrino energy for each neutrino may be known at the sub-percent level. These data may allow direct measurements of the neutrino energy reconstruction bias and the energy dependence of cross sections~\cite{Acerbi:2025wzo,Pinheiro:2026pul}.

NuSTORM~\cite{nuSTORM:2012jbd, nuSTORM:2022div} is another proposal for a dedicated facility to make neutrino cross-section measurements. NuSTORM uses neutrinos produced during the decay of muons in a storage ring, which have a well understood neutrino energy spectrum. Muon storage rings are not yet technically feasible, but have been discussed recently in the context of renewed interest in a muon collider (see Ref.~\cite{Accettura:2023ked} for a recent review).

Another alternative beam design was proposed in Ref.~\cite{Angelico:2019gyi} in which the difference in arrival times of neutrinos at a detector is used to subdivide them by neutrino energy\footnote{Alternatively, the same principle applies to decay muons, which has been explored in Ref.~\cite{Frisch:2022vwc}.}. This relies on the relationship between the energy of the focussed secondary pions and of the neutrinos produced by their decay. Although all daughter neutrinos will travel at close to the speed of light, the differences in the secondary hadron velocity and its average decay position introduce neutrino energy dependent differences in the average time between the proton bunch striking the primary target and a neutrino arriving at the detector. This technique relies on having a neutrino detector with $\mathcal{O}$(100 ps) timing resolution, and a similarly narrow bunch structure of the primary proton beam. The method has been further explored in the context of DUNE and various Fermilab beamlines in Refs.~\cite{Frisch:2022vwc, Ganguly:2024lqh}, but has not been implemented anywhere.

\subsection{Mobile near detectors}
\label{sec:prism}
Another approach, known as the Precision Reaction-Independent Spectrum Measurement (\textit{PRISM}), first described in Ref.~\cite{nuPRISM:2014mzw}, is being pursued as part of both the DUNE ND~\cite{DUNE:2021tad} and Hyper-K IWCD~\cite{Hyper-Kamiokande:2025asb} detector designs. 
PRISM exploits the change in the neutrino energy distribution at different angles away from the meson beam's central axis for neutrinos that are produced in highly-boosted meson decays. 
The neutrino energy distribution becomes narrower, and peaks at a lower energy, as the detector moves further away from the beam axis (or, further \textit{off axis}), as can be seen in~\autoref{fig:PRISMXSec}, using a DUNE example. 
By using a moveable ND to make measurements at a variety of off-axis angles, and therefore over a variety of different neutrino energy distributions with differing contributions of dominant interaction channels, PRISM helps break the flux--cross section degeneracy and may provide a means to infer the energy dependence of the cross section~\cite{DUNE:2025lvs} or provide novel flux constraints~\cite{Gehrlein:2025lbj}.
Moreover, models that successfully predict the observed data across a range of different neutrino energy distributions are much less likely to mismodel the relationship between true and reconstructed neutrino energy. Off-axis ND measurements can therefore identify critical mismodelling that could be misattributed by an on-axis only ND analysis, and thus highlight when the ND to FD extrapolation is not robust.

\begin{figure}[tb]
\centering
\includegraphics[width=0.48\textwidth]{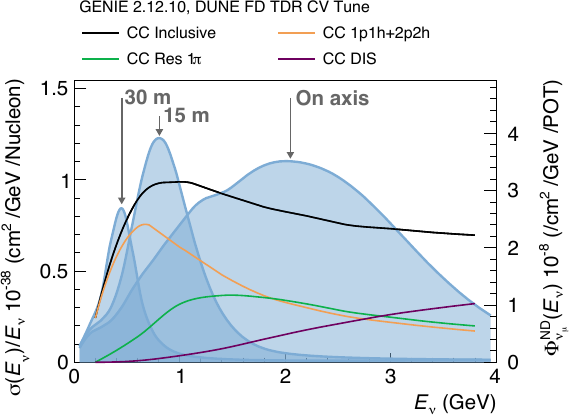}
\caption{The predicted muon neutrino flux  at the DUNE ND is shown for three different off-axis positions. As the DUNE ND is situated 574 m from the proton beam target, the equivalent off-axis angles are approximately $1.5^\circ$ and $3^\circ$ for 15~m and 30~m off axis, respectively. Overlaid is the GENIE-predicted CC cross section. Reproduced from Ref.~\cite{DUNE:2021tad}.} 
\label{fig:PRISMXSec}
\end{figure}

An alternate analysis approach enabled by PRISM is to construct linear combinations of samples from different off-axis positions to synthesise ND measurements in oscillated fluxes. 
The key departure from traditional methods is that the effect of the oscillation hypothesis is included in the linear combination, rather than by weighting simulated FD observations as a function of true energy. 
This removes the implicit (or explicit) need to directly connect the true neutrino energy to the observed event kinematics for signal modelling that is central to traditional approaches. As a result, this approach enables a much more direct ND to FD comparison, mitigating the first-order dependence on the signal channel interaction model.

Although promising, there are challenges for the linear combination approach that introduce explicit model-dependent corrections: the backgrounds discussed in \autoref{subsec:samplemigration} are not directly constrained; it is challenging to synthesise the shape of an oscillated flux beyond the peak energy of the on-axis flux\footnote{This is more problematic for DUNE where the FD is on axis. For Hyper-K, where the FD is $2.5^\circ$ off axis, the mobile ND can sample higher peak energies than the FD.}; 
and differences in ND and FD design and acceptance significantly complicate the analysis. Due to both the low intrinsic $\nu_{e}^{\bracketbar}$ content and the minimal variation of the $\nu_{e}^{\bracketbar}$ energy spectra as a function of off axis angle, it is likely to be infeasible to model the FD $\nu_{e}^{\bracketbar}$-appearance spectra using linear combinations of the ND $\nu_{e}^{\bracketbar}$ flux. Instead, $\nu_{e}^{\bracketbar}$ predictions at the FD are likely to use linear combinations of ND $\nu_{\mu}^{\bracketbar}$ measurements, and be subject to $\nu^{\bracketbar}_{e}/\nu^{\bracketbar}_{\mu}$ modelling uncertainties. At further off-axis positions the relative intrinsic $\nu^{\bracketbar}_{e}$ content in the beam increases due to the three-body kinematics of $\nu^{\bracketbar}_{e}$-producing kaon decays. This feature may enable improved \textit{in situ} constraints of the $\nu^{\bracketbar}_{e}$ and $\nu^{\bracketbar}_{e}/\nu^{\bracketbar}_{\mu}$ cross sections~\cite{nuPRISM:2014mzw,DUNE:2021tad}. However, the low statistical power of such measurements is likely to render them a cross-check of \textit{a priori} theory uncertainties rather than sources of dominant constraints on the cross section ratio.

The linear combination approach also relies on more details of the neutrino flux modelling than fixed ND measurements. It is highly sensitive to how the flux varies as a function of off-axis position, which depends on the beamline geometry, meson-beam focussing system, and hadron production modelling. Precise beam monitoring and accurate flux modelling will be crucial for linear combination analyses.

\section{Challenges for next-generation experiments}
\label{sec:prospects}

It is clear from \autoref{sec:issuesforoa} and \autoref{sec:impactonoa} that the potential mismodelling of neutrino interaction cross sections poses a critical challenge for next-generation neutrino oscillation experiments. \autoref{sec:constraints} demonstrates that the path to sufficient model improvement using current neutrino cross section measurements is unclear and, whilst \autoref{sec:novelapproach} shows there are a variety of approaches to reduce the impact of systematic uncertainties, there is no way to entirely remove cross-section model dependence from DUNE and Hyper-K oscillation measurements.
Given the context provided by the rest of this review, here we provide a non-exhaustive summary of the key neutrino interaction modelling challenges faced by Hyper-K and DUNE, as well as the prospects for overcoming them.

\subsection{Hyper-K}
Hyper-K benefits greatly from 15 years of active model development, motivated by the T2K experiment, which has focused theoretical efforts on understanding cross-section modelling uncertainties in the same $E_{\nu}^{\mathrm{peak}} \approx 0.6$ GeV beam, with the same $E_{\nu}^{\mathrm{QE}}$ energy reconstruction approach. Hyper-K also significantly benefits from having a core analysis that is predominantly sensitive to the relatively well understood CC0$\pi$ cross section as a function of outgoing lepton kinematics. There have been significant developments in the modelling of the dominant contributing interaction channels (CCQE and 2p2h) including neutrino--nucleon cross-section calculations from LQCD, as well as detailed characterisations of nuclear effects using mean field and \textit{ab initio} approaches. Some of these calculations have been implemented in neutrino event generators, improving the baseline simulations used and facilitating comparisons between models to inspire or validate systematic uncertainty treatments. 

In addition to theory developments, the last decade has seen a wealth of cross-section measurements of muon neutrino and antineutrino interactions on hydrocarbon and water targets in the relevant energy range for Hyper-K, although many are limited by large statistical errors or constrained detector acceptances. However, as discussed in \autoref{sec:constraints}, using these measurements to improve or constrain inputs for oscillation analyses is challenging as currently available models are unable to describe many datasets in isolation, let alone together. Attempts to use neutrino cross-section measurements on nuclear targets to develop a consistent model for T2K have met challenges, including limitations in experimental data releases and apparent inconsistencies between model--prediction agreement across different measurements. Overcoming these challenges will be a significant but essential task to meet Hyper-K's precision measurement goals.

As well as benefitting from the cross-section model development work that has been inspired by T2K, Hyper-K has the benefit of directly building on the past decade of T2K oscillation analyses which has identified some of the key modelling challenges and taken the first steps to mitigate them. However, significant work is left in order to meet Hyper-K's precision aims. In particular, there are a number of outstanding issues in neutrino interaction modelling which need to be addressed for Hyper-K:
\begin{itemize}
  \item {\it Electron neutrino and muon neutrino differences}--- Current sensitivity studies from Hyper-K show that the uncertainty on the double ratio of electron to muon and neutrino to antineutrino cross sections, $(\nue/\numu)/(\nueb/\numub)$, will be a dominant systematic uncertainty on measurements of $\delta_{\mathrm{CP}}$ and CP violation. This is challenging to address directly with ND measurements or via measurements by other experiments because the \nue and \nueb component in unoscillated conventional accelerator neutrino beams is small and follows a very different energy spectrum than the oscillated \nue or \nueb signal of interest. However, the source of the uncertainties (specific nuclear effects or radiative corrections) may be targeted via tailored muon (anti)neutrino measurements or theory calculation campaigns. 
  \item {\it The low energy transfer CCQE cross section}--- Recent oscillation analyses from T2K have found that uncertainties designed to cover variations of the CCQE cross section at low energy transfer are significant. Many measurements of the CC0$\pi$ cross section show large discrepancies with model predictions in the kinematic region most sensitive to this ($\cos\theta_{\mu} \gtrsim 0.95$), but the significant overlapping contributions from 2p2h and pion absorption FSI confuse interpretation. More sophisticated theory calculations for both the low energy transfer CCQE and the non-QE contributions, with realistic associated uncertainties, may help lift degeneracies between them. A new targeted measurement campaign of this kinematic region reporting the outgoing hadronic energy in addition to lepton kinematics may also be needed. 
  \item {\it Non-CCQE contributions to the oscillation signal}--- Although CCQE interactions dominate the Hyper-K oscillation signal, contributions from 2p2h and RPP interactions constitute significant contributions ($\sim$15\% each) to the primary CC0$\pi$ signal and drive much of the bias in kinematic neutrino energy reconstruction. RPP contributes the majority of the interactions in Hyper-K's secondary oscillation samples (which make up $\sim$10\% of the $\nue$ appearance signal), and a large fraction of the wrong-flavour and NC contributions to all samples. Whilst the last few years have seen a variety of new models, 2p2h and RPP interaction cross sections are significantly more complicated than the CCQE case and, as they are less important for T2K's statistics-limited measurements, they have also not received the same degree of community attention. RPP models in particular are challenging to adequately constrain with ND data. Moreover, the contribution from RPP to CC0$\pi$ comes via pion-absorption FSI which dramatically increases the difficulty of robustly predicting its size and shape. New pion--nucleus scattering measurements may help improve FSI modelling, provided sufficient theoretical work has been undertaken to relate {\it intranuclear} and {\it extranuclear} pion scattering.
  \item {\it Carbon--oxygen CC0$\pi$ cross-section differences}--- To fully exploit the major hydrocarbon target components of the Hyper-K ND, the experiment needs models that can robustly extrapolate carbon CC0$\pi$ cross-section measurements to oxygen CC0$\pi$ cross-section predictions. None of the current models available in neutrino event generators convincingly meet this requirement, as they fail to describe measured cross-section differences on different nuclear targets and predict significantly different carbon--oxygen CC0$\pi$ cross-section ratios. Overcoming this challenge requires new theory calculations which can both better relate carbon and oxygen cross sections and provide reliable uncertainties on their differences.
\end{itemize}

The Hyper-K ND, consisting of T2K's upgraded ND280 detector and a new PRISM-capable IWCD, is designed to address many of these challenges. IWCD PRISM measurements aim to directly constrain the electron (anti)neutrino cross section at relevant energies and provide high-statistics muon (anti)neutrino measurements on water to reduce the importance of modelling carbon--oxygen differences. However, this relies on tight control over flux and background uncertainties, which has not yet been publicly demonstrated. When coupled with sufficient theory inputs, the wide acceptance and calorimetric capabilities of the upgraded ND280 may be able to partially unpick CCQE modelling issues from those of 2p2h or RPP. Additional upgrades of ND280 are under consideration which may be able to provide additional sensitivity to outstanding modelling issues.

Despite the capable ND complex, the route to constraining neutrino interaction models sufficiently for Hyper-K's precision goals is not straightforward. Hyper-K will rely on continued developments from the theory community, particularly concerning the low-energy transfer region and non-CCQE contributions to the oscillation signal, as well as potentially from some external measurements such as pion--nucleus scattering data. Translating these inputs into improvements within neutrino event generators used for oscillation measurements, and validating them against ND data, is necessarily iterative. Improving on one facet of interaction modelling will reveal the importance of others, which can themselves be refined with theory developments and new measurements. Hyper-K benefits from numerous prior iterations of this cycle in support of T2K. However, sustaining sufficient support for both the theoretical and experimental strands of this effort through further iterations remains essential for Hyper-K to realise its full physics reach.

\subsection{DUNE}

DUNE's broad-band flux and calorimetric energy reconstruction approach give it unique sensitivity to the shape of the oscillation pattern across a wide range of neutrino energies, offering powerful handles on many oscillation parameters simultaneously. However, realising this potential places correspondingly broad demands on neutrino interaction modelling. Having roughly equal contributions of interactions from CCQE+2p2h, RPP and the SIS/DIS transition region, DUNE requires accurate modelling of a range of interaction channels. Moreover, the accuracy of DUNE’s calorimetric neutrino energy reconstruction is sensitive to the outgoing lepton \textit{and} hadron kinematics, making oscillation analyses reliant on some of the most challenging facets of interaction modelling, including hadronisation and FSI. DUNE still benefits from much of the theoretical development work carried out in support of few-GeV accelerator neutrino experiments generally, although the targeted theoretical program needed to support this broader kinematic reach and calorimetric approach is still developing. It is also notable that much of the theoretical work and experimental measurements supporting the few-GeV program has focused on lighter targets than argon. On the other hand, DUNE's higher energy leaves it significantly less sensitive to uncertainties on electron and muon (anti)neutrino cross-section differences.

An experimental program producing neutrino--argon cross-section measurements is well underway with the SBN program in the BNB beam. MicroBooNE has already produced a number of neutrino--argon scattering measurements, and higher-statistics results are anticipated from SBND. These measurements of CCQE+2p2h and $\Delta(1232)$-dominated RPP cross sections for neutrino--argon scattering may be valuable for benchmarking models of these processes. However, they only cover the lower-energy portion ($\lesssim$2.5 GeV) of DUNE's region of interest (0.5--10 GeV). Additionally, the limited prospects for the BNB to run in antineutrino-enhanced mode will limit the model-constraining power of SBND relevant for DUNE, given that DUNE explicitly gains sensitivity to oscillation parameters by running in both neutrino- and antineutrino-enhanced modes.

Whilst there is some prospect of neutrino--argon measurements at higher energies through the off-axis NuMI exposure of ICARUS, these measurements will have limited statistics, are unable to cleanly separate neutrinos from antineutrinos, and will not cover the full kinematic phase space relevant for DUNE. The 2x2 DUNE ND prototype is on axis in the NuMI medium energy beam, providing very relevant neutrino--argon scattering data where no other neutrino--argon measurements exist, but its small size, limited acceptance, and lack of a magnetic field severely restricts the measurements it can make. MINERvA's extensive measurements, primarily on hydrocarbon but including heavier targets (although not argon) provide a complementary input at higher energies, but, as described in \autoref{sec:constraints}, the path to resolving the large discrepancies between models and MINERvA measurements is not clear. Measurements of electron--argon interactions at DUNE energies are part of the e4nu program and may provide useful information on the relationship between cross sections on different targets, FSI, and high-W processes, although the translation of precision cross section predictions from electron scattering measurements to neutrino scattering remains a challenge.

To fully exploit its physics reach, DUNE will need focused progress in several areas:
\begin{itemize}
\item {\it Modelling of RPP processes}--- More sophisticated calculations of neutrino--nucleon and neutrino--nucleus RPP processes are in development, but there are large uncertainties on many aspects of pion production which are challenging theoretically, and which are difficult to disentangle experimentally. Unknowns in the neutrino--nucleon cross section and the variety of nuclear effects and contributions of other channels through FSI to single pion final states make developing predictive models for both leptonic and hadronic kinematics in RPP difficult. A mismatch between the measurable pion production topologies at the ND and impactful topologies at the FD add to the impact of this challenge. Outgoing hadron kinematics from RPP are particularly poorly understood, even at the neutrino--nucleon level. Recent work from the LQCD and nuclear theory communities targetting calculations of transition form factors and nuclear effects are promising and may help overcome \textit{some} of these modelling challenges provided the work is sufficiently supported. 

\item {\it SIS/DIS and hadronisation}--- Approximately $\sim$1/3 of DUNE events are expected to be from the SIS/DIS transition region, where there are large uncertainties on the cross section, even as a function of neutrino energy or lepton kinematics. The situation is even more challenging for hadron kinematics, where currently-used hadronisation models rely on PYTHIA in a regime where it is not applicable. This leaves the relevant SIS/DIS cross sections for DUNE and their impact on the mapping between reconstructed and true neutrino energy poorly constrained. This kinematic region has not received the sustained experimental and theoretical focus that lower energy transfer processes have, especially for argon targets, and so currently there is no clear path to overcoming this challenge.

\item {\it FSI in argon}--- The energy lost to neutrons, the number of pions in the final state, and the distribution of energy between particles are all significantly modified by FSI, making this a critical challenge for DUNE's calorimetric approach to neutrino energy reconstruction. Currently, there is little hadron--argon scattering data available with which to constrain or inform models. However, ProtoDUNE is beginning to provide some relevant measurements, although these may not cover all of DUNE's needs, and generator simulations of FSI are undergoing substantial improvements with the introduction of more sophisticated models. Current theoretical efforts to understand the impact of FSI in CCQE interactions using microscopic models are promising and have allowed some aspects of simulation to be benchmarked by robust theory, but this has typically focused on lighter nuclei than argon and does not yet cover pion production interactions that are most relevant to DUNE.
\end{itemize}

The DUNE ND complex is designed to provide data to address many of these challenges directly once it begins operation. The ND will deploy state-of-the-art argon-based detectors to collect $\mathcal{O}$(10--100M) neutrino interactions per year, providing the potential for unrivalled \textit{in situ} constraints. However, fully benefitting from this requires a neutrino interaction model with associated uncertainties that can describe the data the DUNE ND collects. Since current models are far from this standard, a sustained and significant investment in focused theoretical work aimed at building new models relevant for DUNE is essential. As described for the Hyper-K case, this must be accompanied by an iterative model improvement cycle based on close collaboration between experimentalists and theorists. For DUNE, this process is still in its early stages. Beginning these iterations now, using inputs from relevant neutrino and non-neutrino scattering data, will be critical to ensuring that DUNE is well positioned to continue this cycle once ND data become available, enabling precision oscillation analyses as soon as possible after operations begin.

Given the challenges in neutrino interaction modelling faced by DUNE, analysis techniques which significantly reduce model dependence are likely to be very important. PRISM analyses offer a new capability which aim to lessen the reliance on cross-section modelling in the ND-to-FD extrapolation, though a full assessment of its associated systematic uncertainties has not yet been publicly demonstrated.

\section{Conclusions and outlook}
\label{sec:concl}

Few-GeV accelerator-neutrino oscillation physics is entering a new and ambitious phase. The next-generation experiments, DUNE and Hyper-K, are designed to characterise CP violation in the lepton sector, determine the neutrino mass ordering, and test the completeness of the three-flavour PMNS mixing paradigm with a precision that will take the field firmly out of the statistically-limited era. Whether that precision can be realised in practice depends critically on how well neutrino--nucleus interactions can be modelled and constrained. This review has attempted to trace this dependence systematically, covering: the underlying interaction physics and its simulation in event generators; the ways in which modelling uncertainties propagate into oscillation analyses; the current state of experimental constraints; and the prospects for improvements. There are several clear conclusions.

First, neutrino interaction systematic uncertainties are not a peripheral concern for next-generation experiments but rather may be the limit to their physics reach. Current systematic uncertainties on FD event rates are already at the 4--9\% level for T2K and NOvA. A dramatic reduction is required for DUNE and Hyper-K, which will collect event samples which are an order of magnitude larger. The near detectors of both experiments are essential to providing the \textit{in situ} constraints that may make this reduction possible, but ND measurements alone cannot solve the problem: the extrapolation from near to far detector depends on interaction model predictions that must be robust. It will not be possible to fully exploit higher-statistics ND data without a reliable baseline neutrino interaction model, accompanied by a well-motivated set of systematic uncertainties. 

Second, no currently available model meets the required standards. Dedicated reviews of model-measurement comparisons, and the comparisons presented in this work, consistently find that no generator configuration provides a good description of global neutrino scattering measurements. 
Disagreements are pervasive across interaction topologies, nuclear targets, and neutrino energies. The disagreements are often larger for variables built from hadronic kinematics, on which calorimetric neutrino energy reconstruction at DUNE directly depends. The spread between generator predictions for quantities that oscillation analyses rely on, including cross-section energy dependence, flavour ratios, nuclear target scaling, and visible hadronic energy, consistently suggests that neutrino interaction modelling is not controlled at the level of precision that next-generation experiments require.

Third, the path to reducing neutrino interaction uncertainties for DUNE and Hyper-K is different, as described in \autoref{sec:prospects}. Hyper-K inherits targeted theoretical and experimental development motivated by T2K, operating at the same range of neutrino energies using the same target materials. Some of the most important systematic uncertainties have been identified, and recent work has provided a credible path toward controlling many of them at the level needed. However, some significant challenges remain.
Conversely, DUNE has a broader neutrino energy spectrum in which roughly equal fractions of its event rate will come from QE, RPP, and SIS/DIS interactions, each carrying distinct modelling uncertainties. The focused theoretical and experimental development that has characterised improvements in CC0$\pi$ modelling has not yet been matched for the pion production channels. Moreover, the calorimetric energy reconstruction used in DUNE oscillation analyses amplifies sensitivity to hadronic modelling, particularly to pion multiplicities and to the fraction of energy carried by neutrons. Despite these challenges, there is an advantage in the fact that DUNE and Hyper-K are most affected by a distinct set of model uncertainties, and comparisons between their oscillation results will be a stringent test of their systematic uncertainty models.
As an attempt to confront some of these challenges, the past decade has seen a major investment in producing neutrino--nucleus cross-section measurements. However, although valuable, most measurements are sensitive to a wide variety of nuclear processes and so have a limited utility for constraining any specific piece of the complex neutrino interaction modelling picture. 
Additionally, there is currently no agreed upon approach for
minimising the potential for model dependence in neutrino interaction measurements, or for
testing for its impact on published results, although this is an active area of research. This makes the task of benchmarking models against the growing body of cross-section measurements challenging due to the circularity that arises with building model constraints from model-dependent measurements.
Continued investment in developing and applying rigorous statistical methods for cross-section extraction and model testing, building on recent progress in model-independent measurement techniques and open data releases, is essential to ensure that future measurements translate into genuine improvements in oscillation analyses.

Despite the general challenges with neutrino--nucleus data, the community has made progress with new approaches, such as lepton--hadron correlation measurements, or by pursuing non-neutrino measurements such as electron--nucleus scattering, in order to provide useful model constraints. In some cases, most notably at lower-energy transfers, experiment and theory have worked together iteratively, with new theory being developed to describe experimental observation, and with subsequently refined measurements providing more stringent probes. However, overall, the focus of the field has been measurement-heavy, leading to a situation where experimental analyses have long since outstripped the ability of theoretical models to describe the results. Ensuring the theory community has the support and resources needed to keep up with progress in experimental measurements is among the most important steps the field can take to safeguard the physics reach of next-generation experiments.
Both Hyper-K and DUNE have powerful near detectors, which will provide vital information for resolving some aspects of the modelling issues described above. Additionally, the alternative approaches described in ~\autoref{sec:novelapproach} may partially mitigate some of these challenges. However, none provide a means to completely remove the fundamental and significant dependence of oscillation measurements on neutrino--nucleus interaction modelling. As has been the case for T2K and NOvA, DUNE and Hyper-K will need to develop the cross-section models used in their analyses in order to mitigate the inevitable data--simulation discrepancies at the near detector, and in response to an evolving understanding of which uncertainties introduce the most potential for bias in their analyses. In order to avoid serious delays to producing credible and robust oscillation results, both DUNE and Hyper-K require a significant and coordinated community effort to improve the modelling of few-GeV neutrino cross-sections, before they become operational.

The experimental programme that will take us into the precision era of neutrino oscillation physics is underway. The extent to which it succeeds will depend largely on how the neutrino physics community faces the challenge of understanding neutrino--nucleus interactions.

\ack{
S.D. would like to thank L. Munteanu and R. Gran for helpful discussions related to this work. We would like to thank U. Mosel for providing details about the GiBUU event generator and for discussions on its predictions. We additionally thank O. Tomalak for feedback on our manuscript and discussions related to tau neutrino cross sections and radiative corrections. We further acknowledge useful discussions during the NuXTract workshop at CERN and workshops hosted by ECT*, supported by EURO-LABS, which received funding from the European Union’s Horizon Europe Research and Innovation programme under grant agreement No 101057511.  C.Wret and L.P. are supported by The Royal Society under awards URF\textbackslash R1\textbackslash 241892 and URF\textbackslash R1\textbackslash 211661 respectively, and the United Kingdom Science and Technologies Facilities Council. C.W. is supported by the U.S. Department of Energy (DOE), Office of Science under contract DE-AC02-05CH11231. P.S. is supported by a United Kingdom Science and Technologies Facilities Council Consolidated Grant award. This research used resources of the National Energy Research Scientific Computing Center (NERSC), a U.S. Department of Energy Office of Science User Facility located at Lawrence Berkeley National Laboratory, operated under Contract No. DE-AC02-05CH11231 using NERSC award HEP-ERCAP0035522.
}

\section*{Simulation availability}

The outputs of event generators used in this work have been made available via a publicly accessible directory on the NERSC computing centre: \url{https://portal.nersc.gov/project/nuisance/MC_IOP_review/}.

\section*{Affiliation disclaimer}

All authors are active members of the T2K collaboration.
L.P., C.W., and S.D. are active members of the DUNE collaboration and C.Wret was a member until 2022. C.Wret and P.S are active members of the Hyper-K collaboration. P.S. is an active member of the Super-K collaboration. S.D. is an active member of the ICARUS and nuSCOPE collaborations. C.Wret was a member of the MINERvA collaboration between 2018 and 2022.

All authors are core developers of the NUISANCE framework. L.P. and C.Wret are lead developers of the NEUT neutrino interaction simulation program library. C.W. has contributed significant model development efforts to NEUT. S.D. has contributed significant model development efforts to both GENIE and NEUT.
\vspace{5mm}

\bibliography{main.bib}
\bibliographystyle{unsrturl.bst}

\appendix

\section{Further details on neutrino event generators}
\label{appen:gen}

This appendix provides details of the neutrino interaction generator configurations used within this work, expanding the summary given in~\autoref{tab:models}. 

\subsection{NEUT}
The primary neutrino event generator used by the T2K and Super-Kamiokande collaborations, NEUT is described in Refs.~\cite{Hayato:2002sd,Hayato:2009,Hayato:2021heg}. For this work, we generate events with NEUT 5.8.0 in two different configurations which differ in their choice of RPP model. In both cases, the CCQE model uses an LFG nuclear ground state and a cross section based on Ref.~\cite{Nieves:2011pp} with a custom approach to the treatment of nuclear removal energy~\cite{Bourguille:2020bvw}. A dipole axial form-factor is used with $\maqe=1.05$ GeV. NEUT uses the 2p2h model from the Valencia group~\cite{gran2013neutrino,Nieves:2011pp}, maintaining the theory calculation of the nucleon pair content resulting in $\approx$70\% neutron--proton initial state pairs (although the pair content is dependent on the interaction kinematics). Interaction channels other than CCQE and 2p2h use an RFG nuclear ground state model with no removal energy. For SIS and DIS interactions, NEUT employs modifications from GRV98 DIS cross sections using Bodek--Yang corrections~\cite{Bodek:2003wc} at low four-momentum transfer. For hadronisation, PYTHIA5~\cite{Sjostrand:1992mh} is used at high invariant mass (W$>$2 GeV), whilst a custom model is used between $1.3 \leq W \leq 2.0$~GeV, based on KNO scaling (see Sec.~V.C of Ref.~\cite{NuSTEC:2020nsl}). Double counting with RPP interactions for $W<2$~GeV is prevented by modelling only interactions with more than one pion (before FSI). FSI are included through a custom intranuclear cascade model~\cite{Hayato:2021heg} tuned to pion scattering measurements~\cite{PinzonGuerra:2018rju}. 

In predictions labelled \textit{NEUT}, RPP uses the Rein--Sehgal resonant model~\cite{Rein:1980wg}, with improvements to the nucleon axial form factors~\cite{Graczyk:2014dpa,Graczyk:2007bc} and the inclusion of final-state lepton mass effects~\cite{Berger:2007rq,Graczyk:2007xk,Kuzmin:2003ji}, henceforth referred to as the Berger--Sehgal model. For predictions labelled \textit{NEUT DCC} employs the recently-implemented dynamically coupled channels model~\cite{Nakamura:2015rta}. In both cases, NEUT includes resonances with masses up to 2~GeV as well as interferences between them. 

Importantly for predictions on non-isoscalar targets (such as argon), during the course of this work, an apparent issue was found with the NEUT CCQE model used, in which the cross section scales approximately linearly as a function of the number of neutrons in the target nucleus for both neutrino and antineutrino interactions. This means that for targets with more neutrons than protons, the cross section is too large for antineutrino interactions. This issue has been reported to the NEUT developers.

\subsection{NuWro}
This work uses two versions of the NuWro event generator~\cite{Golan:2012wx,Golan2012nuwro}. Predictions labelled \textit{NuWro 19} use the well-established version 19.02.2, used by many experiments in their data--simulation comparisons, whilst those labelled \textit{NuWro 25} use the newer version 25.03.1\cite{Prasad:2025efg}. The configurations used also differ in their choice of 2p2h and RPP model. 

Both versions of NuWro use an LFG-based ground state model for all interaction channels. CCQE interactions are generated according to the Llewellyn-Smith approach~\cite{LLEWELLYNSMITH1972261} with custom corrections~\cite{Graczyk:2003ru} combined with an effective nuclear momentum-dependent potential~\cite{Juszczak:2003zw}. A dipole axial form-factor is used with $\maqe=1.03$ GeV. For DIS interactions, cross sections are calculated using GRV98~\cite{Gluck:1998xa} and hadronisation is simulated using PYTHIA6~\cite{Sjostrand:2006za}. FSI are modelled using a custom cascade model~\cite{Golan:2012wx,Dytman:2021ohr,ershova2022study, Niewczas:2019fro}. 

For 2p2h interactions, \textit{NuWro 19} uses the Valencia model~\cite{gran2013neutrino,Nieves:2011pp}, with a fixed fraction of initial state nucleon--nucleon pairs (85\% proton--neutron). \textit{NuWro 25} uses an updated version of the Valencia 2p2h model~\cite{Sobczyk:2020dkn}, including a more realistic approach to generating nucleon kinematics and a 3p3h contribution~\cite{Prasad:2024gnv}.

For RPP interactions, \textit{NuWro 19} employs the Adler model~\cite{Adler:1975mt, ADLER1968189}, considering only the $\Delta$(1232) resonance. A linear transition between RPP and DIS (the SIS region) is used from $1.4\leq W \leq 1.6$~GeV. \textit{NuWro 25} uses the Hybrid RPP model~\cite{Yan:2024kkg}, which includes some higher mass resonances (up to S$_{11}$(1535)\footnote{Often referred to as the second resonance region.}) and interferences with the non-resonant background based on the approach of Ref.~\cite{Hernandez:2007qq}. It also considers a Regge-based description at higher invariant masses. A linear transition between the RPP and DIS models is used from 1.6 $\leq W \leq 1.9$~GeV.

\subsection{GiBUU}
In this work we use version 2025p1 of the GiBUU theory framework~\cite{Buss:2011mx}. GiBUU works with a stable groundstate, modelled by placing the nucleons into a momentum-dependent mean-field nuclear potential with the momentum distribution given by that of a local Fermi gas. This potential enters into the calculations of both the initial and the final-state cross sections, ensuring a consistent description of all interaction channels~\cite{Leitner:2008ue,Gallmeister:2016dnq}. Although GiBUU does not explicitly account for RPA corrections, its precise handling of bound nucleons implies that a much weaker RPA correction would be required compared to most other models~\cite{Lalakulich:2012hs}.
For 2p2h interactions, GiBUU employs a phenomenological model guided by electron scattering data~\cite{OConnell:1972edu,Gallmeister:2016dnq,Bodek:2022gli}. When extended to neutrino interactions, the model applies a normalisation scaling that varies linearly with the difference between the proton and neutron content of the target nucleon. This approach has been benchmarked against cross-section measurements, and leads to a notably larger 2p2h prediction than other models for an argon nuclear target~\cite{Dolan:2018sbb, Wilkinson:2022dyx}. GiBUU uses a sophisticated custom approach to simulating FSI, as described in \autoref{sec:nuclear}.

The GiBUU single pion production model includes baryon resonances up to 2~GeV$/c^2$ in mass. Details of the model and evaluations against data can be found in Refs.~\cite{Mosel:2023zek,Mosel:2017nzk,Yan:2025aau}.
The treatment of the $\Delta(1232)$ resonance and non-resonant background use theory models based on Refs.~\cite{Hernandez:2007qq,Lalakulich:2010ss,Lalakulich:2005cs,Leitner:2009zz} and higher resonances are described phenomenologically~\cite{Leitner:2008ue,Leitner:2009zz}.
The vector form factors for the resonances that are not the $\Delta(1232)$ follow the MAID 2007 analysis~\cite{Drechsel:2007if}.
Non-resonant contributions from $1\pi$ and $2\pi$ final states are obtained by converting the Bosted--Christy parametrization~\cite{Christy:2007ve} of electron-scattering data using approximate relations between the electron and neutrino structure functions.
GiBUU's SIS and DIS description starts at $W=2~\text{GeV}/c^2$, where between $2<W<3~\text{GeV}/c^2$
the cross sections come from the Bosted-Christy parametrization mentioned above, and the particle kinematics are generated by Pythia. Pythia is exclusively used for the pure DIS region, defined as $W>3~\text{GeV}/c^2$.

Unlike other event generators, GiBUU does not generate events with a probability proportional to their cross section. Instead, GiBUU events are produced with a weight which must be applied event-by-event in order to correctly model the shape of the cross section. The typical weight for an event is dependent on its kinematics. As a result, the statistical uncertainty in GiBUU simulations does not scale simply with the number of events generated, and achieving small statistical fluctuations for some regions of phase space requires very large event samples.
Indeed, our experience has been that as the weights generated can span many orders of magnitude, a high-weight tail can dominate the statistical uncertainty for some configurations, no matter how many events have been generated.
In this work, we generate 10$^8$ events per neutrino flavour and nuclear target when using experiment-based fluxes, and more than 10$^9$ events when producing cross sections as a function of neutrino energy. Despite this, statistical fluctuations in GiBUU predictions of the cross-section ratios (or double ratios) shown in \autoref{sec:issuesforoa} are non-negligible, as is remarked in the text.

GiBUU also requires the user to decide when to stop propagating particles through the FSI simulation, and which particles from the FSI should be placed in the final state of the neutrino interaction. 
As described in \autoref{sec:nuclear}, GiBUU propagates ensembles of hadrons in the nucleus together in steps of time. The propagation stops once a user-defined number of time steps is exceeded, but there remains some ambiguity in which nucleons escape the nucleus, especially for low-energy transfer interactions containing relatively low-momentum nucleons. 
In this work, we consider 120 time steps where each step size is 0.2~fm. We assert that all particles exceeding a radial distance of 6~fm from the centre of the nucleus have escaped with the exception of interactions on hydrogen (for which there is no such cut), independent of the nucleus\footnote{From correspondence with GiBUU author and developer U.~Mosel.}.
For heavy targets, such as iron or lead, the distance cut should be adapted to each nuclear target, but we expect such tuning to have only a small impact for the relatively light nuclei considered in this work as the nuclear density for $^{12}\text{C}$ and $^{40}\text{Ar}$ are similar above 6~fm (both approaching zero).
This leaves a portion of interactions, which are almost exclusively for CCQE and 2p2h, with no final-state hadrons, the amount of which depends on the species and the flux, shown in \autoref{fig:app_gibuu_blocking}. 
For experiments at lower energy, which have large contributions from events with low-energy transfer (such as T2K and MicroBooNE) the fraction of lepton-only neutrino interactions is relatively large ($\approx$9\%), and is even larger for antineutrinos ($\approx$18\%). This is approximately 25\% of the CCQE events, and 10\% of the 2p2h events. 
For higher-energy experiments, like the NuMI LE and ME configurations, the fraction is a mere $\approx$3\%. The fraction of events in which no hadrons escape the nucleus peaks around $q_0\sim0~\text{GeV}$ and is virtually zero for $q_0 \gtrsim 0.2~\text{GeV}$.
Specifically in \autoref{sec:constraints}, these events are excluded from the analysis as the events cause an unphysical peak at exactly zero in the available energy in ~\autoref{fig:minerva-ccinc}, but are included elsewhere. Their exclusion makes no appreciable difference to any other figures in \autoref{sec:constraints}.
\begin{figure}[hbtp]
    \centering
    \includegraphics[width=0.49\linewidth,page=1,clip,trim=0mm 0mm 0mm 15mm]{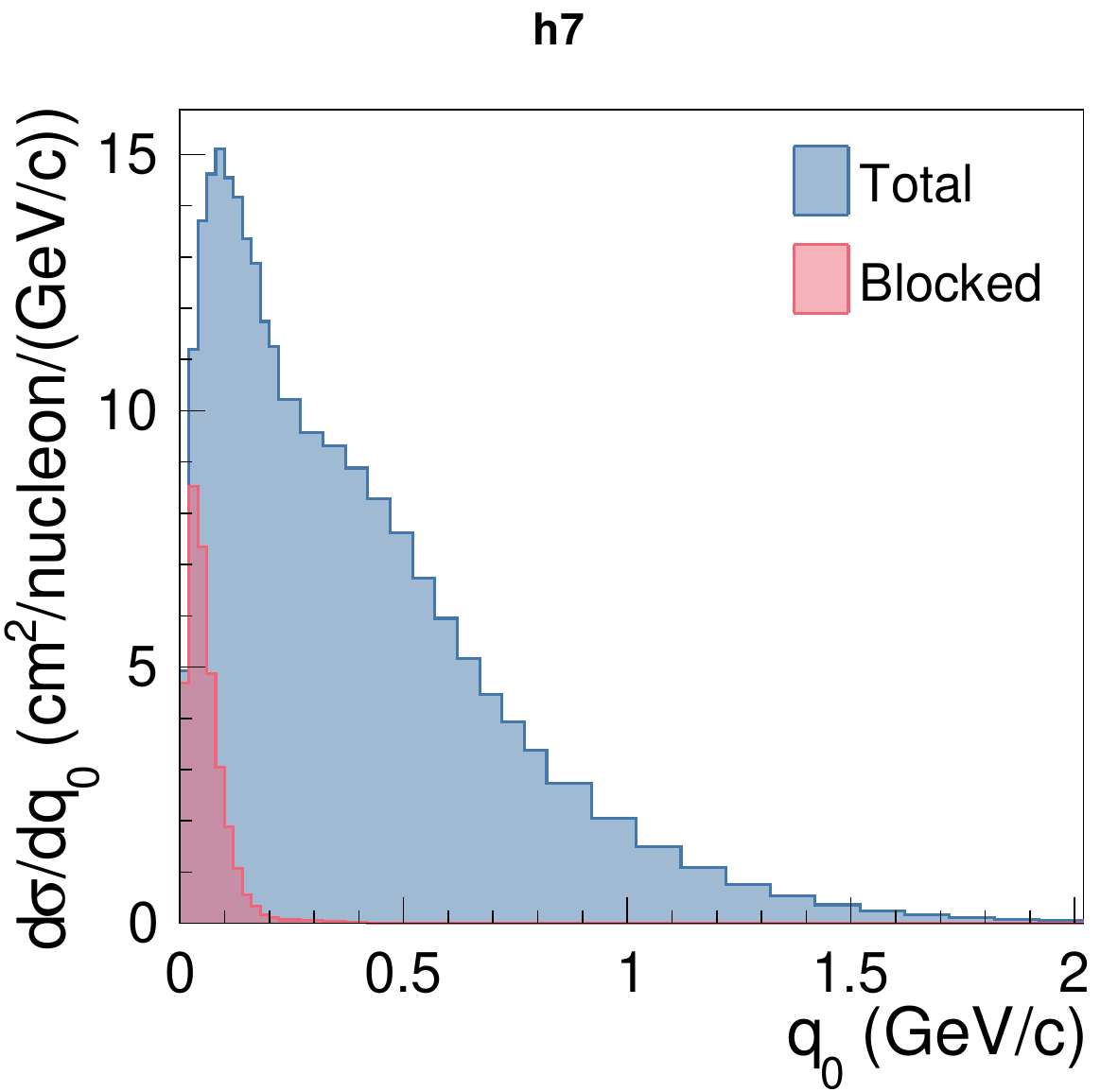}
    \includegraphics[width=0.49\linewidth,page=1,clip,trim=0mm 0mm 0mm 15mm]{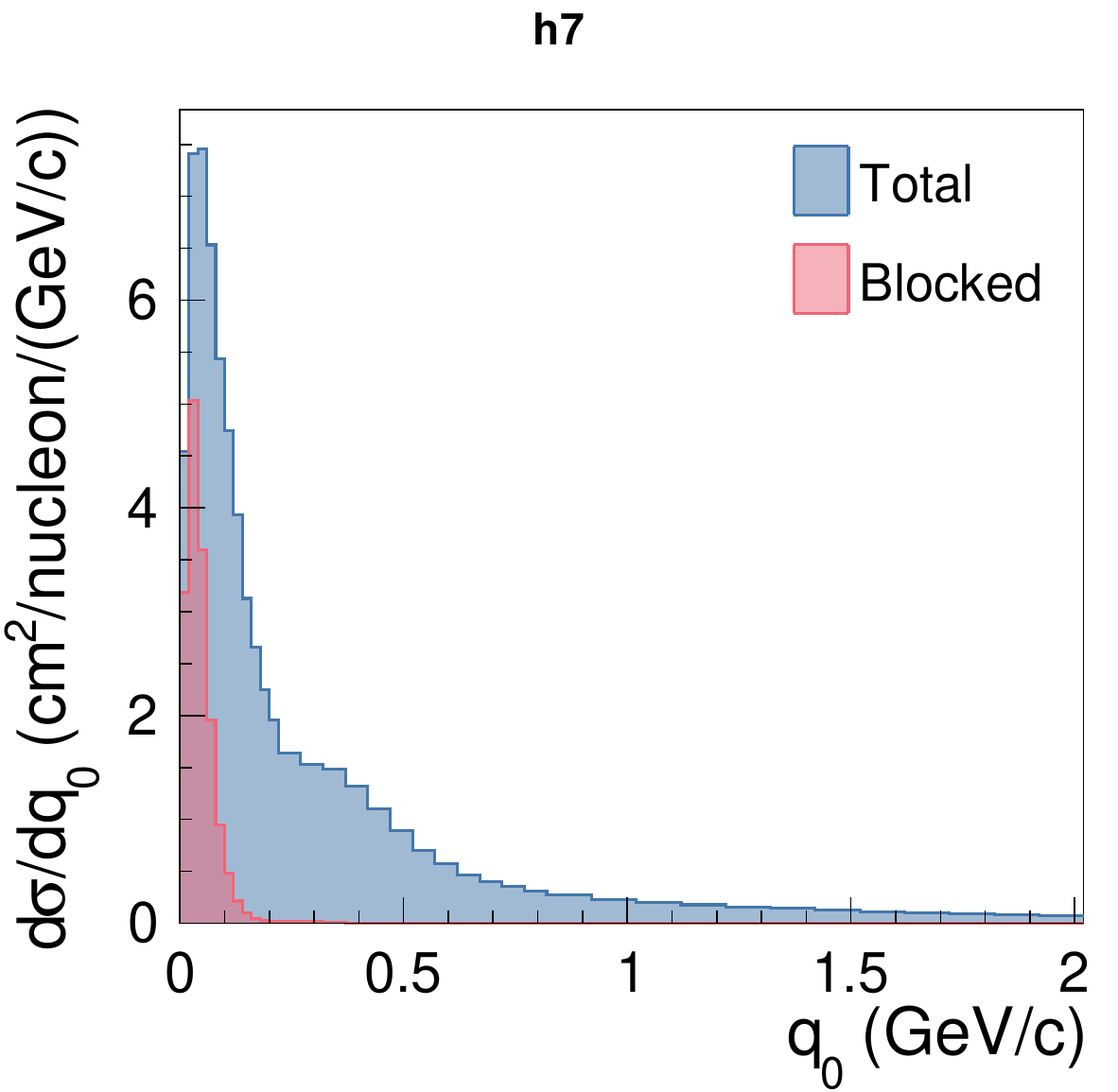}

    \includegraphics[width=0.49\linewidth,page=1,clip,trim=0mm 0mm 0mm 15mm]{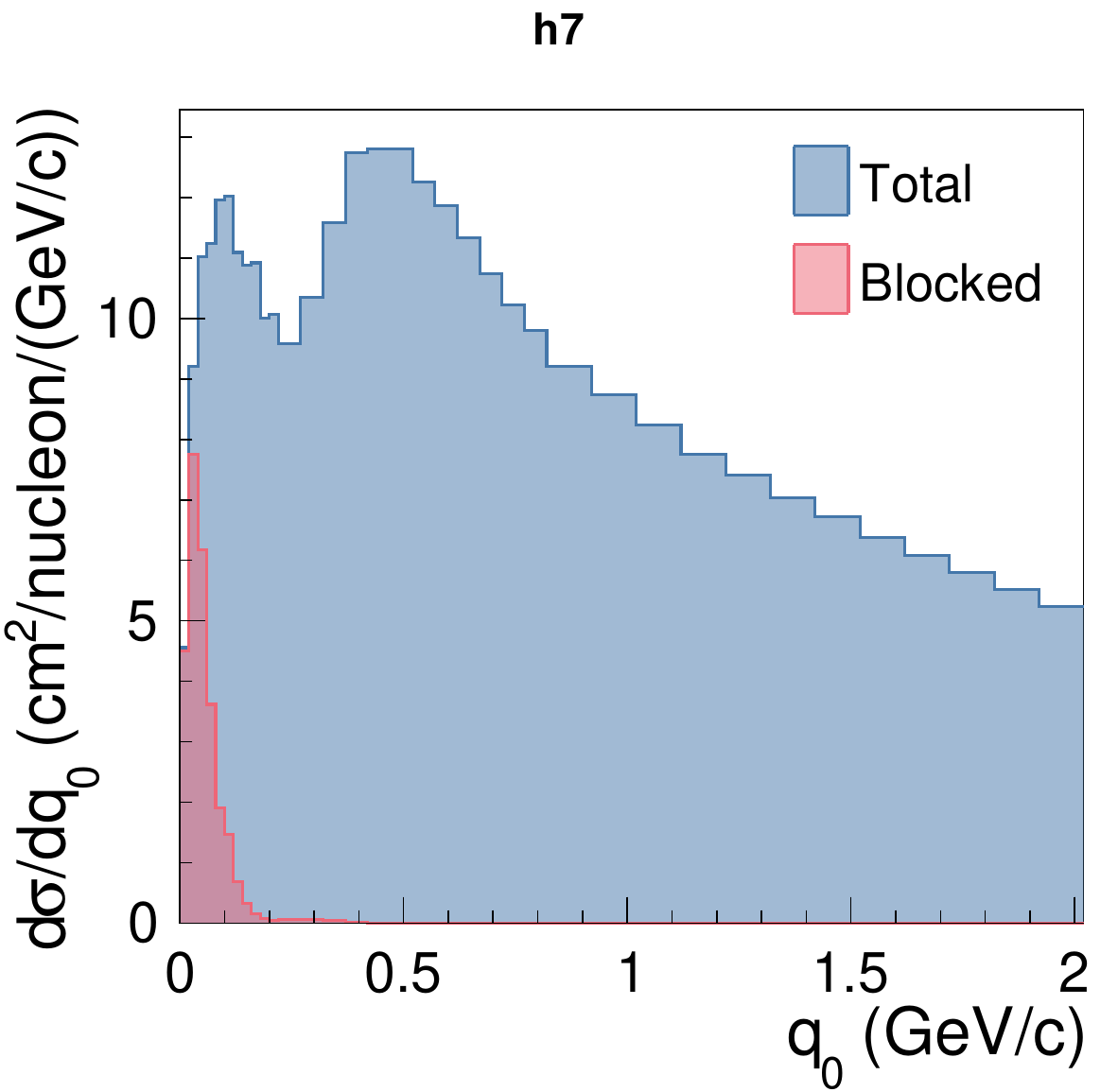}
    \includegraphics[width=0.49\linewidth,,page=1,clip,trim=0mm 0mm 0mm 15mm]{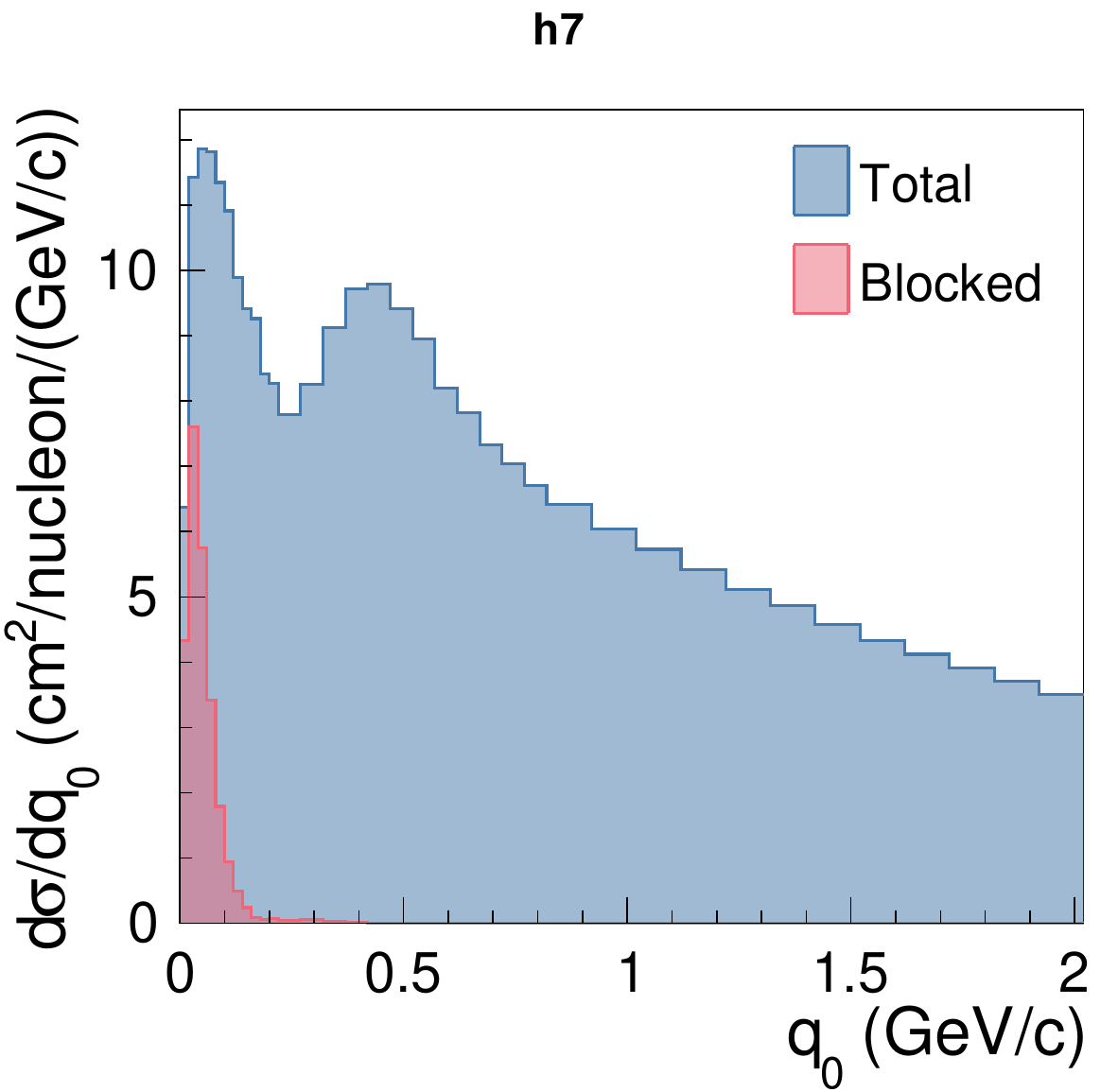}
    \caption{The total CCINC cross section as a function of energy transfer, $q_0$, as predicted by GiBUU, shown in blue, for the MicroBooNE $\nu_\mu$--\argon (top left), T2K $\bar{\nu}_\mu$--C$_8$H$_8$ (top right), NuMI LE $\nu_\mu$--C$_8$H$_8$ (bottom left), and NuMI ME $\bar{\nu}_\mu$--C$_8$H$_8$ (bottom right) fluxes. In each case, the \textit{blocked} contribution of events, in which no hadrons escape the nucleus, is overlaid in red.}
    \label{fig:app_gibuu_blocking}
\end{figure}

\subsection{GENIEv3}
The GENIE neutrino interaction event generator \cite{Andreopoulos:2009rq, Andreopoulos:2015wxa} is the primary generator used by most Fermilab neutrino-beam experiments. In this work we consider a variety of model configurations in GENIE version 3.02.00 based on publicly released GENIE tunes \cite{GENIE:2021zuu, GENIE:2021wox, GENIE:2022qrc}. These configurations alter the CCQE, 2p2h and FSI models, whilst the modelling of other interaction modes remains the same. CCQE interactions use a dipole axial form-factor with $\maqe=0.99$ GeV. RPP is simulated similarly to NEUT, employing the Berger--Sehgal model and considering a variety of resonances up to $W = 1.7~\text{GeV}$ but neglecting interferences between them. DIS and SIS cross sections are described similarly to NEUT, but using a different custom \textit{AGKY} hadronisation model~\cite{Yang:2009zx} for $W \leq 2.3~\text{GeV}$, whilst PYTHIA6~\cite{Sjostrand:2006za} is used for $W > 3.0~\text{GeV}$\footnote{As of GENIE v3.4.0, PYTHIA8 is available as an alternative option, and is expected to become the default in future versions of GENIE.}. A linear transition is considered between them.

\paragraph{\textit{10a, 10b, 10c}:}

GENIE configurations 10a and 10b differ only in their treatment of FSI. GENIE 10b uses the \textit{hN} intranuclear cascade model, similar to NEUT and NuWro, whilst 10a uses the \textit{hA} empirical approach in which the overall \textit{fate} of hadrons in the cascade is decided in a single interaction rather than in a stepped process~\cite{Dytman:2021ohr}. GENIE 10c employs a new implementation of the INCL intranuclear cascade model~\cite{boudard2013new,mancusi2014extension,rodriguez2017improvement,hirtz2018parametrization}. It should be noted that this INCL implementation is expected to be updated in future GENIE versions.

Like NEUT, the QE and 2p2h models used in GENIEv3 10a, 10b and 10c are implementations of the Valencia group's calculations~\cite{Nieves:2011pp}. With respect to NEUT, a more simplistic version of the nuclear ground state is used in which the nuclear removal energy is fixed rather than evolving with the initial state nucleon momentum.

\paragraph{\textit{CRPA}:}

The CRPA GENIE configuration uses the same models as GENIE 10a except for CCQE and 2p2h. For CCQE interactions it uses the GENIE implementation of the CRPA model from the Ghent group~\cite{Jachowicz:2002rr,Pandey:2014tza}, as detailed in Ref.~\cite{Dolan:2021rdd}. 
Additionally, the version of CRPA used in this work has a fast transition to the \textit{SuSAv2} CCQE cross section model~\cite{Megias:2016lke,Gonzalez-Jimenez:2014eqa,Dolan:2019bxf} at momentum transfers between $\sim$500--700~MeV/c~\cite{Dolan:2019bxf}, as the latter uses an unregularised nucleon--nucleon interaction within the RPA calculation~\cite{Nikolakopoulos:2020alk,Jachowicz:2021ieb}, which may be better suited to low energy and momentum transfer interactions. 
As in GENIE 10a, the hA FSI model is used to simulate the propagation of final-state hadrons through the nucleus. The CRPA QE model includes the role of FSI via a distorted wavefunction approach on the outgoing lepton kinematics and so FSI is considered using two separate approaches. 

The GENIE CRPA model uses the SuSAv2 2p2h model, as described in Ref.~\cite{Dolan:2019bxf}, which makes various different theory assumptions to the Valencia model resulting in a significantly different 2p2h prediction. The pair contributions in SuSAv2 2p2h are approximately 80\% proton--neutron and, like the Valencia model case, also dependent on the interaction kinematics.

\section{Differences in electron and muon neutrino--nucleon RPP interactions}
\label{appen:rppnuenumu}

When considering a fixed incoming neutrino energy, lepton mass differences imply that electron and muon (anti)neutrinos have different allowed ranges of energy and momentum, thereby changing the cross section. 
Whilst such effects have been studied in some detail for CCQE interactions~\cite{Dieminger:2023oin,Nikolakopoulos:2019qcr,Ankowski:2017yvm,Martini:2023kem}, less literature is available for other channels, despite this potentially being an important effect, especially when the energy is near the process threshold. 
This can be particularly important for resonant pion interactions where, for a given fixed neutrino energy, the differences in kinematically allowed interaction phase space renders electron and muon (anti)neutrinos sensitive to different resonant (and non-resonant) structures.  
This leads to large differences in the ratio of their cross sections near the threshold energy of the prominent resonances (for example, $\Delta(1232)$). This is demonstrated in \autoref{tab:1pi_nue_numu}, which shows the ratios of the total neutrino--nucleon cross section for electron and muon (anti)neutrinos in RPP interactions considering for four different models of single pion production.

The four models shown in \autoref{tab:1pi_nue_numu} are: the Rein--Sehgal model~\cite{Rein:1980wg}, which has been implemented with variations in both GENIE and NEUT; the \textit{MK} model which refers to the model by Monireh Kabirnezhad~\cite{Kabirnezhad:2017jmf}; the \textit{DCC} model which refers to the dynamically coupled channels model by Nakamura et al~\cite{Nakamura:2015rta}; and the \textit{HNV} model which refers to the Hernandez, Nieves, and Vicente-Vacas model~\cite{Sobczyk:2018ghy} (occasionally called the \textit{Valencia single pion model}).

The Rein--Sehgal model is the simplest, and attempts to describe a wide range of resonances up to a hadronic mass of 2.0 GeV$/c^2$ (15--20 resonances in total) using the quark-harmonic oscillator~\cite{Feynman:1971wr}, which has been subject to a variety of criticisms, including violation of unitarity, but continues to be in use by experiments~\cite{Hayato:2021heg, geniev304TN}.
It has a simplistic recipe for the non-resonant background, and includes resonance--resonance interference, but not interference between the non-resonant and resonant contributions.
The HNV model simplifies the structure of the resonances considerably, focussing on the dominant $\Delta(1232)$ and $D_{13}(1520)$ resonances and a non-resonant background based on a non-linear sigma model which is fixed from chiral symmetries. It includes their interferences and partially restores unitarity.
The MK model uses ingredients similar to the HNV model, such as aspects of the non-resonant background and the method to restore unitarity, and attempts to describe a broader range of hadronic masses (similar to the Rein--Sehgal model) by using a meson-dominance model that respects quantum chromodynamics.
The DCC model similarly targets a wide range of resonances, and uniquely accounts for that resonance interactions can happen through a number of channels which are strongly coupled (for instance $\pi N$, $\pi \pi N$, $\eta N$, $K \Lambda$, $K \Sigma$). All of these couplings need to be modelled if unitarity is to be properly conserved, which is what the DCC model addresses. 
Importantly, the HNV, MK and DCC models have been extensively benchmarked against electron-scattering data, and the latter two have additionally been benchmarked against pion and photon induced pion production data.
Generally speaking, all the models are expected to see similar overall behaviour, for instance $\Delta(1232)$ dominance, but the details of that dominance, such as the \textit{turn-on} of the interaction in hadronic mass which the threshold region is sensitive to, differ considerably. 

\autoref{tab:1pi_nue_numu} shows that at 0.63 GeV of neutrino energy, the neutrino--nucleon cross section for electron neutrinos is 23--29\% larger than muon neutrinos depending on model, which decreases to 2--3\% at 2 GeV, and 1\% at 5 GeV. Thus, experiments at higher energy are less sensitive to the threshold behaviour for the resonant processes, but are instead more sensitive to the threshold behaviour of higher energy processes, such as DIS.
This is a relatively little-studied effect that could impact current and future neutrino oscillation experiments. 
\begin{table*}[h]
\centering
\begin{tabular}{l | c | c | c | c }
     Neutrino & \multicolumn{4}{c }{$\sigma_{\nu_e}/\sigma_{\nu_\mu}$ $\left(\sigma_{\bar{\nu}_e}/\sigma_{\bar{\nu}_\mu}\right)$}  \\
     energy (GeV) & RS & MK & DCC & HNV \\
     \hline
     0.63  & 1.29 (1.60) & 1.29 (1.19) & 1.23 (1.42) & 1.26 (1.47) \\
     0.88  & 1.09 (1.21) & 1.10 (1.11) & 1.07 (1.17) & 1.07 (1.16) \\
     1.00  & 1.07 (1.16) & 1.08 (1.09) & 1.05 (1.13) & 1.06 (1.12) \\
     1.50  & 1.03 (1.07) & 1.04 (1.05) & 1.03 (1.06) & 1.03 (1.05) \\
     2.00  & 1.02 (1.04) & 1.03 (1.03) & 1.02 (1.04) & 1.02 (1.03) \\
     5.00  & 1.01 (1.01) & 1.01 (1.01) & 1.01 (1.01) & 1.01 (1.01) \\
\end{tabular}
\caption{Ratios of the total neutrino--nucleon cross section for electron and muon (anti)neutrinos in the CC$1\pi^+1p$ (CC$1\pi^-1n$) channel, for four different models of single pion production for different fixed neutrino energies.}
\label{tab:1pi_nue_numu}
\end{table*}

\end{document}